\documentclass[12pt]{ociamthesis}  % default square logo 

\usepackage{titlesec}
\titlespacing*{\subsubsection}
{0pt}{5pt plus 1pt minus 1pt}{5pt plus 1pt}

\usepackage[usenames, dvipsnames]{xcolor}
\usepackage{graphicx, float, array, xspace, amscd, amsthm, amssymb, latexsym, longtable, bbm, fancyhdr,slashed,pifont}
\usepackage[letterpaper, body={6.5in, 9.25in}, top=1.0in, left=1.5in]{geometry}
\usepackage[bbgreekl]{mathbbol}
\usepackage[utf8]{inputenc}
\usepackage[sort&compress, comma, square, numbers]{natbib}
\usepackage[bottom]{footmisc}
\usepackage{subcaption}
\usepackage{bm}
\usepackage{mathtools}
\usepackage{tikz}
\usepackage{tikz-cd}
\usepackage[pdfpagelabels=false]{hyperref}

\usepackage{comment}
\usepackage{adjustbox}

\usepackage{tabularx, caption, boldline}
\usepackage{cellspace}
\DeclareSymbolFontAlphabet{\mathbb}{AMSb}
\DeclareSymbolFontAlphabet{\mathbbl}{bbold}

\let\nordico=\o % save Nordic o
\let\SS=\S % save \S before it is redefined

\renewcommand{\d}{\delta}

\newcommand{\m}{\mu}

\newcommand{\p}{\pi}

\renewcommand{\S}{\Sigma}

\renewcommand{\o}{\omega}

\DeclareFontFamily{OT1}{pzc}{}
\DeclareFontShape{OT1}{pzc}{m}{it}{<-> s * [1.200] pzcmi7t}{}
\DeclareMathAlphabet{\mathpzc}{OT1}{pzc}{m}{it}

\newcommand{\cA}{\mathcal{A}}
\newcommand{\cB}{\mathcal{B}}
\newcommand{\cC}{\mathcal{C}}

\newcommand{\cE}{\mathcal{E}}
\newcommand{\cF}{\mathcal{F}}

\newcommand{\cI}{\mathcal{I}}

\newcommand{\cK}{\mathcal{K}}

\newcommand{\cM}{\mathcal{M}}

\newcommand{\cS}{\mathcal{S}}

\DeclareFontFamily{U}{bbold}{}
\DeclareFontShape{U}{bbold}{m}{n}
{  <-5.5> s*[1.05] bbold5
	<5.5-6.5> s*[1.05] bbold6
	<6.5-7.5> s*[1.05] bbold7
	<7.5-8.5> s*[1.05] bbold8
	<8.5-9.5> s*[1.05] bbold9
	<9.5-11.5> s*[1.05] bbold10
	<11.5-16> s*[1.05] bbold12
	<16-> s*[1.05] bbold17
}{}

\newcommand{\IC}{\mathbbl{C}}

\newcommand{\IF}{\mathbbl{F}}

\newcommand{\IH}{\mathbbl{H}}

\newcommand{\IN}{\mathbbl{N}}

\newcommand{\IP}{\mathbbl{P}}
\newcommand{\IQ}{\mathbbl{Q}}
\newcommand{\IR}{\mathbbl{R}}

\newcommand{\IW}{\mathbbl{W}}

\newcommand{\IY}{\mathbbl{Y}}
\newcommand{\IZ}{\mathbbl{Z}}

\font\elevenrmfromseventeenrm = cmr17 at 11pt
\newcommand{\inbar}{\vrule height6.9pt depth-0.2pt width0.35pt}
\newcommand{\zero}{\hbox{{\elevenrmfromseventeenrm 0}\kern-3.5pt\inbar\kern1pt\inbar\kern2pt}}

\font\eightrmfromseventeenrm = cmr17 at 8pt
\newcommand{\ssinbar}{\vrule height5pt depth-0.1pt width0.3pt}
\newcommand{\sszero}{\hbox{{\eightrmfromseventeenrm 0}\kern-2.53pt\ssinbar\kern0.7pt\ssinbar\kern2pt}}

\newcommand{\one}{\mathbbm{1}}

\newcommand{\fC}{\mathfrak{C}}

\newcommand{\fh}{\mathfrak{h}}

\newcommand{\fK}{\mathfrak{K}}

\newcommand{\fp}{\mathfrak{p}}

\newcommand{\fs}{\mathfrak{s}}

\newcommand{\2}{\mathbf{2}}
\newcommand{\1}{\mathbf{1}}

\newcommand{\beq}{\begin{equation}}
\newcommand{\eeq}{\end{equation}}
\newcommand{\beqnn}{\begin{equation*}}
\newcommand{\eeqnn}{\end{equation*}}
\newcommand{\bea}{\begin{eqnarray}}
\newcommand{\eea}{\end{eqnarray}}
\newcommand{\bean}{\begin{eqnarray*}}
	\newcommand{\eean}{\end{eqnarray*}}

\newcommand{\cicy}[2]{\begin{matrix} #1\end{matrix}\!\left[\begin{matrix}#2 \end{matrix}\right]}

\newcommand{\defineas}{\buildrel\rm def\over =}

\newcommand{\place}[3]{\vbox to0pt{\kern-\parskip\kern-7pt
		\kern-#2in\hbox{\kern#1in #3}
		\vss}\nointerlineskip}
\newcommand{\capt}[3]{\parbox{#1}{\renewcommand{\baselinestretch}{1.0}
		\caption{\label{#2}\small\it #3}}}

\newcommand{\+}{\phantom{-}}
\renewcommand{\=}{\;=\;}

\newcommand{\fref}[1]{figure~\ref{#1}}
\newcommand{\tref}[1]{table~\ref{#1}}
\newcommand{\sref}[1]{\SS\ref{#1}}
\newcommand{\cref}[1]{Chapter \ref{#1}}

\DeclareFontFamily{U}{wncy}{}
\DeclareFontShape{U}{wncy}{m}{n}{<->wncyr10}{}
\DeclareSymbolFont{mcy}{U}{wncy}{m}{n}
\DeclareMathSymbol{\sha}{\mathord}{mcy}{"58}
\renewcommand{\Im}{\text{Im}}
\renewcommand{\Re}{\text{Re}}

\newcommand{\CS}{{\IC\text{S}}}
\newcommand{\KC}{{\IC\text{K}}}
\newcommand{\AD}{{\text{AD}}}

\newcommand{\ee}{\text{e}}

\newcommand{\ii}{\text{i}}
\newcommand{\dd}{\text{d}}
\newcommand{\cym}{Calabi-Yau manifold\xspace}

\newcommand{\wt}[1]{\widetilde{#1}}

\makeatletter
\g@addto@macro\bfseries{\boldmath}

\def\blindfootnote{\xdef\@thefnmark{}\@footnotetext}
\makeatother

\newcommand{\zeroFtwo}[3]{ {}_0F_2\left(#1,\,#2;~#3\right) }

\newcommand{\zeroFfour}[5]{ {}_0F_4\left(#1,\,#2,\,#3,\,#4;~#5\right) }

\newcommand{\ip}{\amalg}

\newcommand{\Peq}{\;\stackrel{\mathclap{\normalfont\mbox{P}}}{=}\;}

\usepackage{amssymb}
\usepackage{amsmath}
\usepackage{epigraph}
\usepackage{setspace}
\renewcommand{\arraystretch}{0.75}
\usepackage{lscape}
\usepackage{tikz}

\usepackage[OT2,OT1]{fontenc}
\newcommand\textcyr[1]{{\fontencoding{OT2}\fontfamily{wncyr}\selectfont#1}}
\usepackage{booktabs}
\usepackage{pgfplotstable}
\usepackage{comment}
\usepackage{adjustbox}

\title{Investigations in Calabi-Yau\\[1ex]     %your thesis title,
        Modularity and Mirror Symmetry}   %note \\[1ex] is a line break in the title

\author{Joseph McGovern\\[1ex]}             %your name
\college{The House or College of Scholars of Merton \\[1ex] in the University of Oxford \\[1ex]\textit{and} \\[1ex] Hertford College\\[3ex]}  %your college

\degree{Doctor of Philosophy}     %the degree
\degreedate{Trinity 2023}         %the degree date

\begin{document}

%this baselineskip gives sufficient line spacing for an examiner to easily
%markup the thesis with comments
%\baselineskip=18pt plus1pt

%set the number of sectioning levels that get number and appear in the contents
\setcounter{secnumdepth}{3}
\setcounter{tocdepth}{3}

\begin{singlespace}
\maketitle                  % create a title page from the preamble info
\begin{dedication}
Dedicated to my\\[5pt]
sister and brother,\\[5pt]
Charlotte and Stewart.
\end{dedication}        % include a dedication.tex file
\begin{abstract}

This thesis lays out a number of different projects, linked by the common thread of Calabi-Yau manifolds.

We produce tables of instanton numbers for various multiparameter Calabi-Yau manifolds. Studying those tables reveals, in some cases, a Coxeter group of symmetries that act on sets of instanton numbers. Instanton numbers are constant over the orbits of group actions on the curve classes, and these groups can be infinite. 

We investigate to what extent this symmetry can be used to constrain the holomorphic ambiguity that arises in higher genus topological string free energy computations. In the example we study, at genus 2, combining this Coxeter constraint with the set of vanishing numbers fixes the holomorphic ambiguity.

A new class of solutions is provided to the supersymmetric flux vacuum equations, which have been conjectured elsewhere to give weight-two modular manifolds.  

Another search, complementary to those that have already been carried out, is made for rank-two attractors in the AESZ list. Two novel examples are found, both of which belong to moduli spaces with two points of maximal unipotent monodromy each. For one of these operators, the additional MUM point corresponds to another derived equivalent geometry. For the other operator, we compute \emph{nonintegral} values of the triple intersection number and genus 0 instanton numbers, and so a geometric interpretation is lacking for the second MUM point.

We provide several instances of summation identities that express a ratio of critical L-values as an infinite sum, whose terms contain the Gromov-Witten invariants of a Calabi-Yau manifold $Y$. In one example there is no manifold $Y$, only an operator, but a set of invariants can nonetheless be computed for this and a summation identity is found. The L-functions are associated to a modular manifold $X$ which is mirror to $Y$. These sums can be divergent, but Pad\'e resummation cures this.

In addition to these results, we briefly review informing aspects of Calabi-Yau geometry, black holes in 4d $\mathcal{N}=2$ supergravity, flux compactifications, topological string theory, and number theory.

The contents of the tables of this thesis are available to download at \cite{mcgovern2023a}.

\end{abstract}          % include the abstract

\begin{originality}
Parts of this thesis are based on joint work with Philip Candelas, Xenia de la Ossa, and Pyry Kuusela.

Chapters One and Two serve as an introduction and review of relevant background material, neither contains new results.

Chapter Three describes ongoing work, building on ideas that appeared in \cite{Candelas:2021lkc} where the appearance of the Coxeter groups was noted.

Chapter Four also describes ongoing work. Parts of the discussion of flux vacua appear in \cite{Candelas:2023yrg}. The series solution method for the IIA attractor equations was first given in \cite{Candelas:2021mwz}.

\end{originality}
\begin{acknowledgementslong}
\thispagestyle{empty}
\vskip-10pt\small{I will always be grateful to my supervisor, Philip Candelas, for affording me the opportunity to do this degree. He has been patient, kind, generous with his time, and shared many fascinating things with me. Xenia de la Ossa, who began cosupervising me in my second year, has also been of immense help and I feel very lucky to have worked with her, sharing in the beautiful world of her work. It has been very interesting to work with Pyry Kuusela, my collaborator and former fellow graduate student. Our projects together were accelerated by his technical wizardry and fearsome work ethic. I thank him for his willingness to share and discuss his ideas, and his friendship. Mohamed Elmi, as a former student of this working group, has been very helpful to me, in Oxford during my early days and later on during his postdoc. We have benefited from interesting conversations with Duco van Straten.

My course through life greatly benefited from the supervision of Kolya Gromov during my undergraduate degree in KCL. I was extremely lucky to find a mentor in him, and working with him was a very happy period. For our meetings and their help during those times, I thank Andre\'a Cavaglia, Fedya Levkovich-Maslyuk, and Julius. Also from KCL, I thank Nadav Drukker, J\"urgen Berndt, Sameer Murthy, Alexander Pushnitski, Paul Cook, and Dionysios Anninos, for the varied help that they gave me at different times.

I would like here to belatedly thank the anonymous JHEP referee of my first paper on Wilson loop insertions. In my naivety I neglected to do so in the acknowledgements section of that paper, in spite of their helpful comments.

I am grateful to James Sparks and Fernando Alday for giving their time to critically assess my work at stages of this degree. For this I also thank Erik Panzer, as well as for some additional inspirational conversations. I am also very glad that Kilian B\"onisch was able to visit, and offer me some helpful advice on technical matters which greatly helped in \sref{sect:SUGRA_GWS}. I also thank my college advisor Dawid Kielak for some helpful conversations.

I expect that regularly seeing Sujay Nair in our office will be the part of Oxford that I most come to miss. I feel that our time together has shaped me,  and I look forward to discussing graphene with him one day. I know that he will do very well, and probably without making a big deal of it.

I will miss Carmen Jorge-Diaz, who was an excellent colleague, housemate, and friend. By transitivity I will also miss Lockheed Martin Purkiss. I could not legally submit this thesis without first thanking Doctor Mega Ultra Mateo $G_{2}$ Galdea\~{n}o Feliz ``Solans" Huesca. We had many spooky adventures. Even after he left he was still being helpful, because I then had use of his desk, desktop, and sword. Thank you Diego Berdeja Suarez, first for introducing me to the group as my DPhil buddy, and then for being my friend. James Hefford has been a wonderful person to know, he gave me a lot more food than I gave him. Also from Hertford, I thank Anastasia Sigutina for her sense of humour and the heartfelt effort she puts into her friendships. I appreciated Manuel Schmigdall's approach to 214b. 

From my time living in Mansfield, I am happy to have known Tom McAuliffe, David Rytz, Thomas Han, and the chief Sigma males Armands Strikis and Lily Mott (Armands is the favourite of those two). Thank you Alex Cliffe for not only being my friend, but for organising happy hours.

Living in Norham Gardens was an unexpected blessing, even more so for the characters within. Rick Longley, you are hilarious and thank you for your sincerity. Thank you Esraa Shaban, Tom Anglim Lagones, Ryan O'Sullivan, Facundo Herrero, Katie Hearne, Christine Arndt, Janko Ondras, Ben Ta, Insaif Bakeer Markar, Pamela, Philip, Sarah, and everyone from the Labyrinth.

My final year in Oxford was enhanced by the chance to work in Merton, which I enjoyed greatly. I am grateful for their decision to employ me, and I thank the many interesting people I met there, and those who I worked with. I thank all of those who I had the opportunity to teach, particularly those at Merton in this last year, who were a most impressive collection.

My year spent playing chess in the City Club was memorable, so thank you to them, in particular Phil Hayward and Will van Noordt.

My friends from KCL changed my view of what was possible, thank you Jack Heimrath and Mary Czarnecka for among so much else hosting us in Poland, Marcello Harle-Cowan, Louis Yudowitz, Yusuf Broujerdi, Sil Linskens, Maite Arcos-Enriquez, Andreas Sandbakken, Vlad Petrica, Enzo Desiage. My time in Falmouth Road with Jonathan Gorard and Jan Stanczuk was special.

I thank Keir Crawley and the Music Academy, for making my life so much richer. Time spent with Chris Williams, Michael Papasavva, Warren Alexander, Moeed Majeed, Terry Murphy, Kyriacos Papasavva, Shiv Sharma, and Alex Kempson has been wonderful.

I pay my respects to the brave warriors in the Tiger Mafia and associated entities, for 2 min.

I am grateful to the organisers of the GLSMs@30 conference in the SCGP, Long Island, for facilitating interesting conversations. I also thank the organisers of the Elliptics and Beyond 2023 conference, in ETH Zurich, for the opportunity to talk about my work and for their hospitality. 

I now proceed to thank my favourite members of the vast number of MP DPhil students/postdocs (and affine extensions). Thank you Horia Magureanu for your wisdom. I thank Marieke and Pieter Tunstall van Beest, and Pietro Ferrero, most recently for driving me around Long Island. Our group was extremely fortunate to have so often as hosts Andrea Boido, Enrico Marchetto, and Palash Singh, their generosity has made for many enjoyable evenings.  I thank Jay Swar for the variety of conversations, and also just the variety. I thank Federico Zerbini for his help. I formally rethank Julius Eckhard. Max H\"ubner is pretty good. Spending time with Maria Nocchi and Adam Kmec has been entertaining and joyful. I thank Giulia Albonico for her kindness, Guillermo Arias Tamargo for his Smith-Morra, Christopher Raymond for his explanation of loose units. Filipo Revello, Jean Garret, Johan Henriksson, Murat Kologlu, Connor Behan, Matteo Sacchi, Atul Sharma, Giuseppe Bogna, Lea Bottini, Dario Ascari, the mysterious Dewi Gould, Izar Alonso Lorenzo, Evyatar Sabag, Chris Couzens, Zhenghao Zhong, Dylan Butson, and everyone else who I forgot.

None of the good things in my life would be possible without the love and provision of my mother Helen Clark. Thank you for all that you have done for me all through my life, for working so hard and giving me all that you have.

I also thank my grandfather Kenneth Clark for his support, my stepfather Richard Starnes, and Midna, Barney, Martin, Charlotte and Stewart.

My father Vincent McGovern has done so much for me. Thank you for your support, the adventures across Ireland, and your belief in me. Your loving presence in my life means a great deal to me.

I thank also Patrick McGovern, Katrina McLoughlin, my cousins Malachy, Emer and Brian McGovern, my grandparents Annie and Joe McGovern, Tyke McGovern, and Sunita Ramasawmy.

I finish by thanking Stephen Wallman, Ehsan Pedram, Alison Wilding, Robert Cowen, Alan Godfrey, Despo Speel, and Thore Lindahl Itturiate.

I apologise to the people that I have forgotten in writing this.}
\thispagestyle{empty}
\end{acknowledgementslong}  % include an acknowledgements.tex file
\end{singlespace}

\begin{romanpages}          % start roman page numbering
\tableofcontents            % generate and include a table of contents
%\listoffigures              % generate and include a list of figures
\end{romanpages}            % end roman page numbering

%now include the files of latex for each of the chapters etc
%\input{basics}
%\input{chapter1}
%\input{chapter2}
%\include{conclusions}

\chapter{Introduction}\label{Chapter:Introduction}
\setlength\epigraphwidth{.68\textwidth}\epigraph{I must go down to the seas again, for the call of the running tide\\
Is a wild call and a clear call that may not be denied;}{John Masefield, \textit{Sea Fever}}
\section{Motivation}
It is difficult to overstate the wealth of interesting mathematics attached to Calabi-Yau manifolds as they appear in physics. This thesis documents an attempt to look into these geometries, specifically compact threefolds, to see what can be found. This approach is informed by two broad ideas, both of which have seen major recent advancement on which we rely. These are \emph{modularity} and \emph{mirror symmetry}. 

Modularity has a good modern physics pedigree, appearing for instance as a highly constraining symmetry in CFT partition functions. The modular group also appears as a symmetry of type IIB string theory, which is vital for the geometric uplift to F-theory by way of elliptic fibrations. The ``modular'' here refers to the modular group $\Gamma=\text{PSL}(2,\IZ)$. 

This group $\Gamma$ has a deep significance to the arithmetic of elliptic curves. Arithmetic, by which we mean the study of varieties over finite fields, is perhaps less familiar to physicists of some schools than the aforementioned role played by $\Gamma$. It is a fact that the set of point counts for an elliptic curve, considered as a variety over a finite field, are collected into a generating function that transforms in a highly constrained way under a subgroup of $\Gamma$. Namely, this function is a weight-two modular newform for a congruence subgroup of $\Gamma$. This is the modularity theorem for elliptic curves \cite{Wiles,WilesTaylor,BCDT}. We will better explain that claim in \sref{sect:Essen_CY_Modularity}, but for now press on to make the point that this can, rarely, happen for Calabi-Yau threefolds \cite{Yui2013}. 

Whether or not this happens can change as the Calabi-Yau is subject to a complex structure deformation, so that the question of ``Which Calabi-Yau threefolds are modular?" becomes the question "Which families contain modular members, and where in the moduli space are they?". The reason that the physicist should care about this question lies in two putative answers, each reliant on some conjectures \cite{Moore:1998pn,Moore:1998zu,Kachru:2020abh,Kachru:2020sio}:
\begin{itemize}
\item Rank-two attractors, in the context of four dimensional $\mathcal{N}=2$ supergravity, are weight four modular.
\item Supersymmetric vacua in IIB flux compactifications are weight-two modular.
\end{itemize} 
At the most extreme end of optimism then, one could hope that a better understanding of arithmetic modularity could shed light on black hole physics and string model building. This would see a well-developed area of mathematics being brought to bear on some persistent problems in physics. We come far short of that in this thesis, but these aspirations, and other recent works realising these possibilities, serve as the ultimate motivation. For example, the works \cite{Bonisch:2022mgw,Candelas:2019llw} were able to express D-brane masses and semiclassical black hole entropies using Mellin transforms of the modular forms associated to modular geometries.

The other raison d'\^{e}tre of this thesis is mirror symmetry, specifically two-dimensional mirror symmetry. This is a duality between certain quantum field theories, specifically twisted $\mathcal{N}=(2,2)$ nonlinear sigma models with Calabi-Yau target space. Any duality between two quantum field theories should be squeezed for all that can be learnt from it, and this duality comes with a particularly nice geometric picture. It is possible to obtain, suitably defined, counts of holomorphic embeddings of genus-$g$ curves of various degrees into a Calabi-Yau threefolds $Y$ by studying the complex-structure periods of a mirror geometry $X$.

There is a captivating problem in carrying out this computation at higher genera. There is a procedure for doing this, BCOV recursion \cite{Bershadsky:1993cx},\cite{Bershadsky:1993ta}, which at each genus suffers from a holomorphic ambiguity. This is a rational function that must be fixed, up to which the complete B-model free energy at genus $g$ is determined by the lower genus results. This ambiguity can be fixed by a number of considerations, for instance \cite{Huang:2006hq}, but these do not suffice to reach arbitrarily high genus. A full solution to this stubborn problem could involve a strong handle on patterns in the sets of instanton numbers, which motivates our work along these lines. 

Additionally, the mirror symmetry computation is understood to give higher derivative stringy correction terms to the supergravity action \cite{Gopakumar:1998ii,Gopakumar:1998jq}. This duality then, is offering one way to approach the problem of formulating string theory nonperturbatively. Topological string theory is a simplified version of string theory, and so it is a natural expectation that a full understanding of the latter will be helped by first understanding the topological theory, as discussed recently in \cite{Bah:2022wot}.

Mirror symmetry gives a duality between supergravity compactifications. These are low energy limits of string theories, with there conjecturally being a duality between two full string compactifications that exchanges the nonperturbative contents of both theories. This supergravity duality gives one a means of studying the interplay between mirror symmetry and the modularity that can be present in supergravity compactifications. One can consider extremal black holes from the IIB or IIA perspectives. It is in IIB that we will first consider rank two attractors. In the IIA description the enumerative invariants, computed by mirror symmetry techniques, enter as quantum corrections to the supergravity action. The upshot is that the mirror map provides identities relating the enumerative geometry of one space to the modularity of another, and these are naturally encountered in the course of investigating black hole solutions.

\section{Thesis overview}
We have throughout this thesis strived to denote Calabi-Yau manifolds by $X$ and $Y$, and always have it so that B-model computations are performed on $X$ while A-model computations are reserved for $Y$. We will consider IIB on $X$ and IIA on $Y$. Where $Y$ is a CICY, $X$ is a CICY mirror. We will always refer to instanton numbers of $Y$.

\subsubsection*{Chapter \sref{Chapter:Essentials}}
This is a brief collection of some preliminary results and terminology. We define Calabi-Yaus, and introduce the favourable CICY set. These will recur throughout the thesis as a sufficiently varied set of examples for our purposes. We go on to set up the period computation that we will use frequently. We explain how, for CICY mirrors $X$, the CICY matrix defines the topological quantities and periods of $X$. We also review the splitting process for CICYs, and all of these will bear relation to discussion in \sref{Chapter:Miroirs}.

We go on to give a very basic account of arithmetic modularity, sufficient for the purposes of searching for rational rank-two attractors later on in \sref{Chapter:SUGRA}. We will meet the zeta function, which provides the means of testing for modularity. The zeta function is first defined by suitably collecting point counts into a power series, but remarkably this is resummed into a rational function. For modular manifolds, the numerator of this function factorises. It is from these factors that the relevant modular form is read off. 

\subsubsection*{Chapter \sref{Chapter:Miroirs}}
Here we detail our studies in mirror symmetry. We compute genus 0 numbers for manifolds with $h^{1,1}$ as high as 7. We compute genus one numbers for some geometries with lower $h^{1,1}$, but cannot do this for the 6 and 7 parameter models. The genus 1 computations requires a piece of data not strictly necessary at genus 0, the manifold's discriminant locus. We compute a number of examples of these, and comment on some patterns in this data set, in the hopes of eventually returning to the 6 and 7 parameter models.

It is already known that the process of splitting a Calabi-Yau partitions its instanton numbers, so obtaining the instanton numbers of the new manifold. We showcase some examples of this, for a wider class of models than simply CICYs.

The data from our tables reveals a surprising fact, that sometimes sets of instanton numbers can be symmetric under the action of a Coxeter group on the manifold's second homology. These are groups that are generated by reflections, and are encoded by diagrams similar to Dynkin diagrams. The Weyl group associated to a Dynkin diagram is also a Coxeter group, but Coxeter groups are a much larger set with a different classification theory. We go on to tabulate some Coxeter groups that appear in our data, giving their diagrams and their representation on the curve classes of the Calabi-Yaus to which they are associated. 

Armed with Coxeter symmetry, we address the higher genus computation. We review the construction of the propagators out of which higher genus B-model free energies can be built, and cover BCOV recursion. We work an example at genus 2, presented in such a way that we hope is valuable to anybody hoping to carry out this computation. When it comes time to fix the genus-2 holomorphic ambiguity, we apply the Coxeter symmetry and observe that this does not fully fix the ambiguity. We discuss what other information is required. 

\subsubsection*{Chapter \sref{Chapter:SUGRA}}
This chapter opens by recalling some textbook information on Calabi-Yau supergravity compactifications. Then, we undertake a substantial review of the attractor mechanism's derivation, including the recasting of the attractor equations into Strominger's form for the fixed points of the flow. With the attractor mechanism explained for general 4d $\mathcal{N}=2$ supergravities, we then discuss its details where Calabi-Yau compactifications are concerned.

Next, we depart from black holes and instead look at flux compactifications. We review both the IIB construction and its lift into F-theory \'a la Sen, to see how precisely the IIB axiodilaton profile is realised as the modulus for the elliptic fibre in a Calabi-Yau fourfold.

At this point in the thesis, we make a small digression. We discuss possible sources of corrections to the potential in a Calabi-Yau flux compactification, and explain how the no-scale relation is violated by a nonzero Euler characteristic $\chi$, or by worldsheet instantons if $\chi=0$. 

Then, we return to flux compactifications and explain our solution method for the equations defining supersymmetric vacua. We provide a number of examples, the crux of these being their axiodilaton's dependence on the complex structure moduli. This allows for the F-theory curve to be identified.

Next we turn our sights back to black holes, and use zeta function computations to identify two new rank-two attractors. There exists a way, given a rank-two attractor, to construct a summation identity relating L value ratios to Gromov-Witten invariants. We display a number of these. Not all of these sums converge, but the divergent ones can be treated with Pad\'e resummation. 

The identities so given bring together critical L-values and instanton numbers, in some examples distinct sets of instanton numbers, for different spaces that are derived equivalent. We discuss some of the geometric aspects of these derived equivalent partners.

\chapter{Some essentials}\label{Chapter:Essentials}
\setlength\epigraphwidth{.3\textwidth}\epigraph{This is where the fun begins.}{Anakin Skywalker}
A manifold is \textit{Calabi-Yau} if it is complex, K\"ahler, and has vanishing first Chern class. In this thesis we will be concerned with compact Calabi-Yau manifolds. They possess a twofold utility in string theory. One side of this is that the two-dimensional $\mathcal{N}=(2,2)$ nonlinear sigma model is conformal exactly when the target space is Calabi-Yau. The other side is that an equivalent characterisation of a Calabi-Yau $n$-fold is that the manifold has $\text{SU}(n)$ special holonomy, and so possesses a covariantly constant spinor. This allows for supersymmetry to be preserved in compactifications of supergravity or string theory on these manifolds. Specifically, Calabi-Yau compactifications preserve one half of the supercharges found in the ten-dimensional theory. They saw introduction to the theoretical physics literature in \cite{Candelas:1985en}.

In complex dimension 1, the Calabi-Yau condition is met only by elliptic curves. These can be realised as quotients of the complex plane $\IC$ by a lattice $\Lambda_{\tau}$, with complex structure parameter the lattice parameter $\tau$. The choices $\tau$ and $-\tau$ give isomorphic elliptic curves. Two lattices $\Lambda_{\tau}$, $\Lambda_{\tau'}$ are equal if $\tau$ and $\tau'$ are related by an $\text{SL}(2,\IZ)$ transformation, and so the elliptic curves obtained as quotients $\IC/\Lambda_{\tau}$ and $\IC/\Lambda_{\tau'}$ are also equal. Therefore, the moduli space of one-dimensional Calabi-Yau manifolds is given by the upper half plane modulo $\text{SL}(2,\IZ)$ transformations. 

In complex dimension two, the set of complex K{\"a}hler manifolds with vanishing first Chern class consists of products of elliptic curves, Abelian surfaces, and K3 surfaces (see \cite{Aspinwall:1996mn,FilipK3}). Only K3 surfaces have holonomy $SU(2)$, and they all lie in one complex analytic, but not algebraic, family. 

The situation in three dimensions and higher is starkly different, and does not currently possess a full characterisation. There are distinct families, unrelated by complex structure deformation, and it is not known if the number of such families in any given dimension is finite. Our studies will concern the three-dimensional manifolds. It was shown in \cite{Candelas:1990pi} that the moduli space of a Calabi-Yau manifold $X$ factorises locally, into a product of the space of K\"ahler structures on $X$, $\cM_{K}(X)$, with the space of complex structures on $X$, $\cM_{\IC S}(X)$. The first of these spaces can be thought of as the coefficients of $X$'s K\"ahler form when expanded in a basis of $H^{2}(X,\IZ)$ while the second parametrises complex structure deformations of $X$, which in the simplest cases are variations of the coefficients of the polynomials defining $X$ as a complete intersection. Recall that the spaces of the $n^{th}$ cohomology of a complex manifold admit a Hodge decomposition into $(p,q)$-forms, with $p+q=n$, with each of these spaces having dimension equal to the Hodge number $h^{p,q}$. 

We will only consider examples with full $SU(3)$ holonomy and not a subgroup. This is the case for manifolds with $h^{1,0}=h^{2,0}=0$, because it can be shown that the existence of a holomorphic $(1,0)$ or $(2,0)$ form reduces the holonomy from $SU(3)$ to a subgroup \cite{Yau_Survey}. In so doing we exclude the product of three elliptic curves, and the product of an elliptic curve with a K3 surface. All Hodge numbers of $X$ are fixed once $h^{1,1}$ and $h^{1,2}$ are specified. This is because complex conjugation equates $h^{p,q}$ to $h^{q,p}$, while Hodge duality equates $h^{p,q}$ to $h^{3-q,3-p}$. We have stated that $\text{Dim }\cM_{K}(X)= \text{Dim }H^{2}(X,\IZ)$, which in turn equals $h^{1,1}$ because $h^{2,0}=h^{0,2}=0$. It can also be shown that $\text{Dim }\cM_{\IC S}(X)=h^{2,1}(X)$ \cite{Candelas:1990pi}.

\section{Periods and the triple intersection form}\label{sect:Essen_Periods_and_Yijk}

Up to scale, Calabi-Yau manifolds possess a unique holomorphic $(3,0)$ form denoted $\Omega$. After expanding $\Omega$ in a basis of $H^{3}(X,\IZ)$, one obtains the period vector of $X$. The $b_{3}=2h^{1,2}+2$ components of this vector vary as the complex structure of $X$ is modified, and satisfy the Picard-Fuchs system of $X$. For manifolds with a one-dimensional space of complex structures (so $h^{2,1}=1$), the Picard-Fuchs equation is a fourth order Fuchsian differential equation (that is, a fourth order ODE, with coefficients that are rational functions in the dependent variable, with regular singular points). There are a number of choices of the basis in which one expands $\Omega$, but the two that we will most frequently see are the Frobenius basis and integral symplectic basis. The Frobenius period vector, denoted $\mathbf{\varpi}$, arises as the first set of solutions obtained by the method of Frobenius. The integral symplectic basis sees the most use in physical applications. Recall that linear systems of differential equations possess monodromies, whereby continuing a vector of solutions analytically around a singularity effects a linear transformation on the solution vector. Importantly, the moduli space $\cM_{\IC S}$ possesses a point of maximal unipotent monodromy, or MUM point, around which the monodromy matrix $\text{M}$ obeys $(\text{M}-\one)^{b_{3}}=0$ but $(\text{M}-\one)^{b_{3}-1}\neq 0$. 

We shall write $\varphi^{i}$, $i=1,\,...\,,\,h^{1,2}$ for the complex structure moduli coordinates of $X$, or $\bm{\varphi}$ to denote the list of all coordinates. The Picard-Fuchs system admits as a solution a power series $\varpi_{0}$ about a MUM point, which we shall take to be at $\bm{\varphi}=\bm{0}$, with radius of convergence given by the distance from the MUM point to the nearest singularity of the system. By normalising this function $\varpi_{0}$ to equal 1 at the MUM point we uniquely specify it. Once an analytic expression for the coefficients in the power series defining $\varpi_{0}$ in an expansion about $\bm{\varphi}=\bm{0}$ is known (as a ratio of factorials), along with some topological data that we will explain, it is possible to obtain the other periods using the method of \cite{Hosono:1993qy,Hosono:1994ax} as follows.

First, introduce dependence on formal parameters $\epsilon_{i}$ into $\varpi_{0}$ as follows:
\begin{equation}\label{eq:varpi0}
\varpi_{0}(\bm{\varphi})=\sum_{\bm{m}\geq0}c(\bm{m})\bm{\varphi}^{\bm{m}}\qquad \to \qquad \bm{\varpi}^{\bm{\epsilon}}=\sum_{\bm{m}\geq0}\frac{c(\bm{m}+\bm{\epsilon})}{c(\bm{\epsilon})}\bm{\varphi}^{\bm{m}+\bm{\epsilon}}~.
\end{equation}
Our choice of coordinates $\bm{\varphi}$ is discussed in Appendix \sref{sect:Appendix_coordinates}. The function $c(\bm{m})$ is extended to a function of non-integer argument above by replacing factorials with Gamma functions, $n!\mapsto\Gamma(1+n)$. $h^{2,1}$ periods with logarithmic singularities at the MUM point can then be found by taking $\epsilon$-derivatives:
\begin{align}\label{eq:varpiother}
\varpi_{1,i}&=\partial_{\epsilon_{i}}\varpi^{\bm{\epsilon}}\bigg\vert_{\bm{\epsilon}=0}~.
\intertext{We will shortly introduce and define a set of constants $Y_{ijk}$. This can be used to construct a further $h^{2,1}$ solutions with logarithm-squared singularities at the MUM point like so:}
\varpi_{2,i}&=\frac{1}{2}Y_{ijk}\partial_{\epsilon_{j}}\partial_{\epsilon_{k}}\varpi^{\bm{\epsilon}}\bigg\vert_{\bm{\epsilon}=0}~.
\intertext{The last period, which has logarithm-cubed behaviour near $\bm{\varphi}=0$, is obtained via}
\varpi_{3}&=\frac{1}{6}Y_{ijk}\partial_{\epsilon_{i}}\partial_{\epsilon_{j}}\partial_{\epsilon_{k}}\varpi^{\bm{\epsilon}}\bigg\vert_{\bm{\epsilon}=0}~.
\end{align}
These functions are then collected into the Frobenius period vector,
\begin{equation}
\varpi=\begin{pmatrix}\varpi_{0}\\\varpi_{1,i}\\\varpi_{2,i}\\\varpi_{3}\end{pmatrix}~.
\end{equation}
Let us draw attention to a notational point. Where one-parameter manifolds have been studied in the literature, it has sometimes been common to use the symbols $\varpi_{2}$ and $\varpi_{3}$ to denote the above functions divided by $Y_{111}$, i.e. it has been common to have $\varpi_{2}$ and $\varpi_{3}$ with expansions beginning respectively with $\log(\varphi)^{2}$ and $\log(\varphi)^{3}$. In this thesis we will consider one-parameter manifolds on par with the multiparameter geometries, and so will always use the above contraction with $Y_{ijk}$.

The quantities $Y_{ijk}$ are topological quantities not on $X$, but on a \emph{mirror manifold} to $X$. If $Y$ denotes this mirror family, then $Y_{ijk}$ gives the triple intersection form of two-cycles on $Y$. That is, if $e_{i}$ form a basis of the second cohomology of $Y$,
\begin{equation}\label{eq:Y_ijk}
Y_{ijk}=\int_{Y}e_{i}\wedge e_{j}\wedge e_{k}~.
\end{equation}
This quantity is an integer, equalling the intersection number of the four cycles that are Poincar\'e dual to each $e_{i}$.

Having explained how to form a Frobenius basis of solutions from the data of $\varpi_{0}$ and $Y_{ijk}$, it remains to determine analytic expressions for the coefficients in $\varpi_{0}$'s series expansion. This can be done by analysing the Picard-Fuchs equations directly, but then one must first write these down, which is not always straightforward. $\varpi_{0}$ can be determined by certain residue integrals, as in \cite{Berglund:1993ax}. The method that we will adopt (outside of the one-parameter cases where we have the Picard-Fuchs operator), is to use Liebgober's formula that relates the sought coefficient $c(\bm{m})$ to the total Chern class of $Y$, when $X$ and $Y$ are intersections in toric varieties \cite{libgober1998chern}. The total Chern class can be formally written as a rational function of generators of the second cohomology $e_{i}$, with a numerator and denominator that factorise into linear factors. By effecting for each such factor the replacement 
\begin{equation}
\left(1+\sum_{i=1}^{h^{2,1}(X)}a_{i}e_{i}\right)\qquad\to\qquad\Gamma\left(1+\sum_{i=1}^{h^{2,1}(X)}a_{i}m_{i}\right)^{-1}~,
\end{equation}
we obtain $c(\bm{m})$ as a ratio of Gamma functions.

When we turn to studying string-theory compactifications, we shall find a need to express period vectors in an integral symplectic basis, as we now explain. Consider a path in moduli space that begins and ends at some value $\varphi_{*}$, and goes around a singularity. Encircling singularities in the complex structure moduli space effects monodromy transformations of the period vector. Since such an encircling returns us to the same point $\varphi_{*}$ in moduli space that we started at, these monodromy transformations must be a symmetry of the theory. By choosing to expand the period vector in a basis such that monodromy transformations are integral symplectic matrices, we can pair the monodromy transformations of the period vector with symplectic transformations of the charge vector, to have a manifest monodromy invariance of quantities like D-brane masses or black hole central charges (these are the symplectic inner products of charge vectors with period vectors). Indeed, $\mathcal{N}=2$ Maxwell theory possesses a symplectic duality symmetry. We explain here how to make the change of basis from Frobenius to integral symplectic.

First, one divides each Frobenius period $\varpi$ by an appropriate power of $2\pi\ii$, since upon circling $0$ anticlockwise we have $\frac{1}{2\pi\ii}\log(\varphi)\mapsto\frac{1}{2\pi\ii}\log(\varphi)+1$:
\begin{equation}
\widehat{\varpi}_{0}=\varpi_{0}~,\qquad\widehat{\varpi}_{1,i}=\frac{1}{2\pi\ii}\varpi_{1,i},\qquad\widehat{\varpi}_{2,i}=\frac{1}{(2\pi\ii)^{2}}\varpi_{2,i}~,\qquad\widehat{\varpi}_{3}=\frac{1}{(2\pi\ii)^{3}}\varpi_{3}~.
\end{equation}
The integral symplectic period vector $\Pi$ is then computed from
\begin{equation}
\Pi=\rho\,\widehat{\varpi}~,\qquad\rho=\begin{pmatrix}
-\frac{1}{3} Y_{000} & -\frac{1}{2} \mathbf{Y}_{00}^T & \+\mathbf{0}^T & \+1\\[3pt]
- \frac{1}{2} \mathbf{Y}_{00} & -\IY_{0}  & -  \one \+ & \+\mathbf{0} \\[3pt]
1 & \+ \mathbf{0}^T & \+\mathbf{0}^T & \+0 \\[3pt]
\mathbf{0} & \one & \, \zero & \+ \bm 0
\end{pmatrix}~.
\end{equation}
We have already seen the triple intersection numbers $Y_{ijk}$, with $i$ running from 1 to $h^{2,1}(X)$. it will turn out to be natural to consider these as members of a larger set $Y_{abc}$ with indices running from 0 to $h^{2,1}$. The vector $\mathbf{Y}_{00}$ has components $Y_{00i}$, which are defined in terms of the mirror manifold $Y$'s second Chern class $c_{2}$ as
\begin{equation}\label{eq:Y_00i}
Y_{00i}=-\frac{1}{12}\int_{Y}c_{2}\wedge e_{i}~.
\end{equation}
The quantity $Y_{000}$ is
\begin{equation}\label{eq:Y_000}
Y_{000}=\frac{3\zeta(3)}{(2\pi\ii)^{3}}\chi(X)=-\frac{3\zeta(3)}{(2\pi\ii)^{3}}\chi(Y)~,
\end{equation}
where in the last equality we have used that the mirror pairing exchanges the complex structure and complexified K{\"a}hler structure of $X$ and $Y$, and so their Hodge numbers $h^{1,1}$ and $h^{2,1}$ are interchanged, and thereby $X$ and $Y$ have Euler characteristics that differ by a sign.

The quantities $Y_{0ij}$, which we collect into the matrix $\IY_{0}$, can be taken to have values $0$ or $1/2$. Symplectic transformations allow for shifts by integer values, so we make the choice to take values in $\{0,1/2\}$. As was explained in \cite{Hosono:1994ax}, exactly which number each $Y_{0ij}$ equals is determined by demanding that the monodromy matrix that acts on $\Pi$ after encircling the MUM point at $\bm{\varphi}=0$ is integral and symplectic. This condition relates the $Y_{0ij}$ to the $Y_{ijk}$, and one can compute in the $1,\,2,\,3$ modulus cases that
\begin{equation}\label{eq:Y_0ij}
Y_{0ij}=\frac{Y_{ijj}\text{ Mod }2 }{2}~.
\end{equation}

\section{CICYs}\label{sect:Essen_CICYs}
A set that admits a particularly simple characterisation are the \emph{Complete Intersection Calabi-Yaus}, or CICYs \cite{Candelas:1987kf}. These are the intersection of hypersurfaces in products of projective space. Such a manifold $Y$ can be represented by a configuration matrix, a rectangular array where for each projective space factor in the ambient space there is a row. Each column represents one of the defining equations. The $(i,j)$-entry of the array gives the degree of the $j^{th}$ equation in the $i^{th}$ projective space. Three illustrative examples are
\begin{equation}
\cicy{\IP^{1}\\\IP^{1}\\\IP^{2}}{2\\2\\3}~,\qquad \cicy{\IP^{5}}{3&3}~,\qquad \cicy{\IP^{1}\\\IP^{4}}{1&1\\4&1}~.
\end{equation}
The first of these represents a hypersurface (the vanishing set of a single equation) in ${\IP^{1}\times\IP^{1}\times\IP^{2}}$ defined by a polynomial with degrees $(2,2,3)$ in the homogeneous coordinates on each projective space. The second represents the intersection of two degree-3 hypersurfaces in $\IP^{5}$. The third represents the intersection of two hypersurfaces in the space $\IP^{1}\times\IP^{4}$. The general configuration matrix, which we display to fix notation in this section, is 
\begin{equation}
\cicy{\IP^{n_{1}}\\\vdots\\\IP^{n_{K}}}{d^{\,(1)}_{1}&...&d^{\,(1)}_{C}\\\vdots&...&\vdots\\d^{\,(K)}_{1}&...&d^{\,(K)}_{C}}~.
\end{equation}
For the purposes of this section, $K$ will be the number of projective space factors in the ambient space and $C$ will be the number of equations defining the intersection.

Not any configuration matrix yields a Calabi-Yau. One must verify that the first Chern class $c_{1}(TY)$ vanishes. This is an exercise in classical algebraic geometry: one can form the total Chern class from the entries of the configuration matrix, which is Taylor expanded to get the Chern classes $c_{1}$, $c_{2}$ and $c_{3}$. The first of these needs to vanish, while the third can be used to compute the Euler characteristic $\chi(Y)$. The space $Y$ is an intersection in an ambient space $\cA=\IP^{n_{i}}\times\,...\,\times\IP^{n_{K}}$, and there is a split exact sequence
\begin{equation}
0\to TY\to T\cA \to NY \to 0~,
\end{equation}
where the tangent bundle $TY$ of $Y$ injects into the tangent bundle of $\cA$, which in turn projects onto the normal bundle $NY$ of $Y$. Stated in more pedestrian terms, the following decomposition (which is non-canonical, depending on the choice of the embedding of $NY$ as a subbundle of $T\mathcal{A}$) holds at points on $Y$:
\begin{equation}
T\cA=TY\oplus NY~.
\end{equation}
The Chern class of the vector sum of two spaces is the product of their individual Chern classes, and so
\begin{equation}
\text{Ch}(T\cA)=\text{Ch}(TY)\cdot\text{Ch}(NY)~,\qquad \text{whereby}\qquad \text{Ch}(TY)=\frac{\text{Ch}(T\cA)}{\text{Ch}(NY)}~.
\end{equation}
The total Chern class of $\cA$ is the product of the Chern classes of the factors $\IP^{n_{i}}$. Since $\text{Ch}(\IP^{n_{i}})=(1+E_{i})^{n_{i}+1}$, where $E_{i}$ generates the second cohomology of $\IP^{n_{i}}$, we get
\begin{equation}
\text{Ch}(T\cA)=\prod_{i=1}^{K}\left(1+E_{i}\right)^{n_{i}+1}~.
\end{equation}
It is through the total Chern class of $NY$ that the entries of the configuration matrix enter. One has that
\begin{equation}
\text{Ch}(NY)=\prod_{j=1}^{C}\left(1+\sum_{i=1}^{K}d^{\,(i)}_{j}E_{i}\right)~.
\end{equation}
The total Chern class then is 
\begin{equation}\label{eq:total_chern_class}
\text{Ch}(TY)=\frac{\prod_{i=1}^{K}\left(1+E_{i}\right)^{n_{i}+1}}{\prod_{j=1}^{C}\left(1+\sum_{i=1}^{K}d^{\,(i)}_{j}E_{i}\right)}~,
\end{equation}
which we expand to linear order in the cohomology generators $E_{i}$ to find the first Chern class $c_{1}$. This gives
\begin{equation}
c_{1}(TY)=\sum_{i=1}^{K}\left(1+n_{i}-\sum_{j=1}^{C}d^{\,(i)}_{j}\right)E_{i}~,
\end{equation}
whereupon we learn that the Calabi-Yau condition is met precisely when the sum of the configuration matrix's $i^{th}$ row equals $n_{i}+1$. Since we are interested in threefolds, we must have $c+3=\sum_{i=1}^{K}n_{i}$.

There is a computational problem in enumerating all possible rectangular matrices with nonnegative entries that satisfy the above two conditions. However, there are a number of possible redundancies which mean that a straightforward listing of all possible configuration matrices overcounts the number of CICY threefolds. For example, for $m$-vectors $\mathbf{a},\,\mathbf{b}$, an $n\times m$ matrix $\text{M}$, and a `vector' $P$ of $n$ projective spaces,  there is the identity
\begin{equation}
\cicy{\IP^{1}\\\IP^{1}\\ P}{1&\mathbf{a}^{T}\\1&\mathbf{b}^{T}\\\mathbf{0}&\text{M}}=\cicy{\IP^{1}\\ P}{\mathbf{a}^{T}+\mathbf{b}^{T}\\\text{M}}~,\qquad \text{for example} \qquad \cicy{\IP^{1}\\\IP^{1}\\ \IP^{1}\\\IP^{2}}{1&1\\1&1\\0&2\\0&3}=\cicy{\IP^{1}\\\IP^{1}\\\IP^{2}}{2\\2\\3}~.
\end{equation}
This follows from the fact that
\begin{equation}
\cicy{\IP^{1}\\\IP^{1}}{1\\1}\quad=\quad\IP^{1}~.
\end{equation}
Upon removing such redundant descriptions as above, one arrives at a naive number of 7890 CICY threefolds. This is still not the full story, as additional nontrivial redundancies exist beyond those used to find this number. This list is held in many places, we have in this work made use of Lukas's website \cite{LukasWebsite}, from which one can for instance find the redundancy
\begin{equation}
\cicy{\IP^{1}\\\IP^{1}\\\IP^{1}\\\IP^{1}\\\IP^{1}}{1&1\\1&1\\1&1\\1&1\\1&1}\=\cicy{\IP^{1}\\\IP^{1}\\\IP^{1}\\\IP^{1}\\\IP^{1}}{0&2\\1&1\\1&1\\1&1\\1&1}~.
\end{equation}
This is a family to which we shall give some attention in later sections, preferring to use the more symmetric description on the left-hand side.

More generally, one could consider intersections in toric varieties (of which CICYs are but a special case). Hypersurfaces in toric varieties are in bijection with reflexive four-dimensional polytopes, which were tabulated in \cite{Kreuzer:2000xy}. The enumeration of more general intersections of hypersurfaces in toric varieties of dimension greater than 4 has not been completed. Using gauge theory constructions, intersections in toric varieties are realised as phases of an abelian gauged linear sigma model. By dropping this abelian condition, one could reach a wider-still set of manifolds. Examples of this construction include the R{\nordico}dland manifold \cite{rodland2000pfaffian} and the two-parameter model of Knapp and Hori \cite{Hori:2016txh}.

As we turn to the study of particular problems, we shall lean on the CICY threefolds as our examples. However, we stress that a number of the properties that we will observe can be expected to hold in more general examples.

There is an important partition of the CICY manifolds into two sets: those that are \emph{favourable} and those that are not. A complete intersection $Y$ is favourable if its second cohomology is spanned by the pullbacks to $Y$ of the generators for each projective space's second cohomology. For example, 
\begin{equation}
\cicy{\IP^{5}}{3,3} \qquad \text{and} \qquad \cicy{\IP^{1}\\\IP^{4}}{1&1\\1&4}
\end{equation}
respectively have one- and two-dimensional second cohomologies. In the first example this is spanned by the pullback of the K{\"a}hler form for the ambient $\IP^{5}$, while in the second the pullbacks of the K{\"a}hler forms for the ambient $\IP^{1}$ and $\IP^{4}$ span the manifold's second cohomology. Note that a favourable manifold necessarily has $h^{1,1}=K$. This is to be contrasted with non-favourable CICYs, for which $h^{1,1}>K$ (we will always take $K$ to be the number of projective space factors in the representative among redundant configurations with the smallest number of rows). An example of a CICY with this feature is 
\begin{equation}
\cicy{\IP^{4}\\\IP^{4}}{0&0&2&2&1\\2&2&0&0&1}~,
\end{equation}
which has $h^{1,1}=12$. The pullbacks of the K{\"a}hler form for each $\IP^{4}$ factor each appear in the manifold's second cohomology, but there are ten other generators. To see where these come from, we should recall that the (non-Calabi-Yau) intersection $\cicy{\IP^{4}}{2&2}$ is the Segre surface, isomorphic to the projective plane blown up in 5 points. In addition to the hyperplane class, this space has an additional 5 generators in the second cohomology dual to cycles wrapping each blown up point. So by viewing the above CICY manifold as an intersection in the product of two Segre surfaces, we can account for $h^{1,1}=12$ by pulling back to $X$ the six independent generators in each Segre surface's cohomology. More generally, the $h^{1,1}$ of a non-favourable manifold can be explained in terms of the cohomology classes on del Pezzo surfaces that arise from the configuration matrix in a similar manner as above.

To compute $h^{2,1}$, it suffices to compute $h^{1,1}$ as in the previous paragraph and then use the total Chern class the compute $\chi=\int_{Y}c_{3}$. Recall that the Euler characteristic of $Y$ is the alternating sum of Betti numbers, $\chi(Y)=\sum_{i=0}^{6}(-1)^{i}b_{i}$. Now, $b_{0}=b_{6}=1$ as our manifolds each have a single connected component. Further, $b_{1}=b_{5}=0$ (in light of full $\text{SU}(3)$ holonomy). We have that $b_{2}=b_{4}=h^{1,1}$ and $b_{3}=2+2h^{1,2}$, so that 
\begin{equation}
\chi=2\left(h^{1,1}-h^{1,2}\right)~.
\end{equation}
We see that $h^{2,1}$ can be determined via $h^{2,1}=h^{1,1}-\frac{1}{2}\chi$, with $\chi$ determined from the configuration matrix via the total Chern class formula. Note that for a CICY $Y$, $\chi(Y)\leq0$.

Following \cite{Candelas:1987kf}, the triple intersection numbers of a favourable CICY $Y$ can be computed by noting that each two-form $e_{i}$ on $Y$ is the pullback, with respect to the inclusion map $\iota:Y\mapsto\cA$, of the K{\"a}hler forms $E_{i}$ of each $\IP^{n_{i}}$. That is,
\begin{equation}
e_{i}=\iota^{*}(E_{i}).
\end{equation}
As a result of this
\begin{equation}\label{eq:tripleint}
Y_{ijk}=\int_{Y}e_{i}e_{j}e_{k}=\int_{\cA}E_{i}E_{j}E_{k}\prod_{j=1}^{C}\left(1+\sum_{i=1}^{K}d^{\,(i)}_{j}E_{i}\right)~,
\end{equation}
where the product inserted in the right-hand integral is the Poincar\'e dual to $Y$ in the cohomology of $\cA$. The above integral is performed by singling out the coefficient of the volume form on $\cA$, which is the product $\prod_{i=1}^{K}E_{i}^{n_{i}}$.

We have now provided the total Chern class and triple intersection numbers of a favourable CICY $Y$. We can now use \eqref{eq:total_chern_class} and \eqref{eq:tripleint} to compute the periods of mirror manifolds $X$ as in \eqref{eq:varpi0}. 

\section{Splittings and the conifold transition}\label{sect:Essen_Splits_and_Conifolds}
The data of a CICY matrix gives only the degrees of the polynomials defining the Calabi-Yau variety, and says nothing about the coefficients of these polynomials. Generic choices of these polynomials, related by variation of complex structure, will be smooth and be of the same topological type (in particular they will have the same Hodge numbers and triple intersection form). Nonetheless, for nongeneric choices of these coefficients the geometry will acquire singularities. Deformation (varying the polynomial coefficients again) will return a smooth manifold in the generic family, but one could also resolve these singularities via a blowup to obtain some smooth member of a different family of manifolds, with different topological properties. The latter family obtained from the singular space by resolution is said to be a `split' of the former family obtained from the singular space via deformation.

Let us be more concrete, and explain the splittings via operations on CICY matrices. Let us begin with the family $Y$, with configuration matrix
\begin{equation}
Y:\qquad\cicy{\mathbf{P}}{\mathbf{v}&\text{M}}~,
\end{equation}
with $\mathbf{P}$ denoting a list of projective spaces, $\mathbf{v}$ being a vector of nonnegative integers, and M being a matrix. Now it may be the case that for some nongeneric choice of coefficients, the first equation of the above configuration can be expressed as a determinant. That is to say, the equation read off from $\mathbf{P}[\mathbf{v}]$ is
\begin{equation}
\text{Det}(\text{C}^{a}_{\+ b})\=0~,
\end{equation}
where the square matrix C, of some dimension $(n+1)\times (n+1)$, has entries that are polynomials in the coordinates of $\mathbf{P}$. The set of such equations as the polynomials are varied may have members defining singular spaces, in which case the splitting we go on to describe is said to be \emph{effective} and yields a manifold of a new topological type. Alternatively we may be describing an \emph{ineffective} splitting if the family of vanishing loci of $\text{det}(\text{C})$ has no singularities, so that we are not changing the topology of the family considered, instead just giving a different redundant description.

Vanishing of C's determinant implies the existence of a nullvector for C, a nonzero $n+1$ vector $x^{b}$ such that
\begin{equation}
\text{C}^{a}_{\+ b}\,x^{b}\=0~.
\end{equation}
There are $n+1$ such equations for each choice of the index $a$. Promoting $x^{b}$ to homogeneous coordinates on $\IP^{n}$, The above equation then is some element of the set of polynomials read off from
\begin{equation}
\cicy{\IP^{n}\\\mathbf{P}}{1 & 1 &...& 1\\\mathbf{u}_{1}&\mathbf{u}_{2}&...&\mathbf{u}_{n+1}}~,
\end{equation}
where $\mathbf{v}=\sum_{i=1}^{n+1}\mathbf{u}_{i}$~. We learn that the family $X$ can be split to a family $X'$ with configuration
\begin{equation}
Y':\qquad\cicy{\IP^{n}\\\mathbf{P}}{1 & 1 &...& 1&\mathbf{0}^{T}\\\mathbf{u}_{1}&\mathbf{u}_{2}&...&\mathbf{u}_{n+1}&\text{M}}~,
\end{equation}
which is guaranteed to be Calabi-Yau by the fact that $Y$ was Calabi-Yau (one can check that the rows sum to appropriate values). Some examples of pairs $(A,B)$ where $B$ splits $A$ are
\begin{equation}
\left(\cicy{\IP^{4}}{5}\;,\;\cicy{\IP^{1}\\\IP^{4}}{1&1\\1&4}\right)~,\qquad\left(\cicy{\IP^{4}}{5}\;,\;\cicy{\IP^{4}\\\IP^{4}}{1&1&1&1&1\\1&1&1&1&1}\right)~,\qquad\left(\cicy{\IP^{1}\\\IP^{1}\\\IP^{1}\\\IP^{1}}{2\\2\\2\\2}\;,\;\cicy{\IP^{1}\\\IP^{1}\\\IP^{1}\\\IP^{1}\\\IP^{1}}{1&1\\1&1\\1&1\\1&1\\1&1}\right)~.
\end{equation}
Note that a manifold can have distinct splits, as exemplified by the Quintic $\cicy{\IP^{4}}{5}$ above.

\section{Calabi-Yau modularity}\label{sect:Essen_CY_Modularity}
We turn now to explaining what is meant by a \emph{modular} Calabi-Yau manifold. Having done this, we will go on to see how Calabi-Yau modularity is related to supersymmetric physics.

The conventional physicist's view of a Calabi-Yau manifold is that it is an algebraic variety over the field of complex numbers $\IC$. From a wider perspective, one is free to consider Calabi-Yau manifolds over any field. Of particular relevance to modularity are the finite fields $\IF_{p^{n}}$, where in the subscript $p$ is a prime and $n$ a positive integer. This number $p^{n}$ is the order of the finite field, which is necessarily the power of a prime. 

Given a polynomial with rational coefficients, we can multiply through by the least common multiple of the coefficient's denominators in order to get a polynomial with integer coefficients. This in turn can be reduced modulo a prime.

Alternatively, suppose now that a Calabi-Yau $X$'s complex structure moduli take rational values $m/n$. $X$ is then an algebraic variety over $\IQ$, for which one writes $X/\IQ$. The integers $m$, $n$ have representatives in $\IF_{p}$, simply found as $m \text{ Mod }p$ and $n \text{ Mod } p$. Now, $\IF_{p}$ is a field and so for each integer $n$ such that $p\nmid n$, there is an element $1/n$. So the product $m\cdot\frac{1}{n}$ can be found in $\IF_{p}$. This allows for the equations defining $X$ as an algebraic variety over $\IC$, assumed to have rational coefficients, to be replaced with equations with coefficients in $\IF_{p}$. All of that is to say, we can consider the variety $X$ over the field $\IF_{p}$, denoted $X/\IF_{p}$. 

The variety $X/\IF_{p}$ is the set of solutions to these equations, where the indeterminates (coordinates on the ambient space containing $X$) take values in $\IF_{p}$. So, this variety $X/\IF_{p}$ is the union of a finite number of points in the ambient space. 

The finite fields $\IF_{p^n}$ can be realised as the field extension of $\IF_{p}$ by the root of an irreducible polynomial with order $n$. That is to say, we can realise $\IF_{p^{n}}$ by taking
\begin{equation}
\IF_{p^{n}}\cong\IF_{p}(\alpha)~,\qquad \text{with $\alpha$ a root of an irreducible polynomial $Q$ with coefficients in $\IF_{p}$.}
\end{equation}
 There is a standard field embedding 
\begin{equation}
\IF_{p}\hookrightarrow\IF_{p^{n}}
\end{equation}
which simply sends the field element $r\in\IF_{p}$ to $r\in\IF_{p}(\alpha)\cong\IF_{p^{n}}$.

Since $\IF_{p}$ embeds this way into $\IF_{p^{n}}$, we can also take the equations defining the variety $X/\IF_{p}$ to define another variety $X/\IF_{p^{n}}$, which consists of another union of a finite number of coordinate vectors with elements in $\IF_{p^{n}}$ so that these equations are solved.

Crucially, the information contained in each of these finite sets `fits together' in a way that we shall now describe.

\subsection{The local zeta function}
Fix a prime $p$, and a set of algebraic moduli $\varphi$ for the variety $X_{\varphi}$, which we shall subscript with $\varphi$ in this subsection so as to make dependencies clear.  We will write $N_{p^n}(\varphi)$ for the number of points constituting the variety $X_{\varphi}/\IF_{p^{n}}$.

Now we introduce a formal variable $T$ and collect these point counts into the following generating function:
\begin{equation}\label{eq:zeta_exponential}
\zeta_{p}\left(X_{\varphi};T\right)=\exp\left(\sum_{n=1}^{\infty}\frac{1}{n}\,N_{p^{n}}\,(\varphi)T^{n}\right)~.
\end{equation}
This is the local zeta function of $X_{\varphi}/\IF_{p}$~. The three Weil conjectures \cite{Weil}, proved by Dwork \cite{Dwork}, Grothendieck \cite{Grothendieck}, and Deligne \cite{DeligneRiemann}~, describe the behaviour of this function. We shall state them on the assumption that $X_{\varphi}/\IC$ is a $d$-complex dimensional projective variety, and not necessarily a Calabi-Yau threefold.

\subsubsection*{(1) Rationality}
The zeta function is a rational function of $T$, with the numerator and denominator factorising over $\IZ$ into the form
\begin{equation}\label{eq:zeta_rational}
\zeta_{p}\left(X_{\varphi};T\right)=\frac{P_{1}\left(X_{\varphi};T\right)P_{3}\left(X_{\varphi};T\right)\,...\,P_{2d-1}\left(X_{\varphi};T\right)}{P_{0}\left(X_{\varphi};T\right)P_{2}\left(X_{\varphi};T\right)\,...\,P_{2d}\left(X_{\varphi};T\right)}~,
\end{equation}
where each polynomial $P_{i}\left(X_{\varphi};T\right)$ has degree equal to the Betti number $b_{i}$~. In addition, 
\begin{equation}
P_{0}=1-T\qquad\text{ and }\qquad P_{2d}=1-p^{d}T~.
\end{equation}
\subsubsection*{(2) Functional Equation}
This function obeys the following transformation of argument property:
\begin{equation}\label{eq:zeta_fuctional_equation}
\zeta_{p}\left(X_{\varphi};\frac{1}{p^{d}T}\right)=\pm p^{\frac{d\chi}{2}}T^{\chi}
\zeta_{p}\left(X_{\varphi};T\right)~,
\end{equation}
with $\chi$ the Euler characteristic of $X_{\varphi}$.
\subsubsection*{(3) The Riemann hypothesis}
Each polynomial $P_{i}$ factorises over $\IC$,
\begin{equation}
P_{i}\left(X_{\varphi};T\right)=\prod_{j=1}^{b_{i}}\left(1-\lambda_{ij}(X_{\varphi}) T\right)~,
\end{equation}
and each $\lambda_{ij}$ is an algebraic integer. That is, each $\lambda_{ij}$ is a root in $\IC$ of a monic polynomial with integer coefficients. Further, each $\lambda_{ij}$ has absolute value $p^{i/2}$.

Note that the rational function \eqref{eq:zeta_rational} is further simplified when $X_{\varphi}$ is a Calabi-Yau threefold. Then $d=3$, and the Betti numbers are $b_{0}=b_{6}=1$, $b_{2}=b_{4}=h^{1,1}$, $b_{1}=b_{5}=0$, and $b_{3}=2h^{2,1}+2$. The form of \eqref{eq:zeta_rational} is further constrained by the functional equation \eqref{eq:zeta_fuctional_equation}, so that for a Calabi-Yau threefold we get
\begin{equation}
\zeta_{p}\left(X_{\varphi};T\right)=\frac{R_{p}\left(X_{\varphi};T\right)}{\left(1-T\right)\left(1-pT\right)^{h^{1,1}}\left(1-p^{2}T\right)^{h^{1,1}}\left(1-p^{3}T\right)}~,
\end{equation}
with all $\varphi$ dependence captured in the numerator $R_{p}\left(X_{\varphi};T\right)$, which is a degree $b_{3}=2h^{2,1}+2$ polynomial in $T$.

Let us stress the following point: for generic moduli $\varphi$ we obtain some typically irreducible polynomial $R_{p}\left(X_{\varphi};T\right)$, but for particular choices of moduli $R_{p}\left(X_{\varphi};T\right)$ can factorise for each $p$. When this happens, the coefficients of quadratic factors are conjecturally related to modular forms. We shall now explain this in greater detail.

\subsection{Modular forms and modularity of elliptic curves}
For elliptic curves $E_{\varphi}$ (Calabi-Yau onefolds) over $\IQ$, the zeta function numerator $Q_{p}$ is a quadratic function of $T$ of the form
\begin{equation}
Q_{p}\left(X_{\varphi};T\right)=1-c_{p}(\varphi)pT+p^{3}T^{2}~.
\end{equation}
We remark that the Weil conjectures for elliptic curves were originally proved by Hasse \cite{Hasse}. The modularity theorem for elliptic curves over $\IQ$, proved in work by Breuil, Conrad, Diamond, Taylor, and Wiles \cite{Wiles,WilesTaylor,BCDT} guarantees that the $c_{p}$ so defined are the coefficients of $q^{p}$ in the Fourier expansion of a weight-two modular form (in fact, a newform) for a congruence subgroup $\Gamma_{0}(N)$, where $N$ is the conductor of the elliptic curve $E_{\varphi}$. This number $N$ is divisible by each prime $p$ that has bad reduction, which means that the elliptic curve $E_{\varphi}/\IF_{p}$ is singular. 

The modular group $\Gamma\cong\text{PSL}(2,\IZ)$ of $2\times2$ matrices with integer components and determinant $\pm1$ acts on the upper half plane $\IH$ (complex numbers with positive imaginary part) by the M{\"o}bius action:
\begin{equation}
\gamma(\tau)=\frac{a\tau+b}{c\tau+d}~,\qquad \text{where the matrix $\gamma$ has components}\qquad \gamma=\begin{pmatrix}a&b\\c&d\end{pmatrix}~.
\end{equation}
We shall always take an elements $\gamma$ of $\Gamma$ to have matrix components $a,\,b,\,c,\,d$ as above. A modular form of weight $k\geq0$ is a holomorphic function $f$ on the upper half plane with the following transformation properties under the modular group:
\begin{equation}\label{eq:modular_form_transformation}
f\left(\gamma(\tau)\right)=\left(c\tau+d\right)^{k}f(\tau)~.
\end{equation}
There are functions that satisfy the above property not for $\Gamma$, but for certain subgroups. Specifically, \emph{congruence subgroups}, which we now define. The principal congruence subgroup of level $N$ is the group
\begin{equation}
\Gamma(N)=\left\{\left(\begin{matrix}a&b\\c&d\end{matrix}\right)\in \Gamma:\ a,d=1\text{ and }b,c=0\text{ mod }N \right\}.
\end{equation}
A congruence subgroup is any subgroup of $\Gamma$ that contains a $\Gamma(N)$ as a subgroup. The examples that we shall make frequent return to, appearing in the statement of the modularity theorem for elliptic curves that we gave at the start of this subsection, are the groups
\begin{equation}
\Gamma_{0}(N)=\left\{\left(\begin{matrix}a&b\\c&d
\end{matrix}\right)\in \Gamma:\ c=0\text{ mod }N \right\}.
\end{equation}
A modular form of level $N\in\IN$ and weight $k\geq0$ is a holomorphic function obeying \eqref{eq:modular_form_transformation} for all $\gamma\in\Gamma_{0}(N)$. These functions form a finite-dimensional vector space over $\IC$, and newforms are particular basis elements of this space. These newforms are tabulated in the L-Functions and Modular Forms Database, LMFDB, \cite{LMFDB}. Newforms are in particular cusp forms, which have 0 constant term in their Fourier expansions. We shall make frequent use of the LMFDB labels when discussing examples in later sections.

\subsubsection*{L functions}
The Mellin transform of a modular form $f$ yields the associated $L$-function:
\begin{equation}
L(s)=\frac{(2\pi)^{s}}{\Gamma(s)}\int_{0}^{\infty}\dd y\,f(\ii\, y)\,y^{s-1}~.
\end{equation}
Naturally, $\Gamma(s)$ is the Gamma function and not a group. In practice, one usually has the first few hundred (or more) terms in a Fourier expansion of $f$, which can be inserted into the above integral so as to compute values $L(s)$. The accuracy improves with more terms in $f$'s Fourier expansion, but a manipulation of the integrand using transformation properties of $f$ allows for a quicker evaluation to a given accuracy. To do this, first break up the integral like so:
\begin{equation}\label{eq:L(s)TWO}
L(s)=\frac{(2\pi)^{s}}{\Gamma(s)}\left(\int_{\frac{1}{\sqrt{N}}}^{\infty}\,f(\ii\, y)\,y^{s-1}\,\dd y\;+\;\int_{0}^{\frac{1}{\sqrt{N}}}\,f(\ii\, y)\,y^{s-1}\,\dd y\right)~.
\end{equation}
Because the lower range of integration on the left-hand integral is nonzero, the approximations to this term obtained by truncating $f$'s Fourier expansion are much better than for the right-hand integral. To better approximate the right-hand integral we will make use of the Fricke involution, a property of cusp forms for $\Gamma_{0}(N)$. Such functions, with weight $k$, admit the following transformation (Fricke involution):
\begin{equation}\label{eq:Fricke}
f\left(-\frac{1}{N\tau}\right)=\epsilon N^{k/2}\tau^{k}f(\tau)~.
\end{equation}
The quantity $\epsilon$ is the Fricke sign of the form $f$, which is equal to plus or minus one. We now effect a change of variables $\tau\mapsto-\frac{1}{N\tau}$ in the right-hand integral of \eqref{eq:L(s)TWO}, and then apply the formula \eqref{eq:Fricke}, to obtain
\begin{equation}
L(s)=\frac{(2\pi)^{s}}{\Gamma(s)}\int_{\frac{1}{\sqrt{N}}}^{\infty}\dd y\,f(\ii\, y)\left(y^{s-1}+(-1)^{k/2}\epsilon N^{k/2-s}y^{k-1-s}\right)~.
\end{equation}
This expression is well-approximated upon replacing $f$ by a truncation of its Fourier expansion.

\subsection{Modularity of Calabi-Yau threefolds}
Whereas the modularity of elliptic curves over $\IQ$ is a proven fact for all such curves, the same is not true of threefolds. The topic of Calabi-Yau modularity was reviewed in \cite{Yui2013}. 

A Calabi-Yau is rigid if it has a pointlike space of complex structures, i.e. $h^{2,1}=0$. Then $b_{3}=2$ and the zeta function numerators $R_{p}(X;T)$ are, by the Weil conjectures, automatically quadratic. It was proven by Gouvea and Yui \cite{gouvea2011rigid} that such threefolds are indeed modular, of weight-four. For these Rigid threefolds
\begin{equation}
R_{p}(X;T)=1-\beta_{p} T+p^{3}T^{2}
\end{equation}
and the integers $\beta_{p}$ are the coefficients of the $p^{th}$ terms in the Fourier expansion of a weight-four modular form.

This is in contrast to manifolds $X_{\varphi}$ with $h^{2,1}\neq0$. The most general approach here, which we do not take, is to recognise in the zeta function of $X_{\varphi}$ the coefficients of an automorphic form for a symplectic group's Langlands dual, in the style of the Langlands program \cite{Frenkel:2005pa,Gelbart,Goresky}. The problem that we turn to is, for which moduli $\varphi$ is $X_{\varphi}$ modular? This means that the zeta function numerator $R_{p}(X_{\varphi};T)$ possesses for all good primes $p$ a quadratic factor from which we can read off a modular form. 

There are a number of threefolds proven to be weight-four modular. Certain members of a family associated to the $A_{4}$ lattice were shown to be so by Hulek and Verrill \cite{HulekVerrill}, Schoen proved modularity of singular quintic threefolds in \cite{Schoen1986}, and in \cite{Bonisch:2022mgw} a number of conjecturally modular manifolds were displayed, and among them it was proven that the mirror of $\cicy{\IP^{7}}{2&2&2&2}$ was modular at the conifold. In addition to these proven cases, a number of manifolds are conjectured to be modular with evidence given by an extensive computation of the zeta function. Such cases are included in the works \cite{Candelas:2019llw,Bonisch:2022slo}.

There are two problems in compactifications of type IIB superstring theory on $X_{\varphi}$, solved by restrictions on the moduli $\varphi$ to values $\varphi_{*}$, so that conjecturally the manifolds $X_{\varphi_{*}}$ are modular. The first of these is the identification of \emph{attractor points of rank-two}, a program instigated by Moore in \cite{Moore:1998zu,Moore:1998pn}. The relevant manifolds here are conjecturally weight-four modular. The second such set of moduli are those that give \emph{supersymmetric flux vacua}. The manifolds here should be weight-two, as conjectured by Kachru, Nally, and Yang \cite{Kachru:2020sio,Kachru:2020abh}. We will in later sections exposit both of these, and display new results in both directions. Here, we discuss some general features of modular Calabi-Yau threefolds that bears relevance to both physical setups. 

\subsubsection*{The Frobenius map}
We review this background material following \cite{Candelas:2019llw,Candelas:2000fq,Candelas:2004sk,Candelas:2021tqt}. Fermat's little theorem  gives, for integer $c$,
\begin{equation}
c^{p}=c\qquad\text{ Mod } p~.
\end{equation}
Further, when working modulo $p$ we have a simplified form of the binomial theorem for $c$ and $d$ in $\IF_{p^k}$:
\begin{equation}
(c+d)^{p}=c^{p}+d^{p}\qquad\text{ Mod } p~.
\end{equation}
When considering varieties over finite fields, we refer to the vanishing set of some polynomials. Note that in light of the above two identities, we get a result for polynomials $Q$ that have coefficients in $\IF_{p}$ and indeterminates in any $\IF_{p^{k}}$ with $k\in\IN$. This is
\begin{equation}
Q(x)=0\implies Q(x)^{p}=0\implies Q(x^{p})=0~.
\end{equation}
So, if we act on the coordinates $x$ of the ambient space containing $X_{\varphi}$ by the Frobenius map
\begin{equation}
\text{Frob}_{p}:x\mapsto x^{p}~,
\end{equation}
we leave invariant the vanishing locus $X_{\varphi}$. In light of this, the Frobenius map is an automorphism of any variety $X_{\varphi}/\IF_{p^{k}}$ with coefficients in $\IF_{p}$.

The fixed points of this automorphism obey $x^{p}=x$, which means that $x$ is defined in $\IF_{p}\subset\IF_{p^{k}}$. Fixed points of $\left(\text{Frob}_{p}\right)^{2}$ have $x^{p^{2}}=x$, and so are defined in $\IF_{p^{2}}\subset\IF_{p^{k}}$, and so on for each higher field.

So one has that the zeroes of the defining polynomials of $X_{\varphi}$, which we aim to count, are fixed points of a continuous automorphism of $X_{\varphi}$. By providence, the number of fixed points of such an automorphism appears in Lefschetz's trace formula:
\begin{equation}
\sum_{i=0}^{6}(-1)^{i}\text{Tr}\left((\text{Frob}_{p})^{k}\vert H^{i}(X_{\varphi})\right)=N_{p^{k}}(\varphi)~.
\end{equation}
Here, one uses the $p$-adic cohomology $H^{i}(X_{\varphi})$ (we do not discuss this point further, instead referring to \cite{Candelas:2021tqt}). The 6 in the above formula is the real dimension of the threefolds we study, for a complex $n$-fold one would replace 6 by $2n$.

Using this formula, working from the exponential form \eqref{eq:zeta_exponential} one can show that the polynomials $P_{k}$ appearing in the rational form of the zeta function \eqref{eq:zeta_rational} are in fact determinants:
\begin{equation}
P_{k}(X_{\varphi};T)=\det\left(1-T\,\text{Frob}_{p}^{\+-1}\vert H^{k}(X_{\varphi}\right)~.
\end{equation}
It should be noted that our notation for $P_{k}$ suppresses their dependence on the prime $p$.

Counting the $\IF_{p^{k}}$-points of the varieties $X_{\varphi}$, as done for example in \cite{Kadir:2004zb}, is a laborious process. In \cite{Candelas:2021tqt} a method was developed, using the above determinant formula, to more efficiently compute zeta function numerators. 

%It is this appearance of the cohomology groups that makes a connection to string theory compactifications possible, because the equations in the latter often involve relations between the holomorphic threeform $\Omega$ and charge vectors (which are elements of the cohomology groups of the threefold). In the above  

\chapter{Reflections in the mirror}\label{Chapter:Miroirs}
\setlength\epigraphwidth{.85\textwidth}\epigraph{And I saw how the stars of Heaven come out, and counted the Gates out of which they come, and wrote down all their outlets, for each one, individually, according to their number.  And their names, according to their constellations, their positions, their times, and their months, as the Angel Uriel, who was with me, showed me.}{Enoch 33.3}
Given a complex manifold $M$, a quantity of foremost interest in classical algebraic geometry is the number of holomorphic embeddings of a Riemann surface with given genus and degree in $M$. This statement requires some delicacy, for such curves can lie in continuous families and so a more refined notion of `counting' is required. A major success of string theory is the means it provides of obtaining these curve counts when $M$ is Calabi-Yau. This is done by studying the partition function of two distinct topological string theories, the A and B models. The A-model bears a direct relation to these curve counts, the partition function is their generating function. A direct computation of the A-model partition function is infeasible, but the B-model partition function can be computed from the period functions of $M$ (there are however complications with increasing genus). The two models are related by a string duality, \emph{mirror symmetry}, which means that once either partition function is computed, both are known. And so the `easy' B-model computation yields the sought curve counts. 

To elaborate, as covered in the textbook \cite{Hori:2003ic} the A and B models are topological twists of the two-dimensional $\mathcal{N}=(2,2)$ nonlinear sigma model with target space $M$. The R-symmetry group of a 2d $\mathcal{N}=(2,2)$ theory is $\text{U}(1)_{A}\times\text{U}(1)_{V}$, the product of axial and vector $\text{U}(1)$ symmetries. The vector $\text{U}(1)_{V}$ symmetry is never anomalous, while the axial $\text{U}(1)_{A}$ is anomalous unless $M$ is Calabi-Yau. The $A$ model is obtained by twisting the theory with respect to the $U(1)_{V}$ R-symmetry, while the B-model comes from twisting with respect to $U(1)_{A}$. The bosonic field content of the nonlinear sigma model includes the map $\phi$ that gives the coordinates of an embedding of the worldsheet $\Sigma$ into the target space $M$,
\begin{equation}
\phi:\Sigma\mapsto M~.
\end{equation}
The path integral of a supersymmetric quantum field theory can be computed from the field configurations where the supersymmetry variations of the fermions vanish. This is the principle of supersymmetric localisation. %The quantum-mechanical path integral is a sum over these configurations of $\ee^{-S}$, with $S$ being the action.

The characterisation of such field configurations is different in the A and B twists, and is covered in the textbook \cite{Hori:2003ic}. The upshot is that in the A model the fermionic variation vanishes for field configurations where the map $\phi$ is holomorphic, i.e.
\begin{equation}
\partial_{\bar{z}}\phi=0
\end{equation}
where $(z,\bar{z})$ are complex coordinates on the Riemann surface $\Sigma$. For such a field configuration, $\phi$ embeds the string \emph{holomorphically} into the target space. Since the closed string worldsheet is a genus-$g$ Riemann surface, the partition function will serve as a generating function for counts of holomorphic embeddings, of various degrees, of genus-$g$ Riemann surfaces in the target space.

On the other hand, in the B-model the path integral localises onto \emph{constant} maps. Very roughly speaking, the space of constant maps to $X$ is $X$, and so the B-model partition function is computed by integrals over the target space, using Hodge theory. We will now give a computationally-oriented account of these identities.

\section{The genus 0 prepotential}\label{sect:Miroirs_genus0}
We will denote $m=h^{2,1}(X)=h^{1,1}(Y)$. In terms of the Frobenius periods $\varpi$ as detailed in \eqref{eq:varpiother}, the B-model prepotential at genus 0 is
\begin{equation}\label{eq:B_FE_0}
\cF^{(0)}(\mathbf{\varphi})=\frac{1}{2}\left(-\varpi_{0}\varpi_{3}+\sum_{i=1}^{m}\varpi_{1,i}\varpi_{2,i}\right)~,
\end{equation}
with each of the periods a function of the complex structure variables $\mathbf{\varphi}$. %Note that the B-model free energy is a section, undergoing changes in scale when K{\"a}hler transformations are made, and the above formula is only one representative among in the orbit of these scale transformations. 

The A-model prepotential at genus 0 admits the expansion
\begin{equation}\label{eq:A_FE_0}
F^{(0)}=\frac{1}{6}Y_{ijk}t^{i}t^{j}t^{k}+\frac{1}{(2\pi\ii)^{3}}\sum_{\mathbf{b}\geq0}n^{(0)}_{\mathbf{b}}\text{Li}_{3}\left(\ee^{2\pi\ii\,\mathbf{b}\cdot\mathbf{t}}\right)~,
\end{equation}
with the $Y_{ijk}$ being the triple-intersection numbers on the manifold $Y$~. The quantity $\mathbf{b}\cdot\mathbf{t}$ should be understood as the integral of the K{\"a}hler form over a two-cycle which, in the basis dual to canonical generators of the K{\"a}hler cone, has nonnegative components $\mathbf{b}$. The summation over $\mathbf{b}\in\IZ^{h^{1,1}(Y)}_{\geq0}$ gives a sum over all possible degree vectors that a rational curve embedded in $Y$ can have (neglecting cases with torsion classes), and $\mathbf{b}\cdot\mathbf{t}$ is the area of such a curve.  

Since $\cF^{(0)}$ is given in terms of solutions to differential equations (which we can find as explicit power series), the above pair of formulae provide the means of determining the genus 0 instanton numbers (integers also nomenclated genus 0 Gopakumar-Vafa invariants) $n^{(0)}_{\beta}$~of~$Y$. It only remains to find a suitable transformation of coordinates 
\begin{equation}
\bm{\varphi}\to\mathbf{t}
\end{equation}
and make a compatible choice of gauge. 

The complex structure moduli space is itself K{\"a}hler, with K{\"a}hler potential $K=-\log\left(\ii\overline{\Pi}^{T}\Sigma\Pi\right)$ and metric $G_{ij}=\partial_{\varphi^{i}}\partial_{\varphi^{j}}K$. The necessary choice of coordinates is the flat one, in which the Christoffel symbols for the metric $G$ vanish. This is
\begin{equation}\label{eq:mirror_map1}
t^{i}=\frac{1}{2\pi\ii}\frac{\varpi_{1,i}(\bm{\varphi)}}{\varpi_{0}(\bm{\varphi})}~.
\end{equation}
If $\varphi_{i}$ is moved in a circle about the MUM point $0$, we get a monodromy transformation $t^{i}\mapsto t^{i}+1$. This is a symmetry of the theory, and leaves the free energy $F^{(0)}$ invariant when the symplectic transformations of the $Y_{ijk}$ are taken into account (this amounts to a change of basis for $H_{2}(Y,\IZ)$). A convenient repackaging of the $\mathbf{t}$ in a new coordinate $\mathbf{q}$ makes this manifest:
\begin{equation}
q^{i}=\exp\left(2\pi\ii \,t^{i}\right)\equiv \exp\left(\frac{\varpi_{1,i}(\bm{\varphi)}}{\varpi_{0}(\bm{\varphi})}\right)~.
\end{equation}

The appropriate change of scale is to multiply \eqref{eq:B_FE_0} by $\frac{1}{(2\pi\ii)^{3}\varpi_{0}^{2}}$, and so the numbers $n^{(0)}_{\mathbf{b}}$ are fixed by
\begin{equation}\label{eq:genus_0_prescription}
\begin{aligned}
(\2\pi\ii)^{3}\,F^{(0)}(\mathbf{q})&=\frac{1}{2\varpi_{0}^{2}}\left(-\varpi_{0}\varpi_{3}+\sum_{i=1}^{m}\varpi_{1,i}\varpi_{2,i}\right)\bigg\vert_{\mathbf{\varphi}=\mathbf{\varphi}(\mathbf{q})}\\[5pt]&=\frac{1}{6}Y_{ijk}\log(q^{i})\log(q^{j})\log(q^{k})+\sum_{\mathbf{b}\geq0}n^{(0)}_{\mathbf{b}}\,\text{Li}_{3}\left(\mathbf{q}^{\mathbf{b}}\right)~.
\end{aligned}
\end{equation}
By $\mathbf{q}^{\mathbf{b}}$ we mean the product $\prod_{i}\left(q^{i}\right)^{b_{i}}$. This form of $F^{(0)}(\mathbf{q})$ holds for manifolds whose homology groups lack torsion. Manifolds with torsion homology classes were considered in \cite{Braun:2008sf,Braun:2007xh,Braun:2007vy}~. The problem of efficiently performing this computation was addressed first in the package INSTANTON \cite{Klemm:2004km}, and more recently in the library CYtools \cite{Demirtas:2022hqf,Demirtas:2023als}.

\subsection*{Performing the computation efficiently}

Let us make three points on the efficient computation of the genus 0 invariants. 

One is that instead of replacing $\mathbf{\varphi}$ with the $q$-series $\varphi(\mathbf{q})$ in the first line of \eqref{eq:genus_0_prescription}, the appearances of $q^{i}$ in the second line can be replaced with $\mathbf{\varphi}$-series. This means that, if only genus 0 numbers are sought, the inverse mirror map $\mathbf{\varphi}(\mathbf{q})$ does not need to be computed.

Secondly, if $\mathbf{q}$ is replaced with $\mathbf{q}(\mathbf{\varphi})$ in the bottom line of \eqref{eq:genus_0_prescription}, the $\log(q^{i})$ terms can be expanded and one finds (as necessitated) that the terms proportional to $\log(\varphi^{i})$ in the top line (coming from the logarithmic parts of the Frobenius periods) can be subtracted from both sides, so we need only work with the power series that appear in each Frobenius period. Said another way, we can set up the computation so that the logarithmic terms in each $\varpi$ are removed from the start. 

Finally, we draw attention to the use of symmetric polynomials in suitable examples. Computing the instanton numbers up to degree vectors $\mathbf{b}$ with total $n$ involves finding the $n^{th}$ order Taylor series of the prepotential, which can be found from the $n^{th}$ order series parts of each Frobenius period. For one-parameter manifolds the number of terms in this series is~$n$. If the number of parameters is greater than 1 ($h^{1,1}(Y)\geq 2$), then these truncated Taylor series become very large much faster as $n$ is increased, with a number of new terms at order $n$ equal to the number of partitions of $n$ with length $h^{1,1}(Y)$. The process of multiplying the large intermediate polynomials is computationally expensive, and prevents one from straightforwardly computing instanton numbers to very high degrees for multiparameter manifolds.

This latter problem, of an exponential growth in the number of terms in each power series, can to some extent be mitigated in examples with a symmetry in their parameters. Suppose three (or more) of the parameters enter on symmetric footing, for example the case where $Y$ is a CICY with three identical rows. Then $\varpi_{0}$ and $F^{(0)}$ are symmetric functions of the three parameters. 

A symmetric polynomial in $n$ variables $a_{i}$ can be uniquely expressed in terms of the elementary symmetric polynomials
\begin{equation}\label{eq:elem_sym_pol}
A_{i}=\sum_{\substack{L\text{ a length-$i$}\\ \text{subset of \{1,...,n\}}}}\;\prod_{k\in L}a_{k}~,
\end{equation}
with a convention $A_{0}=1$.

The utility here is that the number of distinct new monomials in the polynomials $A_{i}$ at each degree is smaller than the number of new monomials in the $a_{i}$, because $A_{i}$ has degree $i$. The asymptotic growth is still exponential, but with a smaller base.

Now, a number of the functions involved in the genus 0 computation will not be symmetric in these variables. However, if the coordinates $\varphi^{j},\,\varphi^{k}$ enter the problem symmetrically then $\varpi_{1,k},\,\varpi_{2,k}$ can be obtained from $\varpi_{1,j},\,\varpi_{2,j}$ by effecting a swap $\varphi^{j}\leftrightarrow\varphi^{k}$. So, the number of computations to perform has decreased.

Moreover, identities between the $A_{i}$ and $a_{i}$ allow for further simplification of the less symmetric functions like $\varpi_{1,i}$. We have that, for each $k$,
\begin{equation}
\sum_{i=1}^{m}(-a_{k})^{i}A_{m-i}=0~.
\end{equation}
This means that, for the power series of a function like $\varpi_{1,k}$ where $a_{k}$ breaks the symmetry, we can uniquely express a truncation to a finite degree as a polynomial in the $A_{i}$ and $a_{k}$ with degree less than $m$ in $a_{k}$. This decreases the number of terms in the expression.

Using these ideas, and the ARC supercomputing resources \cite{richards_2015_22558}, it was possible in \cite{Candelas:2021lkc} to compute genus 0 instanton numbers up to degree 29 for the five-parameter Mirror Hulek-Verrill model
\begin{equation}\label{eq:CICY_MHV}
\cicy{\IP^{1}\\\IP^{1}\\\IP^{1}\\\IP^{1}\\\IP^{1}}{1&1\\1&1\\1&1\\1&1\\1&1}~,
\end{equation}
which possesses an $S_{5}$ symmetry corresponding to permuting rows of this matrix.
 
To illustrate our claims anew here, we present in \tref{tab:Instanton_Numbers_6_36} genus 0 numbers for the family\footnote{We will, as and when useful in this section but not throughout the thesis, write CICY matrices with a superscript displaying the Hodge numbers $(h^{1,1},h^{2,1})$ and a subscript displaying the Euler characteristic.}
\begin{equation}\label{eq:CICY_6_36}
\cicy{\IP^{1}\\\IP^{1}\\\IP^{1}\\\IP^{1}\\\IP^{1}\\\IP^{4}}{1&0&0&0&0&1\\0&1&0&0&0&1\\0&0&1&0&0&1\\0&0&0&1&0&1\\0&0&0&0&1&1\\1&1&1&1&1&0}^{(6,36)}_{-60}~,
\end{equation}
which similarly possesses an $S_{5}$ symmetry originating in the freedom to permute the first five rows. These numbers were computed on a 6-CPU desktop machine with 16GB of RAM, and so we do not reach so high a degree as was done for the five parameter model \eqref{eq:CICY_MHV} in~\cite{Candelas:2021lkc}. 

More can be done, and we also present in \tref{tab:Instanton_Numbers_7_27} the genus 0 numbers up to degree 13 for the family
\begin{equation}\label{eq:CICY_7_27}
\cicy{\IP^{1}\\\IP^{1}\\\IP^{1}\\\IP^{1}\\\IP^{1}\\\IP^{4}\\\IP^{4}}{1&0&0&0&0&1&0&0&0&0\\0&1&0&0&0&0&1&0&0&0\\0&0&1&0&0&0&0&1&0&0\\0&0&0&1&0&0&0&0&1&0\\0&0&0&0&1&0&0&0&0&1\\1&1&1&1&1&0&0&0&0&0\\0&0&0&0&0&1&1&1&1&1}^{(7,27)}_{-40}~.
\end{equation}

Another nice example is found in the family of tetraquadrics
\begin{equation}\label{eq:CICY_4_68}
\cicy{\IP^{1}\\\IP^{1}\\\IP^{1}\\\IP^{1}}{2\\2\\2\\2}^{(4,68)}_{-128}~.
\end{equation}
We present the genus 0 numbers up to degree 20 for this model in \tref{tab:Instanton_Numbers_4_68}.

We conclude this run of examples with a model previously studied by Hosono and Takagi, and for which we will have more to say in our later discussions on higher genus invariants and modularity. This is the 2-parameter family
\begin{equation}\label{eq:CICY_2_52}
\cicy{\IP^{4}\\\IP^{4}}{1&1&1&1&1\\1&1&1&1&1}^{(2,52)}_{-100}~.
\end{equation}
Genus 0 invariants up to total degree 18 were already given in \cite{hosono2014mirror}. We provide them to degree 37 in \tref{tab:Instanton_Numbers_2_52}.

One motivation for studying these families is their presence in \cite{Candelas:2008wb}, along with the mirror Hulek-Verrill manifold. These geometries are related by splittings, and all possess certain symmetries that allow for quotient constructions with small Hodge numbers. 

The content of the following tables is available in electronic form \cite{mcgovern2023a}.

\begin{table}[H]
\centering
\def\arraystretch{1.05}
\begin{adjustbox}{width=\columnwidth,center}
\tiny{\pgfplotstabletypeset[
col sep=space,
white space chars={]},
ignore chars={\ },
every head row/.style={before row=\hline,after row=\hline\hline},
every first row/.style={before row=\vrule height12pt width0pt depth0pt},
every last row/.style={before row=\vrule height0pt width0pt depth6pt,after row=\hline},
%after row=\hline,    %  Uncomment this to get back lines between all rows.
columns={A,B,A,B,A,B,A,B},
display columns/0/.style={select equal part entry of={0}{4},column type = {|l},column name=\vrule height12pt width0pt depth6pt \hfil $\bm p$},
display columns/1/.style={select equal part entry of={0}{4},column type = {|l|},column name=\hfil $n^{(0)}_{\bm p}$},
display columns/2/.style={select equal part entry of={1}{4},column type = {|l},column name=\hfil $\bm p$},
display columns/3/.style={select equal part entry of={1}{4},column type = {|l|},column name=\hfil $n^{(0)}_{\bm p}$},
display columns/4/.style={select equal part entry of={2}{4},column type = {|l},column name=\hfil $\bm p$},
display columns/5/.style={select equal part entry of={2}{4},column type = {|l|},column name=\hfil $n^{(0)}_{\bm p}$},
display columns/6/.style={select equal part entry of={3}{4},column type = {|l},column name=\hfil $\bm p$},
display columns/7/.style={select equal part entry of={3}{4},column type = {|l|},column name=\hfil $n^{(0)}_{\bm p}$},
string type]
{
A ] B

{0, 0, 0, 0, 0, 1}]10
{0, 0, 0, 0, 1, 0}]12
{0, 0, 0, 0, 1, 1}]12
{0, 0, 0, 1, 1, 0}]1
{0, 0, 0, 1, 1, 1}]22
{0, 0, 0, 1, 1, 2}]1
{0, 0, 1, 1, 1, 1}]56
{0, 0, 1, 1, 1, 2}]56
{0, 0, 1, 1, 2, 1}]1
{0, 0, 1, 1, 2, 2}]22
{0, 0, 1, 1, 2, 3}]1
{0, 0, 1, 2, 2, 2}]12
{0, 0, 1, 2, 2, 3}]12
{0, 0, 2, 2, 2, 2}]10
{0, 0, 2, 2, 2, 3}]60
{0, 0, 2, 2, 2, 4}]10
{0, 0, 2, 2, 3, 3}]12
{0, 0, 2, 2, 3, 4}]12
{0, 0, 2, 3, 3, 3}]1
{0, 0, 2, 3, 3, 4}]22
{0, 0, 2, 3, 3, 5}]1
{0, 0, 3, 3, 3, 4}]56
{0, 0, 3, 3, 3, 5}]56
{0, 0, 3, 3, 4, 4}]1
{0, 0, 3, 3, 4, 5}]22
{0, 1, 1, 1, 1, 1}]174
{0, 1, 1, 1, 1, 2}]756
{0, 1, 1, 1, 1, 3}]174
{0, 1, 1, 1, 2, 1}]12
{0, 1, 1, 1, 2, 2}]540
{0, 1, 1, 1, 2, 3}]540
{0, 1, 1, 1, 2, 4}]12
{0, 1, 1, 1, 3, 1}]-2
{0, 1, 1, 1, 3, 2}]10
{0, 1, 1, 1, 3, 3}]96
{0, 1, 1, 1, 3, 4}]10
{0, 1, 1, 1, 3, 5}]-2
{0, 1, 1, 2, 2, 2}]474
{0, 1, 1, 2, 2, 3}]1852
{0, 1, 1, 2, 2, 4}]474
{0, 1, 1, 2, 3, 2}]12
{0, 1, 1, 2, 3, 3}]540
{0, 1, 1, 2, 3, 4}]540
{0, 1, 1, 2, 3, 5}]12
{0, 1, 1, 2, 4, 3}]1
{0, 1, 1, 2, 4, 4}]22
{0, 1, 1, 2, 4, 5}]1
{0, 1, 1, 3, 3, 3}]174
{0, 1, 1, 3, 3, 4}]756
{0, 1, 1, 3, 3, 5}]174
{0, 1, 1, 3, 4, 4}]56
{0, 1, 1, 3, 4, 5}]56
{0, 1, 1, 4, 4, 4}]1
{0, 1, 1, 4, 4, 5}]22
{0, 1, 2, 2, 2, 2}]540
{0, 1, 2, 2, 2, 3}]6708
{0, 1, 2, 2, 2, 4}]6708
{0, 1, 2, 2, 2, 5}]540
{0, 1, 2, 2, 3, 2}]22
{0, 1, 2, 2, 3, 3}]2796
{0, 1, 2, 2, 3, 4}]8860
{0, 1, 2, 2, 3, 5}]2796
{0, 1, 2, 2, 3, 6}]22
{0, 1, 2, 2, 4, 3}]56
{0, 1, 2, 2, 4, 4}]1344
{0, 1, 2, 2, 4, 5}]1344
{0, 1, 2, 2, 4, 6}]56
{0, 1, 2, 2, 5, 4}]1
{0, 1, 2, 2, 5, 5}]22
{0, 1, 2, 3, 3, 3}]1344
{0, 1, 2, 3, 3, 4}]13968
{0, 1, 2, 3, 3, 5}]13968
{0, 1, 2, 3, 3, 6}]1344
{0, 1, 2, 3, 4, 3}]22
{0, 1, 2, 3, 4, 4}]2796
{0, 1, 2, 3, 4, 5}]8860
{0, 1, 2, 3, 5, 4}]12
{0, 1, 2, 4, 4, 4}]540
{0, 1, 3, 3, 3, 3}]798
{0, 1, 3, 3, 3, 4}]25662
{0, 1, 3, 3, 3, 5}]69528
{0, 1, 3, 3, 4, 3}]12
{0, 1, 3, 3, 4, 4}]6708
{0, 2, 2, 2, 2, 2}]756
{0, 2, 2, 2, 2, 3}]25662
{0, 2, 2, 2, 2, 4}]69516
{0, 2, 2, 2, 2, 5}]25662
{0, 2, 2, 2, 2, 6}]756
{0, 2, 2, 2, 3, 2}]56
{0, 2, 2, 2, 3, 3}]13968
{0, 2, 2, 2, 3, 4}]103000
{0, 2, 2, 2, 3, 5}]103000
{0, 2, 2, 2, 3, 6}]13968
{0, 2, 2, 2, 4, 3}]756
{0, 2, 2, 2, 4, 4}]25662
{0, 2, 2, 2, 4, 5}]69516
{0, 2, 2, 2, 5, 4}]540
{0, 2, 2, 3, 3, 2}]1
{0, 2, 2, 3, 3, 3}]8860
{0, 2, 2, 3, 3, 4}]176251
{0, 2, 2, 3, 3, 5}]425712
{0, 2, 2, 3, 4, 3}]540
{0, 2, 2, 3, 4, 4}]54504
{0, 2, 2, 4, 4, 3}]10
{0, 2, 3, 3, 3, 3}]6708
{0, 2, 3, 3, 3, 4}]342924
{0, 2, 3, 3, 4, 3}]474
{0, 3, 3, 3, 3, 3}]6204
{1, 1, 1, 1, 1, 1}]700
{1, 1, 1, 1, 1, 2}]8900
{1, 1, 1, 1, 1, 3}]8900
{1, 1, 1, 1, 1, 4}]700
{1, 1, 1, 1, 2, 1}]66
{1, 1, 1, 1, 2, 2}]8892
{1, 1, 1, 1, 2, 3}]27492
{1, 1, 1, 1, 2, 4}]8892
{1, 1, 1, 1, 2, 5}]66
{1, 1, 1, 1, 3, 1}]-24
{1, 1, 1, 1, 3, 2}]180
{1, 1, 1, 1, 3, 3}]9444
{1, 1, 1, 1, 3, 4}]9444
{1, 1, 1, 1, 3, 5}]180
{1, 1, 1, 1, 3, 6}]-24
{1, 1, 1, 1, 4, 1}]3
{1, 1, 1, 1, 4, 2}]-22
{1, 1, 1, 1, 4, 3}]59
{1, 1, 1, 1, 4, 4}]1024
{1, 1, 1, 1, 4, 5}]59
{1, 1, 1, 1, 4, 6}]-22
{1, 1, 1, 1, 4, 7}]3
{1, 1, 1, 2, 2, 2}]10792
{1, 1, 1, 2, 2, 3}]95648
{1, 1, 1, 2, 2, 4}]95648
{1, 1, 1, 2, 2, 5}]10792
{1, 1, 1, 2, 3, 2}]276
{1, 1, 1, 2, 3, 3}]43326
{1, 1, 1, 2, 3, 4}]125676
{1, 1, 1, 2, 3, 5}]43326
{1, 1, 1, 2, 3, 6}]276
{1, 1, 1, 2, 4, 2}]-48
{1, 1, 1, 2, 4, 3}]432
{1, 1, 1, 2, 4, 4}]22320
{1, 1, 1, 2, 4, 5}]22320
{1, 1, 1, 2, 4, 6}]432
{1, 1, 1, 2, 5, 2}]3
{1, 1, 1, 2, 5, 3}]-22
{1, 1, 1, 2, 5, 4}]59
{1, 1, 1, 2, 5, 5}]1024
{1, 1, 1, 3, 3, 3}]22680
{1, 1, 1, 3, 3, 4}]194664
{1, 1, 1, 3, 3, 5}]194664
{1, 1, 1, 3, 3, 6}]22680
{1, 1, 1, 3, 4, 3}]276
{1, 1, 1, 3, 4, 4}]43326
{1, 1, 1, 3, 4, 5}]125676
{1, 1, 1, 3, 5, 3}]-24
{1, 1, 1, 3, 5, 4}]180
{1, 1, 1, 4, 4, 4}]10792
{1, 1, 2, 2, 2, 2}]15902
{1, 1, 2, 2, 2, 3}]369646
{1, 1, 2, 2, 2, 4}]920760
{1, 1, 2, 2, 2, 5}]369646
{1, 1, 2, 2, 2, 6}]15902
{1, 1, 2, 2, 3, 2}]596
{1, 1, 2, 2, 3, 3}]210748
{1, 1, 2, 2, 3, 4}]1399712
{1, 1, 2, 2, 3, 5}]1399712
{1, 1, 2, 2, 3, 6}]210748
{1, 1, 2, 2, 4, 2}]-176
{1, 1, 2, 2, 4, 3}]3558
{1, 1, 2, 2, 4, 4}]360222
{1, 1, 2, 2, 4, 5}]964648
{1, 1, 2, 2, 5, 2}]36
{1, 1, 2, 2, 5, 3}]-400
{1, 1, 2, 2, 5, 4}]2560
{1, 1, 2, 2, 6, 2}]-4
{1, 1, 2, 2, 6, 3}]30
{1, 1, 2, 3, 3, 2}]2
{1, 1, 2, 3, 3, 3}]139918
{1, 1, 2, 3, 3, 4}]2449206
{1, 1, 2, 3, 3, 5}]5705460
{1, 1, 2, 3, 4, 3}]2676
{1, 1, 2, 3, 4, 4}]766084
{1, 1, 2, 3, 5, 3}]-376
{1, 1, 2, 4, 4, 3}]20
{1, 1, 3, 3, 3, 3}]112196
{1, 1, 3, 3, 3, 4}]4878540
{1, 1, 3, 3, 4, 3}]2775
{1, 2, 2, 2, 2, 2}]27600
{1, 2, 2, 2, 2, 3}]1585020
{1, 2, 2, 2, 2, 4}]8537376
{1, 2, 2, 2, 2, 5}]8537376
{1, 2, 2, 2, 2, 6}]1585020
{1, 2, 2, 2, 3, 2}]1344
{1, 2, 2, 2, 3, 3}]1097052
{1, 2, 2, 2, 3, 4}]14667648
{1, 2, 2, 2, 3, 5}]32044488
{1, 2, 2, 2, 4, 2}]-552
{1, 2, 2, 2, 4, 3}]26364
{1, 2, 2, 2, 4, 4}]4907940
{1, 2, 2, 2, 5, 2}]198
{1, 2, 2, 2, 5, 3}]-3960
{1, 2, 2, 2, 6, 2}]-48
{1, 2, 2, 3, 3, 2}]24
{1, 2, 2, 3, 3, 3}]878696
{1, 2, 2, 3, 3, 4}]28577520
{1, 2, 2, 3, 4, 2}]-2
{1, 2, 2, 3, 4, 3}]24120
{1, 2, 3, 3, 3, 3}]823008
{2, 2, 2, 2, 2, 2}]52740
{2, 2, 2, 2, 2, 3}]7545520
{2, 2, 2, 2, 2, 4}]80109420
{2, 2, 2, 2, 2, 5}]166265920
{2, 2, 2, 2, 3, 2}]2928
{2, 2, 2, 2, 3, 3}]6157524
{2, 2, 2, 2, 3, 4}]154063572
{2, 2, 2, 2, 4, 2}]-1338
{2, 2, 2, 2, 4, 3}]178908
{2, 2, 2, 2, 5, 2}]660
{2, 2, 2, 3, 3, 2}]142
{2, 2, 2, 3, 3, 3}]5734779
{2, 2, 2, 3, 4, 2}]-24

}
}
\end{adjustbox}
\vskip10pt
\capt{6.5in}{tab:Instanton_Numbers_6_36}{The genus~0 instanton numbers of total degree $\leqslant 15$ for the family \eqref{eq:CICY_6_36}. The numbers not in this list are either zero, or given by those in the table after a permutation of the first five indices. The sixth index \textbf{cannot} be exchanged with the others.}
\end{table}

\begin{table}[H]
\centering
\def\arraystretch{1.1}
\begin{adjustbox}{width=\columnwidth,center}
\tiny{\pgfplotstabletypeset[
col sep=space,
white space chars={]},
ignore chars={\ },
every head row/.style={before row=\hline,after row=\hline\hline},
every first row/.style={before row=\vrule height12pt width0pt depth0pt},
every last row/.style={before row=\vrule height0pt width0pt depth6pt,after row=\hline},
%after row=\hline,    %  Uncomment this to get back lines between all rows.
columns={A,B,A,B,A,B},
display columns/0/.style={select equal part entry of={0}{3},column type = {|l},column name=\vrule height12pt width0pt depth6pt \hfil $\bm p$},
display columns/1/.style={select equal part entry of={0}{3},column type = {|l|},column name=\hfil $n^{(0)}_{\bm p}$},
display columns/2/.style={select equal part entry of={1}{3},column type = {|l},column name=\hfil $\bm p$},
display columns/3/.style={select equal part entry of={1}{3},column type = {|l|},column name=\hfil $n^{(0)}_{\bm p}$},
display columns/4/.style={select equal part entry of={2}{3},column type = {|l},column name=\hfil $\bm p$},
display columns/5/.style={select equal part entry of={2}{3},column type = {|l|},column name=\hfil $n^{(0)}_{\bm p}$},
string type]
{
A ] B

{0, 0, 0, 0, 0, 0, 1}]10
{0, 0, 0, 0, 1, 0, 0}]6
{0, 0, 0, 0, 1, 0, 1}]6
{0, 0, 0, 0, 1, 1, 1}]6
{0, 0, 0, 1, 1, 0, 1}]1
{0, 0, 0, 1, 1, 1, 1}]20
{0, 0, 0, 1, 1, 1, 2}]1
{0, 0, 1, 1, 1, 1, 1}]28
{0, 0, 1, 1, 1, 1, 2}]28
{0, 0, 1, 1, 1, 2, 2}]28
{0, 0, 1, 1, 2, 1, 2}]1
{0, 0, 1, 1, 2, 2, 2}]20
{0, 0, 1, 1, 2, 2, 3}]1
{0, 0, 1, 2, 2, 2, 2}]6
{0, 0, 1, 2, 2, 2, 3}]6
{0, 0, 1, 2, 2, 3, 3}]6
{0, 0, 2, 2, 2, 2, 3}]10
{0, 0, 2, 2, 2, 3, 3}]40
{0, 0, 2, 2, 2, 3, 4}]10
{0, 0, 2, 2, 3, 3, 3}]6
{0, 1, 1, 1, 1, 1, 1}]24
{0, 1, 1, 1, 1, 1, 2}]126
{0, 1, 1, 1, 1, 1, 3}]24
{0, 1, 1, 1, 1, 2, 2}]504
{0, 1, 1, 1, 1, 2, 3}]126
{0, 1, 1, 1, 1, 3, 3}]24
{0, 1, 1, 1, 2, 1, 2}]6
{0, 1, 1, 1, 2, 1, 3}]6
{0, 1, 1, 1, 2, 2, 2}]264
{0, 1, 1, 1, 2, 2, 3}]264
{0, 1, 1, 1, 2, 2, 4}]6
{0, 1, 1, 1, 2, 3, 3}]264
{0, 1, 1, 1, 2, 3, 4}]6
{0, 1, 1, 1, 3, 1, 3}]-2
{0, 1, 1, 1, 3, 2, 3}]10
{0, 1, 1, 1, 3, 3, 3}]80
{0, 1, 1, 1, 3, 3, 4}]10
{0, 1, 1, 2, 2, 2, 2}]72
{0, 1, 1, 2, 2, 2, 3}]330
{0, 1, 1, 2, 2, 2, 4}]72
{0, 1, 1, 2, 2, 3, 3}]1192
{0, 1, 1, 2, 2, 3, 4}]330
{0, 1, 1, 2, 3, 2, 3}]6
{0, 1, 1, 2, 3, 2, 4}]6
{0, 1, 1, 2, 3, 3, 3}]264
{0, 1, 2, 2, 2, 2, 2}]6
{0, 1, 2, 2, 2, 2, 3}]264
{0, 1, 2, 2, 2, 2, 4}]264
{0, 1, 2, 2, 2, 3, 3}]3090
{0, 1, 2, 2, 3, 2, 3}]1
{0, 2, 2, 2, 2, 2, 3}]126
{1, 1, 1, 1, 1, 1, 1}]20
{1, 1, 1, 1, 1, 1, 2}]330
{1, 1, 1, 1, 1, 1, 3}]330
{1, 1, 1, 1, 1, 1, 4}]20
{1, 1, 1, 1, 1, 2, 2}]4120
{1, 1, 1, 1, 1, 2, 3}]4120
{1, 1, 1, 1, 1, 2, 4}]330
{1, 1, 1, 1, 1, 3, 3}]4120
{1, 1, 1, 1, 1, 3, 4}]330
{1, 1, 1, 1, 1, 4, 4}]20
{1, 1, 1, 1, 2, 1, 2}]15
{1, 1, 1, 1, 2, 1, 3}]36
{1, 1, 1, 1, 2, 1, 4}]15
{1, 1, 1, 1, 2, 2, 2}]1692
{1, 1, 1, 1, 2, 2, 3}]5478
{1, 1, 1, 1, 2, 2, 4}]1692
{1, 1, 1, 1, 2, 2, 5}]15
{1, 1, 1, 1, 2, 3, 3}]16464
{1, 1, 1, 1, 2, 3, 4}]5478
{1, 1, 1, 1, 3, 1, 3}]-12
{1, 1, 1, 1, 3, 1, 4}]-12
{1, 1, 1, 1, 3, 2, 3}]90
{1, 1, 1, 1, 3, 2, 4}]90
{1, 1, 1, 1, 3, 3, 3}]4644
{1, 1, 1, 1, 4, 1, 4}]3
{1, 1, 1, 2, 2, 2, 2}]480
{1, 1, 1, 2, 2, 2, 3}]4916
{1, 1, 1, 2, 2, 2, 4}]4916
{1, 1, 1, 2, 2, 3, 3}]42908
{1, 1, 1, 2, 3, 2, 3}]54
{1, 1, 2, 2, 2, 2, 2}]80
{1, 1, 2, 2, 2, 2, 3}]3293
{1, 2, 2, 2, 2, 2, 2}]6

}
}
\end{adjustbox}
\vskip10pt
\capt{6.5in}{tab:Instanton_Numbers_7_27}{The genus~0 instanton numbers of total degree $\leqslant 13$ for the family \eqref{eq:CICY_7_27}. The numbers not in this list are either zero, or given by those in the table after permutations of the first five and last two indices. The sixth and seventh indices \textbf{cannot} be exchanged with the first five.}
\end{table}

\begin{table}[H]
\centering
\def\arraystretch{1.005}
\begin{adjustbox}{width=\columnwidth,center}
\tiny{\pgfplotstabletypeset[
col sep=space,
white space chars={]},
ignore chars={\ },
every head row/.style={before row=\hline,after row=\hline\hline},
every first row/.style={before row=\vrule height12pt width0pt depth0pt},
every last row/.style={before row=\vrule height0pt width0pt depth6pt,after row=\hline},
%after row=\hline,    %  Uncomment this to get back lines between all rows.
columns={A,B,A,B,A,B},
display columns/0/.style={select equal part entry of={0}{3},column type = {|l},column name=\vrule height12pt width0pt depth6pt \hfil $\bm p$},
display columns/1/.style={select equal part entry of={0}{3},column type = {|l|},column name=\hfil $n^{(0)}_{\bm p}$},
display columns/2/.style={select equal part entry of={1}{3},column type = {|l},column name=\hfil $\bm p$},
display columns/3/.style={select equal part entry of={1}{3},column type = {|l|},column name=\hfil $n^{(0)}_{\bm p}$},
display columns/4/.style={select equal part entry of={2}{3},column type = {|l},column name=\hfil $\bm p$},
display columns/5/.style={select equal part entry of={2}{3},column type = {|l|},column name=\hfil $n^{(0)}_{\bm p}$},
string type]
{
A ] B
{1, 0, 0, 0}]48
{1, 1, 0, 0}]160
{1, 1, 1, 0}]2432
{1, 1, 1, 1}]86016
{2, 1, 0, 0}]48
{2, 1, 1, 0}]5056
{2, 1, 1, 1}]518784
{2, 2, 0, 0}]128
{2, 2, 1, 0}]34640
{2, 2, 1, 1}]7037184
{2, 2, 2, 0}]507904
{2, 2, 2, 1}]171165840
{2, 2, 2, 2}]6547900416
{3, 1, 1, 0}]2432
{3, 1, 1, 1}]899072
{3, 2, 0, 0}]48
{3, 2, 1, 0}]61824
{3, 2, 1, 1}]28639616
{3, 2, 2, 0}]2089008
{3, 2, 2, 1}]1297013760
{3, 2, 2, 2}]80605022416
{3, 3, 0, 0}]160
{3, 3, 1, 0}]308352
{3, 3, 1, 1}]241754112
{3, 3, 2, 0}]17677056
{3, 3, 2, 1}]17200647552
{3, 3, 2, 2}]1557963029504
{3, 3, 3, 0}]252810752
{3, 3, 3, 1}]352652451840
{3, 3, 3, 2}]43431181213824
{3, 3, 3, 3}]1646607181615104
{4, 1, 1, 0}]160
{4, 1, 1, 1}]518784
{4, 2, 1, 0}]34640
{4, 2, 1, 1}]44662400
{4, 2, 2, 0}]3265280
{4, 2, 2, 1}]4042603552
{4, 2, 2, 2}]426305384448
{4, 3, 0, 0}]48
{4, 3, 1, 0}]508160
{4, 3, 1, 1}]789382400
{4, 3, 2, 0}]57521568
{4, 3, 2, 1}]94371041920
{4, 3, 2, 2}]13002686858864
{4, 3, 3, 0}]1424625024
{4, 3, 3, 1}]3031512413312
{4, 3, 3, 2}]527889850255360
{4, 3, 3, 3}]27526058324060160
{4, 4, 0, 0}]128
{4, 4, 1, 0}]2089008
{4, 4, 1, 1}]4985888864
{4, 4, 2, 0}]360824832
{4, 4, 2, 1}]871452197440
{4, 4, 2, 2}]166647022068736
{4, 4, 3, 0}]13391051328
{4, 4, 3, 1}]39844405855232
{4, 4, 3, 2}]9199812356321968
{4, 4, 3, 3}]627205272329861504
{4, 4, 4, 0}]188051928064
{4, 4, 4, 1}]741058165634496
{4, 4, 4, 2}]217057048330727424
{4, 4, 4, 3}]18647281489625577504
{4, 4, 4, 4}]699794636853614635008
{5, 1, 1, 1}]86016
{5, 2, 1, 0}]5056
{5, 2, 1, 1}]28639616
{5, 2, 2, 0}]2089008
{5, 2, 2, 1}]5834714624
{5, 2, 2, 2}]1108729141920
{5, 3, 1, 0}]308352
{5, 3, 1, 1}]1154723840
{5, 3, 2, 0}]84056832
{5, 3, 2, 1}]249714336000
{5, 3, 2, 2}]54663940002816
{5, 3, 3, 0}]3808483584
{5, 3, 3, 1}]12944311033856
{5, 3, 3, 2}]3295227059205504
{5, 3, 3, 3}]240139807447941120
{5, 4, 0, 0}]48
{5, 4, 1, 0}]3265984
{5, 4, 1, 1}]14129065088
{5, 4, 2, 0}]1017784976
{5, 4, 2, 1}]3874428076928
{5, 4, 2, 2}]1074177329475184
{5, 4, 3, 0}]59821118208
{5, 4, 3, 1}]260287936814720
{5, 4, 3, 2}]82385600163793920
{5, 4, 3, 3}]7471070109015318656
{5, 4, 4, 0}]1268723299376
{5, 4, 4, 1}]6894039650704512
{5, 4, 4, 2}]2643331064630855200
{5, 4, 4, 3}]291051848701788699648
{5, 4, 4, 4}]13862614815470765167056
{5, 5, 0, 0}]160
{5, 5, 1, 0}]11746432
{5, 5, 1, 1}]73320878080
{5, 5, 2, 0}]5243776704
{5, 5, 2, 1}]27965750549248
{5, 5, 2, 2}]10380731579151872
{5, 5, 3, 0}]432161466624
{5, 5, 3, 1}]2545321829650432
{5, 5, 3, 2}]1042877238127097984
{5, 5, 3, 3}]120304302520611954688
{5, 5, 4, 0}]12709250594240
{5, 5, 4, 1}]90222224188616064
{5, 5, 4, 2}]43400958203031979648
{5, 5, 4, 3}]5917889244182186813696
{5, 5, 4, 4}]347399539693965379619648
{5, 5, 5, 0}]176223512332928
{5, 5, 5, 1}]1575672398623395840
{5, 5, 5, 2}]922075476759511908864
{5, 5, 5, 3}]151692725887218685190144
{5, 5, 5, 4}]10726096697730587601447552
{5, 5, 5, 5}]399456970261757309912334336
{6, 1, 1, 1}]2432
{6, 2, 1, 0}]48
{6, 2, 1, 1}]7037184
{6, 2, 2, 0}]507904
{6, 2, 2, 1}]4042603552
{6, 2, 2, 2}]1514352078848
{6, 3, 1, 0}]61824
{6, 3, 1, 1}]789382400
{6, 3, 2, 0}]57521568
{6, 3, 2, 1}]342823329792
{6, 3, 2, 2}]125771915722896
{6, 3, 3, 0}]5243776704
{6, 3, 3, 1}]30008993843584
{6, 3, 3, 2}]11575415612576768
{6, 3, 3, 3}]1203234568961692800
{6, 4, 1, 0}]2089008
{6, 4, 1, 1}]19804670208
{6, 4, 2, 0}]1424616960
{6, 4, 2, 1}]9173232641616
{6, 4, 2, 2}]3846287763468288
{6, 4, 3, 0}]141793830384
{6, 4, 3, 1}]938779827042688
{6, 4, 3, 2}]419700792333851408
{6, 4, 3, 3}]51546803704385918976
{6, 4, 4, 0}]4650991239168
{6, 4, 4, 1}]35938232545114592
{6, 4, 4, 2}]18502170216819978240
{6, 4, 4, 3}]2653048822718080612368
{6, 4, 4, 4}]161629518930438998396928
{6, 5, 0, 0}]48
{6, 5, 1, 0}]17677056
{6, 5, 1, 1}]188011416064
{6, 5, 2, 0}]13391051328
{6, 5, 2, 1}]107087122937856
{6, 5, 2, 2}]55530466229981712
{6, 5, 3, 0}]1651763971584
{6, 5, 3, 1}]13704006681994624
{6, 5, 3, 2}]7493044028521115648
{6, 5, 3, 3}]1120675249090550046720
{6, 5, 4, 0}]69188204803344
{6, 5, 4, 1}]661446106005076224
{6, 5, 4, 2}]409473371380697240000
{6, 5, 4, 3}]70283944963932592401152
{6, 5, 4, 4}]5129348718881933358532960
{6, 5, 5, 0}]1335543626995200
{6, 5, 5, 1}]15470217373819587584
{6, 5, 5, 2}]11284911987863253245440
{6, 5, 5, 3}]2276112606495732006202112
{6, 5, 5, 4}]195588673772290431107673344
{6, 6, 0, 0}]128
{6, 6, 1, 0}]57521568
{6, 6, 1, 1}]845112783616
{6, 6, 2, 0}]59821080576
{6, 6, 2, 1}]645375311627376
{6, 6, 2, 2}]435566317842866176
{6, 6, 3, 0}]9911375239392
{6, 6, 3, 1}]108118212077621248
{6, 6, 3, 2}]74937314945510009952
{6, 6, 3, 3}]13961655375523769282816
{6, 6, 4, 0}]549828820220928
{6, 6, 4, 1}]6736033668350881952
{6, 6, 4, 2}]5160552306598346379264
{6, 6, 4, 3}]1080443741007758375555472
{6, 6, 4, 4}]95469681182710628821921792
{6, 6, 5, 0}]13962620228584512
{6, 6, 5, 1}]201987366061445211648
{6, 6, 5, 2}]178213464042928851755872
{6, 6, 5, 3}]42979893686208656793686016
{6, 6, 6, 0}]191827050005069824
{6, 6, 6, 1}]3375664486259351631456
{6, 6, 6, 2}]3521325683147770386726912
{7, 2, 1, 1}]518784
{7, 2, 2, 0}]34640
{7, 2, 2, 1}]1297013760
{7, 2, 2, 2}]1108729141920
{7, 3, 1, 0}]2432
{7, 3, 1, 1}]241754112
{7, 3, 2, 0}]17677056
{7, 3, 2, 1}]249714336000
{7, 3, 2, 2}]165296580806656
{7, 3, 3, 0}]3808483584
{7, 3, 3, 1}]39529406119936
{7, 3, 3, 2}]24136954745543808
{7, 3, 3, 3}]3672469607846903808
{7, 4, 1, 0}]508160
{7, 4, 1, 1}]14129065088
{7, 4, 2, 0}]1017784976
{7, 4, 2, 1}]12163074768640
{7, 4, 2, 2}]8102781267426048
{7, 4, 3, 0}]188051992832
{7, 4, 3, 1}]1984799528343552
{7, 4, 3, 2}]1294128535920058368
{7, 4, 3, 3}]219681085367035373952
{7, 4, 4, 0}]9911375239392
{7, 4, 4, 1}]112338620588205056
{7, 4, 4, 2}]79594696896349282688
{7, 4, 4, 3}]15110868372965862621696
{7, 4, 4, 4}]1189045428745766422260768
{7, 5, 1, 0}]11746432
{7, 5, 1, 1}]255547666432
{7, 5, 2, 0}]18178255616
{7, 5, 2, 1}]233839970568832
{7, 5, 2, 2}]176514847485662208
{7, 5, 3, 0}]3600925105536
{7, 5, 3, 1}]43715576259518464
{7, 5, 3, 2}]32805238780261668736
{7, 5, 3, 3}]6480417642541472432128
{7, 5, 4, 0}]221844802863872
{7, 5, 4, 1}]2932079676057142912
{7, 5, 4, 2}]2387966888748874810112
{7, 5, 4, 3}]523912487781610422477568
{7, 5, 4, 4}]47981652609161486365088768
{7, 5, 5, 0}]6036801603750144
{7, 5, 5, 1}]92621472022310780928
{7, 5, 5, 2}]85897234681261670146944
{7, 5, 5, 3}]21540883712581909578244096
{7, 6, 0, 0}]48
{7, 6, 1, 0}]84056832
{7, 6, 1, 1}]2010040031104
{7, 6, 2, 0}]141793830384

}
}
\end{adjustbox}
\vskip10pt
\capt{6.5in}{tab:Instanton_Numbers_4_68}{The genus~0 instanton numbers of total degree $\leqslant 20$ for the family \eqref{eq:CICY_4_68}. The numbers not in this list are either zero, or given by those in the table after permuting indices.}
\end{table}

\begin{table}[H]\notag
\centering
\def\arraystretch{1.005}
\begin{adjustbox}{width=\columnwidth,center}
\tiny{\pgfplotstabletypeset[
col sep=space,
white space chars={]},
ignore chars={\ },
every head row/.style={before row=\hline,after row=\hline\hline},
every first row/.style={before row=\vrule height12pt width0pt depth0pt},
every last row/.style={before row=\vrule height0pt width0pt depth6pt,after row=\hline},
%after row=\hline,    %  Uncomment this to get back lines between all rows.
columns={A,B,A,B,A,B},
display columns/0/.style={select equal part entry of={0}{3},column type = {|l},column name=\vrule height12pt width0pt depth6pt \hfil $\bm p$},
display columns/1/.style={select equal part entry of={0}{3},column type = {|l|},column name=\hfil $n^{(0)}_{\bm p}$},
display columns/2/.style={select equal part entry of={1}{3},column type = {|l},column name=\hfil $\bm p$},
display columns/3/.style={select equal part entry of={1}{3},column type = {|l|},column name=\hfil $n^{(0)}_{\bm p}$},
display columns/4/.style={select equal part entry of={2}{3},column type = {|l},column name=\hfil $\bm p$},
display columns/5/.style={select equal part entry of={2}{3},column type = {|l|},column name=\hfil $n^{(0)}_{\bm p}$},
string type]
{
A ] B

{7, 6, 2, 1}]2208595900084224
{7, 6, 2, 2}]2021884379122016048
{7, 6, 3, 0}]33789217322496
{7, 6, 3, 1}]503495581721955456
{7, 6, 3, 2}]455218569908331757568
{7, 6, 3, 3}]107719074363142059583360
{7, 6, 4, 0}]2566483940906640
{7, 6, 4, 1}]41379145350412949504
{7, 6, 4, 2}]40123559676936131346720
{7, 6, 4, 3}]10410193126310745548984320
{7, 6, 5, 0}]87046894784712960
{7, 6, 5, 1}]1607540052866407577728
{7, 6, 5, 2}]1749693070862210564431872
{7, 6, 6, 0}]1576912001012723760
{7, 6, 6, 1}]34462954830661856456704
{7, 7, 0, 0}]160
{7, 7, 1, 0}]252810752
{7, 7, 1, 1}]8087692763136
{7, 7, 2, 0}]567589415680
{7, 7, 2, 1}]11584959011321216
{7, 7, 2, 2}]13491384176760961024
{7, 7, 3, 0}]176223512332928
{7, 7, 3, 1}]3369477175185817600
{7, 7, 3, 2}]3792683314958877584384
{7, 7, 3, 3}]1099277650161884440166400
{7, 7, 4, 0}]17175761255142144
{7, 7, 4, 1}]348295401563531685632
{7, 7, 4, 2}]412395106599944043177984
{7, 7, 5, 0}]740932627227834624
{7, 7, 5, 1}]16875996286670980751360
{7, 7, 6, 0}]16999688846773790208
{8, 2, 1, 1}]5056
{8, 2, 2, 0}]128
{8, 2, 2, 1}]171165840
{8, 2, 2, 2}]426305384448
{8, 3, 1, 1}]28639616
{8, 3, 2, 0}]2089008
{8, 3, 2, 1}]94371041920
{8, 3, 2, 2}]125771915722896
{8, 3, 3, 0}]1424625024
{8, 3, 3, 1}]30008993843584
{8, 3, 3, 2}]30743146145765888
{8, 3, 3, 3}]7079306295276204288
{8, 4, 1, 0}]34640
{8, 4, 1, 1}]4985888864
{8, 4, 2, 0}]360824832
{8, 4, 2, 1}]9173232641616
{8, 4, 2, 2}]10353839703716352
{8, 4, 3, 0}]141793830384
{8, 4, 3, 1}]2538964516666880
{8, 4, 3, 2}]2508525839463392816
{8, 4, 3, 3}]602927912105481193728
{8, 4, 4, 0}]12709250181888
{8, 4, 4, 1}]219342311907269104
{8, 4, 4, 2}]219737543096535650304
{8, 4, 4, 3}]56255611905947864806512
{8, 4, 4, 4}]5785403639953201842686976
{8, 5, 1, 0}]3265984
{8, 5, 1, 1}]188011416064
{8, 5, 2, 0}]13391051328
{8, 5, 2, 1}]302207140853120
{8, 5, 2, 2}]347926332689829040
{8, 5, 3, 0}]4650991497216
{8, 5, 3, 1}]86320414207198848
{8, 5, 3, 2}]91592941667768938752
{8, 5, 3, 3}]24378159414883707592448
{8, 5, 4, 0}]439063136203008
{8, 5, 4, 1}]8247835975441925056
{8, 5, 4, 2}]9033581369124138593840
{8, 5, 4, 3}]2573204027887680476759296
{8, 5, 5, 0}]17175761255142144
{8, 5, 5, 1}]356496506323192211328
{8, 5, 5, 2}]428090722447387987896384
{8, 6, 1, 0}]57521568
{8, 6, 1, 1}]2668638725632
{8, 6, 2, 0}]188051928064
{8, 6, 2, 1}]4533354437642480
{8, 6, 2, 2}]5860852677002366976
{8, 6, 3, 0}]69188204803344
{8, 6, 3, 1}]1462074329960965376
{8, 6, 3, 2}]1771264367116721723008
{8, 6, 3, 3}]542307422558027394152960
{8, 6, 4, 0}]7456149925576704
{8, 6, 4, 1}]162132923614385349360
{8, 6, 4, 2}]203054957441346394726912
{8, 6, 5, 0}]343945562331791376
{8, 6, 5, 1}]8263402711793164324608
{8, 6, 6, 0}]8277365987776740864
{8, 7, 0, 0}]48
{8, 7, 1, 0}]360828928
{8, 7, 1, 1}]18129723322752
{8, 7, 2, 0}]1268723299376
{8, 7, 2, 1}]36298079877646592
{8, 7, 2, 2}]55987725237356145792
{8, 7, 3, 0}]549828822442752
{8, 7, 3, 1}]14005384303898305024
{8, 7, 3, 2}]20171041734451334944000
{8, 7, 4, 0}]71290241748070896
{8, 7, 4, 1}]1864208769268632504064
{8, 7, 5, 0}]3988612138839842816
{8, 8, 0, 0}]128
{8, 8, 1, 0}]1017784976
{8, 8, 1, 1}]66748924807328
{8, 8, 2, 0}]4650991239168
{8, 8, 2, 1}]170418886235419696
{8, 8, 2, 2}]328173467451404489600
{8, 8, 3, 0}]2566483940906640
{8, 8, 3, 1}]82211057906371385344
{8, 8, 4, 0}]417176124089319424
{9, 2, 2, 1}]7037184
{9, 2, 2, 2}]80605022416
{9, 3, 1, 1}]899072
{9, 3, 2, 0}]61824
{9, 3, 2, 1}]17200647552
{9, 3, 2, 2}]54663940002816
{9, 3, 3, 0}]252810752
{9, 3, 3, 1}]12944311033856
{9, 3, 3, 2}]24136954745543808
{9, 3, 3, 3}]8791828001566988288
{9, 4, 1, 0}]160
{9, 4, 1, 1}]789382400
{9, 4, 2, 0}]57521568
{9, 4, 2, 1}]3874428076928
{9, 4, 2, 2}]8102781267426048
{9, 4, 3, 0}]59821118208
{9, 4, 3, 1}]1984799528343552
{9, 4, 3, 2}]3120841582270747392
{9, 4, 3, 3}]1094190143621388343680
{9, 4, 4, 0}]9911375239392
{9, 4, 4, 1}]273504607073902144
{9, 4, 4, 2}]400065629964453039984
{9, 4, 4, 3}]141106319957374906048512
{9, 5, 1, 0}]308352
{9, 5, 1, 1}]73320878080
{9, 5, 2, 0}]5243776704
{9, 5, 2, 1}]233839970568832
{9, 5, 2, 2}]435172594319284224
{9, 5, 3, 0}]3600925105536
{9, 5, 3, 1}]108024514779267072
{9, 5, 3, 2}]167816512674800089472
{9, 5, 3, 3}]61564588462323994361856
{9, 5, 4, 0}]549828822442752
{9, 5, 4, 1}]15172240325024688384
{9, 5, 4, 2}]22894439058514947334656
{9, 5, 5, 0}]31786723861681536
{9, 5, 5, 1}]913426712060194627584
{9, 6, 1, 0}]17677056
{9, 6, 1, 1}]2010040031104
{9, 6, 2, 0}]141793830384
{9, 6, 2, 1}]5744002412760576
{9, 6, 2, 2}]10966073503132221968
{9, 6, 3, 0}]87594475420800
{9, 6, 3, 1}]2738079662657568640
{9, 6, 3, 2}]4572692498129982156800
{9, 6, 4, 0}]13962620228584512
{9, 6, 4, 1}]420332107728557286912
{9, 6, 5, 0}]895919580397273344
{9, 7, 1, 0}]252810752
{9, 7, 1, 1}]23624438407168
{9, 7, 2, 0}]1651763971584
{9, 7, 2, 1}]70930323016569984
{9, 7, 2, 2}]150750588373893541888
{9, 7, 3, 0}]1071736381056384
{9, 7, 3, 1}]37743466871928029184
{9, 7, 4, 0}]191827050036601856
{9, 8, 0, 0}]48
{9, 8, 1, 0}]1424625024
{9, 8, 1, 1}]142582803775232
{9, 8, 2, 0}]9911375239392
{9, 8, 2, 1}]497106580728095936
{9, 8, 3, 0}]7456149932740608
{9, 9, 0, 0}]160
{9, 9, 1, 0}]3808483584
{9, 9, 1, 1}]487860526727168
{9, 9, 2, 0}]33789217322496
{10, 2, 2, 1}]34640
{10, 2, 2, 2}]6547900416
{10, 3, 1, 1}]2432
{10, 3, 2, 0}]48
{10, 3, 2, 1}]1297013760
{10, 3, 2, 2}]13002686858864
{10, 3, 3, 0}]17677056
{10, 3, 3, 1}]3031512413312
{10, 3, 3, 2}]11575415612576768
{10, 3, 3, 3}]7079306295276204288
{10, 4, 1, 1}]44662400
{10, 4, 2, 0}]3265280
{10, 4, 2, 1}]871452197440
{10, 4, 2, 2}]3846287763468288
{10, 4, 3, 0}]13391051328
{10, 4, 3, 1}]938779827042688
{10, 4, 3, 2}]2508525839463392816
{10, 4, 3, 3}]1332539766738727936000
{10, 4, 4, 0}]4650991239168
{10, 4, 4, 1}]219342311907269104
{10, 4, 4, 2}]487713798985291233280
{10, 5, 1, 0}]5056
{10, 5, 1, 1}]14129065088
{10, 5, 2, 0}]1017784976
{10, 5, 2, 1}]107087122937856
{10, 5, 2, 2}]347926332689829040
{10, 5, 3, 0}]1651763971584
{10, 5, 3, 1}]86320414207198848
{10, 5, 3, 2}]204999947699695388672
{10, 5, 4, 0}]439063136203008
{10, 5, 4, 1}]18557339208009894656
{10, 5, 5, 0}]38952597058723520
{10, 6, 1, 0}]2089008
{10, 6, 1, 1}]845112783616
{10, 6, 2, 0}]59821080576
{10, 6, 2, 1}]4533354437642480
{10, 6, 2, 2}]13486794785676976128
{10, 6, 3, 0}]69188204803344
{10, 6, 3, 1}]3368380766765580288
{10, 6, 4, 0}]17175761244601344
{10, 7, 1, 0}]84056832
{10, 7, 1, 1}]18129723322752
{10, 7, 2, 0}]1268723299376
{10, 7, 2, 1}]88462669038163968
{10, 7, 3, 0}]1335543626995200
{10, 8, 1, 0}]1017784976
{10, 8, 1, 1}]182970871579264
{10, 8, 2, 0}]12709250181888
{10, 9, 0, 0}]48
{10, 9, 1, 0}]5243776704
{10, 10, 0, 0}]128
{11, 2, 2, 2}]171165840
{11, 3, 2, 1}]28639616
{11, 3, 2, 2}]1557963029504
{11, 3, 3, 0}]308352
{11, 3, 3, 1}]352652451840
{11, 3, 3, 2}]3295227059205504
{11, 3, 3, 3}]3672469607846903808
{11, 4, 1, 1}]518784
{11, 4, 2, 0}]34640
{11, 4, 2, 1}]94371041920
{11, 4, 2, 2}]1074177329475184

}
}
\end{adjustbox}
\caption*{\tref{tab:Instanton_Numbers_4_68} continued.}
\end{table}

\begin{table}[H]\notag
\centering
\def\arraystretch{1.005}
\begin{adjustbox}{width=\columnwidth,center}
\tiny{\pgfplotstabletypeset[
col sep=space,
white space chars={]},
ignore chars={\ },
every head row/.style={before row=\hline,after row=\hline\hline},
every first row/.style={before row=\vrule height12pt width0pt depth0pt},
every last row/.style={before row=\vrule height0pt width0pt depth6pt,after row=\hline},
%after row=\hline,    %  Uncomment this to get back lines between all rows.
columns={A,B,A,B,A,B,A,B},
display columns/0/.style={select equal part entry of={0}{4},column type = {|l},column name=\vrule height12pt width0pt depth6pt \hfil $\bm p$},
display columns/1/.style={select equal part entry of={0}{4},column type = {|l|},column name=\hfil $n^{(0)}_{\bm p}$},
display columns/2/.style={select equal part entry of={1}{4},column type = {|l},column name=\hfil $\bm p$},
display columns/3/.style={select equal part entry of={1}{4},column type = {|l|},column name=\hfil $n^{(0)}_{\bm p}$},
display columns/4/.style={select equal part entry of={2}{4},column type = {|l},column name=\hfil $\bm p$},
display columns/5/.style={select equal part entry of={2}{4},column type = {|l|},column name=\hfil $n^{(0)}_{\bm p}$},
display columns/6/.style={select equal part entry of={3}{4},column type = {|l},column name=\hfil $\bm p$},
display columns/7/.style={select equal part entry of={3}{4},column type = {|l|},column name=\hfil $n^{(0)}_{\bm p}$},
string type]
{
A ] B

{11, 4, 3, 0}]1424625024
{11, 4, 3, 1}]260287936814720
{11, 4, 3, 2}]1294128535920058368
{11, 4, 4, 0}]1268723299376
{11, 4, 4, 1}]112338620588205056
{11, 5, 1, 1}]1154723840
{11, 5, 2, 0}]84056832
{11, 5, 2, 1}]27965750549248
{11, 5, 2, 2}]176514847485662208
{11, 5, 3, 0}]432161466624
{11, 5, 3, 1}]43715576259518464
{11, 5, 4, 0}]221844802863872
{11, 6, 1, 0}]61824
{11, 6, 1, 1}]188011416064
{11, 6, 2, 0}]13391051328
{11, 6, 2, 1}]2208595900084224
{11, 6, 3, 0}]33789217322496
{11, 7, 1, 0}]11746432
{11, 7, 1, 1}]8087692763136
{11, 7, 2, 0}]567589415680
{11, 8, 1, 0}]360828928
{12, 2, 2, 2}]507904
{12, 3, 2, 1}]61824
{12, 3, 2, 2}]80605022416
{12, 3, 3, 0}]160
{12, 3, 3, 1}]17200647552
{12, 3, 3, 2}]527889850255360
{12, 4, 1, 1}]160
{12, 4, 2, 1}]4042603552
{12, 4, 2, 2}]166647022068736
{12, 4, 3, 0}]57521568
{12, 4, 3, 1}]39844405855232
{12, 4, 4, 0}]188051928064
{12, 5, 1, 1}]28639616
{12, 5, 2, 0}]2089008
{12, 5, 2, 1}]3874428076928
{12, 5, 3, 0}]59821118208
{12, 6, 1, 0}]48
{12, 6, 1, 1}]19804670208
{12, 6, 2, 0}]1424616960
{12, 7, 1, 0}]508160
{13, 3, 2, 2}]1297013760
{13, 3, 3, 1}]241754112
{13, 4, 2, 1}]44662400
{13, 4, 3, 0}]508160
{13, 5, 1, 1}]86016
{13, 5, 2, 0}]5056

}
}
\end{adjustbox}
\caption*{\tref{tab:Instanton_Numbers_4_68} continued.}
\end{table}

\begin{table}[H]
\centering
\def\arraystretch{1.15}
\begin{adjustbox}{width=\columnwidth,center}
\tiny{\pgfplotstabletypeset[
col sep=space,
white space chars={]},
ignore chars={\ },
every head row/.style={before row=\hline,after row=\hline\hline},
every first row/.style={before row=\vrule height12pt width0pt depth0pt},
every last row/.style={before row=\vrule height0pt width0pt depth6pt,after row=\hline},
%after row=\hline,    %  Uncomment this to get back lines between all rows.
columns={A,B,A,B,A,B},
display columns/0/.style={select equal part entry of={0}{3},column type = {|l},column name=\vrule height12pt width0pt depth6pt \hfil $\bm p$},
display columns/1/.style={select equal part entry of={0}{3},column type = {|l|},column name=\hfil $n^{(0)}_{\bm p}$},
display columns/2/.style={select equal part entry of={1}{3},column type = {|l},column name=\hfil $\bm p$},
display columns/3/.style={select equal part entry of={1}{3},column type = {|l|},column name=\hfil $n^{(0)}_{\bm p}$},
display columns/4/.style={select equal part entry of={2}{3},column type = {|l},column name=\hfil $\bm p$},
display columns/5/.style={select equal part entry of={2}{3},column type = {|l|},column name=\hfil $n^{(0)}_{\bm p}$},
string type]
{
A ] B
{1, 0}]50
{1, 1}]650
{2, 1}]1475
{2, 2}]29350
{3, 1}]650
{3, 2}]148525
{4, 1}]50
{3, 3}]3270050
{4, 2}]250550
{4, 3}]24162125
{5, 2}]148525
{4, 4}]545403950
{5, 3}]75885200
{6, 2}]29350
{5, 4}]5048036025
{6, 3}]110273275
{7, 2}]1475
{5, 5}]114678709000
{6, 4}]22945154050
{7, 3}]75885200
{6, 5}]1231494256550
{7, 4}]55531376500
{8, 3}]24162125
{6, 6}]27995704239850
{7, 5}]7175800860250
{8, 4}]74278763500
{9, 3}]3270050
{7, 6}]334030085380350
{8, 5}]24352783493100
{9, 4}]55531376500
{10, 3}]148525
{7, 7}]7584889119913750
{8, 6}]2329042266808650
{9, 5}]50034381769600
{10, 4}]22945154050
{11, 3}]650
{8, 7}]97887416945961075
{9, 6}]10084936612321850
{10, 5}]63477362571125
{11, 4}]5048036025
{8, 8}]2218998811196105750
{9, 7}]782114760930236000
{10, 6}]28126794522576400
{11, 5}]50034381769600
{12, 4}]545403950
{9, 8}]30429684503634827875
{10, 7}]4075722566708421875
{11, 6}]51642034298930775
{12, 5}]24352783493100
{13, 4}]24162125
{9, 9}]688579463588598857500
{10, 8}]270605922599775866950
{11, 7}]14322131205924119500
{12, 6}]63157566038079800
{13, 5}]7175800860250
{14, 4}]250550
{10, 9}]9910287533252141060075
{11, 8}]1632070561204989561850
{12, 7}]34680269311023701250
{13, 6}]51642034298930775
{14, 5}]1231494256550
{15, 4}]50
{10, 10}]223872593965525056524000
{11, 9}]96088214450066089180650
{12, 8}]6879784715166845894000
{13, 7}]58660895139129344250
{14, 6}]28126794522576400
{15, 5}]114678709000
{11, 10}]3351508924925685769008400
{12, 9}]652460665889917943662000
{13, 8}]20691735554891324819375
{14, 7}]69837157468295256300
{15, 6}]10084936612321850
{16, 5}]5048036025
{11, 11}]75590773394298108275641400
{12, 10}]34897260129076709170702250
{13, 9}]3187856349624109686861750
{14, 8}]45029496161343522802000
{15, 7}]58660895139129344250
{16, 6}]2329042266808650
{17, 5}]75885200
{12, 11}]1169273852253720661047855850
{13, 10}]261422628471008778452895000
{14, 9}]11425471345666372778573625
{15, 8}]71553701937328489430500
{16, 7}]34680269311023701250
{17, 6}]334030085380350
{18, 5}]148525
{12, 12}]26334146932509192721297606250
{13, 11}]12925735995730366743674988500
{14, 10}]1443695931340763707228964750
{15, 9}]30455096636986392995454400
{16, 8}]83451873384004446556500
{17, 7}]14322131205924119500
{18, 6}]27995704239850
{13, 12}]418749190393922926454264339775
{14, 11}]105192221331381719953056030200
{15, 10}]5983442985255839039237107500
{16, 9}]60960783194314781836252175
{17, 8}]71553701937328489430500
{18, 7}]4075722566708421875
{19, 6}]1231494256550
{13, 13}]9418685010993246523213147309050
{14, 12}]4871005171529900672353774747800
{15, 11}]644225116339930387252675093800
{16, 10}]18858521597882017598430587450
{17, 9}]92220396289894438953276250
{18, 8}]45029496161343522802000
{19, 7}]782114760930236000
{20, 6}]22945154050
{14, 13}]153346515556322207530993642061375
{15, 12}]42551462569251226858792953023050
{16, 11}]3018482957837620078514600231550
{17, 10}]45641749489863534673387285075
{18, 9}]105822944845470819411145650
{19, 8}]20691735554891324819375
{20, 7}]97887416945961075
{21, 6}]110273275
{14, 14}]3445063666410127667138401567917450
{15, 13}]1863829015771384379547064210671650
}
}
\end{adjustbox}
\vskip10pt
\capt{6.5in}{tab:Instanton_Numbers_2_52}{The genus~0 instanton numbers of total degree $\leqslant 37$ for the family \eqref{eq:CICY_2_52}. The numbers not in this list are either zero, or given by those in the table after permuting indices.}
\end{table}

\begin{table}[H]\notag
\centering
\def\arraystretch{1.15}
\begin{adjustbox}{width=\columnwidth,center}
\tiny{\pgfplotstabletypeset[
col sep=space,
white space chars={]},
ignore chars={\ },
every head row/.style={before row=\hline,after row=\hline\hline},
every first row/.style={before row=\vrule height12pt width0pt depth0pt},
every last row/.style={before row=\vrule height0pt width0pt depth6pt,after row=\hline},
%after row=\hline,    %  Uncomment this to get back lines between all rows.
columns={A,B,A,B},
display columns/0/.style={select equal part entry of={0}{2},column type = {|l},column name=\vrule height12pt width0pt depth6pt \hfil $\bm p$},
display columns/1/.style={select equal part entry of={0}{2},column type = {|l|},column name=\hfil $n^{(0)}_{\bm p}$},
display columns/2/.style={select equal part entry of={1}{2},column type = {|l},column name=\hfil $\bm p$},
display columns/3/.style={select equal part entry of={1}{2},column type = {|l|},column name=\hfil $n^{(0)}_{\bm p}$},
string type]
{
A ] B

{16, 12}]284748690462403703849327898236400
{17, 11}]10956684127545224855242712323800
{18, 10}]85415237718946312165876907100
{19, 9}]92220396289894438953276250
{20, 8}]6879784715166845894000
{21, 7}]7584889119913750
{22, 6}]29350
{15, 14}]57245804146829441141855075686421600
{16, 13}]17311291067627079301898110530922200
{17, 12}]1482101901841193380170774085390975
{18, 11}]31105645711336079302777239697100
{19, 10}]124182064589288217124451139225
{20, 9}]60960783194314781836252175
{21, 8}]1632070561204989561850
{22, 7}]334030085380350
{15, 15}]1284701496853760180092631899781959250
{16, 14}]722903234302220208838481980961898100
{17, 13}]125101067315642934987038992528141500
{18, 12}]6071452882998806212881647096382400
{19, 11}]69561055478912921075310779919000
{20, 10}]140638929443068626672454410250
{21, 9}]30455096636986392995454400
{22, 8}]270605922599775866950
{23, 7}]7175800860250
{16, 15}]21731435735419411130551512688193970350
{17, 14}]7084069788870929694102349973792379125
{18, 13}]713360358445986060308899284173421250
{19, 12}]19755724290719957639765719857823000
{20, 11}]123171917767680954127111199469800
{21, 10}]124182064589288217124451139225
{22, 9}]11425471345666372778573625
{23, 8}]30429684503634827875
{24, 7}]55531376500
{16, 16}]487219827311876979652523699533587321150
{17, 15}]283802089997587090105839672112883654150
{18, 14}]54760100433813765734980222782145239000
{19, 13}]3245572721974319554949303862415944250
{20, 12}]51423738528601615934359979687767300
{21, 11}]173303334724056406174320632632850
{22, 10}]85415237718946312165876907100
{23, 9}]3187856349624109686861750
{24, 8}]2218998811196105750
{25, 7}]75885200
{17, 16}]8371985898150331724830984679777619591325
{18, 15}]2915743577162092867565871545455088037975
{19, 14}]338273641516001759928541421158949393625
{20, 13}]11885728626129063303448250398108564900
{21, 12}]107659493684116349217356173601610375
{22, 11}]194151779259881472757738870486275
{23, 10}]45641749489863534673387285075
{24, 9}]652460665889917943662000
{25, 8}]97887416945961075
{26, 7}]1475
{17, 17}]187534310501258236886420065561355215672850
{18, 16}]112636401191272489352275415887745870490500
{19, 15}]23921213512060867824603015906296302482550
{20, 14}]1687381561533043508236480530888841957000
{21, 13}]35280439004752280205968167087919443350
{22, 12}]182005705950504851032041523111610550
{23, 11}]173303334724056406174320632632850
{24, 10}]18858521597882017598430587450
{25, 9}]96088214450066089180650
{26, 8}]2329042266808650
{18, 17}]3267643622168623942896097738545068717322400
{19, 16}]1206865467127098649476297800122829258555125
{20, 15}]158613328551370248346752880868799119011250
{21, 14}]6853883162475990624130911021611665107450
{22, 13}]85346765578856941178005038357338030125
{23, 12}]249140684221048318682166766371696025
{24, 11}]123171917767680954127111199469800
{25, 10}]5983442985255839039237107500
{26, 9}]9910287533252141060075
{27, 8}]24352783493100
{18, 18}]73136802934944577315316758502397717744613800
{19, 17}]45145186660990785137737197513727989291648750
{20, 16}]10440390902521863107237190843251830811038400
{21, 15}]858330748648887482867022141394382548080200
{22, 14}]22822795902093633748803493742442397239000
{23, 13}]168970246230273909199356496016178585000
{24, 12}]276580864121735152873483395276585800
{25, 11}]69561055478912921075310779919000
{26, 10}]1443695931340763707228964750
{27, 9}]688579463588598857500
{28, 8}]74278763500
{19, 18}]1290315264071203979827100216377332535448538000
{20, 17}]502246809291235370794554217924188280591690325
{21, 16}]73738372223409643492419300864401743615326225
{22, 15}]3821053219977680771955490543838867117725025
{23, 14}]62641363244422442148059754716581069033750
{24, 13}]274635658050542709324085728636859818125
{25, 12}]249140684221048318682166766371696025
{26, 11}]31105645711336079302777239697100
{27, 10}]261422628471008778452895000
{28, 9}]30429684503634827875
{29, 8}]24162125

}
}
\end{adjustbox}
\vskip10pt
\caption*{\tref{tab:Instanton_Numbers_2_52} continued.}
\end{table}

\section{Genus 1 mirror symmetry: Counting elliptic curves}\label{sect:Miroirs_genus1}
The B-model prepotential was determined in \cite{Bershadsky:1993ta} and is given in the topological limit ${\bar{t}\rightarrow-\ii\infty}$ by
\begin{equation}\label{eq:B_genus1}
\cF^{(1)}=\log\left[\varpi_{0}^{-\left(3+h^{2,1}(X)+\chi(X)/12\right)/2}\text{det}\left(\frac{\partial \bm{\varphi}}{\partial \mathbf{t}}\right)^{1/2}f(\bm{\varphi})\right]~,
\end{equation}
with the function $f$ giving the correct behaviour at singularities of the moduli space, to be fixed after imposing a boundary condition that imposes consistency with the A-model expansion. The large complex structure expansion of the A-model genus 1 prepotential is
\begin{equation}\label{eq:A_genus1}
F^{(1)}=\frac{1}{2}Y_{00i}t^{i}+\sum_{\mathbf{b}\geq0}\left(n^{(1)}_{\mathbf{b}}+\frac{n^{(0)}_{\mathbf{b}}}{12}\right)\,\text{Li}_{1}\left(\mathbf{q}^{\mathbf{b}}\right)~.
\end{equation}
The function $f$ should have a factor that gives zeroes at the roots of the discriminant locus $\Delta$, the set of moduli $\bm{\varphi}$ for which the manifold $X$ becomes singular. Additionally, $f$ should vanish at the large complex structure\footnote{so named because this point is taken by the mirror map to the large volume point $t=\ii\infty$.} (LCS) point $\bm{\varphi}=\bm{0}$, with the order of vanishing such that the leading behaviour of $F^{(1)}$ is recovered. These considerations fix $f$ to be of the form
\begin{equation}\label{eq:Not_Hol_Amb}
f=\frac{1}{\Delta^{c}\prod_{i=1}^{h^{2,1}(X)}\left(\varphi^{i}\right)^{(1-Y_{00i})/2}}~.
\end{equation}
We remark that the determinant in \eqref{eq:B_genus1} goes like $\left(\varphi^{i}\right)^{1/2}$ as $\varphi^{i}\to0$, hence the exponents $\frac{1-Y_{00i}}{2}$ above. The exponent $c$ is specified by the kind of singularities encoded by $\Delta$. For a conifold singularity, $c$ equals $1/12$. More complicated behaviours are possible at orbifold singularities, see for instance the computation of genus 1 numbers on quotient manifolds in~\cite{Candelas:2019llw}, informed by \cite{Gopakumar:1997dv}. For our purposes, $\Delta$ is always a conifold locus and we work with $c=1/12$. For clarity, we remark that while $f$ encapsulates a number of singularities, there is only one $\cF^{(1)}$, and this behaves appropriately at each different singularity through $f$'s appearance in \eqref{eq:B_genus1}.

The polynomial $\Delta$ is in the variables $\varphi^{i}$, and we explain this choice of complex structure coordinates in Appendix \sref{sect:Appendix_coordinates}. $\Delta$ encodes singularities not of the CICYs that we display, but of their mirrors $X$. Appendix \sref{sect:Appendix_coordinates} discusses the construction of those mirror families $X$, and their parametrisation in terms of $\varphi^{i}$.

We shall display some genus 1 numbers for two families, the Tetraquadric \eqref{eq:CICY_4_68} and the maximally split Quintic \eqref{eq:CICY_2_52}, whose mirrors respectively have discriminants
\begin{equation}
\begin{aligned}
\cicy{\IP^{1}\\\IP^{1}\\\IP^{1}\\\IP^{1}}{2\\2\\2\\2}^{(4,68)}_{-128}~,&\qquad \Delta=\prod_{\epsilon_{i}\in\{1,-1\}}\left(1+2\epsilon_{1}\sqrt{\varphi^{1}}+2\epsilon_{2}\sqrt{\varphi^{2}}+2\epsilon_{3}\sqrt{\varphi^{3}}+2\epsilon_{4}\sqrt{\varphi^{4}}\right)~;\\[30pt]
\cicy{\IP^{4}\\\IP^{4}}{1&1&1&1&1\\1&1&1&1&1}^{(2,52)}_{-100}~,&\qquad \Delta=\prod_{1\leq n_{1},n_{2}\leq5}\left(1-\ee^{\frac{2\pi\ii}{5}n_{1}}\left(\varphi^{1}\right)^{1/5}-\ee^{\frac{2\pi\ii}{5}n_{2}}\left(\varphi^{2}\right)^{1/5}\right)~.
\end{aligned}
\end{equation}
The $q$-series that we have used for obtaining genus-1 curve counts $n^{(1)}_{\mathbf{b}}$, formula \eqref{eq:A_genus1}, is that given by the Gopakumar-Vafa prescription \cite{Gopakumar:1998ii,Gopakumar:1998jq}. In earlier work by Bershadsky, Cecotti, Ooguri and Vafa \cite{Bershadsky:1993ta}, the prescription differed slightly (counting what they called ``primitive elliptic curves") and so some care is needed when comparing curve counts across the literature. In either case the B-model prepotential is the same, and the two sets of data are equivalent.

The content of the following tables is available in electronic form \cite{mcgovern2023a}.

\begin{table}[H]
\centering
\def\arraystretch{1.005}
\begin{adjustbox}{width=\columnwidth,center}
\tiny{\pgfplotstabletypeset[
col sep=space,
white space chars={]},
ignore chars={d,[,\ },
every head row/.style={before row=\hline,after row=\hline\hline},
every first row/.style={before row=\vrule height12pt width0pt depth0pt},
every last row/.style={before row=\vrule height0pt width0pt depth6pt,after row=\hline},
%after row=\hline,    %  Uncomment this to get back lines between all rows.
columns={A,B,A,B,A,B},
display columns/0/.style={select equal part entry of={0}{3},column type = {|l},column name=\vrule height12pt width0pt depth6pt \hfil $\bm p$},
display columns/1/.style={select equal part entry of={0}{3},column type = {|l|},column name=\hfil $n^{(1)}_{\bm p}$},
display columns/2/.style={select equal part entry of={1}{3},column type = {|l},column name=\hfil $\bm p$},
display columns/3/.style={select equal part entry of={1}{3},column type = {|l|},column name=\hfil $n^{(1)}_{\bm p}$},
display columns/4/.style={select equal part entry of={2}{3},column type = {|l},column name=\hfil $\bm p$},
display columns/5/.style={select equal part entry of={2}{3},column type = {|l|},column name=\hfil $n^{(1)}_{\bm p}$},
string type]
{
A ] B
d[{2, 2, 0, 0}]]4
d[{2, 2, 1, 0}]]-96
d[{2, 2, 1, 1}]]1984
d[{2, 2, 2, 0}]]-10272
d[{2, 2, 2, 1}]]269280
d[{2, 2, 2, 2}]]93388992
d[{3, 2, 1, 0}]]-320
d[{3, 2, 1, 1}]]10496
d[{3, 2, 2, 0}]]-69568
d[{3, 2, 2, 1}]]2765056
d[{3, 2, 2, 2}]]1812768672
d[{3, 3, 1, 0}]]-4864
d[{3, 3, 1, 1}]]112640
d[{3, 3, 2, 0}]]-1046656
d[{3, 3, 2, 1}]]54265088
d[{3, 3, 2, 2}]]57695471872
d[{3, 3, 3, 0}]]-25362432
d[{3, 3, 3, 1}]]1683640320
d[{3, 3, 3, 2}]]2698701579264
d[{3, 3, 3, 3}]]173800712052736
d[{4, 2, 1, 0}]]-96
d[{4, 2, 1, 1}]]20096
d[{4, 2, 2, 0}]]-124616
d[{4, 2, 2, 1}]]9983872
d[{4, 2, 2, 2}]]12011885472
d[{4, 3, 1, 0}]]-10112
d[{4, 3, 1, 1}]]241408
d[{4, 3, 2, 0}]]-4386240
d[{4, 3, 2, 1}]]353758784
d[{4, 3, 2, 2}]]621519800480
d[{4, 3, 3, 0}]]-188206400
d[{4, 3, 3, 1}]]17552569088
d[{4, 3, 3, 2}]]43026117967872
d[{4, 3, 3, 3}]]3851778847706880
d[{4, 4, 0, 0}]]4
d[{4, 4, 1, 0}]]-69568
d[{4, 4, 1, 1}]]-1580992
d[{4, 4, 2, 0}]]-38445312
d[{4, 4, 2, 1}]]3987616000
d[{4, 4, 2, 2}]]10558546637760
d[{4, 4, 3, 0}]]-2397896960
d[{4, 4, 3, 1}]]286009826304
d[{4, 4, 3, 2}]]999142624314816
d[{4, 4, 3, 3}]]117575052436736704
d[{4, 4, 4, 0}]]-45623190048
d[{4, 4, 4, 1}]]6666610324992
d[{4, 4, 4, 2}]]31633436508163392
d[{4, 4, 4, 3}]]4709891230826927232
d[{4, 4, 4, 4}]]239082475099257093312
d[{5, 2, 1, 1}]]10496
d[{5, 2, 2, 0}]]-69568
d[{5, 2, 2, 1}]]15218816
d[{5, 2, 2, 2}]]35189119872
d[{5, 3, 1, 0}]]-4864
d[{5, 3, 1, 1}]]286720
d[{5, 3, 2, 0}]]-6904192
d[{5, 3, 2, 1}]]1021728256
d[{5, 3, 2, 2}]]3034052551424
d[{5, 3, 3, 0}]]-578600960
d[{5, 3, 3, 1}]]83573229568
d[{5, 3, 3, 2}]]317438354195968
d[{5, 3, 3, 3}]]40174633448570880
d[{5, 4, 1, 0}]]-124608
d[{5, 4, 1, 1}]]-14325760
d[{5, 4, 2, 0}]]-127854976
d[{5, 4, 2, 1}]]19566085376
d[{5, 4, 2, 2}]]80961860336992
d[{5, 4, 3, 0}]]-12801118336
d[{5, 4, 3, 1}]]2116485979904
d[{5, 4, 3, 2}]]10738139481866240
d[{5, 4, 3, 3}]]1691948070016716544
d[{5, 4, 4, 0}]]-372907450720
d[{5, 4, 4, 1}]]71025289697408
d[{5, 4, 4, 2}]]466267539878814528
d[{5, 4, 4, 3}]]89381762755780427776
d[{5, 4, 4, 4}]]5783834578826599864896
d[{5, 5, 1, 0}]]-631296
d[{5, 5, 1, 1}]]-193064960
d[{5, 5, 2, 0}]]-831856256
d[{5, 5, 2, 1}]]156514373632
d[{5, 5, 2, 2}]]951193768367744
d[{5, 5, 3, 0}]]-114273286144
d[{5, 5, 3, 1}]]23771145744384
d[{5, 5, 3, 2}]]165492284557610240
d[{5, 5, 3, 3}]]33247249326979840000
d[{5, 5, 4, 0}]]-4598298755008
d[{5, 5, 4, 1}]]1075385950542336
d[{5, 5, 4, 2}]]9348192801960947328
d[{5, 5, 4, 3}]]2224378492069768184832
d[{5, 5, 4, 4}]]177840753331297192191808
d[{5, 5, 5, 0}]]-78791963521792
d[{5, 5, 5, 1}]]21862949555982336
d[{5, 5, 5, 2}]]243574478608897658880
d[{5, 5, 5, 3}]]70047199737530626793472
d[{5, 5, 5, 4}]]6756842515156306030742016
d[{5, 5, 5, 5}]]310274010609799385697632256
d[{6, 2, 1, 1}]]1984
d[{6, 2, 2, 0}]]-10272
d[{6, 2, 2, 1}]]9983872
d[{6, 2, 2, 2}]]49952569216
d[{6, 3, 1, 0}]]-320
d[{6, 3, 1, 1}]]241408
d[{6, 3, 2, 0}]]-4386240
d[{6, 3, 2, 1}]]1445166080
d[{6, 3, 2, 2}]]7576252775296
d[{6, 3, 3, 0}]]-831856256
d[{6, 3, 3, 1}]]205607705344
d[{6, 3, 3, 2}]]1238547915076352
d[{6, 3, 3, 3}]]226786087463462912
d[{6, 4, 1, 0}]]-69568
d[{6, 4, 1, 1}]]-25295744
d[{6, 4, 2, 0}]]-188206960
d[{6, 4, 2, 1}]]48618551008
d[{6, 4, 2, 2}]]323138860266848
d[{6, 4, 3, 0}]]-33378648000
d[{6, 4, 3, 1}]]8232411529408
d[{6, 4, 3, 2}]]61783631376376704
d[{6, 4, 3, 3}]]13300730517821622784
d[{6, 4, 4, 0}]]-1541331810048
d[{6, 4, 4, 1}]]404052229218240
d[{6, 4, 4, 2}]]3723977589477458176
d[{6, 4, 4, 3}]]935179511020340807456
d[{6, 4, 4, 4}]]77803599946908880718016
d[{6, 5, 1, 0}]]-1046656
d[{6, 5, 1, 1}]]-734128384
d[{6, 5, 2, 0}]]-2397896960
d[{6, 5, 2, 1}]]626843119104
d[{6, 5, 2, 2}]]5792732588373728
d[{6, 5, 3, 0}]]-497787993344
d[{6, 5, 3, 1}]]139578486511360
d[{6, 5, 3, 2}]]1362193196667713536
d[{6, 5, 3, 3}]]356453634018444596992
d[{6, 5, 4, 0}]]-28770544038624
d[{6, 5, 4, 1}]]8683903357985664
d[{6, 5, 4, 2}]]101579932691307744640
d[{6, 5, 4, 3}]]30534454347465167308416
d[{6, 5, 4, 4}]]3044826118116035108486016
d[{6, 5, 5, 0}]]-692429114488064
d[{6, 5, 5, 1}]]237977798080726528
d[{6, 5, 5, 2}]]3452295445836449045632
d[{6, 5, 5, 3}]]1220175570920202576920832
d[{6, 5, 5, 4}]]143356009888676753465771648
d[{6, 6, 0, 0}]]4
d[{6, 6, 1, 0}]]-4386240
d[{6, 6, 1, 1}]]-5444761792
d[{6, 6, 2, 0}]]-12801129824
d[{6, 6, 2, 1}]]3899397189472
d[{6, 6, 2, 2}]]52718694526710016
d[{6, 6, 3, 0}]]-3510327932032
d[{6, 6, 3, 1}]]1211618774875136
d[{6, 6, 3, 2}]]15814322874224859712
d[{6, 6, 3, 3}]]5160195188934834782144
d[{6, 6, 4, 0}]]-267545068581728
d[{6, 6, 4, 1}]]98216197349510400
d[{6, 6, 4, 2}]]1487879821243945966432
d[{6, 6, 4, 3}]]546260879195635250051616
d[{6, 6, 4, 4}]]66049985769182760347092224
d[{6, 6, 5, 0}]]-8478485684300928
d[{6, 6, 5, 1}]]3465741933227797120
d[{6, 6, 5, 2}]]63508834166133397153792
d[{6, 6, 5, 3}]]26872551390971959525026816
d[{6, 6, 6, 0}]]-136960390214771328
d[{6, 6, 6, 1}]]64863757395907531392
d[{6, 6, 6, 2}]]1466437377855662330826240
d[{7, 2, 2, 0}]]-96
d[{7, 2, 2, 1}]]2765056
d[{7, 2, 2, 2}]]35189119872

d[{7, 3, 1, 1}]]112640
d[{7, 3, 2, 0}]]-1046656
d[{7, 3, 2, 1}]]1021728256
d[{7, 3, 2, 2}]]10223759052288
d[{7, 3, 3, 0}]]-578600960
d[{7, 3, 3, 1}]]276199923712
d[{7, 3, 3, 2}]]2739373720985856
d[{7, 3, 3, 3}]]748137868720717824
d[{7, 4, 1, 0}]]-10112
d[{7, 4, 1, 1}]]-14325760
d[{7, 4, 2, 0}]]-127854976
d[{7, 4, 2, 1}]]65476152960
d[{7, 4, 2, 2}]]723323393538432
d[{7, 4, 3, 0}]]-45623185536
d[{7, 4, 3, 1}]]18148629097216
d[{7, 4, 3, 2}]]206208348590980608
d[{7, 4, 3, 3}]]62079704012640705280
d[{7, 4, 4, 0}]]-3510327932032
d[{7, 4, 4, 1}]]1335576781170688
d[{7, 4, 4, 2}]]17562256414556246528
d[{7, 4, 4, 3}]]5884952965650513045248
d[{7, 4, 4, 4}]]636300785891465292898752
d[{7, 5, 1, 0}]]-631296
d[{7, 5, 1, 1}]]-1115140096
d[{7, 5, 2, 0}]]-3380586368
d[{7, 5, 2, 1}]]1393481688832
d[{7, 5, 2, 2}]]20022577101423360
d[{7, 5, 3, 0}]]-1166173618944
d[{7, 5, 3, 1}]]469840676702208
d[{7, 5, 3, 2}]]6554880932704430336
d[{7, 5, 3, 3}]]2281737394795361083392
d[{7, 5, 4, 0}]]-100934478032512
d[{7, 5, 4, 1}]]41094723073773568
d[{7, 5, 4, 2}]]656136930137156080768
d[{7, 5, 4, 3}]]253382111085115792710400
d[{7, 5, 4, 4}]]31839291301539012129070080
d[{7, 5, 5, 0}]]-3469762473325056
d[{7, 5, 5, 1}]]1531922193657643008
d[{7, 5, 5, 2}]]29300021039056213398016
d[{7, 5, 5, 3}]]12921015476002583834605568
d[{7, 6, 1, 0}]]-6904192
d[{7, 6, 1, 1}]]-16517906944
d[{7, 6, 2, 0}]]-33378648000
d[{7, 6, 2, 1}]]13324886863872
d[{7, 6, 2, 2}]]271114676514738784
d[{7, 6, 3, 0}]]-13264263786752
d[{7, 6, 3, 1}]]5997956867302144
d[{7, 6, 3, 2}]]106978781273942999040
d[{7, 6, 3, 3}]]44494307569970341498112
d[{7, 6, 4, 0}]]-1392349606298016
d[{7, 6, 4, 1}]]649011301399128064
d[{7, 6, 4, 2}]]12931687111454709126208
d[{7, 6, 4, 3}]]5900151322919540445706752
d[{7, 6, 5, 0}]]-59273706883766400
d[{7, 6, 5, 1}]]29832962488756485888
d[{7, 6, 5, 2}]]699633783726226960540672
d[{7, 6, 6, 0}]]-1270394426831547360
d[{7, 6, 6, 1}]]719306357328985736192
d[{7, 7, 1, 0}]]-25362432
d[{7, 7, 1, 1}]]-93294921728
d[{7, 7, 2, 0}]]-154245226112
}
}
\end{adjustbox}
\vskip10pt
\capt{6.5in}{tab:Instanton_Numbers_4_68_genus1}{The genus~1 instanton numbers of total degree $\leqslant 20$ for the family \eqref{eq:CICY_4_68}. The numbers not in this list are either zero, or given by those in the table after permuting indices.}
\end{table}
\begin{table}[H]\notag
\centering
\def\arraystretch{1.005}
\begin{adjustbox}{width=\columnwidth,center}
\tiny{\pgfplotstabletypeset[
col sep=space,
white space chars={]},
ignore chars={d,[,\ },
every head row/.style={before row=\hline,after row=\hline\hline},
every first row/.style={before row=\vrule height12pt width0pt depth0pt},
every last row/.style={before row=\vrule height0pt width0pt depth6pt,after row=\hline},
%after row=\hline,    %  Uncomment this to get back lines between all rows.
columns={A,B,A,B,A,B},
display columns/0/.style={select equal part entry of={0}{3},column type = {|l},column name=\vrule height12pt width0pt depth6pt \hfil $\bm p$},
display columns/1/.style={select equal part entry of={0}{3},column type = {|l|},column name=\hfil $n^{(1)}_{\bm p}$},
display columns/2/.style={select equal part entry of={1}{3},column type = {|l},column name=\hfil $\bm p$},
display columns/3/.style={select equal part entry of={1}{3},column type = {|l|},column name=\hfil $n^{(1)}_{\bm p}$},
display columns/4/.style={select equal part entry of={2}{3},column type = {|l},column name=\hfil $\bm p$},
display columns/5/.style={select equal part entry of={2}{3},column type = {|l|},column name=\hfil $n^{(1)}_{\bm p}$},
string type]
{
A ] B

d[{7, 7, 2, 1}]]67556863891200
d[{7, 7, 2, 2}]]2038282247914091008
d[{7, 7, 3, 0}]]-78791963521792
d[{7, 7, 3, 1}]]42886489670017024
d[{7, 7, 3, 2}]]1004665124573369920512
d[{7, 7, 3, 3}]]512053569223265100480512
d[{7, 7, 4, 0}]]-10569489713182080
d[{7, 7, 4, 1}]]5915250606113727488
d[{7, 7, 4, 2}]]149886488946568628294656
d[{7, 7, 5, 0}]]-572125762297261056
d[{7, 7, 5, 1}]]340591525223538673664
d[{7, 7, 6, 0}]]-15558927658117177088
d[{8, 2, 2, 0}]]4
d[{8, 2, 2, 1}]]269280
d[{8, 2, 2, 2}]]12011885472
d[{8, 3, 1, 1}]]10496
d[{8, 3, 2, 0}]]-69568
d[{8, 3, 2, 1}]]353758784
d[{8, 3, 2, 2}]]7576252775296
d[{8, 3, 3, 0}]]-188206400
d[{8, 3, 3, 1}]]205607705344
d[{8, 3, 3, 2}]]3556448533895424
d[{8, 3, 3, 3}]]1507518654558092544
d[{8, 4, 1, 0}]]-96
d[{8, 4, 1, 1}]]-1580992
d[{8, 4, 2, 0}]]-38445312
d[{8, 4, 2, 1}]]48618551008
d[{8, 4, 2, 2}]]942605235833872
d[{8, 4, 3, 0}]]-33378648000
d[{8, 4, 3, 1}]]23534563926272
d[{8, 4, 3, 2}]]418143162721593312
d[{8, 4, 3, 3}]]181018160115934668416
d[{8, 4, 4, 0}]]-4598298768072
d[{8, 4, 4, 1}]]2691518925004800
d[{8, 4, 4, 2}]]51542075860983487936
d[{8, 4, 4, 3}]]23528568383214020784160
d[{8, 4, 4, 4}]]3349905544323212030802336
d[{8, 5, 1, 0}]]-124608
d[{8, 5, 1, 1}]]-734128384
d[{8, 5, 2, 0}]]-2397896960
d[{8, 5, 2, 1}]]1809944501504
d[{8, 5, 2, 2}]]41383783201391104
d[{8, 5, 3, 0}]]-1541331760640
d[{8, 5, 3, 1}]]955800484656896
d[{8, 5, 3, 2}]]19486899779598178176
d[{8, 5, 3, 3}]]9230145138666014389504
d[{8, 5, 4, 0}]]-210165664436480
d[{8, 5, 4, 1}]]120646194440285120
d[{8, 5, 4, 2}]]2670176886160388673088
d[{8, 5, 4, 3}]]1347917974198473929771648
d[{8, 5, 5, 0}]]-10569489713182080
d[{8, 5, 5, 1}]]6203840748464569344
d[{8, 5, 5, 2}]]158349506217985346295936
d[{8, 6, 1, 0}]]-4386240
d[{8, 6, 1, 1}]]-23596592896
d[{8, 6, 2, 0}]]-45623190048
d[{8, 6, 2, 1}]]27081415941696
d[{8, 6, 2, 2}]]840745818190013632
d[{8, 6, 3, 0}]]-28770544038624
d[{8, 6, 3, 1}]]18089515225187456
d[{8, 6, 3, 2}]]449402422813104411520
d[{8, 6, 3, 3}]]243302732065835955711744
d[{8, 6, 4, 0}]]-4345925673694464
d[{8, 6, 4, 1}]]2674825644762726656
d[{8, 6, 4, 2}]]71095410940695469670768
d[{8, 6, 5, 0}]]-254128031458899136
d[{8, 6, 5, 1}]]162347641909980632448
d[{8, 6, 6, 0}]]-7295475375082553200
d[{8, 7, 1, 0}]]-38443520
d[{8, 7, 1, 1}]]-249012852224
d[{8, 7, 2, 0}]]-372907450720
d[{8, 7, 2, 1}]]201277836622720
d[{8, 7, 2, 2}]]9200312079392193664
d[{8, 7, 3, 0}]]-267545068649344
d[{8, 7, 3, 1}]]185891068420360448
d[{8, 7, 3, 2}]]5838261073702086594688
d[{8, 7, 4, 0}]]-47957368939626656
d[{8, 7, 4, 1}]]33506015096031958400
d[{8, 7, 5, 0}]]-3380856350018782208
d[{8, 8, 0, 0}]]4
d[{8, 8, 1, 0}]]-127854976
d[{8, 8, 1, 1}]]-1181782901504
d[{8, 8, 2, 0}]]-1541331810048
d[{8, 8, 2, 1}]]844741431788416
d[{8, 8, 2, 2}]]59548327861798208892
d[{8, 8, 3, 0}]]-1392349606298016
d[{8, 8, 3, 1}]]1140095263765735424
d[{8, 8, 4, 0}]]-311703356855240960
d[{9, 2, 2, 1}]]1984
d[{9, 2, 2, 2}]]1812768672
d[{9, 3, 2, 0}]]-320
d[{9, 3, 2, 1}]]54265088
d[{9, 3, 2, 2}]]3034052551424
d[{9, 3, 3, 0}]]-25362432
d[{9, 3, 3, 1}]]83573229568
d[{9, 3, 3, 2}]]2739373720985856
d[{9, 3, 3, 3}]]1899437094349107200
d[{9, 4, 1, 1}]]241408
d[{9, 4, 2, 0}]]-4386240
d[{9, 4, 2, 1}]]19566085376
d[{9, 4, 2, 2}]]723323393538432
d[{9, 4, 3, 0}]]-12801118336
d[{9, 4, 3, 1}]]18148629097216
d[{9, 4, 3, 2}]]527901584611872640
d[{9, 4, 3, 3}]]340169678300685140992
d[{9, 4, 4, 0}]]-3510327932032
d[{9, 4, 4, 1}]]3390956750776448
d[{9, 4, 4, 2}]]97201242671174343424
d[{9, 4, 4, 3}]]61931896048581461894144
d[{9, 5, 1, 0}]]-4864
d[{9, 5, 1, 1}]]-193064960
d[{9, 5, 2, 0}]]-831856256
d[{9, 5, 2, 1}]]1393481688832
d[{9, 5, 2, 2}]]52565098296996352
d[{9, 5, 3, 0}]]-1166173618944
d[{9, 5, 3, 1}]]1207709952303104
d[{9, 5, 3, 2}]]37014997608595211776
d[{9, 5, 3, 3}]]24481664831474093735936
d[{9, 5, 4, 0}]]-267545068649344
d[{9, 5, 4, 1}]]227415315078834432
d[{9, 5, 4, 2}]]7109119409066563817856
d[{9, 5, 5, 0}]]-20339595410138112
d[{9, 5, 5, 1}]]16444600976022953984
d[{9, 6, 1, 0}]]-1046656
d[{9, 6, 1, 1}]]-16517906944
d[{9, 6, 2, 0}]]-33378648000
d[{9, 6, 2, 1}]]34157599278208
d[{9, 6, 2, 2}]]1634935031609724864
d[{9, 6, 3, 0}]]-37107163518144
d[{9, 6, 3, 1}]]34589813440859136
d[{9, 6, 3, 2}]]1221567939348425063424
d[{9, 6, 4, 0}]]-8478485684300928
d[{9, 6, 4, 1}]]7170055316812400128
d[{9, 6, 5, 0}]]-699282768851250048
d[{9, 7, 1, 0}]]-25362432
d[{9, 7, 1, 1}]]-342488981504
d[{9, 7, 2, 0}]]-497787993344
d[{9, 7, 2, 1}]]377316002640896
d[{9, 7, 2, 2}]]26201449192835298304
d[{9, 7, 3, 0}]]-547093483682304
d[{9, 7, 3, 1}]]513891279029764096
d[{9, 7, 4, 0}]]-136960390218830336
d[{9, 8, 1, 0}]]-188206400
d[{9, 8, 1, 1}]]-2888853456640
d[{9, 8, 2, 0}]]-3510327932032
d[{9, 8, 2, 1}]]2188585915040384
d[{9, 8, 3, 0}]]-4345925673853440
d[{9, 9, 1, 0}]]-578600960
d[{9, 9, 1, 1}]]-12088456933376
d[{9, 9, 2, 0}]]-13264263786752
d[{10, 2, 2, 1}]]-96
d[{10, 2, 2, 2}]]93388992
d[{10, 3, 2, 1}]]2765056
d[{10, 3, 2, 2}]]621519800480
d[{10, 3, 3, 0}]]-1046656
d[{10, 3, 3, 1}]]17552569088
d[{10, 3, 3, 2}]]1238547915076352
d[{10, 3, 3, 3}]]1507518654558092544
d[{10, 4, 1, 1}]]20096
d[{10, 4, 2, 0}]]-124616
d[{10, 4, 2, 1}]]3987616000
d[{10, 4, 2, 2}]]323138860266848
d[{10, 4, 3, 0}]]-2397896960
d[{10, 4, 3, 1}]]8232411529408
d[{10, 4, 3, 2}]]418143162721593312
d[{10, 4, 3, 3}]]419022629321006302208
d[{10, 4, 4, 0}]]-1541331810048
d[{10, 4, 4, 1}]]2691518925004800
d[{10, 4, 4, 2}]]119867722557338716800
d[{10, 5, 1, 1}]]-14325760
d[{10, 5, 2, 0}]]-127854976
d[{10, 5, 2, 1}]]626843119104
d[{10, 5, 2, 2}]]41383783201391104
d[{10, 5, 3, 0}]]-497787993344
d[{10, 5, 3, 1}]]955800484656896
d[{10, 5, 3, 2}]]45752405840566558208
d[{10, 5, 4, 0}]]-210165664436480
d[{10, 5, 4, 1}]]280378657159159808
d[{10, 5, 5, 0}]]-25243927618418816
d[{10, 6, 1, 0}]]-69568
d[{10, 6, 1, 1}]]-5444761792
d[{10, 6, 2, 0}]]-12801129824
d[{10, 6, 2, 1}]]27081415941696
d[{10, 6, 2, 2}]]2036173541408295680
d[{10, 6, 3, 0}]]-28770544038624
d[{10, 6, 3, 1}]]42837663308834816
d[{10, 6, 4, 0}]]-10569489712041568
d[{10, 7, 1, 0}]]-6904192
d[{10, 7, 1, 1}]]-249012852224
d[{10, 7, 2, 0}]]-372907450720
d[{10, 7, 2, 1}]]463223659993600
d[{10, 7, 3, 0}]]-692429114488064
d[{10, 8, 1, 0}]]-127854976
d[{10, 8, 1, 1}]]-3867646677632
d[{10, 8, 2, 0}]]-4598298768072
d[{10, 9, 1, 0}]]-831856256
d[{10, 10, 0, 0}]]4
d[{11, 2, 2, 2}]]269280
d[{11, 3, 2, 1}]]10496
d[{11, 3, 2, 2}]]57695471872
d[{11, 3, 3, 0}]]-4864
d[{11, 3, 3, 1}]]1683640320
d[{11, 3, 3, 2}]]317438354195968
d[{11, 3, 3, 3}]]748137868720717824
d[{11, 4, 2, 0}]]-96

}
}
\end{adjustbox}
\vskip10pt
\caption*{\tref{tab:Instanton_Numbers_4_68_genus1} continued.}
\end{table}
\begin{table}[H]\notag
\centering
\def\arraystretch{1.005}
\begin{adjustbox}{width=\columnwidth,center}
\tiny{\pgfplotstabletypeset[
col sep=space,
white space chars={]},
ignore chars={d,[,\ },
every head row/.style={before row=\hline,after row=\hline\hline},
every first row/.style={before row=\vrule height12pt width0pt depth0pt},
every last row/.style={before row=\vrule height0pt width0pt depth6pt,after row=\hline},
%after row=\hline,    %  Uncomment this to get back lines between all rows.
columns={A,B,A,B,A,B},
display columns/0/.style={select equal part entry of={0}{3},column type = {|l},column name=\vrule height12pt width0pt depth6pt \hfil $\bm p$},
display columns/1/.style={select equal part entry of={0}{3},column type = {|l|},column name=\hfil $n^{(1)}_{\bm p}$},
display columns/2/.style={select equal part entry of={1}{3},column type = {|l},column name=\hfil $\bm p$},
display columns/3/.style={select equal part entry of={1}{3},column type = {|l|},column name=\hfil $n^{(1)}_{\bm p}$},
display columns/4/.style={select equal part entry of={2}{3},column type = {|l},column name=\hfil $\bm p$},
display columns/5/.style={select equal part entry of={2}{3},column type = {|l|},column name=\hfil $n^{(1)}_{\bm p}$},
string type]
{
A ] B

d[{11, 4, 2, 1}]]353758784
d[{11, 4, 2, 2}]]80961860336992
d[{11, 4, 3, 0}]]-188206400
d[{11, 4, 3, 1}]]2116485979904
d[{11, 4, 3, 2}]]206208348590980608
d[{11, 4, 4, 0}]]-372907450720
d[{11, 4, 4, 1}]]1335576781170688
d[{11, 5, 1, 1}]]286720
d[{11, 5, 2, 0}]]-6904192
d[{11, 5, 2, 1}]]156514373632
d[{11, 5, 2, 2}]]20022577101423360
d[{11, 5, 3, 0}]]-114273286144
d[{11, 5, 3, 1}]]469840676702208
d[{11, 5, 4, 0}]]-100934478032512
d[{11, 6, 1, 0}]]-320
d[{11, 6, 1, 1}]]-734128384
d[{11, 6, 2, 0}]]-2397896960
d[{11, 6, 2, 1}]]13324886863872
d[{11, 6, 3, 0}]]-13264263786752
d[{11, 7, 1, 0}]]-631296
d[{11, 7, 1, 1}]]-93294921728
d[{11, 7, 2, 0}]]-154245226112
d[{11, 8, 1, 0}]]-38443520
d[{12, 2, 2, 2}]]-10272
d[{12, 3, 2, 1}]]-320
d[{12, 3, 2, 2}]]1812768672
d[{12, 3, 3, 1}]]54265088
d[{12, 3, 3, 2}]]43026117967872
d[{12, 4, 2, 1}]]9983872
d[{12, 4, 2, 2}]]10558546637760
d[{12, 4, 3, 0}]]-4386240
d[{12, 4, 3, 1}]]286009826304
d[{12, 4, 4, 0}]]-45623190048
d[{12, 5, 1, 1}]]10496
d[{12, 5, 2, 0}]]-69568
d[{12, 5, 2, 1}]]19566085376
d[{12, 5, 3, 0}]]-12801118336
d[{12, 6, 1, 1}]]-25295744
d[{12, 6, 2, 0}]]-188206960
d[{12, 7, 1, 0}]]-10112
d[{13, 3, 2, 2}]]2765056
d[{13, 3, 3, 1}]]112640
d[{13, 4, 2, 1}]]20096
d[{13, 4, 3, 0}]]-10112

}
}
\end{adjustbox}
\vskip10pt
\caption*{\tref{tab:Instanton_Numbers_4_68_genus1} continued.}
\end{table}

\begin{table}[H]
\centering
\begin{adjustbox}{width=\columnwidth,center}
\tiny{\pgfplotstabletypeset[
col sep=space,
white space chars={]},
ignore chars={\ },
every head row/.style={before row=\hline,after row=\hline\hline},
every first row/.style={before row=\vrule height12pt width0pt depth0pt},
every last row/.style={before row=\vrule height0pt width0pt depth6pt,after row=\hline},
%after row=\hline,    %  Uncomment this to get back lines between all rows.
columns={A,B,A,B},
display columns/0/.style={select equal part entry of={0}{2},column type = {|l},column name=\vrule height12pt width0pt depth6pt \hfil $\bm p$},
display columns/1/.style={select equal part entry of={0}{2},column type = {|l|},column name=\hfil $n^{(1)}_{\bm p}$},
display columns/2/.style={select equal part entry of={1}{2},column type = {|l},column name=\hfil $\bm p$},
display columns/3/.style={select equal part entry of={1}{2},column type = {|l|},column name=\hfil $n^{(1)}_{\bm p}$},
string type]
{
A ] B
{3, 3}]1475
{4, 3}]29350
{4, 4}]2669500
{5, 3}]148525
{5, 4}]46911250
{6, 3}]250550
{5, 5}]2311178040
{6, 4}]303610050
{7, 3}]148525
{6, 5}]38756326500
{7, 4}]882636150
{8, 3}]29350
{6, 6}]1477879258975
{7, 5}]298784327925
{8, 4}]1249719025
{9, 3}]1475
{7, 6}]24724246516200
{8, 5}]1207298050100
{9, 4}]882636150
{7, 7}]824125289385950
{8, 6}]217335663077200
{9, 5}]2731112702750
{10, 4}]303610050
{8, 7}]13948250904141600
{9, 6}]1103600201154950
{10, 5}]3573290410020
{11, 4}]46911250
{8, 8}]428216622053327300
{9, 7}]135767820281303350
{10, 6}]3417167213249325
{11, 5}]2731112702750
{12, 4}]2669500
{9, 8}]7360276988409757150
{10, 7}]817523002761866550
{11, 6}]6658383337394000
{12, 5}]1207298050100
{13, 4}]29350
{9, 9}]213796802016132371125
{10, 8}]77723709111160034550
{11, 7}]3186381984770132650
{12, 6}]8301844531611000
{13, 5}]298784327925
{10, 9}]3733143718641168532250
{11, 8}]534623718661493240750
{12, 7}]8273823575633968400
{13, 6}]6658383337394000
{14, 5}]38756326500
{10, 10}]104214421442680518762070
{11, 9}]42137416928528774899500
{12, 8}]2489528873573792774625

}
}
\end{adjustbox}
\vskip10pt
\def\arraystretch{1.2}
\capt{6.5in}{tab:Instanton_Numbers_2_52_genus1}{The genus~1 instanton numbers of total degree $\leqslant 37$ for the family \eqref{eq:CICY_2_52}. The numbers not in this list are either zero, or given by those in the table after permuting indices.}
\end{table}
\begin{table}[H]\notag
\centering
\def\arraystretch{1.2}
\begin{adjustbox}{width=\columnwidth,center}
\tiny{\pgfplotstabletypeset[
col sep=space,
white space chars={]},
ignore chars={\ },
every head row/.style={before row=\hline,after row=\hline\hline},
every first row/.style={before row=\vrule height12pt width0pt depth0pt},
every last row/.style={before row=\vrule height0pt width0pt depth6pt,after row=\hline},
%after row=\hline,    %  Uncomment this to get back lines between all rows.
columns={A,B,A,B},
display columns/0/.style={select equal part entry of={0}{2},column type = {|l},column name=\vrule height12pt width0pt depth6pt \hfil $\bm p$},
display columns/1/.style={select equal part entry of={0}{2},column type = {|l|},column name=\hfil $n^{(1)}_{\bm p}$},
display columns/2/.style={select equal part entry of={1}{2},column type = {|l},column name=\hfil $\bm p$},
display columns/3/.style={select equal part entry of={1}{2},column type = {|l|},column name=\hfil $n^{(1)}_{\bm p}$},
string type]
{
A ] B

{13, 7}]14571606313456936800
{14, 6}]3417167213249325
{15, 5}]2311178040
{11, 10}]1846950755735426212125500
{12, 9}]322658009930052499145200
{13, 8}]8054854261423242104500
{14, 7}]17578605828858033550
{15, 6}]1103600201154950
{16, 5}]46911250
{11, 11}]50041665253951501461197500
{12, 10}]22037047145009664053322650
{13, 9}]1732761485763012342489325
{14, 8}]18428114211388322399550
{15, 7}]14571606313456936800
{16, 6}]217335663077200
{17, 5}]148525
{12, 11}]899012021570846648502276300
{13, 10}]184364061855133820169125300
{14, 9}]6680798279094720126093800
{15, 8}]30150777878498691717250
{16, 7}]8273823575633968400
{17, 6}]24724246516200
{12, 12}]23797576472047430629503926275
{13, 11}]11242357248165502750651190625
{14, 10}]1113322592174963231485326725
{15, 9}]18802673258937852338243175
{16, 8}]35503691126837007672225
{17, 7}]3186381984770132650
{18, 6}]1477879258975
{13, 12}]432820127858166659059675434050
{14, 11}]101324272018566859752134278000
{15, 10}]4956305470261096852879492520
{16, 9}]39078728637782455132498000
{17, 8}]30150777878498691717250
{18, 7}]817523002761866550
{19, 6}]38756326500
{13, 13}]11245532977494243857131235680275
{14, 12}]5634805973327046211368971793400
{15, 11}]675189575619716190356450822375
{16, 10}]16520552027852727161441204150
{17, 9}]60438328652210496631088500
{18, 8}]18428114211388322399550
{19, 7}]135767820281303350
{20, 6}]303610050
{14, 13}]206807254359453225476163905375300
{15, 12}]54117569679383890033269008256800
{16, 11}]3390671418197231819920137904000
{17, 10}]41701179917420632714678892850
{18, 9}]69860383953641753591175350
{19, 8}]8054854261423242104500
{20, 7}]13948250904141600
{21, 6}]250550
{14, 14}]5291910264323169346514519871886400
{15, 13}]2787959348321923974606513689960775
{16, 12}]392220183802149653816905782829075
{17, 11}]13018317138588633706128782358825
{18, 10}]80361250809900507464623047700
{19, 9}]60438328652210496631088500
{20, 8}]2489528873573792774625
{21, 7}]824125289385950
{15, 14}]98295022688621627634296681347227000
{16, 13}]28288279315284624001518199057961000
{17, 12}]2182612077022792907314617911510950
{18, 11}]38633710437048284516888316634750
{19, 10}]118872243137920021974764986250
{20, 9}]39078728637782455132498000
{21, 8}]534623718661493240750
{22, 7}]24724246516200
{15, 15}]2483451908782029978483441284953085840
{16, 14}]1366181074996754720323336949015132850
{17, 13}]220457095613704667909688795083550075
{18, 12}]9451168057601953137717939435446150
{19, 11}]89342683428580923415751095264250
{20, 10}]135397354308622811768958599380
{21, 9}]18802673258937852338243175
{22, 8}]77723709111160034550
{23, 7}]298784327925
{16, 15}]46546285854998069136664551148160029750
{17, 14}]14543614745639013135983680186883590850
{18, 13}]1340566237649146039899848536120642500
{19, 12}]32177823874685392221678086500438300
{20, 11}]161956900412845480278959039133500
{21, 10}]118872243137920021974764986250
{22, 9}]6680798279094720126093800
{23, 8}]7360276988409757150
{24, 7}]882636150
{16, 16}]1163420080671892184401603599755019719900
{17, 15}]664598046187475942195256766297105189925
{18, 14}]120763194531086741850020681429354372600
{19, 13}]6439973193384613124797926096685847700
{20, 12}]86828419567453882491964244521715475
{21, 11}]231063040271162785144114394861625
{22, 10}]80361250809900507464623047700
{23, 9}]1732761485763012342489325
{24, 8}]428216622053327300
{25, 7}]148525
{17, 16}]21983863245711864916146007548819920783200
{18, 15}]7380688961068662640476640993756532397450
{19, 14}]793510861046493220781342058077180934050
{20, 13}]24681216007215096706670234574240357500
{21, 12}]186826323241756428747689323102774900
{22, 11}]260054031226217864291042886267450
{23, 10}]41701179917420632714678892850
{24, 9}]322658009930052499145200
{25, 8}]13948250904141600
{17, 17}]544441278609756260678075733228688830971900
{18, 16}]321496661014907630133282804314901041773300
{19, 15}]64811223550225163381933657270162177991375
{20, 14}]4173727947710375348427686143589666021850
{21, 13}]76049417732402225585766755889537693250
{22, 12}]321974061113618057538097390567433500
{23, 11}]231063040271162785144114394861625

}
}
\end{adjustbox}
\vskip10pt
\caption*{\tref{tab:Instanton_Numbers_2_52_genus1} continued.}
\end{table}
\begin{table}[H]\notag
\centering
\def\arraystretch{1.05}
\begin{adjustbox}{width=\columnwidth,center}
\tiny{\pgfplotstabletypeset[
col sep=space,
white space chars={]},
ignore chars={\ },
every head row/.style={before row=\hline,after row=\hline\hline},
every first row/.style={before row=\vrule height12pt width0pt depth0pt},
every last row/.style={before row=\vrule height0pt width0pt depth6pt,after row=\hline},
%after row=\hline,    %  Uncomment this to get back lines between all rows.
columns={A,B},
display columns/0/.style={select equal part entry of={0}{1},column type = {|l},column name=\vrule height12pt width0pt depth6pt \hfil $\bm p$},
display columns/1/.style={select equal part entry of={0}{1},column type = {|l|},column name=\hfil $n^{(1)}_{\bm p}$},
string type]
{
A ] B

{24, 10}]16520552027852727161441204150
{25, 9}]42137416928528774899500
{26, 8}]217335663077200
{18, 17}]10364068253575077523538680193754837859801100
{19, 16}]3707131738581515278187095373395684544431750
{20, 15}]455990757375323038689152386845291780997980
{21, 14}]17736215361969271597308350645891203085550
{22, 13}]189541398071564094191656020501685749700
{23, 12}]445795478920394854284300481376874700
{24, 11}]161956900412845480278959039133500
{25, 10}]4956305470261096852879492520
{26, 9}]3733143718641168532250
{27, 8}]1207298050100
{18, 18}]254628881812169884863144737907071363850384275
{19, 17}]154848648177340961815627818425036135998755750
{20, 16}]34212563460249368657994755599372244965731625
{21, 15}]2598107539139344864070481843220029434149400
{22, 14}]61346528057542695552753204264092617938350
{23, 13}]383896440392157077525881068439189823375
{24, 12}]496770523785734165693381637859440175
{25, 11}]89342683428580923415751095264250
{26, 10}]1113322592174963231485326725
{27, 9}]213796802016132371125
{28, 8}]1249719025
{19, 18}]4879913180567429516866248000617525315434124200
{20, 17}]1846594387973758522093973511561276425586720500
{21, 16}]255797150291719302931481295802475608460495500
{22, 15}]12092961029767718697960992221193131180072800
{23, 14}]173738502202568390450523304606890631973000
{24, 13}]634051294393795149147968219347562400500
{25, 12}]445795478920394854284300481376874700
{26, 11}]38633710437048284516888316634750
{27, 10}]184364061855133820169125300
{28, 9}]7360276988409757150
{29, 8}]29350

}
}
\end{adjustbox}
\vskip10pt
\caption*{\tref{tab:Instanton_Numbers_2_52_genus1} continued.}
\end{table}

\section{Discriminant loci and Yukawa couplings}\label{sect:Miroirs_DiscandYukawa}
We do not give genus one numbers for six and seven parameter families \eqref{eq:CICY_6_36}, \eqref{eq:CICY_7_27}. Performing this computation would be possible if we knew the relevant discriminants $\Delta$, which we do not. To make progress on this, we compute a number of discriminants for other models, in the hopes of recognising a formula that reproduces the known examples. This can be done for a CICY matrix with identical columns, as in this case $\Delta$ takes a factored form as for the two families above.

To compute a model's discriminant, we compute the Yukawa couplings. These are rational functions (in the coordinates defined in Appendix \sref{sect:Appendix_coordinates}), and the discriminant can be identified in the denominators. This process can be run over all two-parameter CICYs, and we also do this for a number of three-parameter CICYs. However, this has not yet led to any good candidates for the models \eqref{eq:CICY_6_36}, \eqref{eq:CICY_7_27}.

The Yukawa couplings $C_{ijk}$ can be computed from the formula
\begin{equation}\label{eq:Yukawas}
\begin{aligned}
C_{ijk}&=\frac{\partial}{\partial\varphi^{i}}\frac{\partial}{\partial\varphi^{j}}\frac{\partial}{\partial\varphi^{k}}\cF^{(0)}=-\Pi^{T}\cdot\Sigma\cdot\frac{\partial}{\partial\varphi^{i}}\frac{\partial}{\partial\varphi^{j}}\frac{\partial}{\partial\varphi^{k}}\Pi
&=-\varpi^{T}\cdot\sigma\cdot\frac{\partial}{\partial\varphi^{i}}\frac{\partial}{\partial\varphi^{j}}\frac{\partial}{\partial\varphi^{k}}\varpi~.
\end{aligned}
\end{equation}
We give again here for ease of reading the matrix $\Sigma$, and also the matrix $\sigma$:
\begin{equation}
\Sigma=\begin{pmatrix}
\+0&\one_{m+1}\\-\one_{m+1}&0
\end{pmatrix}~,\qquad \sigma=\begin{pmatrix}
0&\+0&0&-1\\
0&\+0&\one_{m}&\+0\\
0&-\one_{m}&0&\+0\\
1&\+0&0&\+0
\end{pmatrix}~,\qquad m=h^{2,1}(X)~.
\end{equation}
These functions $C_{ijk}$ are rational function of $\varphi^{i}$, and their denominators all contain a factor of $\Delta$, which equals the polynomial least common multiple of the denominators of $C_{ijk}$. In practice formula \eqref{eq:Yukawas} serves to compute $C_{ijk}$ as power series in $\varpi^{i}$, and one can fit a rational function to this. If the discriminant $\Delta$ is known, then one can form the power series of $\varphi^{i}\varphi^{j}\varphi^{k}\cdot\Delta \cdot C_{ijk}$ to find a polynomial, the numerator of $C_{ijk}$. If $\Delta$ is not known then one must fit a rational function to $C_{ijk}$, which becomes prohibitively complicated at large numbers of parameters. A more detailed discussion of the Yukawa couplings can be found in \cite{CoxKatz,Candelas:1987se}. 

We remark that a rule holds for CICYs with identical columns $(d^{(1)},\,d^{(2)},\,...\,,d^{(k)})^{T}$, so of the form
\begin{equation}\label{eq:simples}
\cicy{\IP^{n_{1}}\\\vdots\\\IP^{n_{k}}}{d^{(1)}&...&d^{(1)}\\\vdots&...&\vdots\\ d^{(k)}&...&d^{(k)}}.
\end{equation}
This rule is
\begin{equation}
\Delta=\prod_{j_{1}=1}^{n_{1}+1}\cdot\cdot\cdot\prod_{j_{k}=1}^{n_{k}+1}\left(1-\sum_{i=1}^{k}d^{(i)}\exp\left(\frac{2\pi\ii\,j_{i}}{n_{i}+1}\right)\varphi_{i}^{\frac{1}{n_{i}+1}}\right)~.
\end{equation}
This does not give the discriminant of the families \eqref{eq:simples}, but of their mirrors. We discuss the parametrisation of the mirror manifolds that we are employing in Appendix \sref{sect:Appendix_coordinates}. We verify that this holds for all CICY configurations of the form \eqref{eq:simples}, so the following 11 families:
\begin{equation}
\begin{gathered}
\cicy{\IP^{4}}{5}~,\qquad\cicy{\IP^{5}}{3&3}~,\qquad \cicy{\IP^{7}}{2&2&2&2}~,
\\[20pt]
\cicy{\IP^{4}\\\IP^{4}}{1&1&1&1&1\\1&1&1&1&1}~,\qquad \cicy{\IP^{2}\\\IP^{2}}{3\\3}~, \qquad \cicy{\IP^{1}\\\IP^{3}}{2\\4}~,
\\[10pt]
\cicy{\IP^{2}\\\IP^{2}\\\IP^{2}}{1&1&1\\1&1&1\\1&1&1}~,\qquad \cicy{\IP^{1}\\\IP^{1}\\\IP^{2}}{2\\2\\3}~,\qquad \cicy{\IP^{1}\\\IP^{1}\\\IP^{3}}{1&1\\1&1\\2&2}~,\qquad \cicy{\IP^{1}\\\IP^{1}\\\IP^{1}\\\IP^{1}}{2\\2\\2\\2}~,\qquad\cicy{\IP^{1}\\\IP^{1}\\\IP^{1}\\\IP^{1}\\\IP^{1}}{1&1\\1&1\\1&1\\1&1\\1&1}~.
\end{gathered}\end{equation}
For the last model above, with five parameters, this discriminant is
\begin{equation}\label{eq:discriminant_HV}
\Delta=\prod_{\epsilon_{i}\in\{1,-1\}}\left(1+\epsilon_{1}\sqrt{\varphi^{1}}+\epsilon_{2}\sqrt{\varphi^{2}}+\epsilon_{3}\sqrt{\varphi^{3}}+\epsilon_{4}\sqrt{\varphi^{4}}+\epsilon_{5}\sqrt{\varphi^{5}}\right)~.
\end{equation}
We proceed to tabulate all CICYs with $h^{1,1}=2$ \cite{LukasWebsite}, together with $\Delta$, which we compute by forming the polynomial least common multiple of the Yukawa coupling denominators \cite{Hosono:1994ax}.

\newpage
\begin{landscape}
\begin{equation}\notag
\tiny{\begin{array}{|l|}
\hline \vrule depth5pt height 10pt width 0pt \text{CICYs with }h^{1,1}=2\text{, with the discriminant polynomial  }\Delta \text{ of the mirror. \cite{mcgovern2023a}} \\
\hline
\hline
\vrule depth5pt height15pt width0pt\cicy{\IP^{2}\\\IP^{5}}{0&0&2&1\\2&2&1&1}\\\vrule depth5pt height15pt width0pt\left(1-4 \varphi _1\right){}^2 \left(1-8 \varphi _1+16 \varphi _1^2-48 \varphi _2-320 \varphi _1 \varphi _2+768 \varphi _2^2+256 \varphi _1 \varphi _2^2-4096 \varphi _2^3\right)\\\hline
\vrule depth5pt height15pt width0pt\cicy{\IP^{4}\\\IP^{4}}{2&0&1&1&1\\0&2&1&1&1}\\\vrule depth5pt height15pt width0pt\left(1-4 \varphi _1\right) \left(1-4 \varphi _2\right) \left(1-12 \varphi _1+48 \varphi _1^2-64 \varphi _1^3-12 \varphi _2-336 \varphi _1 \varphi _2-192 \varphi _1^2 \varphi _2+48 \varphi _2^2-192 \varphi _1 \varphi _2^2-64 \varphi _2^3\right)\\\hline
\vrule depth5pt height15pt width0pt\cicy{\IP^{2}\\\IP^{4}}{0&2&1\\3&1&1}\\\vrule depth5pt height15pt width0pt\left(1-4 \varphi _1\right){}^2 \left(1-8 \varphi _1+16 \varphi _1^2-81 \varphi _2-540 \varphi _1 \varphi _2+2187 \varphi _2^2+729 \varphi _1 \varphi _2^2-19683 \varphi _2^3\right)\\\hline
\vrule depth5pt height15pt width0pt\cicy{\IP^{2}\\\IP^{6}}{0&0&1&1&1\\2&2&1&1&1}\\\vrule depth5pt height15pt width0pt\left(1-\varphi _1\right){}^2 \left(1-3 \varphi _1+3 \varphi _1^2-\varphi _1^3-48 \varphi _2-336 \varphi _1 \varphi _2-48 \varphi _1^2 \varphi _2+768 \varphi _2^2-768 \varphi _1 \varphi _2^2-4096 \varphi _2^3\right)\\\hline
\vrule depth5pt height15pt width0pt\cicy{\IP^{3}\\\IP^{5}}{0&1&1&1&1\\2&1&1&1&1}\\\vrule depth5pt height15pt width0pt\left(1-\varphi _1\right) \left(1-4 \varphi _1+6 \varphi _1^2-4 \varphi _1^3+\varphi _1^4-16 \varphi _2-496 \varphi _1 \varphi _2-496 \varphi _1^2 \varphi _2-16 \varphi _1^3 \varphi _2+96 \varphi _2^2-1984 \varphi _1 \varphi _2^2+96 \varphi _1^2 \varphi _2^2-256 \varphi _2^3-256 \varphi _1 \varphi _2^3+256 \varphi _2^4\right)\\\hline
\vrule depth5pt height15pt width0pt\cicy{\IP^{2}\\\IP^{4}}{0&2&1\\2&1&2}\\\vrule depth5pt height15pt width0pt\left(1-4 \varphi _1\right) \left(1-12 \varphi _1+48 \varphi _1^2-64 \varphi _1^3-48 \varphi _2-696 \varphi _1 \varphi _2+96 \varphi _1^2 \varphi _2+768 \varphi _2^2+384 \varphi _1 \varphi _2^2-432 \varphi _1^2 \varphi _2^2-4096 \varphi _2^3\right)\\\hline
\vrule depth5pt height15pt width0pt\cicy{\IP^{3}\\\IP^{4}}{0&2&1&1\\2&1&1&1}\\\vrule depth5pt height15pt width0pt\left(1-4 \varphi _1\right) \left(1-12 \varphi _1+48 \varphi _1^2-64 \varphi _1^3-16 \varphi _2-664 \varphi _1 \varphi _2-544 \varphi _1^2 \varphi _2+96 \varphi _2^2-1280 \varphi _1 \varphi _2^2+16 \varphi _1^2 \varphi _2^2-256 \varphi _2^3+128 \varphi _1 \varphi _2^3+256 \varphi _2^4\right)\\\hline
\vrule depth5pt height15pt width0pt\cicy{\IP^{4}\\\IP^{4}}{1&1&1&1&1\\1&1&1&1&1}\\\vrule depth5pt height15pt width0pt1-5 \varphi _1+10 \varphi _1^2-10 \varphi _1^3+5 \varphi _1^4-\varphi _1^5-5 \varphi _2-605 \varphi _1 \varphi _2-1905 \varphi _1^2 \varphi _2-605 \varphi _1^3 \varphi _2-5 \varphi _1^4 \varphi _2+10 \varphi _2^2-1905 \varphi _1 \varphi _2^2+1905 \varphi _1^2 \varphi _2^2-10 \varphi _1^3 \varphi _2^2-10 \varphi _2^3-605 \varphi _1 \varphi _2^3-10 \varphi _1^2 \varphi _2^3+5 \varphi _2^4-5 \varphi _1 \varphi _2^4-\varphi _2^5\\\hline
\vrule depth5pt height15pt width0pt\cicy{\IP^{3}\\\IP^{3}}{2&1&1\\1&2&1}\\\vrule depth5pt height15pt width0pt1-16 \varphi _1+96 \varphi _1^2-256 \varphi _1^3+256 \varphi _1^4-16 \varphi _2-1261 \varphi _1 \varphi _2-2968 \varphi _1^2 \varphi _2+304 \varphi _1^3 \varphi _2+96 \varphi _2^2-2968 \varphi _1 \varphi _2^2+69 \varphi _1^2 \varphi _2^2-27 \varphi _1^3 \varphi _2^2-256 \varphi _2^3+304 \varphi _1 \varphi _2^3-27 \varphi _1^2 \varphi _2^3+256 \varphi _2^4\\\hline

\end{array}}
\end{equation}
\begin{equation}\notag
\tiny{\begin{array}{|l|}
\hline \vrule depth5pt height 10pt width 0pt \text{CICYs with }h^{1,1}=2\text{, with the discriminant polynomial  }\Delta \text{ of the mirror. \cite{mcgovern2023a}} \\
\hline
\hline
\vrule depth5pt height15pt width0pt\cicy{\IP^{1}\\\IP^{4}}{0&2\\3&2}\\\vrule depth5pt height15pt width0pt\left(1-4 \varphi _1\right){}^2 \left(1-8 \varphi _1+16 \varphi _1^2-216 \varphi _2-864 \varphi _1 \varphi _2+11664 \varphi _2^2\right)\\\hline
\vrule depth5pt height15pt width0pt\cicy{\IP^{2}\\\IP^{5}}{0&1&1&1\\2&2&1&1}\\\vrule depth5pt height15pt width0pt\left(1-\varphi _1\right) \left(1-4 \varphi _1+6 \varphi _1^2-4 \varphi _1^3+\varphi _1^4-48 \varphi _2-664 \varphi _1 \varphi _2-320 \varphi _1^2 \varphi _2+8 \varphi _1^3 \varphi _2+768 \varphi _2^2-2176 \varphi _1 \varphi _2^2+16 \varphi _1^2 \varphi _2^2-4096 \varphi _2^3\right)\\\hline
\vrule depth5pt height15pt width0pt\cicy{\IP^{2}\\\IP^{5}}{0&1&1&1\\3&1&1&1}\\\vrule depth5pt height15pt width0pt\left(1-\varphi _1\right){}^2 \left(1-3 \varphi _1+3 \varphi _1^2-\varphi _1^3-81 \varphi _2-567 \varphi _1 \varphi _2-81 \varphi _1^2 \varphi _2+2187 \varphi _2^2-2187 \varphi _1 \varphi _2^2-19683 \varphi _2^3\right)\\\hline
\vrule depth5pt height15pt width0pt\cicy{\IP^{3}\\\IP^{4}}{1&1&1&1\\2&1&1&1}\\\vrule depth5pt height15pt width0pt1-5 \varphi _1+10 \varphi _1^2-10 \varphi _1^3+5 \varphi _1^4-\varphi _1^5-16 \varphi _2-905 \varphi _1 \varphi _2-1889 \varphi _1^2 \varphi _2-318 \varphi _1^3 \varphi _2+3 \varphi _1^4 \varphi _2+96 \varphi _2^2-4296 \varphi _1 \varphi _2^2+814 \varphi _1^2 \varphi _2^2-3 \varphi _1^3 \varphi _2^2-256 \varphi _2^3-784 \varphi _1 \varphi _2^3+\varphi _1^2 \varphi _2^3+256 \varphi _2^4\\\hline
\vrule depth5pt height15pt width0pt\cicy{\IP^{1}\\\IP^{5}}{0&1&1\\2&2&2}\\\vrule depth5pt height15pt width0pt\left(1-\varphi _1\right) \left(1-4 \varphi _1+6 \varphi _1^2-4 \varphi _1^3+\varphi _1^4-128 \varphi _2-768 \varphi _1 \varphi _2-128 \varphi _1^2 \varphi _2+4096 \varphi _2^2\right)\\\hline
\vrule depth5pt height15pt width0pt\cicy{\IP^{1}\\\IP^{6}}{0&0&1&1\\2&2&2&1}\\\vrule depth5pt height15pt width0pt\left(1-\varphi _1\right){}^2 \left(1-3 \varphi _1+3 \varphi _1^2-\varphi _1^3-128 \varphi _2-320 \varphi _1 \varphi _2+16 \varphi _1^2 \varphi _2+4096 \varphi _2^2\right)\\\hline
\vrule depth5pt height15pt width0pt\cicy{\IP^{1}\\\IP^{7}}{0&0&0&1&1\\2&2&2&1&1}\\\vrule depth5pt height15pt width0pt\left(1-\varphi _1\right){}^3 \left(1-2 \varphi _1+\varphi _1^2-128 \varphi _2-128 \varphi _1 \varphi _2+4096 \varphi _2^2\right)\\\hline
\vrule depth5pt height15pt width0pt\cicy{\IP^{2}\\\IP^{4}}{1&1&1\\2&2&1}\\\vrule depth5pt height15pt width0pt1-5 \varphi _1+10 \varphi _1^2-10 \varphi _1^3+5 \varphi _1^4-\varphi _1^5-48 \varphi _2-1280 \varphi _1 \varphi _2-1672 \varphi _1^2 \varphi _2-124 \varphi _1^3 \varphi _2-\varphi _1^4 \varphi _2+768 \varphi _2^2-5888 \varphi _1 \varphi _2^2-128 \varphi _1^2 \varphi _2^2-4096 \varphi _2^3\\\hline
\vrule depth5pt height15pt width0pt\cicy{\IP^{1}\\\IP^{5}}{0&0&2\\2&2&2}\\\vrule depth5pt height15pt width0pt\left(1-4 \varphi _1\right){}^2 \left(1-8 \varphi _1+16 \varphi _1^2-128 \varphi _2-512 \varphi _1 \varphi _2+4096 \varphi _2^2\right)\\\hline

\end{array}}
\end{equation}
\begin{equation}\notag
\tiny{\begin{array}{|l|}
\hline \vrule depth5pt height 10pt width 0pt \text{CICYs with }h^{1,1}=2\text{, with the discriminant polynomial  }\Delta \text{ of the mirror. \cite{mcgovern2023a}} \\
\hline
\hline
\vrule depth5pt height15pt width0pt\cicy{\IP^{1}\\\IP^{6}}{0&0&0&2\\2&2&2&1}\\\vrule depth5pt height15pt width0pt\left(1-4 \varphi _1\right){}^3 \left(1-4 \varphi _1-128 \varphi _2+4096 \varphi _2^2\right)\\\hline
\vrule depth5pt height15pt width0pt\cicy{\IP^{2}\\\IP^{3}}{2&1\\1&3}\\\vrule depth5pt height15pt width0pt1-16 \varphi _1+96 \varphi _1^2-256 \varphi _1^3+256 \varphi _1^4-81 \varphi _2-1809 \varphi _1 \varphi _2+440 \varphi _1^2 \varphi _2-400 \varphi _1^3 \varphi _2+2187 \varphi _2^2+1458 \varphi _1 \varphi _2^2-3375 \varphi _1^2 \varphi _2^2+3125 \varphi _1^3 \varphi _2^2-19683 \varphi _2^3\\\hline
\vrule depth5pt height15pt width0pt\cicy{\IP^{1}\\\IP^{5}}{0&1&1\\3&2&1}\\\vrule depth5pt height15pt width0pt\left(1-\varphi _1\right){}^2 \left(1-3 \varphi _1+3 \varphi _1^2-\varphi _1^3-216 \varphi _2-540 \varphi _1 \varphi _2+27 \varphi _1^2 \varphi _2+11664 \varphi _2^2\right)\\\hline
\vrule depth5pt height15pt width0pt\cicy{\IP^{2}\\\IP^{3}}{2&1\\2&2}\\\vrule depth5pt height15pt width0pt1-16 \varphi _1+96 \varphi _1^2-256 \varphi _1^3+256 \varphi _1^4-48 \varphi _2-2656 \varphi _1 \varphi _2-5120 \varphi _1^2 \varphi _2+512 \varphi _1^3 \varphi _2+768 \varphi _2^2-8704 \varphi _1 \varphi _2^2+256 \varphi _1^2 \varphi _2^2-4096 \varphi _2^3\\\hline
\vrule depth5pt height15pt width0pt\cicy{\IP^{2}\\\IP^{4}}{0&2&1\\2&2&1}\\\vrule depth5pt height15pt width0pt\left(1-4 \varphi _1\right) \left(1-12 \varphi _1+48 \varphi _1^2-64 \varphi _1^3-48 \varphi _2-1344 \varphi _1 \varphi _2-768 \varphi _1^2 \varphi _2+768 \varphi _2^2-3072 \varphi _1 \varphi _2^2-4096 \varphi _2^3\right)\\\hline
\vrule depth5pt height15pt width0pt\cicy{\IP^{1}\\\IP^{4}}{1&1\\3&2}\\\vrule depth5pt height15pt width0pt1-5 \varphi _1+10 \varphi _1^2-10 \varphi _1^3+5 \varphi _1^4-\varphi _1^5-216 \varphi _2-2052 \varphi _1 \varphi _2-873 \varphi _1^2 \varphi _2+16 \varphi _1^3 \varphi _2+11664 \varphi _2^2\\\hline
\vrule depth5pt height15pt width0pt\cicy{\IP^{3}\\\IP^{3}}{2&1&1\\2&1&1}\\\vrule depth5pt height15pt width0pt1-16 \varphi _1+96 \varphi _1^2-256 \varphi _1^3+256 \varphi _1^4-16 \varphi _2-1984 \varphi _1 \varphi _2-7936 \varphi _1^2 \varphi _2-1024 \varphi _1^3 \varphi _2+96 \varphi _2^2-7936 \varphi _1 \varphi _2^2+1536 \varphi _1^2 \varphi _2^2-256 \varphi _2^3-1024 \varphi _1 \varphi _2^3+256 \varphi _2^4\\\hline
\vrule depth5pt height15pt width0pt\cicy{\IP^{1}\\\IP^{6}}{0&0&1&1\\3&2&1&1}\\\vrule depth5pt height15pt width0pt\left(1-\varphi _1\right){}^3 \left(1-2 \varphi _1+\varphi _1^2-216 \varphi _2-216 \varphi _1 \varphi _2+11664 \varphi _2^2\right)\\\hline
\vrule depth5pt height15pt width0pt\cicy{\IP^{2}\\\IP^{4}}{1&1&1\\3&1&1}\\\vrule depth5pt height15pt width0pt1-5 \varphi _1+10 \varphi _1^2-10 \varphi _1^3+5 \varphi _1^4-\varphi _1^5-81 \varphi _2-1674 \varphi _1 \varphi _2-1417 \varphi _1^2 \varphi _2+55 \varphi _1^3 \varphi _2-8 \varphi _1^4 \varphi _2+2187 \varphi _2^2-10206 \varphi _1 \varphi _2^2+108 \varphi _1^2 \varphi _2^2-16 \varphi _1^3 \varphi _2^2-19683 \varphi _2^3\\\hline

\end{array}}
\end{equation}
\begin{equation}\notag
\tiny{\begin{array}{|l|}
\hline \vrule depth5pt height 10pt width 0pt \text{CICYs with }h^{1,1}=2\text{, with the discriminant polynomial  }\Delta \text{ of the mirror. \cite{mcgovern2023a}} \\
\hline
\hline
\vrule depth5pt height15pt width0pt\cicy{\IP^{1}\\\IP^{5}}{0&0&2\\3&2&1}\\\vrule depth5pt height15pt width0pt\left(1-4 \varphi _1\right){}^3 \left(1-4 \varphi _1-216 \varphi _2+11664 \varphi _2^2\right)\\\hline
\vrule depth5pt height15pt width0pt\cicy{\IP^{1}\\\IP^{5}}{0&1&1\\2&3&1}\\\vrule depth5pt height15pt width0pt\left(1-\varphi _1\right) \left(1-4 \varphi _1+6 \varphi _1^2-4 \varphi _1^3+\varphi _1^4-216 \varphi _2-864 \varphi _1 \varphi _2+72 \varphi _1^2 \varphi _2-16 \varphi _1^3 \varphi _2+11664 \varphi _2^2\right)\\\hline
\vrule depth5pt height15pt width0pt\cicy{\IP^{1}\\\IP^{4}}{0&2\\2&3}\\\vrule depth5pt height15pt width0pt\left(1-4 \varphi _1\right) \left(1-12 \varphi _1+48 \varphi _1^2-64 \varphi _1^3-216 \varphi _2-2592 \varphi _1 \varphi _2+11664 \varphi _2^2\right)\\\hline
\vrule depth5pt height15pt width0pt\cicy{\IP^{2}\\\IP^{3}}{2&1\\3&1}\\\vrule depth5pt height15pt width0pt1-16 \varphi _1+96 \varphi _1^2-256 \varphi _1^3+256 \varphi _1^4-81 \varphi _2-4968 \varphi _1 \varphi _2-11664 \varphi _1^2 \varphi _2+256 \varphi _1^3 \varphi _2+2187 \varphi _2^2-31347 \varphi _1 \varphi _2^2-19683 \varphi _2^3\\\hline
\vrule depth5pt height15pt width0pt\cicy{\IP^{2}\\\IP^{2}}{3\\3}\\\vrule depth5pt height15pt width0pt1-81 \varphi _1+2187 \varphi _1^2-19683 \varphi _1^3-81 \varphi _2-15309 \varphi _1 \varphi _2-59049 \varphi _1^2 \varphi _2+2187 \varphi _2^2-59049 \varphi _1 \varphi _2^2-19683 \varphi _2^3\\\hline
\vrule depth5pt height15pt width0pt\cicy{\IP^{1}\\\IP^{4}}{1&1\\4&1}\\\vrule depth5pt height15pt width0pt1-5 \varphi _1+10 \varphi _1^2-10 \varphi _1^3+5 \varphi _1^4-\varphi _1^5-512 \varphi _2-2816 \varphi _1 \varphi _2+320 \varphi _1^2 \varphi _2-144 \varphi _1^3 \varphi _2+27 \varphi _1^4 \varphi _2+65536 \varphi _2^2\\\hline
\vrule depth5pt height15pt width0pt\cicy{\IP^{1}\\\IP^{5}}{0&1&1\\4&1&1}\\\vrule depth5pt height15pt width0pt\left(1-\varphi _1\right){}^3 \left(1-2 \varphi _1+\varphi _1^2-512 \varphi _2-512 \varphi _1 \varphi _2+65536 \varphi _2^2\right)\\\hline
\vrule depth5pt height15pt width0pt\cicy{\IP^{1}\\\IP^{3}}{2\\4}\\\vrule depth5pt height15pt width0pt1-16 \varphi _1+96 \varphi _1^2-256 \varphi _1^3+256 \varphi _1^4-512 \varphi _2-12288 \varphi _1 \varphi _2-8192 \varphi _1^2 \varphi _2+65536 \varphi _2^2\\\hline
\vrule depth5pt height15pt width0pt\cicy{\IP^{1}\\\IP^{4}}{0&2\\4&1}\\\vrule depth5pt height15pt width0pt\left(1-4 \varphi _1\right){}^3 \left(1-4 \varphi _1-512 \varphi _2+65536 \varphi _2^2\right)\\\hline

\end{array}}
\end{equation}
\end{landscape}

\section{Contractions and instanton summations}\label{sect:Miroirs_Contractions}
Upon inspecting the CICY matrices, one can see that the (7,27) model\footnote{It is clearest here for us to refer to these manifolds by their Hodge numbers $(h^{1,1},h^{2,1})$. The configurations for the families referred to in this paragraph can be found in formulae \eqref{eq:CICY_7_27}, \eqref{eq:CICY_6_36}, \eqref{eq:CICY_MHV}, \eqref{eq:CICY_4_68}.} is a split of the (6,36) model, which is itself a split of the (5,45) Mirror Hulek Verrill family. Moreover, this (5,45) family is a split of the Tetraquadric with Hodge numbers (4,68). 

Further still, a sequence of five splits from the (2,52) family\footnote{See the configuration \eqref{eq:CICY_2_52}.} yields the (7,27) family. Finally, the (2,52) family is a split of the Quintic $\cicy{\IP^{4}}{5}^{(1,101)}_{-200}$. 

To pass from a manifold $Y$ to its split $\wt{Y}$ involves blowing up some number of exceptional divisors, which are degree-1 rational curves (so genus 0). Now consider a rational curve $\wt{C}$ on the split manifold. Corresponding to this curve is a degree vector $\wt{\textbf{b}}$, giving the curve's homology class. In blowing down, we project out one of the homology classes on $\wt{Y}$. The image of $\wt{C}$ in $Y$ is a rational curve $C$ with degree vector $\textbf{b}$ obtained from $\wt{\mathbf{b}}$ by deleting the entries corresponding to the homology class that was projected out. Let $\rho$ denote the projection operator on degree vectors that deletes the entry corresponding to the projected-out homology class. We expect to be able to recover the instanton numbers $n^{(0)}_{\mathbf{b}}$ for $Y$ from the numbers $n^{(0)}_{\wt{\mathbf{b}}}$ for $\wt{Y}$ via a sum rule
\begin{equation}\label{eq:sumrule}
n^{(0)}_{\mathbf{b}}=\sum_{\wt{\mathbf{b}}|\,\rho\left(\wt{\mathbf{b}}\right)=\mathbf{b}}n^{(0)}_{\wt{\mathbf{b}}}~.
\end{equation}

This can also be seen at the level of the Calabi-Yau genus prepotential $F$, which is closely related to the topological string free energy $F_{0}$, but incorporates the full set of topological quantities $Y_{abc}$ as follows:
\begin{equation}
F=-\frac{1}{6}Y_{abc}t^{a}t^{b}t^{c}-\frac{1}{(2\pi\ii)^{3}}\sum_{\mathbf{b}\geq0}n^{(0)}_{\mathbf{b}}\,\text{Li}_{3}\left(\mathbf{q}^{\mathbf{b}}\right)~.
\end{equation}
Reiterating, $a,\,b,\,c$ run from $0$ to $h^{1,1}(Y)$ while $i,\,j,\,k$ run from $1$ to $h^{1,1}(Y)$. The $t^{i}$ give coefficients in the expansion of $Y$'s complexified K{\"a}hler form in a basis of $H^{2}(Y,\IZ)$. We have said this again in order to set the stage for shrinking two-cycles, a birational map between threefolds realised by sending one of the $t^{i}$ to zero.

Let $Y$ have $h^{1,1}=n$. Take $\wt{Y}$ with $h^{1,1}=n+1$ to be a split of $Y$. The coordinates $t^{i}$, $i=1,...,n+1$ give the complexified K{\"a}hler structure parameters on $\wt{Y}$, which equal the integral of the complexified K{\"a}hler form\footnote{$J$ is the K{\"a}hler form and $B$ is the Kalb-Ramond B-field.} $B+\ii J$ of two-cycles in a basis of $H_{2}(\wt{Y},\IZ)$. Let $t^{n+1}$ be the integral of $B+\ii J$ over the two-cycle wrapped by the collapsing two-spheres. Then we should have
\begin{equation}\label{eq:contraction_derivation}
F_{Y}(t^{1},...,t^{n})=F_{\wt{Y}}(t^{1},...,t^{n},t^{n+1})\bigg\vert_{t^{n+1}=0}~.
\end{equation}
Inspecting \eqref{eq:A_FE_0}, setting $t_{n+1}=0$, and comparing the $q$-expansions of both sides, we recover the sum rule for all degree vectors $\mathbf{b}$ except the zero vector. Equating the terms linear, quadratic, and cubic in $t^{i}$ we get equalities
\begin{equation}
\wt{Y}_{abc}=Y_{abc}~,\qquad0\leq a,b,c\leq n \text{ and }(a,b,c)\neq(0,0,0)~,
\end{equation}
with the left hand side giving topological numbers $\wt{Y}_{abc}$ on $\wt{Y}$ and the right hand side giving the numbers $Y_{abc}$ for $Y$.

The exception to the sum rule is related to the change in Euler characteristic. In light of $\text{Li}_{3}(1)=\zeta(3)$, equating constant terms in \eqref{eq:contraction_derivation} yields
\begin{equation}
\sum_{k}n^{(0)}_{0,...,0,k}=-\frac{1}{2}\bigg[\chi\left(\wt{Y}\right)-\chi\left(Y\right)\bigg]~.
\end{equation}
This formula could have been anticipated more geometrically, as the left hand side counts the curves that are blown down in the birational map $\wt{Y}\mapsto Y$. Since $\chi\left(\IP^{1}\right)=2$, minus twice the above left hand side gives the change in Euler characteristic.

This can be verified for each pair of CICYs where one splits the other, but other examples can be found for suitable $\wt{Y}$. For instance, the genus 0 invariants for the families
\begin{equation}
\cicy{\IP^{3}\\\IP^{3}}{1&1&2\\1&1&2}~,\qquad\cicy{\IP^{3}\\\IP^{3}}{1&2&1\\1&1&2}~,\qquad\cicy{\IP^{1}\\\IP^{3}}{2\\4}~,
\end{equation}
can be summed in this manner to recover invariants for the degree 8 hypersurface in the weighted projective space $\IW\IP^{4}_{(1,1,1,1,4)}$. This is mirror to a manifold with Picard-Fuchs operator labelled AESZ7 in the database \cite{AESZ-db}, a hypergeometric model with indices $\left(\frac{1}{8}~,\,\frac{3}{8}~,\,\frac{5}{8}~,\,\frac{7}{8}\right)$. This family has also seen recent attention in \cite{Bourjaily:2019hmc}, wherein three-loop ``wheel" Feynman diagrams were evaluated in terms of Calabi-Yau periods.

In \cite{Hori:2016txh}, a two-parameter non-Abelian gauged linear sigma model (GLSM) \cite{Witten:1993yc} with gauge group
\begin{equation}
G=\frac{U(1)\times U(1)\times O(2)}{\{\pm1\}\times\{\pm1\}\times\{\pm\one_{2}\}}
\end{equation}
was studied. This model has six phases, and of relevance to our discussion is their phase $\text{I}_{+}$ and phase IV. These were both geometric, in the sense that the GLSM flowed to a nonlinear sigma model on Calabi-Yau spaces with Hodge numbers $(h^{1,1},h^{2,1})=(2,24)$. The phase $\text{I}_{+}$ model was a quotient of a complete intersection in a toric variety, while the phase IV geometry was a determinental hypersurface in another toric variety. We remark that the phase $\text{I}_{+}$ geometry can be contracted in either of two ways, and the instanton sums recover the genus-0 invariants and Euler characteristic of either the Reye congruence \cite{hosono2014mirror,hosono2016double} or a $\IZ_{2}$ quotient of the quadriconic in $\IP^{7}$:
\begin{equation}
\cicy{\IP^{4}\\\IP^{4}}{1&1&1&1&1\\1&1&1&1&1}_{/\IZ_{2}}~,\qquad \IP^{7}[2,2,2,2]_{/\IZ_{2}}~.
\end{equation}

On the other hand, instanton number summations suggest that the phase IV geometry can be contracted to the intersection of two degree four hypersurfaces in $\IW\IP^{5}_{(1,1,1,1,2,2)}$. This manifold is mirror to a geometry with Picard-Fuchs operator AESZ10, a hypergeometric operator with indices $\left(\frac{1}{4}~,\,\frac{1}{4}~,\,\frac{3}{4}~,\,\frac{3}{4}\right)$.

We will briefly mention, but say no more about, the question of connecting all Calabi-Yau threefolds by such transitions (including generalisations, such as where instead of two-spheres, orbifolds shrink), which in the mathematics literature relates to a conjecture of Reid \cite{Reid1987}, see also  \cite{gross1997primitive}~. The conifold transition in string theory was studied in \cite{Candelas:1989ug,Candelas:1989js,Candelas:1988di}. It was shown in \cite{Greene:1995hu} that these transitions could be explained by black hole condensation.

Studying the prepotential $F$ makes clear that the genus 0 instanton numbers on the split manifold can be summed into the genus 0 numbers of the contracted manifold, with the topological quantities and change in Euler character accounted for. At higher genera, setting a $t^{k}$ to zero reproduces the same summation rules for higher genus instanton numbers. So the rule \eqref{eq:sumrule} also holds at higher genus:
\begin{equation}
n^{(g)}_{\mathbf{b}}=\sum_{\wt{\mathbf{b}}|\,\rho\left(\wt{\mathbf{b}}\right)=\mathbf{b}}n^{(g)}_{\wt{\mathbf{b}}}~.
\end{equation}
Such data can serve as useful boundary data in higher genus free energy computations, as was noted in \cite{Klemm:2004km}.

\section{Coxeter groups}\label{sect:Miroirs_Coxeter}
When the tables of genus 0 and 1 instanton numbers in \cite{Candelas:2021lkc} were produced, a number of repetitions of values were noted. A similar phenomenon can be observed in our tables \tref{tab:Instanton_Numbers_4_68} and \tref{tab:Instanton_Numbers_4_68_genus1} for the Tetraquadric. For example, we have equalities
\begin{equation}
\begin{aligned}
n^{(0)}_{(1,1,1,0)}=n^{(0)}_{(6,1,1,0)}=n^{(0)}_{(10,3,1,1)}&=2432\\
n^{(0)}_{(3,3,1,1)}=n^{(0)}_{(7,3,1,1)}=n^{(0)}_{(13,3,3,1)}&=241754112\\
n^{(1)}_{(4,2,1,0)}=n^{(1)}_{(7,2,2,0)}=n^{(1)}_{(8,4,1,0)}&=-96\\
n^{(1)}_{(3,2,2,2)}=n^{(1)}_{(9,2,2,2)}=n^{(1)}_{(12,3,2,2)}&=1812768672~.
\end{aligned}
\end{equation}
It should be noted that when genus 0 numbers $n^{(0)}_{\bm{p}}$ and $n^{(0)}_{\bm{q}}$ are equal, then the genus 1 numbers $n^{(1)}_{\bm{p}}$ and $n^{(1)}_{\bm{q}}$ are equal.

In fact, instanton numbers are equal for degree vectors related by a certain linear transformation:
\begin{equation}
n^{(g)}_{(i,j,k,l)}=n^{(g)}_{-i+2j+2k+2l,j,k,l}~.
\end{equation}
Our tables support this claim for genera 0 and 1, but this is expected to hold at every genus. It is true to the full extent of the tables of genus one numbers, but at genus 0 there is one exception, which is that this operation takes $(1,0,0,0)$ to $(-1,0,0,0)$, yet the instanton number $n^{(0)}_{(1,0,0,0)}=48$ is nonzero. For this particular Tetraquadric family, the above identity (including the mentioned exception) can be proven at every genus, either by inheriting an action from the splitting (5,45) family (which itself has a proven Coxeter symmetry \cite{Candelas:2021lkc}), or by the arguments from flop transitions that appear in \cite{Lukas:2022crp}. As we go on to give more examples of such identities, they will for the purposes of this thesis strictly be conjectures supported by the instanton number computations. Any example where these reflections can be traced to a flop transition (such as CICYs that split another family) would see a rigorous proof along the same lines, and indeed a number of our examples appear in \cite{Lukas:2022crp} with proofs.

We will denote this operation by
\begin{equation}
g_{1}:(i,j,k,l)\mapsto(2d-3i,j,k,l)~,\qquad d=i+j+k+l~.
\end{equation}
This operation is an involution, with $g_{1}^{2}=1$. Additionally, there are the obvious permutation operations that leave instanton numbers invariant, because the mirror Tetraquadric's complex structure moduli $\varphi_{i}$ can be exchanged by an $S_{4}$ symmetry. The group $S_{4}$ has three generators, whose actions on $\IZ^{4}$ are
\begin{equation}
s_{1}:(i,j,k,l)\mapsto(j,i,k,l)~,\quad s_{2}:(i,j,k,l)\mapsto(i,k,j,l)~,\quad s_{3}:(i,j,k,l)\mapsto(i,j,l,k)~.
\end{equation}
We should be clear that there are a few distinct group orbits with the same instanton number, a point that demands further study. Collecting these together, we find the following group presentation
\begin{equation}
\langle g_{1},\,s_{1},\,s_{2},\,s_{3}\,|d_{1}^{2}=s_{1}^{2}=s_{2}^{2}=s_{3}^{2}=1~,\left(d_{1}s_{2}\right)^{2}=\left(d_{1}s_{3}\right)^{2}=1~,\,\left(s_{1}s_{2}\right)^{3}=\left(s_{1}s_{3}\right)^{3}=\left(s_{2}s_{3}\right)^{3}=1\rangle~.
\end{equation}
Note that the product $d_{1}s_{1}$ is a group element of infinite order. This is a group generated by reflections $r_{i}$, and a symmetric matrix $m_{ij}$ (with two formally infinite entries) gives the relations that products of reflections obey:
\begin{equation}
\left(r_{i}r_{j}\right)^{m_{ij}}=1~.
\end{equation}
Groups with such a presentation are termed \emph{Coxeter groups}. For our group, the Coxeter matrix $m_{ij}$ reads
\begin{equation}
m_{ij}=\begin{pmatrix}
1&\infty&2&2\\\infty&1&3&2\\
2&3&1&3\\
2&2&3&1
\end{pmatrix}~.
\end{equation}
The number of generators $r_{i}$ is termed the \emph{rank} of the Coxeter group. The data of a Coxeter matrix can be neatly encoded in a Coxeter diagram. This is a graph with 
a number of nodes equal to the group's rank. If $m_{ij}=2$, then no edges connect the $i,\,j$ nodes. If $m_{ij}=3$ then an unlabelled edge connects nodes $i,\,j$. If $m_{ij}\geq4$ then nodes $i,\,j$ are connected by an edge labelled with the entry $m_{ij}$. The Tetraquadric's Coxeter group has the following diagram:
\begin{figure}[H]
\begin{center}
\begin{tikzpicture}[scale=2.0]
%[every node/.style={circle, draw, scale=.8}, scale=1.0, rotate = 180, xscale = -1]

\node [style={circle, draw, fill=black}] (1) at ( 0.0, 0.0) {};
\node [style={circle, draw, fill=black}] (2) at ( 1.0, 0.0) {};
\node [style={circle, draw, fill=black}] (3) at ( 2.0, 0.0) {};
\node [style={circle, draw, fill=black}] (4) at ( 3.0, 0.0) {};

\draw (1) -- (2) node [midway,above] {$\infty$};
\draw (2) -- (3);
\draw (3) -- (4);

\end{tikzpicture}
\end{center}
\end{figure}

The appearance of Coxeter group actions as symmetries of sets of instanton numbers was also discussed in \cite{Lukas:2022crp,Brodie:2021toe}, and were explained by flop transitions between manifolds in the same complex deformation family. We incorporate permutation symmetries in our discussion, so our presented Coxeter groups differ to those of \cite{Lukas:2022crp}~. The paper \cite{hosono2018movable} explained how an infinite symmetry group could arise from infinite sequences of birational transformations.

We list some CICY matrices for which a Coxeter group symmetry acts on the instanton numbers. Each of these geometries possesses an $S_{h^{1,1}}$ symmetry, which is extended by an operation $g$ for which we give the action on the second homology, to get the group specified by the diagram that we give. When we give the operation $g$, the symbol $d$ refers to to the sum of a vector's elements. For instance in the first example $d=i+j+k+l+m$, while in the third $d=i+j+k$.

\begin{figure}[H]
\begin{minipage}{.5\textwidth}
\begin{equation}\notag
\cicy{\IP^{1}\\\IP^{1}\\\IP^{1}\\\IP^{1}\\\IP^{1}}{1&1\\1&1\\1&1\\1&1\\1&1}~,\qquad g:(i,j,k,l,m)\mapsto(d-2i,j,k,l,m)~,
\end{equation}
\end{minipage}
\begin{minipage}{.5\textwidth}
\vskip60pt
\begin{flushleft}
\begin{tikzpicture}[scale=2.0]
%[every node/.style={circle, draw, scale=.8}, scale=1.0, rotate = 180, xscale = -1]

\node [style={circle, draw, fill=black}] (1) at ( 0.0, 0.0) {};
\node [style={circle, draw, fill=black}] (2) at ( 1.0, 0.0) {};
\node [style={circle, draw, fill=black}] (3) at ( 2.0, 0.0) {};
\node [style={circle, draw, fill=black}] (4) at ( 3.0, 0.0) {};
\node [style={circle, draw, fill=black}] (5) at ( 4.0, 0.0) {};

\draw (1) -- (2) node [midway,above] {$6$};
\draw (2) -- (3);
\draw (3) -- (4);
\draw (4) -- (5);

\end{tikzpicture}
\end{flushleft}
\end{minipage}
\end{figure}

\begin{figure}[H]
\begin{minipage}{.5\textwidth}
\begin{flushleft}
\begin{equation}\notag
\cicy{\IP^{1}\\\IP^{1}\\\IP^{1}\\\IP^{1}}{2\\2\\2\\2}~,\qquad g:(i,j,k,l)\mapsto(2d-3i,j,k,l)~,
\end{equation}
\end{flushleft}
\end{minipage}
\begin{minipage}{.5\textwidth}
\vskip60pt
\begin{flushleft}
\begin{tikzpicture}[scale=2.0]
%[every node/.style={circle, draw, scale=.8}, scale=1.0, rotate = 180, xscale = -1]

\node [style={circle, draw, fill=black}] (1) at ( 0.0, 0.0) {};
\node [style={circle, draw, fill=black}] (2) at ( 1.0, 0.0) {};
\node [style={circle, draw, fill=black}] (3) at ( 2.0, 0.0) {};
\node [style={circle, draw, fill=black}] (4) at ( 3.0, 0.0) {};

\draw (1) -- (2) node [midway,above] {$\infty$};
\draw (2) -- (3);
\draw (3) -- (4);

\end{tikzpicture}
\end{flushleft}
\end{minipage}
\end{figure}

\begin{figure}[H]
\begin{minipage}{.5\textwidth}
\begin{flushleft}
\begin{equation}\notag
\cicy{\IP^{2}\\\IP^{2}\\\IP^{2}}{1&1&1\\1&1&1\\1&1&1}~,\qquad g:(i,j,k)\mapsto(2d-3i,j,k)~,
\end{equation}
\end{flushleft}
\end{minipage}
\begin{minipage}{.5\textwidth}
\vskip60pt
\begin{flushleft}
\begin{tikzpicture}[scale=2.0]
%[every node/.style={circle, draw, scale=.8}, scale=1.0, rotate = 180, xscale = -1]

\node [style={circle, draw, fill=black}] (1) at ( 0.0, 0.0) {};
\node [style={circle, draw, fill=black}] (2) at ( 1.0, 0.0) {};
\node [style={circle, draw, fill=black}] (3) at ( 2.0, 0.0) {};

\draw (1) -- (2) node [midway,above] {$\infty$};
\draw (2) -- (3);

\end{tikzpicture}
\end{flushleft}
\end{minipage}
\end{figure}

\begin{figure}[H]
\begin{minipage}{.5\textwidth}
\begin{flushleft}
\begin{equation}\notag
\cicy{\IP^{4}\\\IP^{4}}{1&1&1&1&1\\1&1&1&1&1}~,\qquad g:(i,j)\mapsto(4d-5i,j)~,
\end{equation}
\end{flushleft}
\end{minipage}
\begin{minipage}{.5\textwidth}
\vskip60pt
\begin{flushleft}
\begin{tikzpicture}[scale=2.0]
%[every node/.style={circle, draw, scale=.8}, scale=1.0, rotate = 180, xscale = -1]

\node [style={circle, draw, fill=black}] (1) at ( 0.0, 0.0) {};
\node [style={circle, draw, fill=black}] (2) at ( 1.0, 0.0) {};

\draw (1) -- (2) node [midway,above] {$\infty$};

\end{tikzpicture}
\end{flushleft}
\end{minipage}
\end{figure}

\begin{figure}[H]
\begin{minipage}{.5\textwidth}
\begin{flushleft}
\begin{equation}\notag
\cicy{\IP^{3}\\\IP^{3}}{1&1&2\\1&1&2}~,\qquad g:(i,j)\mapsto(6d-7i,j)~,
\end{equation}
\end{flushleft}
\end{minipage}
\begin{minipage}{.5\textwidth}
\vskip60pt
\begin{flushleft}
\begin{tikzpicture}[scale=2.0]
%[every node/.style={circle, draw, scale=.8}, scale=1.0, rotate = 180, xscale = -1]

\node [style={circle, draw, fill=black}] (1) at ( 0.0, 0.0) {};
\node [style={circle, draw, fill=black}] (2) at ( 1.0, 0.0) {};

\draw (1) -- (2) node [midway,above] {$\infty$};

\end{tikzpicture}
\end{flushleft}
\end{minipage}
\end{figure}

\begin{figure}[H]
\begin{minipage}{.5\textwidth}
\begin{flushleft}
\begin{equation}\notag
\cicy{\IP^{3}\\\IP^{3}}{1&1&2\\1&2&1}~,\qquad g:(i,j)\mapsto(7d-8i,j)~,
\end{equation}
\end{flushleft}
\end{minipage}
\begin{minipage}{.5\textwidth}
\vskip60pt
\begin{flushleft}
\begin{tikzpicture}[scale=2.0]
%[every node/.style={circle, draw, scale=.8}, scale=1.0, rotate = 180, xscale = -1]

\node [style={circle, draw, fill=black}] (1) at ( 0.0, 0.0) {};
\node [style={circle, draw, fill=black}] (2) at ( 1.0, 0.0) {};

\draw (1) -- (2) node [midway,above] {$\infty$};

\end{tikzpicture}
\end{flushleft}
\end{minipage}
\end{figure}

In the latter three two-parameter examples we see the appearance of the infinite dihedral group. In \sref{sect:Miroirs_Genus2} we will discuss the application of this infinite symmetry to the computation of higher genus invariants.  All of the above examples possess an $S_{h^{1,1}}$ permutation symmetry, but Coxeter group symmetries can also be found for manifolds that do not have this permutation symmetry. Some examples, for which we now give a full set of generators, are as follows:
\begin{figure}[H]
\begin{minipage}{.5\textwidth}
\begin{flushleft}
\begin{equation}\notag
\cicy{\IP^{1}\\\IP^{3}}{2\\4}~,\qquad g:(i,j)\mapsto(4j-i,j)~,
\end{equation}
\end{flushleft}
\end{minipage}
\begin{minipage}{.5\textwidth}
\vskip60pt
\begin{flushleft}
\begin{tikzpicture}[scale=2.0]
%[every node/.style={circle, draw, scale=.8}, scale=1.0, rotate = 180, xscale = -1]

\node [style={circle, draw, fill=black}] (1) at ( 0.0, 0.0) {};

\end{tikzpicture}
\end{flushleft}
\end{minipage}
\end{figure}

\begin{figure}[H]
\begin{minipage}{.5\textwidth}
\begin{flushleft}
\begin{equation}\notag
\begin{aligned}
\cicy{\IP^{1}\\\IP^{1}\\\IP^{3}}{1&1\\1&1\\2&2}~,\qquad &g:(i,j,k)\mapsto(2k+j-i,j,k)~,\\
&s:(i,j,k)\mapsto(j,i,k)
\end{aligned}
\end{equation}
\end{flushleft}
\end{minipage}
\begin{minipage}{.5\textwidth}
\vskip60pt
\begin{flushleft}
\begin{tikzpicture}[scale=2.0]
%[every node/.style={circle, draw, scale=.8}, scale=1.0, rotate = 180, xscale = -1]

\node [style={circle, draw, fill=black}] (1) at ( 0.0, 0.0) {};
\node [style={circle, draw, fill=black}] (2) at ( 1.0, 0.0) {};

\draw (1) -- (2) node [midway,above] {$6$};

\end{tikzpicture}
\end{flushleft}
\end{minipage}
\end{figure}

\begin{figure}[H]
\begin{minipage}{.5\textwidth}
\begin{flushleft}
\begin{equation}\notag
\begin{aligned}
\cicy{\IP^{1}\\\IP^{1}\\\IP^{2}}{2\\2\\3}~,\qquad &g:(i,j,k)\mapsto(3k+2j-i,j,k)~,\\
&s:(i,j,k)\mapsto(j,i,k)
\end{aligned}
\end{equation}
\end{flushleft}
\end{minipage}
\begin{minipage}{.5\textwidth}
\vskip60pt
\begin{flushleft}
\begin{tikzpicture}[scale=2.0]
%[every node/.style={circle, draw, scale=.8}, scale=1.0, rotate = 180, xscale = -1]

\node [style={circle, draw, fill=black}] (1) at ( 0.0, 0.0) {};
\node [style={circle, draw, fill=black}] (2) at ( 1.0, 0.0) {};

\draw (1) -- (2) node [midway,above] {$\infty$};

\end{tikzpicture}
\end{flushleft}
\end{minipage}
\end{figure}

In the first of the above two examples, we have a $\IZ_{2}$ action but no permutation symmetry. This should be contrasted with manifolds that possess a permutation symmetry but have no extra $\IZ_{2}$ symmetry action that changes the sum $i+j$, for example the $h^{1,1}=2$ bicubic.

The second of the above three examples is interesting insofar as the Coxeter group is finite, being the dihedral group of order 12. This is the Weyl group of $G_{2}$. It would be interesting in future work to address the question of whether a Kac-Moody symmetry can be realised in string compactifications on manifolds with Coxeter symmetries of the forms that we describe, and were this the case then this example could be the simplest place to start.

While the instanton numbers are constant across orbits of the Coxeter groups we have given, it is not the case that each such orbit has a distinct associated instanton number. There is additional work to be done in explaining the additional repetitions. For example, the Coxeter group actions that we have given above for the families
\begin{equation}\label{eq:Troublesome_Twosome}
\cicy{\IP^{1}\\\IP^{1}\\\IP^{2}}{2\\2\\3}^{(3,75)}_{-144}~,\qquad\cicy{\IP^{1}\\\IP^{1}\\\IP^{3}}{1&1\\1&1\\2&2}^{(3,55)}_{-104}~,
\end{equation} 
do not change the third component $k$ of the index vector $(i,j,k)$. Nonetheless, the tables of instanton numbers reveal some repetitions of instanton numbers for different values of $k$. We produce those tables to a low order here. Note for instance that for the $\chi=-144$ family
\begin{equation}
n^{(0)}_{(1,0,13)}=n^{(0)}_{(2,0,7)}=5221882080 \qquad \text{and} \qquad n^{(0)}_{(1,0,13)}=n^{(0)}_{(5,0,4)}=2115255492,
\end{equation}
while for the $\chi=-104$ family we observe
\begin{equation}
n^{(0)}_{(1,0,5)}=n^{(0)}_{(3,0,3)}=125824\qquad \text{and} \qquad n^{(0)}_{(3,0,9)}=n^{(0)}_{(5,0,7)}=130181768448~
\end{equation}
Equalities like this are not explained by the flop operations appearing in \cite{Lukas:2022crp}, and for the time being we cannot explain them. It may be of note that the above index vectors have components that are 0.

In another attempt to generalise the results of \cite{Lukas:2022crp}, we can see manifolds with Coxeter symmetries that are not CICYs. We defer discussion of examples with groups larger than $\IZ_{2}$ to future work, and here remark that nontrivial $\IZ_{2}$ symmetries can be found for the following manifolds, which appeared in the tables of \cite{KIM1989108}. Details of Calabi-Yau intersections in weighted projective space that we avail of are found in \cite{Greene:1990du}. Since the instanton numbers that we compute for their $\chi=-112$ families agree, we conjecture that these two families are identical. We also fix a typo in the configuration matrix for the $\chi=-98$ family.
\begin{equation}
\cicy{\IP^{1}\\\IW\IP^{5}_{(1,1,1,1,2,2)}}{0&1&1\\4&2&2}_{\chi=-112}\;\cong\;\cicy{\IP^{1}\\\IW\IP^{4}_{(1,1,1,1,2)}}{0&2\\4&2}_{\chi=-112}~,\qquad g:(i,j)\mapsto(2j-i,j)~.
\end{equation}
\vskip10pt
\begin{equation}
\cicy{\IP^{2}\\\IW\IP^{4}_{(1,1,1,1,2)}}{1&1&1\\2&2&2}_{\chi=-98}~,\qquad g:(i,j)\mapsto(4j-i,j)~.
\end{equation}
Finally, we remark that the instanton numbers for the two-parameter geometry given in Phase IV of the Knapp-Hori model \cite{Hori:2016txh} possess a $\IZ_{2}$ symmetry given by the action $g:(i,j)\mapsto(i,5i-j)$, but this only holds for vectors with $i>0$ at genus 0. Such $\IZ_{2}$ symmetries and their utility in higher genus computations were identified in \cite{Klemm:2004km}.

The content of the following tables is available in electronic form \cite{mcgovern2023a}.
\vskip20pt

\begin{table}[H]
\centering
\begin{adjustbox}{width=\columnwidth,center}
\tiny{\pgfplotstabletypeset[
col sep=space,
white space chars={]},
ignore chars={\ },
every head row/.style={before row=\hline,after row=\hline\hline},
every first row/.style={before row=\vrule height12pt width0pt depth0pt},
every last row/.style={before row=\vrule height0pt width0pt depth6pt,after row=\hline},
%after row=\hline,    %  Uncomment this to get back lines between all rows.
columns={A,B,A,B,A,B},
display columns/0/.style={select equal part entry of={0}{3},column type = {|l},column name=\vrule height12pt width0pt depth6pt \hfil $\bm p$},
display columns/1/.style={select equal part entry of={0}{3},column type = {|l|},column name=\hfil $n^{(0)}_{\bm p}$},
display columns/2/.style={select equal part entry of={1}{3},column type = {|l},column name=\hfil $\bm p$},
display columns/3/.style={select equal part entry of={1}{3},column type = {|l|},column name=\hfil $n^{(0)}_{\bm p}$},
display columns/4/.style={select equal part entry of={2}{3},column type = {|l},column name=\hfil $\bm p$},
display columns/5/.style={select equal part entry of={2}{3},column type = {|l|},column name=\hfil $n^{(0)}_{\bm p}$},
string type]
{
A ] B

{0, 0, 1}]128
{1, 0, 0}]32
{0, 0, 2}]120
{1, 0, 1}]384
{1, 1, 0}]32
{0, 0, 3}]128
{1, 0, 2}]2368
{1, 1, 1}]3072
{2, 0, 1}]128
{0, 0, 4}]104
{1, 0, 3}]10496
{1, 1, 2}]75648
{2, 0, 2}]5056
{2, 1, 1}]3072
{0, 0, 5}]128
{1, 0, 4}]38624
{1, 1, 3}]958464
{2, 0, 3}]70656
{2, 1, 2}]293376
{2, 2, 1}]8000
{3, 0, 2}]2368
{3, 1, 1}]384
{0, 0, 6}]120
{1, 0, 5}]125824
{1, 1, 4}]8368448
{2, 0, 4}]626432
{2, 1, 3}]9744384
{2, 2, 2}]2005824
{3, 0, 3}]125824
{3, 1, 2}]293376
{3, 2, 1}]3072
{4, 0, 2}]120
{0, 0, 7}]128
{1, 0, 6}]373952
{1, 1, 5}]57106432
{2, 0, 5}]4264960
{2, 1, 4}]180973568
{2, 2, 3}]142328576
{3, 0, 4}]2692704
{3, 1, 3}]27253248
{3, 2, 2}]3662848
{3, 3, 1}]3072
{4, 0, 3}]70656
{4, 1, 2}]75648
{4, 2, 1}]128
{0, 0, 8}]104
{1, 0, 7}]1033472
{1, 1, 6}]326085888
{2, 0, 6}]24162944
{2, 1, 5}]2324460544
{2, 2, 4}]5014078976
{3, 0, 5}]36442112
{3, 1, 4}]1082082304
{3, 2, 3}]598081536
{3, 3, 2}]11713408
{4, 0, 4}]4264256
{4, 1, 3}]27253248
{4, 2, 2}]2005824
{4, 3, 1}]384
{5, 0, 3}]10496
{5, 1, 2}]2368
{0, 0, 9}]128
{1, 0, 8}]2692704
{1, 1, 7}]1625063424
{2, 0, 7}]119377920
{2, 1, 6}]23048243712
{2, 2, 5}]111830923392
{3, 0, 6}]368140672
{3, 1, 5}]25668767488
{3, 2, 4}]40106733568
{3, 3, 3}]3737054208
{4, 0, 5}]119377920
{4, 1, 4}]2487784448
{4, 2, 3}]945297408
{4, 3, 2}]11713408
{4, 4, 1}]128
{5, 0, 4}]2692704
{5, 1, 3}]9744384
{5, 2, 2}]293376
{6, 0, 3}]128
{0, 0, 10}]120
{1, 0, 9}]6679296
{1, 1, 8}]7268894432
{2, 0, 8}]529534464
{2, 1, 7}]188236812288
{2, 2, 6}]1803958100480
{3, 0, 7}]3018823680
{3, 1, 6}]428709670912
{3, 2, 5}]1533012331520
{3, 3, 4}]430848150336
{4, 0, 6}]2151740096
{4, 1, 5}]111475558400
{4, 2, 4}]128007090432
{4, 3, 3}]8915561216
{4, 4, 2}]20508560
{5, 0, 5}]174847616
{5, 1, 4}]2487784448
{5, 2, 3}]598081536
{5, 3, 2}]3662848
{6, 0, 4}]626432
{6, 1, 3}]958464
{6, 2, 2}]5056
{0, 0, 11}]128
{1, 0, 10}]15887680
{1, 1, 9}]29762052096
{2, 0, 9}]2151748608
{2, 1, 8}]1320335902720
{2, 2, 7}]22815978525184
{3, 0, 8}]21118162688
{3, 1, 7}]5536125819904
{3, 2, 6}]39458161495040
{3, 3, 5}]26335226884096
{4, 0, 7}]28827891712
{4, 1, 6}]3144631756416
{4, 2, 5}]8535762597376
{4, 3, 4}]1874830230016
{4, 4, 3}]31657668096
{5, 0, 6}]5865465088
{5, 1, 5}]224531369984
{5, 2, 4}]186198155264
{5, 3, 3}]8915561216
{5, 4, 2}]11713408
{6, 0, 5}]119377920
{6, 1, 4}]1082082304
{6, 2, 3}]142328576
{6, 3, 2}]293376
{7, 0, 4}]38624
{7, 1, 3}]10496
{0, 0, 12}]104
{1, 0, 11}]36442112
{1, 1, 10}]113160255872
{2, 0, 10}]8125777088
{2, 1, 9}]8186269737984
{2, 2, 8}]238488938059776
{3, 0, 9}]130181768448
{3, 1, 8}]58621972428800
{3, 2, 7}]755784523825152
{3, 3, 6}]1028951153011840
{4, 0, 8}]310514571520
{4, 1, 7}]63737659456000
{4, 2, 6}]351572541694080
{4, 3, 5}]186761410553984
{4, 4, 4}]11079782988096
{5, 0, 7}]130181768448
{5, 1, 6}]10983601432704
{5, 2, 5}]22728981286912
{5, 3, 4}]3805839345664
{5, 4, 3}]47775759360
{5, 5, 2}]11713408
{6, 0, 6}]8125777088
{6, 1, 5}]224531369984
{6, 2, 4}]128007090432
{6, 3, 3}]3737054208
{6, 4, 2}]2005824
{7, 0, 5}]36442112
{7, 1, 4}]180973568
{7, 2, 3}]9744384
{7, 3, 2}]2368
{8, 0, 4}]104

}
}
\end{adjustbox}
\def\arraystretch{1.15}
\vskip10pt
\capt{6.5in}{tab:Instanton_Numbers_111122}{The genus~0 instanton numbers of total degree $\leqslant 12$ for the $\chi=-104$ family \eqref{eq:Troublesome_Twosome}. The numbers not in this list are either zero, or given by those in the table after permuting the first two indices.}
\end{table}

\begin{table}[H]
\centering
\def\arraystretch{1.005}
\begin{adjustbox}{width=\columnwidth,center}
\tiny{\pgfplotstabletypeset[
col sep=space,
white space chars={]},
ignore chars={\ },
every head row/.style={before row=\hline,after row=\hline\hline},
every first row/.style={before row=\vrule height12pt width0pt depth0pt},
every last row/.style={before row=\vrule height0pt width0pt depth6pt,after row=\hline},
%after row=\hline,    %  Uncomment this to get back lines between all rows.
columns={A,B,A,B,A,B,A,B},
display columns/0/.style={select equal part entry of={0}{4},column type = {|l},column name=\vrule height12pt width0pt depth6pt \hfil $\bm p$},
display columns/1/.style={select equal part entry of={0}{4},column type = {|l|},column name=\hfil $n^{(0)}_{\bm p}$},
display columns/2/.style={select equal part entry of={1}{4},column type = {|l},column name=\hfil $\bm p$},
display columns/3/.style={select equal part entry of={1}{4},column type = {|l|},column name=\hfil $n^{(0)}_{\bm p}$},
display columns/4/.style={select equal part entry of={2}{4},column type = {|l},column name=\hfil $\bm p$},
display columns/5/.style={select equal part entry of={2}{4},column type = {|l|},column name=\hfil $n^{(0)}_{\bm p}$},
display columns/6/.style={select equal part entry of={3}{4},column type = {|l},column name=\hfil $\bm p$},
display columns/7/.style={select equal part entry of={3}{4},column type = {|l|},column name=\hfil $n^{(0)}_{\bm p}$},
string type]
{
A ] B

{{0, 0}, 1}]168
{{1, 0}, 0}]54
{{0, 0}, 2}]168
{{1, 0}, 1}]1080
{{1, 1}, 0}]180
{{0, 0}, 3}]144
{{1, 0}, 2}]9504
{{1, 1}, 1}]22968
{{2, 0}, 1}]1080
{{2, 1}, 0}]54
{{0, 0}, 4}]168
{{1, 0}, 3}]55080
{{1, 1}, 2}]801720
{{2, 0}, 2}]55080
{{2, 1}, 1}]84240
{{2, 2}, 0}]144
{{3, 0}, 1}]168
{{0, 0}, 5}]168
{{1, 0}, 4}]258876
{{1, 1}, 3}]14272344
{{2, 0}, 3}]1045440
{{2, 1}, 2}]9589752
{{2, 2}, 1}]823968
{{3, 0}, 2}]94248
{{3, 1}, 1}]84240
{{3, 2}, 0}]54
{{0, 0}, 6}]144
{{1, 0}, 5}]1045440
{{1, 1}, 4}]169945416
{{2, 0}, 4}]12531888
{{2, 1}, 3}]422121240
{{2, 2}, 2}]212527800
{{3, 0}, 3}]5686200
{{3, 1}, 2}]37017000
{{3, 2}, 1}]2286360
{{3, 3}, 0}]180
{{4, 0}, 2}]55080
{{4, 1}, 1}]22968
{{0, 0}, 7}]168
{{1, 0}, 6}]3781080
{{1, 1}, 5}]1538714160
{{2, 0}, 5}]112746384
{{2, 1}, 4}]10651393728
{{2, 2}, 3}]18704746728
{{3, 0}, 4}]159172380
{{3, 1}, 3}]3897248904
{{3, 2}, 2}]1536760944
{{3, 3}, 1}]14832456
{{4, 0}, 3}]12531888
{{4, 1}, 2}]57195792
{{4, 2}, 1}]2286360
{{4, 3}, 0}]54
{{5, 0}, 2}]9504
{{5, 1}, 1}]1080
{{0, 0}, 8}]168
{{1, 0}, 7}]12531888
{{1, 1}, 6}]11407448232
{{2, 0}, 6}]828397800
{{2, 1}, 5}]185136252912
{{2, 2}, 4}]868185209088
{{3, 0}, 5}]2868991776
{{3, 1}, 4}]196654202136
{{3, 2}, 3}]268467230952
{{3, 3}, 2}]19569181320
{{4, 0}, 4}]828397800
{{4, 1}, 3}]15496835472
{{4, 2}, 2}]4691149344
{{4, 3}, 1}]35294184
{{4, 4}, 0}]144
{{5, 0}, 3}]12531888
{{5, 1}, 2}]37017000
{{5, 2}, 1}]823968
{{6, 0}, 2}]168
{{0, 0}, 9}]144
{{1, 0}, 8}]38713950
{{1, 1}, 7}]72542163168
{{2, 0}, 7}]5221882080
{{2, 1}, 6}]2455050545136
{{2, 2}, 5}]26018243190288
{{3, 0}, 6}]38437207344
{{3, 1}, 5}]6157144423728
{{3, 2}, 4}]22069473542568
{{3, 3}, 3}]5853218557032
{{4, 0}, 5}]29153182176
{{4, 1}, 4}]1560583187460
{{4, 2}, 3}]1709274209400
{{4, 3}, 2}]105124396536
{{4, 4}, 1}]179638056
{{5, 0}, 4}]2115255492
{{5, 1}, 3}]30179989584
{{5, 2}, 2}]6738481008
{{5, 3}, 1}]35294184
{{5, 4}, 0}]54
{{6, 0}, 3}]5686200
{{6, 1}, 2}]9589752
{{6, 2}, 1}]84240
{{0, 0}, 10}]168
{{1, 0}, 9}]112746384
{{1, 1}, 8}]408220124400
{{2, 0}, 8}]29153182176
{{2, 1}, 7}]26462705388768
{{2, 2}, 6}]565228067371704
{{3, 0}, 7}]414019483488
{{3, 1}, 6}]137077369593336
{{3, 2}, 5}]1091242103367168
{{3, 3}, 4}]771574529680320
{{4, 0}, 6}]685227318336
{{4, 1}, 5}]86145995276352
{{4, 2}, 4}]249147504104832
{{4, 3}, 3}]56646795125808
{{4, 4}, 2}]950998199904
{{5, 0}, 5}]147357745992
{{5, 1}, 4}]6324878723688
{{5, 2}, 3}]5553133901424
{{5, 3}, 2}]275408356176
{{5, 4}, 1}]387427104
{{5, 5}, 0}]180
{{6, 0}, 4}]2868991776
{{6, 1}, 3}]30179989584
{{6, 2}, 2}]4691149344
{{6, 3}, 1}]14832456
{{7, 0}, 3}]1045440
{{7, 1}, 2}]801720
{{7, 2}, 1}]1080
{{0, 0}, 11}]168
{{1, 0}, 10}]312318288
{{1, 1}, 9}]2077856570952
{{2, 0}, 9}]147357745992
{{2, 1}, 8}]241891770932622
{{2, 2}, 7}]9578647470994416
{{3, 0}, 8}]3764269848150
{{3, 1}, 7}]2356453861300944
{{3, 2}, 6}]37164898364815152
{{3, 3}, 5}]58494821385825792
{{4, 0}, 7}]12074918985360
{{4, 1}, 6}]3130393529188872
{{4, 2}, 5}]20023868970613584
{{4, 3}, 4}]12074786392584528
{{4, 4}, 3}]815165175453336
{{5, 0}, 6}]6028970554656
{{5, 1}, 5}]618734398390992
{{5, 2}, 4}]1475296893039852
{{5, 3}, 3}]283970290298616
{{5, 4}, 2}]4180709760048
{{5, 5}, 1}]1672396776
{{6, 0}, 5}]414019483488
{{6, 1}, 4}]14260130266464
{{6, 2}, 3}]9866075528304
{{6, 3}, 2}]377099230176
{{6, 4}, 1}]387427104
{{6, 5}, 0}]54
{{7, 0}, 4}]2115255492
{{7, 1}, 3}]15496835472
{{7, 2}, 2}]1536760944
{{7, 3}, 1}]2286360
{{8, 0}, 3}]55080
{{8, 1}, 2}]9504
{{0, 0}, 12}]144
{{1, 0}, 11}]828397800
{{1, 1}, 10}]9721605877056
{{2, 0}, 10}]685227318336
{{2, 1}, 9}]1932024378377232
{{2, 2}, 8}]133082696708836560
{{3, 0}, 9}]29850028039080
{{3, 1}, 8}]33021971994940200
{{3, 2}, 7}]949255452430119360
{{3, 3}, 6}]2949179390777334672
{{4, 0}, 8}]170870441516784
{{4, 1}, 7}]83229565047206400
{{4, 2}, 6}]1047154408278044472
{{4, 3}, 5}]1392090222866615136
{{4, 4}, 4}]259340008844756376
{{5, 0}, 7}]170870441516784
{{5, 1}, 6}]36397140905520432
{{5, 2}, 5}]193464393164553024
{{5, 3}, 4}]100247137259690592
{{5, 4}, 3}]6055793581127544
{{5, 5}, 2}]29817003490128
{{6, 0}, 6}]29850028039080
{{6, 1}, 5}]2540902631155632
{{6, 2}, 4}]5011092898162560
{{6, 3}, 3}]801503420918760
{{6, 4}, 2}]9842930030808
{{6, 5}, 1}]3363048504
{{6, 6}, 0}]144
{{7, 0}, 5}]685227318336
{{7, 1}, 4}]18624092277168
{{7, 2}, 3}]9866075528304
{{7, 3}, 2}]275408356176
{{7, 4}, 1}]179638056
{{8, 0}, 4}]828397800
{{8, 1}, 3}]3897248904
{{8, 2}, 2}]212527800
{{8, 3}, 1}]84240
{{9, 0}, 3}]144
{{0, 0}, 13}]168
{{1, 0}, 12}]2115255492
{{1, 1}, 11}]42320995599600
{{2, 0}, 11}]2966972060160
{{2, 1}, 10}]13783408869528072
{{2, 2}, 9}]1570644007964714736
{{3, 0}, 10}]211313193184296
{{3, 1}, 9}]391767299135571456
{{3, 2}, 8}]19261188790077746538
{{3, 3}, 7}]108334675692791766768
{{4, 0}, 9}]2030806663104960
{{4, 1}, 8}]1732761009324236286
{{4, 2}, 7}]39401939588604883920
{{4, 3}, 6}]102365041178451406320
{{4, 4}, 5}]42886986818729501952
{{5, 0}, 8}]3683509791835230
{{5, 1}, 7}]1475899281351731232
{{5, 2}, 6}]15437538722622496320
{{5, 3}, 5}]17666622173504257920
{{5, 4}, 4}]2952225900540710424
{{5, 5}, 3}]65441894934804480
{{6, 0}, 7}]1358732492843328
{{6, 1}, 6}]243556290929859120
{{6, 2}, 5}]1092271314577190688
{{6, 3}, 4}]485201908448389176
{{6, 4}, 3}]25578337459800960
{{6, 5}, 2}]113389478053344
{{6, 6}, 1}]12857494104
{{7, 0}, 6}]89683487215200
{{7, 1}, 5}]6340790927783952
{{7, 2}, 4}]10262891970293004
{{7, 3}, 3}]1333022580438624
{{7, 4}, 2}]13029988164048
{{7, 5}, 1}]3363048504
{{7, 6}, 0}]54
{{8, 0}, 5}]685227318336
{{8, 1}, 4}]14260130266464
{{8, 2}, 3}]5553133901424
{{8, 3}, 2}]105124396536
{{8, 4}, 1}]35294184
{{9, 0}, 4}]159172380
{{9, 1}, 3}]422121240
{{9, 2}, 2}]9589752
{{9, 3}, 1}]168
{{0, 0}, 14}]168
{{1, 0}, 13}]5221882080
{{1, 1}, 12}]173059142952312
{{2, 0}, 12}]12074918985360
{{2, 1}, 11}]89309343886076376
{{2, 2}, 10}]16165319885559734832
{{3, 0}, 11}]1358732492843328
{{3, 1}, 10}]4045017572706202992
{{3, 2}, 9}]323487707741738306640
{{3, 3}, 8}]3085305331283577953136
{{4, 0}, 10}]20915462494951344
{{4, 1}, 9}]29595710277925790904
{{4, 2}, 8}]1140647229711563099904
{{4, 3}, 7}]5313981987997190937072
{{4, 4}, 6}]4394846243085819779592
{{5, 0}, 9}]64083374604252864
{{5, 1}, 8}]44892419985923962284
{{5, 2}, 7}]846447272733537148032
{{5, 3}, 6}]1887461261292502202904
{{5, 4}, 5}]706177757611379483424
{{5, 5}, 4}]45586829807821571112
{{6, 0}, 8}]44262594615526560
{{6, 1}, 7}]15022136654860343184
{{6, 2}, 6}]133596413514797157576
{{6, 3}, 5}]132683253333213798864
{{6, 4}, 4}]19702612873693234944
{{6, 5}, 3}]400441073987635488
{{6, 6}, 2}]678024552756840
{{7, 0}, 7}]6624537453484920
{{7, 1}, 6}]1009832275947370032
{{7, 2}, 5}]3836114182573117632
{{7, 3}, 4}]1448583132341564928
{{7, 4}, 3}]65054184230486808
{{7, 5}, 2}]246714051981816
{{7, 6}, 1}]24516763128
{{7, 7}, 0}]180
{{8, 0}, 6}]170870441516784
{{8, 1}, 5}]9942236934310944
{{8, 2}, 4}]12999327967495584
{{8, 3}, 3}]1333022580438624
{{8, 4}, 2}]9842930030808
{{8, 5}, 1}]1672396776
{{9, 0}, 5}]414019483488
{{9, 1}, 4}]6324878723688
{{9, 2}, 3}]1709274209400
{{9, 3}, 2}]19569181320
{{9, 4}, 1}]2286360
{{10, 0}, 4}]12531888
{{10, 1}, 3}]14272344
{{10, 2}, 2}]55080
{{0, 0}, 15}]144
{{1, 0}, 14}]12507646968
{{1, 1}, 13}]669793850973648
{{2, 0}, 13}]46536192247248
{{2, 1}, 12}]532443125472289380
{{2, 2}, 11}]148032484883296635024
{{3, 0}, 12}]8041290548966712
{{3, 1}, 11}]37113307255824419664
{{3, 2}, 10}]4635692472845342699712
{{3, 3}, 9}]71228168955194892241824
{{4, 0}, 11}]190978119261948528
{{4, 1}, 10}]428898270234051873648
{{4, 2}, 9}]26635577378628399257880
{{4, 3}, 8}]208542584860846725573792
{{4, 4}, 7}]310728018286366928659728
{{5, 0}, 10}]936802041472321344
{{5, 1}, 9}]1084155094548776246256
{{5, 2}, 8}]34504030323455683365540
{{5, 3}, 7}]137334444454235316481488
{{5, 4}, 6}]100771753799790046258992
{{5, 5}, 5}]15002509184093317799448
{{6, 0}, 9}]1114166808793427904
{{6, 1}, 8}]662747101613708918166
{{6, 2}, 7}]10648710770871871691424
{{6, 3}, 6}]20684312507372786347752
{{6, 4}, 5}]6916476884469525094032
{{6, 5}, 4}]411546921208916198364
{{6, 6}, 3}]3494470176162937224
{{7, 0}, 8}]326486183204225142
{{7, 1}, 7}]95346746262039029904
{{7, 2}, 6}]729068917506240634848
{{7, 3}, 5}]629003708472330915264
{{7, 4}, 4}]82046840116560882948
{{7, 5}, 3}]1485487116716515272
{{7, 6}, 2}]2309426676472032
{{7, 7}, 1}]85286277432
{{8, 0}, 7}]20915462494951344
{{8, 1}, 6}]2715306672487631616
{{8, 2}, 5}]8705702714908296384
{{8, 3}, 4}]2759145061379596596
{{8, 4}, 3}]102967117070019696
{{8, 5}, 2}]318500079686208
{{8, 6}, 1}]24516763128
{{8, 7}, 0}]54
{{9, 0}, 6}]211313193170328
{{9, 1}, 5}]9942236934310944
{{9, 2}, 4}]10262891970293004
{{9, 3}, 3}]801503420918760
{{9, 4}, 2}]4180709760048
{{9, 5}, 1}]387427104
{{10, 0}, 5}]147357745992
{{10, 1}, 4}]1560583187460
{{10, 2}, 3}]268467230952
{{10, 3}, 2}]1536760944
{{10, 4}, 1}]22968
{{11, 0}, 4}]258876
{{11, 1}, 3}]55080

}
}
\end{adjustbox}
\vskip10pt
\capt{6.5in}{tab:Instanton_Numbers_223}{The genus~0 instanton numbers of total degree $\leqslant 15$ for the $\chi=-144$ family \eqref{eq:Troublesome_Twosome}. The numbers not in this list are either zero, or given by those in the table after permuting the first two indices.}
\end{table}

\section{Higher genus mirror symmetry}\label{sect:Miroirs_HigherGenus}
The higher genus B-model prepotentials can be computed via a recursive procedure, originally due to Bershadsky, Cecotti, Ooguri, and Vafa \cite{Bershadsky:1993ta,Bershadsky:1993cx}. This was refined in work by Yamaguchi and Yau \cite{Yamaguchi:2004bt} who demonstrated that the genus-$g$ prepotential was a polynomial in a set of propagators. This was further refined in work by Alim, Laenge, and Scheidegger \cite{Alim:2007qj,Alim:2012ss}, and in \cite{Huang:2015sta} Huang, Katz, and Klemm phrased the recursion as a set of PDEs. This has also been reviewed by Elmi \cite{Elmi:2020jpy} and in the notes \cite{Alim:2012gq}. Work by Klemm, Huang, and Quackenbush \cite{Huang:2006hq} drove the computation to genus 51 for the Quintic, and this has recently been driven higher still by incorporating the modularity of D4-D2-D0 bound states \cite{Alexandrov:2023zjb,Alexandrov:2022pgd}. Additional higher genus results and other developments are found in \cite{Huang:2015sta,Cota:2019cjx}~.

The point of departure for this computation is in realising the complex structure moduli space of $X$ as a complex, K{\"a}hler metric space, with K{\"a}hler potential
\begin{equation}\label{eq:Kahler1}
K(\bm{\varphi},\overline{\bm{\varphi}})=-\log\left(-\ii\Pi^{\dagger}\Sigma\Pi\right)~.
\end{equation}
The metric then is
\begin{equation}
G_{i\bar{j}}=\partial_{i}\partial_{\bar{j}}K~.
\end{equation}
This metric can be used to raise and lower indices, and doing so twice on the complex conjugate of the holomorphic Yukawa coupling gives a quantity
\begin{equation}
\bar{C}_{\bar{i}}^{\+jk}=G^{j\bar{j}}G^{k\bar{k}}\overline{C_{ijk}}~.
\end{equation}
From this, we define non-holomorphic propagators $S^{ij},\,S^{i},\,S$ as potentials:
\begin{equation}
\partial_{\bar{i}}S^{ij}=\bar{C}_{\bar{i}}^{\+ij}~,\qquad\partial_{\bar{i}}S^{j}=G_{i\bar{i}}S^{ij}~,\qquad\partial_{\bar{i}}S=G_{i\bar{i}}S^{i}~.
\end{equation}
Note that $S^{ij}$ is a symmetric tensor, $S^{ij}=S^{ji}$. These functions transform as tensors under coordinate transformations of the moduli space. Under a K{\"a}hler transformation ${K\mapsto K+f+\bar{f}}$ for some holomorphic $f$, each propagator transforms as ${S\mapsto \ee^{2f}S}$, ${S^{i}\mapsto \ee^{2f}S^{i}}$, ${S^{ij}\mapsto \ee^{2f}S^{ij}}$ (one says that they have weight (2,0) under K{\"a}hler transformations).  We display their covariant, with respect to coordinate and K{\"a}hler transformations, derivatives below, and also do so for $K_{i}=\partial_{\varphi^{i}}K$:
\begin{equation}
\begin{aligned}
D_{i}S^{jk}&=\partial_{i}S^{jk}+\Gamma^{j}_{il}S^{lk}+\Gamma^{k}_{il}S^{jl}-2K_{i}S^{jk}~,\\[5pt]
D_{i}S^{j}&=\partial_{i}S^{j}+\Gamma^{j}_{il}S^{l}-2K_{i}S^{j}~,\\[5pt]
D_{i}S&=\partial_{i}S-2K_{i}S~,\\[5pt]
D_{i}K_{j}&=\partial_{i}K_{j}-\Gamma^{l}_{ij}K_{l}~.
\end{aligned}
\end{equation}
The Christoffel symbols $\Gamma^{i}_{jk}$ give the Levi-Civita connection for the metric $G$. We also have provided the covariant derivative $D_{i}K_{j}$ to avoid any confusion in what follows.

For $g\geq2$, the genus-$g$ B-model free energy is a polynomial of degree $3g-3$ in the functions $S^{ij}$ (degree 1), $S^{i}$ (degree 2), $S$ (degree 3), and $K_{i}$ (degree 1). The coefficients of this polynomial are rational functions of the moduli $\bm{\varphi}$, to be determined soon. The K{\"a}hler potential $K$ can be computed from the periods as in \eqref{eq:Kahler1}, and we now explain how to obtain the functions $S^{ij},\,S^{i},\,S$ from the periods. The key insight is the following special geometry relation, which originates in the $tt^{*}$ equations as explained in \cite{Alim:2012gq} (which we avoid discussing):
\begin{equation}\label{eq:SG_Relation}
\begin{aligned}
R_{i\bar{i}\,j}^{\+l}&=\left[\partial_{\bar{i}},D_{i}\right]^{l}_{\+j}\\[5pt]&=\partial_{\bar{i}}\Gamma^{l}_{ij}=\delta^{l}_{i}G_{j\bar{i}}+\delta^{l}_{j}G_{i\bar{i}}-C_{ijk}\bar{C}_{\bar{i}}^{kl}~.
\end{aligned}
\end{equation}
This equation is a total antiholomorphic derivative, with the left and right hand side explicitly given as $\partial_{\bar{i}}$ of quantities that we can either already compute, or of $C_{ijk}S^{kl}$. So we integrate this equation, to get the \emph{integrated special geometry relation} below that we use to fix the first propagator $S^{ij}$:
\begin{equation}\label{eq:ISG}
C_{ijk}S^{kl}=-\Gamma^{l}_{ij}+\delta^{l}_{i}K_{j}+\delta^{l}_{j}K_{i}+s^{l}_{ij}~.
\end{equation}
The quantity $s^{l}_{ij}$ is the holomorphic `constant of integration' which lives in the kernel of $\partial_{\bar{i}}$. It is the first of several `propagator ambiguities' that we will meet. There is some freedom in the choice of $s^{l}_{ij}$. Note that to solve the above equation for $S^{kl}$ we must clear the $C_{ijk}$ from the left hand side. We shall make comments on how to go about this later, and for now press on with reviewing the BCOV procedure.

We now seek to express covariant derivatives of the propagators in terms of the propagators themselves. This involves some algebra. Consider for example $D_{i}S^{jk}$. One hits this with $\partial_{\bar{i}}$, and then expands the expression to get $\partial_{\bar{i}}\left(\partial_{i}S^{jk}+\Gamma^{j}_{il}S^{lk}+\Gamma^{k}_{il}S^{jl}-2K_{i}S^{jk}\right)$. We expand this by the Leibniz rule, with antiholomorphic derivatives of the Christoffel symbols replaced by the special geometry relation \eqref{eq:SG_Relation}. After some manipulation, each term on the right hand side can be written as $\partial_{\bar{i}}$ of something, and this is integrated to get the following:
\begin{equation}\label{eq:BCOV_Derivatives_untilded}
\begin{aligned}
D_{i}S^{jk}&=\delta^{j}_{i}S^{k}+\delta^{k}_{i}S^{j}-C_{imn}S^{mj}S^{nk}+h^{jk}_{i}~,\\[5pt]
D_{i}S^{j}&=2\delta^{j}_{i}S-C_{imn}S^{m}S^{nj}+h^{jk}_{i}K_{k}+h^{j}_{i}~,\\[5pt]
D_{i}S&=-\frac{1}{2}C_{imn}S^{m}S^{n}+\frac{1}{2}h^{mn}_{i}K_{m}K_{n}+h^{j}_{i}K_{j}+h_{i}~,\\[5pt]
D_{i}K_{j}&=-K_{i}K_{j}-C_{ijk}S^{k}+C_{ijk}S^{kl}K_{l}+h_{ij}~.
\end{aligned}
\end{equation}
$h^{jk}_{i},\,h^{j}_{i},\,h_{i},\,h_{ij}$ are all further propagator ambiguities, rational functions of $\bm{\varphi}$ over which there is a degree of choice. It is the first and second of the above equations that are used in practice to compute the propagators $S^{i}$ and $S$. 

The BCOV recursion relation takes its simplest form after a change of variables. One writes the tilded propagators
\begin{equation}
\wt{S}^{ij}=S^{ij}~,\qquad \wt{S}^{i}=S^{i}-S^{ij}K_{j}~,\qquad \wt{S}=S-S^{i}K_{i}+\frac{1}{2}S^{ij}K_{i}K_{j}~,\qquad\wt{K}_{i}=K_{i}~,
\end{equation}
which, we parenthetically remark, obey
\begin{equation}
\partial_{\bar{i}}\wt{S}^{ij}=\bar{C}_{\bar{i}}^{ij}~,\qquad\partial_{\bar{i}}\wt{S}^{i}=-\bar{C}_{\bar{i}}^{ij}K_{j}~,\qquad\partial_{\bar{i}}\wt{S}=\frac{1}{2}\bar{C}_{\bar{i}}^{ij}K_{i}K_{j}~.
\end{equation}
This change of variables allows for a significant simplification: higher genus prepotentials $\cF^{(g\geq2)}$ are degree $3g-3$ polynomials in $\wt{S}^{ij},\,\wt{S}^{i},$ and $\wt{S}$ but \emph{do not explicitly depend on} $\wt{K}$~.

For $g\geq2$, the BCOV recursion relation takes the form \cite{Huang:2015sta}
\begin{equation}\label{eq:BCOV_PDE}
\begin{aligned}
\frac{\partial \cF^{(g)}}{\partial \wt{S}^{ij}}&=\frac{1}{2}\partial_{i}\left(\partial_{j}'\cF^{(g-1)}\right)+\frac{1}{2}\left(C_{ijl}S^{lk}-s^{k}_{ij}\right)\partial_{k}'\cF^{(g-1)}+\frac{1}{2}\left(C_{ijk}S^{k}-h_{ij}\right)c_{g-1}\\[5pt]&\phantom{spooky phantom text}+\frac{1}{2}\sum_{h=1}^{g-1}\left(\partial_{i}'\cF^{(h)}\right)\left(\partial_{j}'\cF^{(g-h)}\right)~,\\[10pt] 
\frac{\partial \cF^{(g)}}{\partial \wt{S}^{i}}&=(2g-3)\partial_{i}'\cF^{(g-1)}+\sum_{h=1}^{g-1}c_{h}\partial_{i}'\cF^{(g-h)}~,\\[10pt]
\frac{\partial \cF^{(g)}}{\partial\wt{S}}&=(2g-3)c_{g-1}+\sum_{h=1}^{g-1}c_{h}c_{g-h}~.
\end{aligned}
\end{equation}
In the above equations we have used 
\begin{equation}\label{eq:BCOV_misc}
\begin{aligned}
c_{g}&=\begin{cases}\frac{\chi(Y)}{24}-1~,&\qquad\qquad\;\, g=1~,\\(2g-2)\cF^{(g)}~,&\qquad\qquad\;\, g>1~,\end{cases}\\[5pt]
\partial_{i}'\cF^{(g)}&=\begin{cases}\frac{1}{2}C_{ijk}S^{jk}+\partial_{i}f^{(1)}~,&\qquad g=1~,\\\partial_{i}\cF^{(g)}~,&\qquad g>1~.\end{cases}
\end{aligned}
\end{equation}
The above symbol $f^{(1)}$ denotes the genus-1 holomorphic ambiguity, but is \emph{not} the function $f$ of \eqref{eq:Not_Hol_Amb} that appeared inside the logarithm of the genus 1 prepotential. Instead,
\begin{equation}
f^{(1)}=-\log\left(\Delta^{c}\prod_{i=1}^{h^{2,1}(X)}\left(\varphi^{i}\right)^{(1-Y_{00i})/2}\overline{\Delta^{c}\prod_{i=1}^{h^{2,1}(X)}\left(\varphi^{i}\right)^{(1-Y_{00i})/2}}\+\right)~,
\end{equation}
with $c=1/12$ in the CICY examples that concern us. We stress that $f^{(1)}$ is a function of $\bm{\varphi}$ and $\overline{\bm{\varphi}}$. We labour this point so as to avoid confusion when we take the topological limit.

The recursive procedure involves working out each $\cF^{(g)}$ by first taking as ansatz the most general polynomial in the tilded propagators of degree $3g-3$, and then comparing the partial derivatives (with respect to the propagators) thereof to the right hand side of \eqref{eq:BCOV_PDE} (which only involves lower-genus prepotentials, guaranteeing a recursion). That right hand side can be written as a polynomial in the tilded propagators by replacing every derivative of a propagator produced by \eqref{eq:BCOV_misc} through the BCOV closure relations \eqref{eq:BCOV_Derivatives_untilded}. For simplicity, we rewrite those closure relations here for the tilded propagators, and expanding the covariant derivatives:
\begin{equation}\label{eq:BCOV_Derivatives_Tilded}
\begin{aligned}
\partial_{i}\wt{S}^{jk}&=C_{imn}\wt{S}^{mj}\wt{S}^{nk}+\delta^{j}_{i}\wt{S}^{k}+\delta^{k}_{i}\wt{S}^{j}-s^{j}_{im}\wt{S}^{mk}-s^{k}_{im}\wt{S}^{mj}+h^{jk}_{i}~,\\[5pt]
\partial_{i}\wt{S}^{j}&=C_{imn}\wt{S}^{mj}\wt{S}^{n}+2\delta^{j}_{i}\wt{S}-s^{j}_{im}\wt{S}^{m}-h_{ik}\wt{S}^{kj}+h^{j}_{i}~,\\[5pt]
\partial_{i}\wt{S}&=\frac{1}{2}C_{imn}\wt{S}^{m}\wt{S}^{n}-h_{ij}\wt{S}^{j}+h_{i}~,\\[5pt]
\partial_{i}K_{j}&=K_{i}K_{j}-C_{ijn}\wt{S}^{mn}K_{m}+s^{m}_{ij}K_{m}-C_{ijk}\wt{S}^{k}+h_{ij}~.
\end{aligned}
\end{equation}
This procedure fixes the polynomial form of $\cF^{(g)}\left(\wt{S}^{ij},\,\wt{S}^{i},\,\wt{S}\right)$ up to the `constant' term $f^{(g)}$, which is a rational function of $\bm{\varphi}$ not fixed by \eqref{eq:BCOV_PDE}. That this function is holomorphic is buried in the derivation of \eqref{eq:BCOV_PDE}, which is a rearrangement of antiholomorphic derivatives that annihilate this \emph{holomorphic ambiguity} $f^{(g)}$. 

Determination of this ambiguity $f^{(g)}$ is the major remaining conceptual problem in topological string theory. It is a rational function, with poles at the zeroes of $\Delta$ and any singularities of the propagators. That the propagators can have singularities outside of $\Delta=0$ is an artefact of how the equations \eqref{eq:ISG} and \eqref{eq:BCOV_Derivatives_untilded} are solved for the propagators. $\cF^{(g)}$ should be nonsingular away from $\Delta=0$, so $f^{(g)}$ must have a residue at these spurious poles so that $\cF^{(g)}$ is regular. 

The singular behaviour of $\cF^{(g)}$ at the conifold locus $\Delta=0$ is fixed by the conifold gap condition \cite{Huang:2006hq}, which shows that as the conifold locus is approached $\cF^{(g)}$ should go like
\begin{equation}
\cF^{(g)}\sim\frac{1}{\Delta^{2g-2}}+\text{regular}
\end{equation}
where ``regular" denotes terms that are nonsingular at $\Delta=0$. The degree of $f^{(g)}$'s numerator should not be too high, so that $\cF^{(g)}$ is regular at $\infty$. 

In this thesis we shall only make it as high as genus 2. To this end we display an expression for the genus 2 prepotential that we will make use of, first derived in \cite{Bershadsky:1993cx}. The authors thereof were able to compute higher-still prepotentials but these do not fit on a page. Note that we now revert to using untilded propagators.
\begin{equation}\label{eq:BCOV_p99}
\begin{aligned}
\cF^{(2)}\=&\frac{1}{2}C^{(1)}_{ij}S^{ij}+\frac{1}{2}C_{i}^{(1)}S^{ij}C_{j}^{(1)}-\frac{1}{8}S^{jk}S^{mn}C_{jkmn}-\frac{1}{2}S^{ij}C_{ijm}S^{mn}C_{n}^{(1)}+\frac{\chi(Y)}{24}S^{i}C_{i}^{(1)}\\[5pt]
&\++\frac{1}{8}S^{ij}C_{ijp}S^{pq}C_{qmn}S^{mn}+\frac{1}{12}S^{ij}S^{pq}S^{mn}C_{ipm}C_{jqn}-\frac{\chi(Y)}{48}S^{i}C_{ijk}S^{jk}\\[5pt]
&\++\frac{\chi(Y)}{24}\left(\frac{\chi(Y)}{24}-1\right)S+f^{(2)}\left(\bm{\varphi}\right)~.
\end{aligned}
\end{equation}
We now fix the above missing pieces of notation:
\begin{equation}
\begin{aligned}
C^{(1)}_{i}&=D_{i}\cF^{(1)}=\partial_{i}\cF^{(1)}~,\qquad C^{(1)}_{ij}=D_{i}D_{j}\cF^{(1)}=\partial_{i}\partial_{j}\cF^{(1)}-\Gamma^{l}_{ij}\partial_{l}\cF^{(1)}~,\\[5pt] 
C_{jklm}&=D_{j}C_{klm}=\partial_{j}C_{klm}-\Gamma^{n}_{jk}C_{nlm}-\Gamma^{n}_{jl}C_{knm}-\Gamma^{n}_{jm}C_{kln}+2K_{j}C_{klm}~.
\end{aligned}
\end{equation}
Note that $C_{jklm}$ is symmetric in its four indices.

In order to obtain curve counts, a change of variables to A-model quantities must be made. The genus-$g$ A model free energy dependence on higher genus instanton numbers $n^{(g)}_{\mathbf{b}}$ is given by the Gopakumar-Vafa formula \cite{Gopakumar:1998ii,Gopakumar:1998jq}:
\begin{equation}
\begin{aligned}
F^{\text{All Genus}}(\lambda,\mathbf{t})&=\sum_{g=0}^{\infty}\lambda^{2g-2}F_{g}(t)
\\&=\frac{c(\mathbf{t})}{\lambda}+l(\mathbf{t})+\sum_{g=0}^{\infty}\sum_{\mathbf{b}\geq0}n^{(g)}_{\mathbf{b}}\sum_{k=1}^{\infty}\frac{1}{k}\left(2\sin\frac{k\lambda}{2}\right)^{2g-2}\exp\left[2\pi\ii\,k\,\mathbf{b}\cdot\mathbf{t}\right]~.
\end{aligned}
\end{equation}
$\lambda$ is the string coupling. $c(\mathbf{t})$ and $l(\mathbf{t})$ are respectively cubic and linear polynomials in $t$, This formula fixes $F_{g}(\mathbf{t}$ at each genus up to a constant term, which we neglect to discuss. This was studied at different genera in the series of papers \cite{Candelas:1990rm,Bershadsky:1993ta,Bershadsky:1993cx,Faber:1998gsw} and nicely presented in the thesis \cite{Elmi:2020jpy}. We give the genus 2 expansion below ahead of making use of it in the next section.
\begin{equation}
F_{2}(\mathbf{t})=\frac{\chi(Y)}{5760}+\sum_{\mathbf{b}\geq0}\left(\frac{n^{(0)}_{\mathbf{b}}}{240}+n^{(2)}_{\mathbf{b}}\right)\text{Li}_{-1}\left(\mathbf{q}^{\mathbf{b}}\right)~.
\end{equation}
One can note that genus 1 numbers $n^{(1)}_{\mathbf{b}}$ do not appear in this formula. Their absence was noted in \cite{Bershadsky:1993cx}, where it was explained as the lack of ``toroidal bubbling" at genus 2.

\subsubsection*{Quantities in the holomorphic limit}
Up until now the propagators have had a dependence on $\mathbf{\varphi}$ and $\overline{\mathbf{\varphi}}$. To compute enumerative invariants, it suffices to work in the topological limit $\bar{t}\mapsto-\ii\infty$, or $\overline{\varphi}\mapsto\infty$. The K{\"a}hler potential can be massaged,
\begin{equation}
K(\bm{\varphi},\overline{\bm{\varphi}})=-\log\left(-\ii\Pi^{\dagger}\Sigma\Pi\right)=\log\left(k(\overline{\mathbf{\varphi}})\right)-\log\left(\varpi_{0}(\mathbf{\varphi})+p\left(\mathbf{\varphi},\bar{\mathbf{\varphi}}\right)\right)~,
\end{equation}
where the function $p\left(\mathbf{\varphi},\overline{\mathbf{\varphi}}\right)$ vanishes when $\overline{\mathbf{\varphi}}\mapsto\infty$. Since it is derivatives of the K{\"a}hler potential that play a role in computing the propagators, we can removed the piece $\log\left(k\left(\overline{\mathbf{\varphi}}\right)\right)$ and retain in this limit
\begin{equation}
K(\mathbf{\varphi})=-\log\left(\varpi_{0}(\mathbf{\varphi})\right)~.
\end{equation}
Similarly, the Christoffel symbols become in this limit
\begin{equation}
\Gamma^{i}_{jk}=\frac{\partial\varphi^{i}}{\partial t^{l}}\frac{\partial^{2}t^{l}}{\partial a^{j} \partial a^{k}}~.
\end{equation}

\section{A worked example at genus 2}\label{sect:Miroirs_Genus2}
We will set about computing genus 2 numbers on the maximally split quintic \eqref{eq:CICY_2_52}~. These were already computed to degree 18 in \cite{hosono2014mirror}, where the holomorphic ambiguity was fixed with an involved implementation of the conifold gap condition. This involves expanding the periods about the conifold locus and imposing the gap condition on the prepotential, expressed in terms of these conifold-adapted periods. 

We instead make the following claim: The genus 2 ambiguity, in this example, can be fixed from two sources of information. These are 
\begin{itemize}
\item The known vanishing instanton numbers~.
\item An infinite Coxeter symmetry~.
\end{itemize}

This example will involve several appearances of the coordinates $\varphi^{i}$ raised to powers. In the name of sanity, we will temporarily write these coordinates with lower indices so that superscripts can be read as exponents. For instance, $\varphi_{2}^{4}$ is $\varphi^{2}$ to the power of four. The only exception to this rule will be when the superscript is an abstract index, as in $\varphi^{i}$, so that tensor contractions behave as they should.

The three-point functions, or Yukawa couplings, are
\begin{equation}
\begin{aligned}
C_{111}&=\frac{5 \left(1+37 \varphi _1+73 \varphi _1^2+14 \varphi _1^3-3 \varphi _2+176 \varphi _1 \varphi _2-73 \varphi _1^2 \varphi _2+3 \varphi _2^2+37 \varphi _1 \varphi _2^2-\varphi _2^3\right)}{\varphi_{1}^{3}\Delta}~,\\[5pt]
C_{112}&=\frac{5\left(2+14 \varphi _1-9 \varphi _1^2-7 \varphi _1^3-\varphi _2-88 \varphi _1 \varphi _2+64 \varphi _1^2 \varphi _2-4 \varphi _2^2-51 \varphi _1 \varphi _2^2+3 \varphi _2^3\right)}{\varphi_{1}^{2}\varphi_{2}\Delta}~,\\[5pt]
C_{122}&=\frac{5\left(2-\varphi _1-4 \varphi _1^2+3 \varphi _1^3+14 \varphi _2-88 \varphi _1 \varphi _2-51 \varphi _1^2 \varphi _2-9 \varphi _2^2+64 \varphi _1 \varphi _2^2-7 \varphi _2^3\right)}{\varphi_{1}\varphi_{2}^{2}\Delta}~,\\[5pt]
C_{222}&=\frac{5\left(1-3 \varphi _1+3 \varphi _1^2-\varphi _1^3+37 \varphi _2+176 \varphi _1 \varphi _2+37 \varphi _1^2 \varphi _2+73 \varphi _2^2-73 \varphi _1 \varphi _2^2+14 \varphi _2^3\right)}{\varphi_{2}^{3}\Delta}~.
\end{aligned}
\end{equation}
The discriminant $\Delta$ is 
\begin{equation}
\begin{aligned}
\Delta&=1-5 \varphi _1+10 \varphi _1^2-10 \varphi _1^3+5 \varphi _1^4-\varphi _1^5-5 \varphi _2-605 \varphi _1 \varphi _2 -1905 \varphi _1^2 \varphi _2-605 \varphi _1^3 \varphi _2\\[5pt]&\+\+\+-5 \varphi
   _1^4 \varphi _2+10 \varphi _2^2-1905 \varphi _1 \varphi _2^2+1905 \varphi _1^2 \varphi _2^2-10 \varphi _1^3 \varphi _2^2-10 \varphi _2^3-605 \varphi _1 \varphi _2^3\\[5pt]&\+\+\+-10 \varphi _1^2
   \varphi _2^3+5 \varphi _2^4-5 \varphi _1 \varphi _2^4-\varphi _2^5~.
\end{aligned}
\end{equation}
The undisplayed $C_{ijk}$ can be obtained from those given by symmetry, e.g. $C_{211}=C_{112}$. 

\subsubsection*{A digression on matrix inversion}
Note that there is a $\IZ_{2}$ symmetry to the example, so that for instance $C_{222}$ can be obtained from $C_{111}$ by effecting a swap $\varphi_{1}\leftrightarrow\varphi_{2}$. This symmetry means that in our final result, the instanton numbers should possess a symmetry $n^{(g)}_{(i,j)}=n^{(g)}_{(j,i)}$. The A and B model prepotentials should both be symmetric functions. 

This means that, if we set up the problem correctly, we can have the ambiguity $f^{(2)}$ be a symmetric function of $\varphi_{1},\,\varphi_{2}$. Let us explain this point. We begin with the determination of $S^{ij}$ from \eqref{eq:ISG}~. In the one-parameter setting, one is free to simply divide both sides of this equation by the quantity $C_{111}$. In multiparameter examples we must invert $C_{ijk}$. The traditional way to do this is to pick a direction in moduli space, say $1$, and then write out the $i=1$ component of \eqref{eq:ISG}:
\begin{equation}\label{eq:ISG_bad}
C_{1jk}S^{kl}=-\Gamma^{l}_{1j}+\delta^{l}_{1}K_{j}+\delta^{l}_{j}K_{1}+s^{l}_{1j}~.
\end{equation}
Now although covariance has been thrown out the window, $C_{1jk}$ is a legitimate matrix that generically has full rank, so that we can find an inverse of it, so that we obtain
\begin{equation}
S^{kl}=-\left(C_{1}^{\+-1}\right)^{kj}\Gamma^{l}_{1j}+\delta^{l}_{1}\left(C_{1}^{\+-1}\right)^{kj}K_{j}+\left(C_{1}^{\+-1}\right)^{kl}K_{1}+\left(C_{1}^{\+-1}\right)^{kj}s^{l}_{1j}~.
\end{equation}
Naively we might say that $s^{l}_{1j}=0$ is the simplest choice, so that we read off $S^{kl}$ directly from the above. This would be wrong, because the $S^{kl}$ so defined does not equal $S^{lk}$. The way to proceed is to write
\begin{equation}\label{eq:amb_sym}
\left(C_{1}^{\+-1}\right)^{kj}s^{l}_{1j}=A^{kl}+E^{kl}~,
\end{equation}
where $A^{kl}$ and $E^{kl}$ are respectively antisymmetric and symmetric in their indices. Our freedom is to redefine the symmetric part $E^{kl}$, and we take this to vanish. Then we get
\begin{equation}
S^{kl}=\frac{1}{2}\left[\left(-\left(C_{1}^{\+-1}\right)^{kj}\Gamma^{l}_{1j}+\delta^{l}_{1}\left(C_{1}^{\+-1}\right)^{kj}K_{j}+\left(C_{1}^{\+-1}\right)^{kl}K_{1}\right)+\left(k\leftrightarrow l\right)\right]~.
\end{equation} 
This choice would be correct, and one could proceed from here to compute all of the other propagators and propagator ambiguities using the $i=1$ components of \eqref{eq:BCOV_Derivatives_untilded}~. For completeness, we give the components of the symmetric matrix $\left(C_{1}^{\+-1}\right)^{ij}$:
\begin{equation}
\begin{aligned}
\left(C_{1}^{\+-1}\right)^{11}&=\frac{-\varphi _1^3 \left(2{-}\varphi _1{-}4 \varphi _1^2{+}3 \varphi _1^3{+}14 \varphi _2{-}88 \varphi _1 \varphi _2{-}51 \varphi _1^2 \varphi _2{-}9 \varphi _2^2{+}64 \varphi _1 \varphi _2^2{-}7 \varphi
   _2^3\right)}{5\left(2-7\varphi_{1}-2\varphi_{2}\right)}~,\\[10pt]
\left(C_{1}^{\+-1}\right)^{12}&=\frac{\varphi_{1}^{2}\varphi_{2}\left(2{+}14 \varphi _1{-}9 \varphi _1^2{-}7 \varphi _1^3{-}\varphi _2{-}88 \varphi _1 \varphi _2{+}64 \varphi _1^2 \varphi _2{-}4 \varphi _2^2{-}51 \varphi _1 \varphi _2^2{+}3 \varphi _2^3\right)}{5\left(2-7\varphi_{1}-2\varphi_{2}\right)}~,\\[10pt]
\left(C_{1}^{\+-1}\right)^{22}&=\frac{-\varphi _1 \varphi _2^2 \left(1{+}37 \varphi _1{+}73 \varphi _1^2{+}14 \varphi _1^3{-}3 \varphi _2{+}176 \varphi _1 \varphi _2{-}73 \varphi _1^2 \varphi _2{+}3 \varphi _2^2{+}37 \varphi _1 \varphi
   _2^2{-}\varphi _2^3\right)}{5\left(2-7\varphi_{1}-2\varphi_{2}\right)}~.\\[5pt]
\end{aligned}
\end{equation}

A problem with this approach is that this breaks the $(1\leftrightarrow2)$ symmetry. For instance, $S^{11}$ does not give $S^{22}$ upon swapping $(\varphi_{1}\leftrightarrow\varphi_{2})$. At the end of the process, $f^{(2)}$ will not be a symmetric function.

This is not such a problem for computations with two-parameter manifolds. However, we are motivated by the power of symmetry evidenced in the genera 0 and 1 computations for manifolds with $\geq3$ parameters to seek another way of determining the propagators, so that we maintain any symmetries that we begin with. The reader who does not care for maintaining this symmetry can use the previous expressions of this subsection for $S^{ij}$. 

\subsubsection*{Resuming the computation}
In what follows, we forget the last five equations and compute $S^{ij}$ anew. To this end, we contract the integrated special geometry relation \eqref{eq:ISG} with $\varphi^{i}$ to obtain
\begin{equation}\label{eq:ISG_Sym}
\varphi^{i}C_{ijk}S^{kl}=-\varphi^{i}\Gamma^{l}_{ij}+\varphi^{l}K_{j}+\delta^{l}_{j}\varphi^{i}K_{i}+\varphi^{i}s^{l}_{ij}~.
\end{equation}
The matrix $\fC_{jk}=\varphi^{i}C_{ijk}$ respects the $(1\leftrightarrow2)$ symmetry, as does its inverse. We introduce more notation,
\begin{equation}
\fC_{jk}=\varphi^{i}C_{ijk}~,\qquad \Upsilon^{l}_{j}=\varphi^{i}\Gamma^{l}_{ij}~,\qquad \fK=\varphi^{i}K_{i}~,\qquad\fs^{l}_{j}=\varphi^{i}s^{l}_{ij}~.
\end{equation}
We shall also use another abuse of notation, whereby we denote the inverse of $\fC_{ij}$ with raised components:
\begin{equation}
\fC^{ij}\fC_{jk}=\delta^{i}_{k}~.
\end{equation}
The components of $\fC_{ij}$ are
\begin{equation}
\begin{aligned}
\fC_{11}&=\frac{5\left(3+51 \varphi _1+64 \varphi _1^2+7 \varphi _1^3-4 \varphi _2+88 \varphi _1 \varphi _2-9 \varphi _1^2 \varphi _2-\varphi _2^2-14 \varphi _1 \varphi _2^2+2 \varphi _2^3\right)}{\varphi_{1}^{2}\Delta}~,\\[10pt]
\fC_{12}&=\frac{5\left(4+13 \varphi _1-13 \varphi _1^2-4 \varphi _1^3+13 \varphi _2-176 \varphi _1 \varphi _2+13 \varphi _1^2 \varphi _2-13 \varphi _2^2+13 \varphi _1 \varphi _2^2-4 \varphi _2^3\right)}{\varphi_{1}\varphi_{2}\Delta}~,\\[10pt]
\fC_{22}&=\frac{5\left(3-4 \varphi _1-\varphi _1^2+2 \varphi _1^3+51 \varphi _2+88 \varphi _1 \varphi _2-14 \varphi _1^2 \varphi _2+64 \varphi _2^2-9 \varphi _1 \varphi _2^2+7 \varphi _2^3\right)}{\varphi_{2}^{2}\Delta}~.
\end{aligned}
\end{equation}
The inverse has components
\begin{equation}
\begin{aligned}
\fC^{11}&=\frac{-\varphi_{1}^{2}\left(3-4 \varphi _1-\varphi _1^2+2 \varphi _1^3+51 \varphi _2+88 \varphi _1 \varphi _2-14 \varphi _1^2 \varphi _2+64 \varphi _2^2-9 \varphi _1 \varphi _2^2+7 \varphi _2^3\right)}{5\left(7-2\varphi_{1}-2\varphi_{2}\right)}~,\\[5pt]
\fC^{12}&=\frac{\varphi_{1}\varphi_{2}\left(4+13 \varphi _1-13 \varphi _1^2-4 \varphi _1^3+13 \varphi _2-176 \varphi _1 \varphi _2+13 \varphi _1^2 \varphi _2-13 \varphi _2^2+13 \varphi _1 \varphi _2^2-4 \varphi _2^3\right)}{5\left(7-2\varphi_{1}-2\varphi_{2}\right)}~,\\[5pt]
\fC^{22}&=\frac{-\varphi_{2}^{2}\left(3+51 \varphi _1+64 \varphi _1^2+7 \varphi _1^3-4 \varphi _2+88 \varphi _1 \varphi _2-9 \varphi _1^2 \varphi _2-\varphi _2^2-14 \varphi _1 \varphi _2^2+2 \varphi _2^3\right)}{5\left(7-2\varphi_{1}-2\varphi_{2}\right)}~.
\end{aligned}
\end{equation}
The denominator appearing above will recur, and so we give it a name:
\begin{equation}
Z=7-2\varphi^{1}-2\varphi^{2}~.
\end{equation}
Analogously to what we did with \eqref{eq:amb_sym}, we have a freedom to change the symmetric part of $\fC^{kj}\fs^{l}_{j}$, and we set this to zero. Then from \eqref{eq:ISG_Sym} we get
\begin{equation}
S^{kl}=\frac{1}{2}\left[\left(-\fC^{kj}\Upsilon^{l}_{j}+\varphi^{l}\fC^{kj}K_{j}+\fK\fC^{kl}\right)+\left(k\leftrightarrow l\right)\right]~.
\end{equation}
The first few terms in $S^{ij}$'s Taylor expansions are
\begin{equation}
\begin{aligned}
S^{11}&=-\frac{3 \varphi _1^2}{35}+\frac{64 \varphi _1^3}{245}-\frac{102}{49} \varphi _1^2 \varphi _2+\frac{79 \varphi _1^4}{1715}-\frac{7654 \varphi _1^3 \varphi _2}{1715}-\frac{7488 \varphi
   _1^2 \varphi _2^2}{1715}+\,...~,\\[10pt]
   S^{12}&=\frac{4 \varphi _1 \varphi _2}{35}+\frac{24}{49} \varphi _1^2 \varphi _2+\frac{24}{49} \varphi _1 \varphi _2^2-\frac{1426 \varphi _1^3 \varphi _2}{1715}-\frac{11182 \varphi _1^2
   \varphi _2^2}{1715}-\frac{1426 \varphi _1 \varphi _2^3}{1715}+\,...~,\\[10pt]
   S^{22}&=-\frac{3 \varphi _2^2}{35}-\frac{102}{49} \varphi _1 \varphi _2^2+\frac{64 \varphi _2^3}{245}-\frac{7488 \varphi _1^2 \varphi _2^2}{1715}-\frac{7654 \varphi _1 \varphi
   _2^3}{1715}+\frac{79 \varphi _2^4}{1715}+\,...~.
\end{aligned}
\end{equation}
Now we turn to computing $S^{i}$, which in the interests of symmetry we do by considering $\varphi^{i}D_{i}S^{jk}$ as in \eqref{eq:BCOV_Derivatives_untilded}. There is another new notation, 
\begin{equation}
\fh^{jk}=\varphi^{i}h^{jk}_{i}~.
\end{equation}
We have the freedom to set $\fh^{1,1}=\fh^{2,2}=0$, and then we can find $S^{1}$ and $S^{2}$ from the $(j,k)=(1,1)$ and $(j,k)=(2,2)$ components of $\varphi^{i}D_{i}S^{jk}$ as below:
\begin{equation}\label{eq:zDS2}
\varphi^{i}D_{i}S^{jk}=\varphi^{j}S^{k}+\varphi^{k}S^{j}-\fC_{mn}S^{mj}S^{nk}+\fh^{jk}~.
\end{equation}
The first few terms in the expansions of the $S^{i}$ so obtained are
\begin{equation}
\begin{aligned}
S^{1}&=-\frac{3 \varphi _1}{70}+\frac{64 \varphi _1^2}{245}-\frac{929 \varphi _1 \varphi _2}{490}+\frac{1389 \varphi _1^3}{1715}-\frac{13539 \varphi _1^2 \varphi _2}{1715}-\frac{7641 \varphi
   _1 \varphi _2^2}{1715}+\,...~,\\[10pt]
S^{2}&=-\frac{3 \varphi _2}{70}-\frac{929 \varphi _1 \varphi _2}{490}+\frac{64 \varphi _2^2}{245}-\frac{7641 \varphi _1^2 \varphi _2}{1715}-\frac{13539 \varphi _1 \varphi
   _2^2}{1715}+\frac{1389 \varphi _2^3}{1715}+\,...~.
\end{aligned}
\end{equation}
We will use the $(j,k)={1,2}$ component of \eqref{eq:zDS2} in order to compute $\fh^{12}$, over which we have no freedom having fixed $\fh^{11}=\fh^{22}=0$. We will need all components of $\fh^{ij}$ when we consider $\varphi^{i}D_{i}S^{j}$, which we will use to compute the remaining propagator $S$~. We find
\begin{equation}
\begin{aligned}
\fh^{12}=\frac{\varphi_{1}\varphi_{2}}{10Z^{2}\Delta}\bigg(98&+679 \varphi _1-5669 \varphi _1^2-6701 \varphi _1^3-29717 \varphi _1^4-29363 \varphi _1^5-4824 \varphi _1^6-2656 \varphi _1^7\\&+21 \varphi _1^8+7 \varphi _1^9+679
   \varphi _2-72532 \varphi _1 \varphi _2-848410 \varphi _1^2 \varphi _2-2858677 \varphi _1^3 \varphi _2\\&-2218750 \varphi _1^4 \varphi _2-307684 \varphi _1^5 \varphi _2+97632 \varphi
   _1^6 \varphi _2+4550 \varphi _1^7 \varphi _2+67 \varphi _1^8 \varphi _2\\&-5669 \varphi _2^2-848410 \varphi _1 \varphi _2^2-1272670 \varphi _1^2 \varphi _2^2+8757690 \varphi _1^3
   \varphi _2^2+3502930 \varphi _1^4 \varphi _2^2\\&-588156 \varphi _1^5 \varphi _2^2+880 \varphi _1^6 \varphi _2^2+280 \varphi _1^7 \varphi _2^2-6701 \varphi _2^3-2858677 \varphi _1
   \varphi _2^3\\&+8757690 \varphi _1^2 \varphi _2^3-13388420 \varphi _1^3 \varphi _2^3+1228780 \varphi _1^4 \varphi _2^3-37094 \varphi _1^5 \varphi _2^3+672 \varphi _1^6 \varphi
   _2^3\\&-29717 \varphi _2^4-2218750 \varphi _1 \varphi _2^4+3502930 \varphi _1^2 \varphi _2^4+1228780 \varphi _1^3 \varphi _2^4-66890 \varphi _1^4 \varphi _2^4\\&+1022 \varphi _1^5
   \varphi _2^4-29363 \varphi _2^5-307684 \varphi _1 \varphi _2^5-588156 \varphi _1^2 \varphi _2^5-37094 \varphi _1^3 \varphi _2^5\\&+1022 \varphi _1^4 \varphi _2^5-4824 \varphi
   _2^6+97632 \varphi _1 \varphi _2^6+880 \varphi _1^2 \varphi _2^6+672 \varphi _1^3 \varphi _2^6-2656 \varphi _2^7\\&+4550 \varphi _1 \varphi _2^7+280 \varphi _1^2 \varphi _2^7+21
   \varphi _2^8+67 \varphi _1 \varphi _2^8+7 \varphi _2^9\bigg)~.
\end{aligned}
\end{equation}
We continue to contract propagator ambiguities with $\varphi^{i}$, and write
\begin{equation}
\fh^{j}=\varphi^{i}h^{j}_{i}
\end{equation}
so that $\varphi^{i}D_{i}S^{j}$ reads
\begin{equation}
\varphi^{i}D_{i}S^{j}=2\varphi^{j}S-\fC_{mn}S^{m}S^{nj}+\fh^{jk}K_{k}+\fh^{j}~.
\end{equation}
We now make another choice, $\frac{1}{\varphi_{1}}\fh^{1}+\frac{1}{\varphi_{2}}\fh^{2}=0$. Then we can get a symmetric $S$ by taking
\begin{equation}
S=\frac{1}{2}\left[\frac{1}{2\varphi_{1}}\left(\varphi^{i}D_{i}S^{1}+\fC_{mn}S^{m}S^{nj}-\fh^{1k}K_{k}\right)+\left(\varphi_{1}\leftrightarrow\varphi_{2}\right)\right]~.
\end{equation}
The first few terms of $S$'s Taylor expansion are
\begin{equation}
S=-\frac{3}{140}-\frac{188 \varphi _1}{245}-\frac{188 \varphi _2}{245}-\frac{5659 \varphi _1^2}{6860}-\frac{34163 \varphi _1 \varphi _2}{3430}-\frac{5659 \varphi _2^2}{6860}+\,...~.
\end{equation}
Having computed each of the propagators, we can work back through the relations \eqref{eq:BCOV_Derivatives_untilded} to obtain all of the propagator ambiguities, which would be necessary to go beyond genus 2. We decline to show those functions here, but remark that all of their denominators are products of powers of $Z$ and $\Delta$. The most complicated propagator ambiguity is $h_{i}$, which has as denominator $40\varphi_{i}Z^{4}\Delta^{3}$, and a numerator of degree 21.

We now plug our propagators into \eqref{eq:BCOV_p99}. It remains to fix the holomorphic ambiguity, which is of the form
\begin{equation}
f^{(2)}=\frac{P_{15}}{Z^{3}\Delta^{2}}~,
\end{equation}
where $P_{15}$ is a degree-15 symmetric polynomial in $\varphi_{1},\,\varphi_{2}$.

We consider the combined set of unknowns, coefficients of $P_{15}$ and the genus 2 instanton numbers $n^{(2)}_{(i,j)}$. We have 73 unknown coefficients of $P_{15}$. Increasing the order to which we work at provides more unknowns $n^{(2)}_{(i,j)}$, and as many equations for these. Some of these new unknowns we can fix in terms of other unknowns by Coxeter symmetry, until with a high enough order we exhaust the independent constraints that Coxeter symmetry offers.

Demanding that the $q$-expansion of $\cF^{(2)}$ possesses infinite dihedral Coxeter symmetry reduces the number of unknowns, but unfortunately these constraints are not all independent and we cannot get by solely by working to a higher order and imposing Coxeter symmetry. Fortunately, for this example, we know ahead of time which instanton numbers should vanish from the tables of Hosono and Takagi \cite{hosono2014mirror}. These are the numbers $n^{(2)}_{(i,j)}$ with $i$ or $j$ less than 4, and also $n^{(2)}_{(4,4)}$. If we include these zeroes as known values, then we get an integer set of solutions for the instanton numbers that agrees with Hosono and Takagi's tables where they overlap, as well as fixing the ambiguity $f^{(2)}$. Curiously, we need \emph{all} of these zeroes. If we leave out $n^{(2)}_{(4,4)}=0$ then we have a solution with one degree of freedom, eliminated by imposing $n^{(2)}_{(4,4)}=0$.

We have taken a barbaric way to know these zeroes, from Hosono and Takagi's computation. Since they used the conifold gap condition, we have not escaped its necessity. Future investigation will determine whether there is a means of fixing all zeroes from another principle, like a Castelnuovo bound (vital for the computation of \cite{Huang:2006hq}, with a possible generalisation discussed in \cite{Alexandrov:2022pgd}), and then we could hope to check how high of a genus Coxeter symmetry will take us. The genus 2 numbers that we compute are tabulated in \tref{tab:Instanton_Numbers_2_52_genus2}~.

In \cite{hosono2014mirror}, Hosono and Takagi display numbers $n^{(g)}_{(i,j)}$ for $0\leq g\leq2$, $0\leq i \leq 12$, $0\leq j \leq 6$. We have in this thesis displayed numbers for this model at the same genera, but for degrees $i+j\leq37$, which includes numbers not displayed before (even when the Coxeter action is taken into account). It would be interesting to return to this problem at genus 3.

The contents of the following table is available in electronic form \cite{mcgovern2023a}.

\begin{table}[H]
\centering
\def\arraystretch{1.1}
\begin{adjustbox}{width=\columnwidth,center}
\tiny{\pgfplotstabletypeset[
col sep=space,
white space chars={]},
ignore chars={\ },
every head row/.style={before row=\hline,after row=\hline\hline},
every first row/.style={before row=\vrule height12pt width0pt depth0pt},
every last row/.style={before row=\vrule height0pt width0pt depth6pt,after row=\hline},
%after row=\hline,    %  Uncomment this to get back lines between all rows.
columns={A,B,A,B},
display columns/0/.style={select equal part entry of={0}{2},column type = {|l},column name=\vrule height12pt width0pt depth6pt \hfil $\bm p$},
display columns/1/.style={select equal part entry of={0}{2},column type = {|l|},column name=\hfil $n^{(2)}_{\bm p}$},
display columns/2/.style={select equal part entry of={1}{2},column type = {|l},column name=\hfil $\bm p$},
display columns/3/.style={select equal part entry of={1}{2},column type = {|l|},column name=\hfil $n^{(2)}_{\bm p}$},
string type]
{
A ] B
{5, 4}]2500
{5, 5}]2238300
{6, 4}]36800
{6, 5}]101188225
{7, 4}]132150
{6, 6}]10885677450
{7, 5}]1276419800
{8, 4}]191850
{7, 6}]321011805475
{8, 5}]6764994000
{9, 4}]132150
{7, 7}]19785981206800
{8, 6}]4007017841650
{9, 5}]17585700200
{10, 4}]36800
{8, 7}]497495418446900
{9, 6}]25512018106050
{10, 5}]24017901850
{11, 4}]2500
{8, 8}]23487690756165150
{9, 7}]6354579702758450
{10, 6}]91091698085900
{11, 5}]17585700200
{9, 8}]546653742258204975
{10, 7}]46325408551373425
{11, 6}]192086807308450
{12, 5}]6764994000
{9, 9}]22058659953217981800
{10, 8}]7207510049560719850
{11, 7}]206305163005291900
{12, 6}]245649059538250
{13, 5}]1276419800
{10, 9}]492220346020866314150
{11, 8}]58488168886336533925
{12, 7}]585085137096464550
{13, 6}]192086807308450
{14, 5}]101188225
{10, 10}]17908353146570145820800
{11, 9}]6703938825093094075300
{12, 8}]307911899250593057650
{13, 7}]1083681563433259200
{14, 6}]91091698085900
{15, 5}]2238300
{11, 10}]389901597341682586245250
{12, 9}]59381136406580839191450
{13, 8}]1088683595004165702950
{14, 7}]1328923645662952325
{15, 6}]25512018106050
{16, 5}]2500
{11, 11}]13181153378953434491429200
{12, 10}]5475299266824539894041150
{13, 9}]356988331876350933661250
{14, 8}]2645148270492863092650
{15, 7}]1083681563433259200
{16, 6}]4007017841650
{12, 11}]282712962376014816940856175
{13, 10}]52176320562765050835877250
{14, 9}]1500298428800313660065325
{15, 8}]4481489156958762453000
{16, 7}]585085137096464550
{17, 6}]321011805475
{12, 12}]9047714570978243256779338500
{13, 11}]4082629478671789440045910650
{14, 10}]349654611337758233664220900
{15, 9}]4499834280177683754617100
{16, 8}]5337842575936922957650
{17, 7}]206305163005291900
{18, 6}]10885677450
{13, 12}]192252132113764705987353896200
{14, 11}]41394311546286990570767724825
{15, 10}]1690595367815132163307836750
{16, 9}]9770462296518547305273800
{17, 8}]4481489156958762453000
{18, 7}]46325408551373425
{19, 6}]101188225
{13, 13}]5896815924361936623044444626800
{14, 12}]2847327956358643190359156085300
{15, 11}]303781402592961107944725255700

}
}
\end{adjustbox}
\vskip10pt
\capt{6.5in}{tab:Instanton_Numbers_2_52_genus2}{The genus~2 instanton numbers of total degree $\leqslant 37$ for the family \eqref{eq:CICY_2_52}. The numbers not in this list are either zero, or given by those in the table after permuting indices.}
\end{table}
\begin{table}[H]\notag
\centering
\def\arraystretch{1.01}
\begin{adjustbox}{width=\columnwidth,center}
\tiny{\pgfplotstabletypeset[
col sep=space,
white space chars={]},
ignore chars={\ },
every head row/.style={before row=\hline,after row=\hline\hline},
every first row/.style={before row=\vrule height12pt width0pt depth0pt},
every last row/.style={before row=\vrule height0pt width0pt depth6pt,after row=\hline},
%after row=\hline,    %  Uncomment this to get back lines between all rows.
columns={A,B},
display columns/0/.style={select equal part entry of={0}{1},column type = {|l},column name=\vrule height12pt width0pt depth6pt \hfil $\bm p$},
display columns/1/.style={select equal part entry of={0}{1},column type = {|l|},column name=\hfil $n^{(2)}_{\bm p}$},
string type]
{
A ] B
{16, 10}]6008644011089373537248656800
{17, 9}]15502680926528679507634750
{18, 8}]2645148270492863092650
{19, 7}]6354579702758450
{20, 6}]36800
{14, 13}]124565089119075700247816681507000
{15, 12}]30450650946759882533741285631525
{16, 11}]1650396599182174240918908567600
{17, 10}]15912299153776396184830406150
{18, 9}]18071193444493417015126625
{19, 8}]1088683595004165702950
{20, 7}]497495418446900
{14, 14}]3693414921590391334562655080548700
{15, 13}]1887312824415684091294137374297850
{16, 12}]241434752646453338262009308998450
{17, 11}]6750850050453875649938033430650
{18, 10}]31701961212371099761950706550
{19, 9}]15502680926528679507634750
{20, 8}]307911899250593057650
{21, 7}]19785981206800
{15, 14}]77734854231576430735155964501483500
{16, 13}]21141578696054339753623844206366900
{17, 12}]1447892872574562822004056745806925
{18, 11}]21056470435268391362146639838000
{19, 10}]47822198309268358643758769375
{20, 9}]9770462296518547305273800
{21, 8}]58488168886336533925
{22, 7}]321011805475
{15, 15}]2242005633475379243138889611530224350
{16, 14}]1202166271275554852096319284514781050
{17, 13}]179193911305392936323515910617613500

}
}
\end{adjustbox}
\caption*{\tref{tab:Instanton_Numbers_2_52_genus2} continued.}
\end{table}
\begin{table}[H]\notag
\centering
\def\arraystretch{1.005}
\begin{adjustbox}{width=\columnwidth,center}
\tiny{\pgfplotstabletypeset[
col sep=space,
white space chars={]},
ignore chars={\ },
every head row/.style={before row=\hline,after row=\hline\hline},
every first row/.style={before row=\vrule height12pt width0pt depth0pt},
every last row/.style={before row=\vrule height0pt width0pt depth6pt,after row=\hline},
%after row=\hline,    %  Uncomment this to get back lines between all rows.
columns={A,B},
display columns/0/.style={select equal part entry of={0}{1},column type = {|l},column name=\vrule height12pt width0pt depth6pt \hfil $\bm p$},
display columns/1/.style={select equal part entry of={0}{1},column type = {|l|},column name=\hfil $n^{(2)}_{\bm p}$},
string type]
{
A ] B

{18, 12}]6669236758606676544482977924485150
{19, 11}]50556113264139731816624429986400
{20, 10}]54823814338848321169711111550
{21, 9}]4499834280177683754617100
{22, 8}]7207510049560719850
{23, 7}]1276419800
{16, 15}]47083170110743580006872424368633531875
{17, 14}]14025908356241673430452097295836041700
{18, 13}]1169983477395354290630422427608175550
{19, 12}]23876076766141789732446125447010225
{20, 11}]94078435233076347832828692471750
{21, 10}]47822198309268358643758769375
{22, 9}]1500298428800313660065325
{23, 8}]546653742258204975
{24, 7}]132150
{16, 16}]1327080986654885164312931683366667358200
{17, 15}]741736053327841335761519208560504382350
{18, 14}]126013267019536739983934798259312856100
{19, 13}]5967526679788804873035522660386672150
{20, 12}]67042622214148592803250933902047750
{21, 11}]136315316135643017078981825252200
{22, 10}]31701961212371099761950706550
{23, 9}]356988331876350933661250
{24, 8}]23487690756165150
{17, 16}]27835064011278355479795425089182885779850
{18, 15}]8971331196558149334377988222442191858000
{19, 14}]885994523265940239167051685255212381600
{20, 13}]24041369335716235988574456740045846350
{21, 12}]148673916874105664720736408737803900
{22, 11}]154206354444180862857899859688400
{23, 10}]15912299153776396184830406150
{24, 9}]59381136406580839191450
{25, 8}]497495418446900
{17, 17}]769459996977013484532529807380612024950900
{18, 16}]445912967342824716142914500096902010818850

}
}
\end{adjustbox}
\caption*{\tref{tab:Instanton_Numbers_2_52_genus2} continued.}
\end{table}
\begin{table}[H]\notag
\centering
\def\arraystretch{1.005}
\begin{adjustbox}{width=\columnwidth,center}
\tiny{\pgfplotstabletypeset[
col sep=space,
white space chars={]},
ignore chars={\ },
every head row/.style={before row=\hline,after row=\hline\hline},
every first row/.style={before row=\vrule height12pt width0pt depth0pt},
every last row/.style={before row=\vrule height0pt width0pt depth6pt,after row=\hline},
%after row=\hline,    %  Uncomment this to get back lines between all rows.
columns={A,B},
display columns/0/.style={select equal part entry of={0}{1},column type = {|l},column name=\vrule height12pt width0pt depth6pt \hfil $\bm p$},
display columns/1/.style={select equal part entry of={0}{1},column type = {|l|},column name=\hfil $n^{(2)}_{\bm p}$},
string type]
{
A ] B

{19, 15}]84847587251331673881762239411974072983050
{20, 14}]4938016593847719974314383381477983741100
{21, 13}]77169654314981775489114851835598126800
{22, 12}]261702486079440770739117430365216750
{23, 11}]136315316135643017078981825252200
{24, 10}]6008644011089373537248656800
{25, 9}]6703938825093094075300
{26, 8}]4007017841650
{18, 17}]16130063331032815277955386421606900094355875
{19, 16}]5569304833367577335205278477402592908516350
{20, 15}]636726757004598625316020644295171791557250
{21, 14}]22041710721238453533303229084611701286825
{22, 13}]198702763601020024146439712249527136275
{23, 12}]366918571341072517525517535457060450
{24, 11}]94078435233076347832828692471750
{25, 10}]1690595367815132163307836750
{26, 9}]492220346020866314150
{27, 8}]6764994000
{18, 18}]438539109875379528721919654529846561401409750
{19, 17}]262362104844021123045694638864495986589287000
{20, 16}]55131178495812759321176609589875814140037750
{21, 15}]3836414759887507545665007708487359916157800
{22, 14}]79449270275648960296194539460339370694050
{23, 13}]412564879312424855910387890012240484400
{24, 12}]410576929861745683208869036387758800
{25, 11}]50556113264139731816624429986400
{26, 10}]349654611337758233664220900
{27, 9}]22058659953217981800
{28, 8}]191850
{19, 18}]9192063299791665753644261196926491990414773625
{20, 17}]3372577227429580157843234196650003432550205075
{21, 16}]438359755343431485741657365115289687329201500
{22, 15}]18737856667070391183320951837536512233506700
{23, 14}]232781528281126665497995449029490453266300
{24, 13}]693400092277019907121874794244775804900
{25, 12}]366918571341072517525517535457060450
{26, 11}]21056470435268391362146639838000
{27, 10}]52176320562765050835877250
{28, 9}]546653742258204975

}
}
\end{adjustbox}
\caption*{\hskip300pt\tref{tab:Instanton_Numbers_2_52_genus2} continued.}
\end{table}

\chapter{Supergravity Compactifications}\label{Chapter:SUGRA}
\setlength\epigraphwidth{.65\textwidth}\epigraph{Rose\phantom{a}\; : I can’t even touch it. Seems to be in a state of flux.\\
Donna : What does that mean?\\
Rose\phantom{a}\; : I don’t know. Sort of thing the Doctor would say.}{Russell T Davies, \textit{Turn Left}}

Our discussion of background material in this section follows the textbooks \cite{Becker:2006dvp,Freedman:2012zz}. If massive degrees of freedom are neglected, then any of the five ten-dimensional superstring theories reduce to ten-dimensional supergravity theories. Of foremost concern to us are the two type II superstring theories (type IIA and IIB), whose low energy effective theories are the two maximal $\mathcal{N}=2$ supergravities (of type IIA and IIB respectively). Either of the maximal supergravities can be compactified on a Calabi-Yau threefold\footnote{Complex dimension three, so real dimension six.} to obtain a four-dimensional $\mathcal{N}=2$ matter coupled supergravity, and at the level of supergravity, mirror symmetry gives a duality between each of the supergravity theories obtained by the compactifications
\begin{equation}
\begin{matrix}\text{Type IIB supergravity}\\\text{on CY threefold X}\end{matrix}\qquad\cong\qquad \begin{matrix}\text{Type IIA supergravity}\\\text{on mirror Y}\end{matrix}~.
\end{equation}
We will review briefly the field content of these 4d theories, in particular recalling that the scalar fields in the matter content include geometric moduli of the threefolds. By comparing supersymmetric black hole solutions to both of the above theories, we can in some cases obtain formulae that relate number-theoretic quantities computed on $X$ to the enumerative invariants of $Y$. 

Moreover, one approach to realising realistic phenomenological models is \emph{flux compactifications}, which we shall always view from the IIB perspective. Here, in addition to compactifying IIB supergravity (or indeed the full string theory) on a Calabi-Yau threefold $X$, one gives nonzero vacuum expectation values to form fields supported on the cohomology of $X$. This breaks supersymmetry further, and the four-dimensional massless theory has $\mathcal{N}=1$ supersymmetry. We will study supersymmetric vacuum configurations, and provide new examples in support of the \emph{flux modularity conjecture} \cite{Kachru:2020sio,Kachru:2020abh}. Such IIB setups can be related to F-theory on an elliptically fibred Calabi-Yau \emph{fourfold} $\wt{X}$, and we will see that (in line with previous conjectures) the elliptic fibre in a supersymmetric vacuum configuration has a surprising relation to the modularity of the threefold $X$.

\section{4d $\mathcal{N}=2$ matter coupled supergravities}\label{sect:SUGRA_4dN=2}
The $\mathcal{N}=2$ gravity multiplet consists of the spin-2 spacetime metric $g_{\mu\nu}$, two spin-$3/2$ gravitini, and a spin-1 vector field dubbed the graviphoton. The CPT conjugates to all of these must be included in the theory as well, so as to have CPT symmetry. Additional vector fields can be found in the matter content, residing in 4d $\mathcal{N}=2$ vector multiplets which each consist of a spin-1 vector field, two spin-$1/2$ gaugini, and two real scalars. These scalars are packaged into one complex scalar. Again, this content is not CPT self-conjugate and so fields with opposite helicities must also be included. Finally, there can be CPT self-conjugate hypermultiplets, which contain two spin-$1/2$ fields, their CPT conugate spin-$-1/2$ fields, and four real scalars.

The attractor mechanism governs the values of the vector multiplet scalars near a stationary black hole solution, so we pause to comment on the possible manifolds in which the vector multiplet scalars of a $\mathcal{N}=2$ theory can live in. With there being $n$ vector multiplets, we have $n+1$ $\text{U}(1)$ vector fields when the graviphoton is incorporated. We will only be concerned with gauge fields for Abelian gauge groups $\text{U}(1)^{n+1}$, i.e. extended Maxwell theory. Just as classical electromagnetism (albeit with electric and magnetic charges) possesses an electromagnetic duality group $\text{SL}(2,\IZ)$ under which the charge vector $(q_{\text{electric}},q_{\text{magnetic}})^{T}$ transforms as a doublet, extended Maxwell theory enjoys a symplectic electromagnetic duality group $\text{Sp}(2n+2,\IZ)$ under which the charge vector $(\mathbf{q}_{\text{electric}},\mathbf{q}_{\text{magnetic}})$ transforms in the vector representation. 

If the 4d theory is to both have these duality transformations and be supersymmetric, then the vector multiplet scalars must be coordinates on a K{\"a}hler manifold with an \\${\text{SP}(2n+2,\IR)\otimes \text{GL}(1,\IC)}$ bundle that possesses a holomorphic symplectic section\footnote{When we turn to studying IIB compactifications on a CY threefold $X$, the bundle will simply be $H^{3}(X,\IC)$ and the section will be the holomorphic three-form $\Omega$.} $\cS$ \cite{Strominger:1990pd}.  The $n$ vector multiplet scalars can be projectivised into $n+1$ scalar fields living on a K{\"a}hler manifold $\cM$. A symplectic duality transformation on the $2n+2$-component charge vector is paired with a symplectic transformation of the section 
\begin{equation}\label{eq:hol_sym_sec}
\cS=\begin{pmatrix}
X^{I}\\\cF_{I}
\end{pmatrix}
\end{equation}
on the scalar manifold $\cM$. We are assuming the existence of such a section $\cS$ in line with \cite{Strominger:1990pd}, and denoting components of $\cS$ by $X_{I}$ and $\cF_{I}$. In this equation and the following, $I,J,K$ run from 0 to $n$. With $\langle\circ,\circ\rangle$ giving the symplectic form, the K{\"a}hler potential of $\cM$ is
\begin{equation}
K=\ii \langle \cS,\bar{\cS}\rangle~.
\end{equation}
There is an additional requirement, that 
\begin{equation}\label{eq:what}
\langle \partial_{I}\cS,\partial_{J}\cS\rangle=0~.
\end{equation}
K{\"a}hler manifolds possessing such a section $\cS$ are termed \emph{special K{\"a}hler}. So long as the matrix $\partial_{I}X^{J}$ is invertible, then there must locally exist a holomorphic function $\cF(X)$, homogeneous of degree 2, such that
 \begin{equation}
 \cF_{I}=\frac{\partial \cF(X)}{\partial X^{I}}.
\end{equation} 
In the absence of fluxes, the Kaluza-Klein reduction (as recounted for instance in the textboook \cite{Becker:2006dvp}) of either maximal 10d supergravity on a Calabi-Yau is a 4d $\mathcal{N}=2$ supergravity with some number of the aforementioned two kinds of matter multiplets. We turn now to describing these theories, with particular mind to the prepotential for their vector multiplet scalars. 

\subsubsection*{Type IIA compactifications}
If type IIA supergravity is compactified on a Calabi-Yau $Y$, the resulting theory possesses $h^{1,1}$ vector multiplets and $h^{2,1}+1$ hypermultiplets. The $h^{1,1}$ vector multiplet's complex scalars parametrise the space of complexified K{\"a}hler classes on $Y$.  The hypermultiplet moduli space contains the space of complex structures on $Y$, in addition to further scalars coming from reduction of the type IIA form fields on $Y$. 

The first half of our symplectic section $\cS$, which we denoted $X^{I}$ in the previous subsection, will be taken to be the projective coordinates $z^{I}$.

The prepotential is closely related to the genus-0 topological A-model free energy,
\begin{equation}\label{eq:IIA_prepotential}
\begin{aligned}
\cF&=-\frac{1}{6z^{0}}Y_{IJK}z^{I}z^{J}z^{K}-\frac{\left(z^{0}\right)^{2}}{(2\pi\ii)^{3}}\sum_{\mathbf{k}\geq0}n^{(0)}_{\mathbf{k}}\,\text{Li}_{3}\left(\exp\left(2\pi\ii\,k_{i}\frac{\bm{z}^{i}}{z^{0}}\right)\right)\\[5pt]
&\equiv-\frac{1}{6z^{0}}Y_{IJK}z^{I}z^{J}z^{K}-\left(z^{0}\right)^{2}\cI\left(\frac{\bm{z}}{z^{0}}\right)~.
\end{aligned}
\end{equation}
We have collected the $z^{i}$ into the vector quantity $\bm{z}$. Here $i$ runs from 1 to $h^{1,1}(Y)$ while $I,\,J,\,K$ run from 0 to $h^{1,1}(Y)$. The $Y_{ijk}$ are the topological numbers that we encountered in \eqref{eq:Y_ijk}, \eqref{eq:Y_00i}, \eqref{eq:Y_0ij} and \eqref{eq:Y_000}.

The projective coordinates $z^{I}$ relate to the usual complexified K{\"a}hler coordinates $t^{i}$ via
\begin{equation}
t^{i}=\frac{z^{i}}{z^{0}}~.
\end{equation}

Note that in \eqref{eq:IIA_prepotential} we have given the prepotential $\cF$ together with its nonperturbative instanton corrections $\left(z^{0}\right)^{2}\cI\left(z^{i}/z^{0}\right)$ and perturbative quantum corrections, those being the terms in $\frac{1}{6 z^{0}}Y_{IJK}z^{I}z^{J}z^{K}$ where any index takes the value 0. We will understand the purely classical value of the prepotential to be
\begin{equation}
\cF_{\text{Classical}}=-\frac{\left(z^{0}\right)^{2}}{6}Y_{ijk}t^{i}t^{j}t^{k}~.
\end{equation}

\subsubsection*{Type IIB compactifications}
Compactifying the type IIB supergravity on a CY threefold $X$ leads to a 4d $\mathcal{N}=2$ theory with $h^{2,1}(X)$ vector multiplets and $h^{1,1}(X)+1$ hypermultiplets. The $h^{2,1}$ vector multiplet scalars parametrise complex structures on $X$, and within the hypermultiplet moduli are $h^{1,1}(X)$ complex scalars parametrising the complexified K{\"a}hler structures on $X$. When $X$ is mirror to $Y$, the field content of IIB on $X$ and IIA on $Y$ are identical. Both theories are actually identical, but this is a more involved duality than simply identifying dimensions by swapping the Hodge numbers, as the hypermultiplet moduli space of either compactification must contain the vector multiplet moduli space of the other. In both compactifications there is a universal hypermultiplet containing the axiodilaton field.

By using the mirror map, so that the $t^{i}$ parametrising $Y$'s K{\"a}hler structures are coordinates on the complex structure moduli space of $X$, we can find the same symplectic section as in the previous subsection on IIA compactifications,
\begin{equation}
\cS=\begin{pmatrix}
z^{I}\\\displaystyle\frac{\partial\cF}{\partial z^{I}}
\end{pmatrix}~,
\end{equation}
with the same $\cF$ as in \eqref{eq:IIA_prepotential}. There is an important point to be made on classical versus nonclassical physics. In the IIA compactification, we have seen that the prepotential $\cF$ is the sum of a classical part and quantum corrections. However, in the IIB frame this quantity is purely classical. Only after changing duality frames does the instanton sum $\cI$ concealed in $\cF$ above take on a nonperturbative interpretation.

The symplectic section $\cS$ has a natural relation to the holomorphic three-form $\Omega$ on $X$. After introducing a symplectic basis $\alpha_{I},\,\beta^{I}$ of $H^{3}(X,\IZ)$ we can write
\begin{equation}\label{eq:Omega_in_ISB}
\Omega=z^{I}\alpha_{I}-\frac{\partial\cF}{\partial z^{I}}\beta^{I}~.%Really?
\end{equation}
Or in other words, the symplectic section is the integral period vector
\begin{equation}
\cS=\Pi~.
\end{equation}

\section{The attractor mechanism}\label{sect:SUGRA_AttMech}
This subsection serves to review some background material. We follow \cite{Freedman:2012zz} and display a derivation of the attractor equations, including the necessary manipulations to obtain Strominger's form of the equations \cite{Strominger:1996kf}. We adopt a set of notations in this section that make the supergravity analysis tractable, but this section is mostly self-contained and included to provide better background to the work on rank-two attractors.

In words, the attractor mechanism fixes the values of the vector multiplet scalars at the horizon of a supersymmetric black hole configuration in terms of the charges of said black hole. We shall at the end of this section explain how in Calabi-Yau compactifications these attractor equations can be interpreted in terms of the internal geometry's cohomology.

We begin with a more concrete account of electromagnetic duality, which was important in arguing for the existence of a holomorphic symplectic section on the scalar manifold. Then we set about describing dilatation-gauge fixing, which is necessary in order to fix the form of the action that we will consider black hole solutions for, and the kinetic terms for the gauge fields.

Subsequently, we will take a spherically symmetric solution ansatz and derive the equations of motion. The scalar field's equations of motions will be recast as a gradient flow equation. Finally we will follow Ferrara and Kallosh \cite{Ferrara:1996dd} to find algebraic expressions for the fixed points of these gradient flows. Note that although we will study attractor points, we never work with the gradient flow equations, so do not find attractors by solving differential equations. We instead will in later sections deal with solutions to the algebraic equations given at the end of this section.

\subsubsection*{Electromagnetic duality}

We shall write the Maxwell fields as $A_{\mu}^{I}$, with $\mu$ a spacetime index and $I$ running from $0$ to $n$ (one of these is the graviphoton and the remaining n are fields in the vector multiplets). The field strengths will then be 

\begin{equation}
F_{\mu\nu}^{\phantom{\mu\nu}I}=\partial_{\mu}A^{I}_{\nu}-\partial_{\nu}A^{I}_{\mu}.
\end{equation}
These field strengths form the first $n+1$ components of a symplectic vector with $2n{+}2$ components. With the action $S$ given below, the quantities 

\begin{equation}
G_{\mu\nu I}=\kappa^{2}\epsilon_{\mu\nu\rho\sigma}\frac{\delta S}{\delta F_{\rho\sigma}^{\phantom{\rho\sigma}I}}
\end{equation}
form the remaining $n+1$ components. If $F_{\mu\nu}^{\phantom{\mu\nu}I}$ and $G_{\mu\nu I}$ solve the gauge field equations of motion then so do $\widehat{F}_{\mu\nu}^{\phantom{\mu\nu}I}$ and $\widehat{G}_{\mu\nu I}$ defined by

\begin{equation}
\left(\begin{matrix}\widehat{F}_{\mu\nu}^{\phantom{\mu\nu}I}\\[5pt]\widehat{G}_{\mu\nu I}\end{matrix}\right)=\mathbf{W}\left(\begin{matrix}F_{\mu\nu}^{\phantom{\mu\nu}I}\\[5pt]G_{\mu\nu I}\end{matrix}\right),
\end{equation}
where $\mathbf{W}$ is a $(2n+2)\times(2n+2)$ symplectic matrix. There is a corresponding transformation law for the gauge kinetic matrix (and hence the scalars, upon which this matrix depends). This is the statement of electromagnetic duality for Maxwell fields.

The electric and magnetic charges $q_{I}$ and $p^{I}$ in a volume $V$ are given by

\begin{equation}\label{eq:charge_integrals}
q_{I}=-\frac{1}{2}\int_{V}\text{d}^{3}x\,\epsilon^{ijk}\partial_{i}G_{jkI}~,\qquad p^{I}=-\frac{1}{2}\int_{V}\text{d}^{3}x\,\epsilon^{ijk}\partial_{i}F_{jk}^{\phantom{jk}I}~.
\end{equation}
Under electromagnetic duality the charge vector $\left(p^{I},q_{I}\right)^{\text{T}}$ is mapped to $\mathbf{W}\left(p^{I},q_{I}\right)^{\text{T}}$. Quantization forces the charges to lie on a lattice:

\begin{equation}
\left(p^{I},q_{I}\right)\left(\begin{matrix}\phantom{-}\mathbf{0}_{n\times n} & \mathbf{1}_{n\times n} \\-\mathbf{1}_{n\times n} & \mathbf{0}_{n\times n}\end{matrix}\right)\left(\begin{matrix}p^{I}\\q_{I}\end{matrix}\right)=2\pi n,\qquad n\in\mathbb{Z},
\end{equation}
and so $\mathbf{W}$ must have integer entries: $\mathbf{W}\in\text{Sp}\left(2n+2,\mathbb{Z}\right)$. 

\subsubsection*{Dilatation gauge fixing and the gauge field kinetic term}
We have already argued that this electromagnetic duality, plus supersymmetry, leads to the requirement that the vector multiplet scalars take values in a special K{\"a}hler manifold with symplectic section $\cS$ and K{\"a}hler potential $K=\ii\langle\cS,\bar{\cS}\rangle$.

With the symplectic section $\cS$ defined in terms of a prepotential $\cF$ as in \eqref{eq:hol_sym_sec}, the gauge field kinetic matrix is 

\begin{equation}
\kappa^{2}\mathcal{N}_{IJ}\left(z,\bar{z}\right)=\bar{F}_{IJ}+\ii\frac{K_{IK}X^{K}\,K_{JL}X^{L}}{K_{MN}X^{M}X^{N}},
\end{equation}
in which all subscripts denote partial derivatives. This couples the scalars to the gauge fields. We write

\begin{equation}\label{eq:reim_N}
\text{\textcyr{R}}_{IJ}=\text{Re}\left[\mathcal{N}_{IJ}\right], \qquad \text{\textcyr{I}}_{IJ}=\text{Im}\left[\mathcal{N}_{IJ}\right].
\end{equation}
The dilatation operator's action on the scalars is to flow them along a Killing vector of this manifold. To gauge fix, one identifies all points in the same orbits of this action. Using the Frobenius theorem one can choose coordinates $y, z^{\alpha}$ so that $\partial_{y}$ is parallel to this Killing~vector. 

We take the quantities $X^{I}$ to be projective by introducing functions $Z^{I}(z)$, that do not depend on $y$, such that $X^{I}=y Z^{I}(z)$. One can then write

\begin{equation}\label{eq:sympvectors}
\cS=\left(\begin{matrix}X^{I}\\[5pt]F_{I}(X)\end{matrix}\right)=y\left(\begin{matrix}Z^{I}(z)\\[5pt]F_{I}\left(Z(z)\right)\end{matrix}\right).
\end{equation}
The dilatation gauge is fixed by choosing a specific value for $y$. This value is chosen to be real and to keep the K{\"a}hler potential $K$ at a fixed constant value of $-\kappa^{-2}$.

Performing this fixing we are left with a \textit{projective special K{\"a}hler} manifold. This has coordinates $z^{\alpha}$, which are the scalars that will appear in the gauge-fixed action. The K{\"a}hler potential of this projective manifold is $\mathcal{K}\left(z,\bar{z}\right)$, where

\begin{equation}
\ee^{-\kappa^{2}\mathcal{K}}=-\kappa^{2}\frac{\partial K}{\partial X^{I} \partial \bar{X}^{\bar{J}}}Z^{I}\bar{Z}^{\bar{J}}=\frac{1}{|y|^{2}}.
\end{equation}
There is some freedom in our choice of scale for $Z^{I}$ and $y$. Had we instead opted for 

\begin{equation}\label{eq:kahl_tr}
\widehat{y}=y\ee^{\kappa^{2}f(z)},\quad \widehat{Z}^{I}=Z^{I}\ee^{-\kappa^{2}f(z)}
\end{equation}
with $f$ an arbitrary holomorphic function, then we would have arrived at a projective potential

\begin{equation}
\widehat{\mathcal{K}}=\mathcal{K}+f+\bar{f}.
\end{equation}
A connection is introduced to give derivatives covariant under the transformations \eqref{eq:kahl_tr}:

\begin{equation}
\begin{aligned}
&\nabla_{\alpha}Z^{I}=\partial_{\alpha}Z^{I}+\kappa^{2}\left(\partial_{\alpha}\mathcal{K}\right)Z^{I},\\[5pt]
&\overline{\nabla}_{\bar{\alpha}}\bar{Z}^{\bar{I}}=\partial_{\bar{\alpha}}\bar{Z}^{\bar{I}}+\kappa^{2}\left(\partial_{\bar{\alpha}}\mathcal{K}\right)\bar{Z}^{\bar{I}},\\[5pt]
&\nabla_{\alpha}\bar{Z}^{\bar{I}}=\partial_{\alpha}\bar{Z}^{\bar{I}}=0,\\[5pt]
&\overline{\nabla}_{\bar{\alpha}}Z^{I}=\partial_{\bar{\alpha}}Z^{I}=0.
\end{aligned}
\end{equation}
The metric on the projective manifold can be written as

\begin{equation}\label{eq:proj_met}
g_{\alpha\bar{\beta}}=\partial_{\alpha}\partial_{\bar{\beta}}\mathcal{K}=\ii\langle\nabla_{\alpha}\cS,\bar{\nabla}_{\bar{\beta}}\bar{\cS}\rangle.
\end{equation}
We end this subsection discussing the scalars by stating the bilinear form $\mathcal{N}_{IJ}$ used in the kinetic term for the gauge fields when a prepotential does not exist. This is

\begin{equation}
\kappa^{2}\bar{\mathcal{N}}_{IJ}=\left(\bar{F}_{I}\; \nabla_{\alpha}F_{I}\right)\left(\bar{X}^{J}\; \nabla_{\alpha}X^{J}\right)^{-1}.
\end{equation}
We will still in this case use the definitions \eqref{eq:reim_N}.

\subsubsection*{Equations of motion}
The bosonic part of the action for this theory is

\begin{equation}\label{eq:Action}
S\!=\frac{1}{2\kappa^{2}}\!\int\!\text{d}^{4}x\left[\sqrt{-g}\left(R{-}2g^{\mu\nu}g_{\alpha\bar{\beta}}\partial_{\mu}z^{\alpha}\partial_{\nu}\bar{z}^{\bar{\beta}}+\frac{1}{2}\text{\textcyr{I}}_{IJ}F_{\mu\nu}^{\phantom{\mu\nu}I}F^{\mu\nu J}\right){-}\frac{1}{4}\text{\textcyr{R}}_{IJ}\,\epsilon^{\mu\nu\rho\sigma}F_{\mu\nu}^{\phantom{\mu\nu}I}F_{\rho\sigma}^{\phantom{\rho\sigma}J} \right].
\end{equation}
One takes the following ansatz for a spherically symmetric, static metric:

\begin{equation}
\dd s^{2}=-\ee^{2U(\tau)}\dd t^{2}+\ee^{-2U(\tau)}\left[\frac{\dd \tau^{2}}{\tau^{4}}+\frac{1}{\tau^{2}}\left(\dd\theta^{2}+\sin^{2}\left(\theta\right)\,\dd\phi^{2}\right)\right].
\end{equation}
The coordinates $\theta$ and $\phi$ are the familiar 2-sphere coordinates. The extremal Reissner-Nordstrom coordinate can be brought to this form, and in this case $\tau$ is $1/r$, where the shifted radial variable $r$ is 0 at the horizon.

In a static, spherically symmetric solution the three-dimensional electric and magnetic fields will only have radial components. Thus our field strengths are of the form

\begin{equation}
F^{I}=F^{\phantom{t\tau} I}_{t\tau}\dd t\wedge\dd\tau+F_{\theta\phi}^{\phantom{\theta\phi}I}\dd\theta\wedge\dd\phi.
\end{equation}
Staticity and spherical symmetry gives

\begin{equation}\nabla_{t}F_{\mu\nu}^{\phantom{\mu\nu}I}=\nabla_{\theta}F_{\mu\nu}^{\phantom{\mu\nu}I}=\nabla_{\Phi}F_{\mu\nu}^{\phantom{\mu\nu}I}=0.
\end{equation}
The latter two of these enter into a Bianchi identity from which one obtains the equation
 
\begin{equation}
\nabla_{\tau}F_{\theta\phi}^{\phantom{\theta\phi}I}=0.
\end{equation}
These equations are solved by
 
\begin{equation}
F^{\phantom{t\tau} I}_{t\tau}=f^{I}(\tau),\qquad F_{\theta\phi}^{\phantom{\theta\phi}I}=-\frac{p^{I}
}{4\pi}\sin\left(\theta\right),
\end{equation}
with the constant of integration $-p^{I}/4\pi$ fixed by the magnetic charge integral in \eqref{eq:charge_integrals}. 

The equation of motion $\delta S/\delta A^{I}_{t}=0$ following from the action \eqref{eq:Action} is 

\begin{equation}\label{eq:an_eom}
\sin\left(\theta\right)\, \partial_{\tau}\left[\ee^{-2U}\text{\textcyr{I}}_{IJ}f^{J}(\tau)-\text{\textcyr{R}}_{IJ}\frac{p^{I}}{4\pi}\right]=0.
\end{equation}
Integration gives $f^{I}(\tau)$.

\begin{equation}
f^{I}(\tau)=\ee^{2U}(\text{\textcyr{I}}^{-1})^{IJ}\left[\text{\textcyr{R}}_{JK}\frac{p^{K}}{4\pi}-\frac{q_{J}}{4\pi}\right].
\end{equation}
Here the constant of integration is obtained by comparing the expression in brackets in \eqref{eq:an_eom} with that of $G_{\mu\nu I}=\kappa^{2}\epsilon_{\mu\nu\rho\sigma}\frac{\delta S}{\delta F_{\rho\sigma}^{\phantom{\rho\sigma}I}}$ and then using the electric charge integral in \eqref{eq:charge_integrals}.

The stress energy tensor $T_{\mu\nu}$ for the above field content is

\begin{equation}\label{eq:setensor}
\kappa^{2}T_{\mu\nu}\!=-\text{\textcyr{I}}_{IJ}\left(F_{\mu\rho}^{\phantom{\mu\rho}I}F_{\nu}^{\phantom{\nu}\rho J}-\frac{1}{4}g_{\mu\nu}F_{\rho\sigma}^{\phantom{\rho\sigma}I}F^{\rho\sigma J}\right)+g_{\alpha\bar{\beta}}\left(\partial_{\mu}z^{\alpha}\partial_{\nu}\bar{z}^{\bar{\beta}}+\partial_{\nu}z^{\alpha}\partial_{\mu}\bar{z}^{\bar{\beta}}-g_{\mu\nu}\partial_{\rho}z^{\alpha}\partial^{\rho}\bar{z}^{\bar{\beta}}\right).
\end{equation}
It is most convenient to work with the following form of the Einstein equations:

\begin{equation}\label{eq:einstein}
R_{\mu\nu}=\kappa^{2}\left(T_{\mu\nu}-\frac{1}{2}g_{\mu\nu}T_{\rho}^{\phantom{\rho}\rho}\right).
\end{equation}
Expanding the right hand side using \eqref{eq:setensor} leads to

\begin{equation}\label{eq:einstein2}
R_{\mu\nu}=-\text{\textcyr{I}}_{IJ}\left(F_{\mu\rho}^{\phantom{\mu\rho}I}F_{\nu}^{\phantom{\nu}\rho J}-\frac{1}{4}g_{\mu\nu}F_{\rho\sigma}^{\phantom{\rho\sigma}I}F^{\rho\sigma J}\right)+g_{\alpha\bar{\beta}}\left(\partial_{\mu}z^{\alpha}\partial_{\nu}\bar{z}^{\bar{\beta}}+\partial_{\nu}z^{\alpha}\partial_{\mu}\bar{z}^{\bar{\beta}}\right).
\end{equation}
A significant simplification is met by introducing the black hole potential $V_{BH}$.

\begin{equation}\label{eq:Vbh}
V_{BH}=\frac{1}{32\pi^{2}}\left(p^{I}\,\,q_{I}\right)\left(\begin{matrix}
-(\text{\textcyr{I}}+\text{\textcyr{R}}\text{\textcyr{I}}^{-1}\text{\textcyr{R}})_{IJ} & (\text{\textcyr{R}}\text{\textcyr{I}}^{-1})_{I}^{\phantom{I}J}\\[5pt]
(\text{\textcyr{I}}^{-1}\text{\textcyr{R}})^{I}_{\phantom{I}J} & -(\text{\textcyr{I}}^{-1})^{IJ}
\end{matrix}\right)\begin{pmatrix}p^{J}\\[5pt]q_{J}\end{pmatrix}.
\end{equation}
There are only two independent Einstein equations. 

\begin{equation}\label{eq:einstein3}
\begin{aligned}
R_{tt}&=\ee^{4U}\tau^{4}\ddot{U}=\ee^{6U}\tau^{4}V_{BH},\\[5pt]
R_{\tau\tau}&=\ddot{U}-2\dot{U}^{2}=-\ee^{2U}V_{BH}+2g_{\alpha\bar{\beta}}\dot{z}^{\alpha}\dot{\bar{z}}^{\bar{\beta}}.
\end{aligned}
\end{equation}
The dot represents differentation with respect to $\tau$. Neatening these gives

\begin{equation}\label{eq:einstein4}
\begin{aligned}
\ddot{U}&=\ee^{2U}V_{BH},\\[5pt]
\dot{U}^{2}&=\ee^{2U}V_{BH}-g_{\alpha\bar{\beta}}\dot{z}^{\alpha}\dot{\bar{z}}^{\bar{\beta}}.
\end{aligned}
\end{equation}
The equations of motion of the scalars are

\begin{equation}\label{eq:eom_scalars}
\frac{\dd}{\dd\tau}\left(g_{\alpha\bar{\beta}}\dot{\bar{z}}^{\bar{\beta}}\right)-\left(\partial_{\alpha}g_{\gamma\bar{\delta}}\right)\dot{z}^{\gamma}\dot{\bar{z}}^{\bar{\delta}}-\ee^{2U}V_{BH}=0,
\end{equation}
plus the complex conjugate of these equations.
\vskip10pt
\subsubsection*{The black hole potential and the central charge}

The central charge $\mathcal{Z}$ is related to the electromagnetic charges and the scalar K{\"a}hler geometry~by 
\begin{equation}\label{eq:cent_ch}
\mathcal{Z}=2\kappa^{-2}\ee^{\frac{\kappa^{2}}{2}\mathcal{K}}Z^{I}\left(q_{I}-\mathcal{N}_{IJ}p^{J}\right)=2\kappa^{-2}\left(X^{I}q_{I}-F_{I}p^{I}\right).
\end{equation}
After some unwinding, one can relate the central charge and the black hole potential.
\begin{equation}
(4\pi)^{2}V_{BH}=\frac{\kappa^{4}}{4}|\mathcal{Z}|^{2}+\frac{\kappa^{2}}{4}\left(\nabla_{\alpha}\mathcal{Z}\right)g^{\alpha\bar{\beta}}\left(\nabla_{\bar{\beta}}\bar{\mathcal{Z}}\right).
\end{equation}
Some formulae from special geometry give further simplification. $\nabla_{\alpha}\bar{\mathcal{Z}}=0$ and  $\nabla_{\alpha}|\mathcal{Z}|^{2}=\partial_{\alpha}|\mathcal{Z}|^{2}$ lead to  
\begin{equation}\label{eq:ccd}
\nabla_{\alpha}\mathcal{Z}=2\sqrt{\frac{\mathcal{Z}}{\bar{\mathcal{Z}}}}\partial_{\alpha}|\mathcal{Z}|.
\end{equation}
This allows for
\begin{equation}\label{eq:vbh_cc}
\frac{V_{BH}}{G^{2}}=|\mathcal{Z}|^{2}+4g^{\alpha\bar{\beta}}\partial_{\alpha}|\mathcal{Z}|\partial_{\bar{\beta}}|\mathcal{Z}|~,\qquad G=\frac{\kappa}{8\pi}\frac{64\pi^{2}}{\kappa^{2}}
\end{equation}

\subsubsection*{The one-dimensional effective action}
The large amount of symmetry has greatly reduced the number of equations. Moreover, the equations \eqref{eq:eom_scalars} and the first of \eqref{eq:einstein4} extremise a single one-dimensional effective action:

\begin{equation}\label{eq:1dea}
S_{1D}\left[U,z,\bar{z}\right]=\int_{0}^{\infty}\dd\tau\left(\dot{U}^{2}+g_{\alpha\bar{\beta}}\dot{z}^{\alpha}\dot{\bar{z}}^{\bar{\beta}}+\ee^{2U}V_{BH}\right).
\end{equation}
Since the above Lagrangian is independent of $\tau$ there is a conserved quantity:

\begin{equation}\label{eq:cons}
\mathcal{E}=\dot{U}^{2}+g_{\alpha\bar{\beta}}\dot{z}^{\alpha}\dot{\bar{z}}^{\bar{\beta}}-\ee^{2U}V_{BH}.
\end{equation}
By fixing $\mathcal{E}=0$ the second equation of \eqref{eq:einstein4} is satisfied. With this constant of integration fixed, this one-dimensional action captures the metric and scalar dynamics. 

One can go further, by inserting the central charge expression \eqref{eq:vbh_cc} into the action \eqref{eq:1dea} the attractor behaviour can be made clear. By completing squares and recognising a total derivative, one can write

\begin{equation}\label{eq:1dea_cc}
\begin{aligned}
S_{1D}\left[U,z,\bar{z}\right]=&\int_{0}^{\infty}\dd\tau\left[\left(\dot{U}+G\ee^{U}|\mathcal{Z}|\right)^{2}+g_{\alpha\bar{\beta}}\left(\dot{z}^{\alpha}+2G\ee^{U}g^{\alpha\bar{\delta}}\partial_{\bar{\delta}}|\mathcal{Z}|\right)\left(\dot{\bar{z}}^{\bar{\beta}}+2G\ee^{U}g^{\gamma\bar{\beta}}\partial_{\gamma}|\mathcal{Z}|\right)\right]\\[5pt]
&-2G\ee^{U}|\mathcal{Z}|\bigg\vert_{\tau=0}^{\tau=\infty}.
\end{aligned}
\end{equation}

\subsubsection*{The attractor equations}
The integral in this action has an integrand that is the sum of two squares. It is thus minimised when both of these squares vanish. This requires that the following attractor equations are satisfied:

\begin{equation}\label{eq:att}
\dot{U}=-G\ee^{U}|\mathcal{Z}|,\qquad \dot{z}^{\alpha}=-2G\ee^{U}g^{\alpha\bar{\beta}}\partial_{\bar{\beta}}|\mathcal{Z}|,
\end{equation}
in addition to the complex conjugate of this last equation.

Note that these minimising equations together directly give an expression for $\dot{U}^{2}+g_{\alpha\bar{\beta}}\dot{z}^{\alpha}\dot{\bar{z}}^{\bar{\beta}}$. This is

\begin{equation}
\dot{U}^{2}+g_{\alpha\bar{\beta}}\dot{z}^{\alpha}\dot{\bar{z}}^{\bar{\beta}}=G^{2}\ee^{2U}\left(|\mathcal{Z}|^{2}+4g^{\alpha\bar{\beta}}\partial_{\alpha}|\mathcal{Z}|\partial_{\bar{\beta}}|\mathcal{Z}|\right)=\ee^{2U}V_{BH}.
\end{equation}
Equation \eqref{eq:vbh_cc} has been used to make the rightmost transition. This implies that $\mathcal{E}=0$ for any solutions to \eqref{eq:att} and so the second equation in \eqref{eq:einstein4} is satisfied.

The formulae \eqref{eq:att} describe the evolution of $U$ and motion of the scalars on their target space as $\tau$ increases.  The second formula implies that under this evolution

\begin{equation}
\frac{\dd}{\dd\tau}|\mathcal{Z}|=-4G\ee^{U}g^{\alpha\bar{\beta}}\partial_{\alpha}|\mathcal{Z}|\partial_{\bar{\beta}}|\mathcal{Z}|.
\end{equation}.

Since $g_{\alpha\bar{\beta}}$ is positive definite this equation informs us that $|\mathcal{Z}|$ is monotonically decreasing along attractor flows, and also that the flows end where $|\mathcal{Z}|$ takes a critical value. Since $|\mathcal{Z}|$ is bounded below (by virtue of being nonnegative), there must be a limiting value for $|\mathcal{Z}|$ as $\tau\rightarrow\infty$. The scalars $z^{\alpha}$ thus flow to the values such that $|\mathcal{Z}|$ takes a critical value. These are the \textit{attractor points} of the scalar target manifold.

\subsubsection*{Saturation of the BPS bound}
One can read off the mass of the black hole from the leading correction to $g_{tt}$ at large $r$. It will be shown that this mass equals the modulus of the central charge at infinity, which is the BPS condition. We seek to find $M$ as per $g_{tt}=-\ee^{2U}=-1+2MG\tau+...~,$ and obtain

\begin{equation}
M=\frac{-1}{2G}\frac{\dd}{\dd\tau}\ee^{2U}\bigg\vert_{\tau=0}.
\end{equation}
The flow equations \eqref{eq:att} facilitate such a computation. Rearranging the equation involving $\dot{U}$, one obtains 

\begin{equation}
\frac{\dd}{\dd\tau}\ee^{-U}=G|\mathcal{Z}|.
\end{equation}
One can proceed by

\begin{equation}
M=\frac{-1}{2G}\frac{\dd}{\dd\tau}\ee^{2U}\bigg\vert_{\tau=0}=\frac{-1}{2G}\frac{\dd}{\dd\tau}\left(\ee^{-U}\right)^{-2}\bigg\vert_{\tau=0}=\frac{1}{G}\ee^{3U}\frac{\dd}{\dd\tau}\ee^{-U}\bigg\vert_{\tau=0}=|\mathcal{Z}|_{\infty}.
\end{equation}

\subsubsection*{Stabilisation equations}

We have seen that that the scalars flow to attractor points where $|\mathcal{Z}|$ is minimised. A set of equations give the locations of these attractor points as a function of the black hole charges $p^{I}$ and $q_{I}$. These were initially derived by Strominger in 1996 \cite{Strominger:1996kf}. Later that year they were derived using a slightly different approach by Ferrara and Kallosh \cite{Ferrara:1996dd}. 

Strominger considered solutions to the field equations with constant scalars (which must necessarily have an attractor point value throughout spacetime). The spacetime geometry in such setups is Reissner-Nordstrom. As these black holes are BPS and have vanishing Fermi fields, the bosonic parts of the fermionic supersymmetry variations necessarily vanish. The gaugino variations give the sought equations for the scalars in terms of the charges.

Ferrara and Kallosh's approach, which we detail here, involves directly extremising the central charge's modulus. The starting point is formula \eqref{eq:cent_ch}, reproduced here:

\begin{equation}\label{eq:cech}
\mathcal{Z}=2\kappa^{-2}\left(X^{I}q_{I}-F_{I}p^{I}\right).
\end{equation}

By formula \eqref{eq:ccd}, $\partial_{\alpha}|\mathcal{Z}|$ is a nonzero multiple of $\nabla_{\alpha}\mathcal{Z}$. Thus the scalars that minimise $|\mathcal{Z}|$ satisfy

\begin{equation}\label{eq:first_deriv}
\frac{\kappa^{2}}{2}\nabla_{\alpha}\mathcal{Z}=\left(\nabla_{\alpha}X^{I}\right)q_{I}-\left(\nabla_{\alpha}F_{I}\right)p^{I}=0.
\end{equation}

The argument will make use of some formulae from special geometry. There are the raising and lowering identities

\begin{equation}\label{eq:raiselower}
F_{I}=\mathcal{N}_{IJ}X^{J},\qquad\bar{F}_{I}=\bar{\mathcal{N}}_{IJ}\bar{X}^{J},\qquad \nabla_{\alpha}F_{I}=\bar{\mathcal{N}}_{IJ}\nabla_{\alpha}X^{J},\qquad \bar{\nabla}_{\bar{\alpha}}\bar{F}_{I}=\mathcal{N}_{IJ}\overline{\nabla}_{\bar{\alpha}}\bar{X}^{J}.
\end{equation}
Using these one arrives at the following identities involving the vector $\cS$ of~\eqref{eq:sympvectors}:

\begin{equation}\label{eq:symp_products}
\begin{aligned}
\langle \cS,\bar{\cS}\rangle&=X^{I}\bar{F}_{I}-\bar{X}^{I}F_{I}=X^{I}\bar{X}^{J}\left(\bar{\mathcal{N}}_{IJ}-\mathcal{N}_{IJ}\right)=-2\ii X^{I}\bar{X}^{J}\text{\textcyr{I}}_{IJ},\\[5pt]
\langle\nabla_{\alpha}\cS,\overline{\nabla}_{\bar{\beta}}\bar{\cS}\rangle&=\left(\nabla_{\alpha}X^{I}\right)\left(\overline{\nabla}_{\bar{\beta}}\bar{F}_{I}\right)-\left(\overline{\nabla}_{\bar{\beta}}\bar{X}^{I}\right)\left(\nabla_{\alpha}F_{I}\right)=2\ii\left(\nabla_{\alpha}X^{I}\right)\left(\overline{\nabla}_{\bar{\beta}}\bar{X}^{J}\right)\text{\textcyr{I}}_{IJ},\\[5pt]
\langle \cS,\overline{\nabla}_{\bar{\alpha}}\bar{\cS}\rangle&=X^{I}\overline{\nabla}_{\bar{\alpha}}\bar{F}^{J}-F_{I}\overline{\nabla}_{\bar{\alpha}}\bar{X}^{I}=X^{I}\mathcal{N}_{IJ}\overline{\nabla}_{\bar{\alpha}}\bar{X}^{J}-X^{J}\mathcal{N}_{JI}\overline{\nabla}_{\bar{\alpha}}\bar{X}^{I}=0,\\[5pt]
\langle\bar{\cS},\nabla_{\alpha}\cS\rangle&=\bar{X}^{I}\nabla_{\alpha}F_{I}-\bar{F}_{I}\nabla_{\alpha}X^{I}=\bar{X}^{I}\bar{\mathcal{N}}_{IJ}\nabla_{\alpha}X^{J}-\bar{X}^{J}\bar{\mathcal{N}}_{IJ}\nabla_{\alpha}X^{I}=0.
\end{aligned}
\end{equation}
The second of these gives a formula for the metric component $g_{\alpha\bar{\beta}}=\ii\langle\nabla_{\alpha}\cS,\overline{\nabla}_{\bar{\beta}}\bar{\cS}\rangle$. The first formula above is an equation for the K{\"a}hler potential of the rigid manifold inside which our projective manifold is embedded\footnote{This embedding is one stage in the construction of supergravity actions by gauge-fixing conformal supergravity, as discussed in the textbook \cite{Freedman:2012zz}.}, $K=\ii\langle \cS,\bar{\cS}\rangle$. In choosing a dilatation gauge we chose values for the coordinate $y$ so that $K$ had a constant value of $-\kappa^{-2}$, but so as to find a general set of stabilisation equations not depending on gauge we do not fix this value here. One should bear in mind that since $\cS$ is holomorphic, $K$ is real.

Noting that by construction $\nabla_{\alpha}y=\nabla_{\alpha}\bar{y}$, all of the equations in \eqref{eq:symp_products} can be packaged into a single $(n+1)\times(n+1)$ matrix equation:

\begin{equation}\label{eq:matrixmess}
\left(\begin{matrix}-K & 0\\[5pt]0 & g_{\alpha\bar{\beta}}\end{matrix}\right)=
-2y\bar{y}\begin{pmatrix}
\bar{Z}^{I}\\[5pt]\nabla_{\alpha}Z^{I}
\end{pmatrix}
\text{\textcyr{I}}
\left(
Z^{I} \,\, \overline{\nabla}_{\bar{\beta}}\bar{Z}^{I}
z\right).
\end{equation}

A comment on the layout of this equation is in order. The matrix on the LHS is read in the obvious way. In the leftmost matrix on the RHS $\bar{Z}^{I}$ is understood as a row vector and $\nabla_{\alpha}Z^{I}$ is an $n\times(n+1)$ matrix. $\text{\textcyr{I}}=\text{Im}[\mathcal{N}]$ is read in the obvious way. In the rightmost matrix on the LHS $Z^{I}$ is a column vector and $\bar{\nabla}_{\bar{\beta}}\bar{Z}^{I}$ is an $(n+1)\times n$ matrix. 

The point of this involved digression is to arrive at a useful formula in the derivation of the stabilisation equations. One can rearrange \eqref{eq:matrixmess} to reach

\begin{equation}\label{eq:inversemess}
\text{\textcyr{I}}^{-1}=-2y\bar{y}\left(\begin{matrix}
Z^{I} & \overline{\nabla}_{\bar{\beta}}\bar{Z}^{I}
\end{matrix}\right)
\left(\begin{matrix}-K^{-1} & 0\\[5pt]0 & g^{\alpha\bar{\beta}}\end{matrix}\right)
\left(\begin{matrix}
\bar{Z}^{I}\\[5pt]\nabla_{\alpha}Z^{I}
\end{matrix}\right).
\end{equation}

The matrix multiplication can be carried out to get an equation for the components of $I^{-1}$.

\begin{equation}\label{eq:inversemess2}
\left(\text{\textcyr{I}}^{-1}\right)^{IJ}=\frac{2}{K}\bar{X}^{I}X^{J}-2g^{\alpha\bar{\beta}}\left(\nabla_{\alpha}X^{I}\right)\left(\overline{\nabla}_{\bar{\beta}}\bar{X}^{J}\right).
\end{equation}
It is now possible to extremise the central charge. First, substitute the third identity from \eqref{eq:raiselower} in \eqref{eq:first_deriv} and use the fact that $\mathcal{N}_{IJ}$ is symmetric to obtain

\begin{equation}
\left(\nabla_{\alpha}X^{I}\right)q_{I}-\bar{\mathcal{N}}_{IJ}\left(\nabla_{\alpha}X^{I}\right)p^{J}=0.
\end{equation}
Next contract with $g^{\alpha\bar{\beta}}\left(\overline{\nabla}_{\bar{\beta}}\bar{X}^{K}\right)$ and use formula \eqref{eq:inversemess2} to replace $g^{\alpha\bar{\beta}}\left(\nabla_{\alpha}X^{I}\right)\left(\overline{\nabla}_{\bar{\beta}}\bar{X}^{K}\right)$.

\begin{equation}\label{eq:contracted}
\left(\frac{1}{K}\bar{X}^{I}X^{K}-\frac{1}{2}\left(\text{\textcyr{I}}^{-1}\right)^{IK}\right)\left(q_{I}-\bar{\mathcal{N}}_{IJ}p^{J}\right)=0.
\end{equation}
One then rearranges this equation,

\begin{equation}\label{eq:simpling}
\begin{aligned}
0&=\frac{1}{K}\left(\bar{X}^{I}q_{I}-\bar{\mathcal{N}}_{IJ}\bar{X}^{I}p^{J}\right)X^{K}-\frac{1}{2}\left(\text{\textcyr{I}}^{-1}\right)^{IK}q_{I}+\frac{1}{2}\left(\text{\textcyr{I}}^{-1}\right)^{IK}\bar{\mathcal{N}}_{IJ}p^{J}\\[5pt]
&=\frac{1}{K}\left(\bar{X}^{I}q_{I}-\bar{F}_{J}p^{J}\right)X^{K}-\frac{1}{2}\left(\text{\textcyr{I}}^{-1}\right)^{IK}q_{I}+\frac{1}{2}\left(\text{\textcyr{I}}^{-1}\right)^{IK}\left(\text{\textcyr{R}}_{IJ}-\ii \text{\textcyr{I}}_{IJ}\right)p^{J}\\[5pt]
&=\frac{\kappa^{2}}{2K}\bar{\mathcal{Z}}X^{K}-\frac{1}{2}\left(\text{\textcyr{I}}^{-1}\right)^{IK}\left(q_{I}-\text{\textcyr{R}}_{IJ}p^{J}\right)-\frac{\ii}{2}p^{K}.
\end{aligned}.
\end{equation}

Passing from \eqref{eq:contracted} to the first line of \eqref{eq:simpling} involves only moving terms around. The passage from the first to the second line makes use of the second lowering identity in \eqref{eq:raiselower} and breaks $\bar{\mathcal{N}}$ into its real and imaginary parts. Passing from the second to the third line necessitates formula \eqref{eq:cech} and multiplying some matrices together. 

Since $\kappa^{2}$, $\text{\textcyr{I}}$, $\text{\textcyr{R}}$, $K$, $p^{I}$ and $q_{I}$ are all real, one can extract the imaginary part of the last equation:

\begin{equation}\label{eq:magstab}
p^{K}=\text{Im}\left[\frac{\kappa^{2}}{K}\bar{\mathcal{Z}}X^{K}\right].
\end{equation}
The real part of \eqref{eq:simpling}, contracted with $\text{\textcyr{I}}_{JK}$, is

\begin{equation}\label{eq:simpreal}
\begin{aligned}
q_{J}&=\text{\textcyr{R}}_{JK}p^{K}+\text{Re}\left[\frac{\kappa^{2}}{K}\bar{\mathcal{Z}}\text{\textcyr{I}}_{JK}X^{K}\right]\\[5pt]
&=\text{Im}\left[\frac{\kappa^{2}}{K}\bar{\mathcal{Z}}\text{\textcyr{R}}_{JK}X^{K}\right]+\text{Im}\left[\frac{\kappa^{2}}{K}\bar{\mathcal{Z}}\ii \text{\textcyr{I}}_{JK}X^{K}\right]\\[5pt]
&=\text{Im}\left[\frac{\kappa^{2}}{K}\bar{\mathcal{Z}}\mathcal{N}_{JK}X^{K}\right]\\[5pt]
&=\text{Im}\left[\frac{\kappa^{2}}{K}\bar{\mathcal{Z}}F_{J}\right].
\end{aligned}
\end{equation}
The quantity $\frac{\kappa^{2}}{K}\bar{\mathcal{Z}}$ depends on the dilatation gauge and the charges. Denoting this quantity's value at an attractor point as $C$, we have reached the stabilisation equations that relate the scalar values at an attractor point to the electromagnetic charges when the central charge does not vanish:

\begin{equation}\label{eq:stab}
\begin{pmatrix}
p^{I}\\q_{I}
\end{pmatrix}
=\text{Im}\left[C
\begin{pmatrix}X^{I}\\F_{I}\end{pmatrix}\right].
\end{equation}

These are $2n+2$ real equations for $2n+4$ real variables --- the real and imaginary parts of $C$ and the $2n+2$ attractor coordinates on the rigid manifold. Fixing a choice of gauge constrains two of these variables, and so the number of equations equals the number of physical variables to solve for.

\subsubsection*{Multi-centred solutions}
In \cite{Denef:2000nb} the attractor mechanism was extended to multi-centered black holes in $\mathcal{N}=2$ supergravity. It was shown that in the presence of $N$ sources located at $\vec{x}_{i}$, $i=1,...,N$, each of charge $\Gamma_{i}$, the vector multiplet scalars take attractor values associated to the charge vector $\Gamma_{i}$ as $\vec{x}_{i}$ is approached.

\section{Calabi-Yau attractors in IIB}\label{sect:SUGRA_CYAtt_IIB}
In a IIB compactification on a Calabi-Yau threefold $X$, a charged black hole has a string-theoretic description as a bound state of D3-branes wrapping special lagrangian cycles in the middle cohomology of $X$. The charge vector $Q$ then is associated to a three-cycle $\gamma$ in $H_{3}(X,\IZ)$. This is dual in cohomology to a threeform $\Gamma$, which we will expand in the same integral symplectic basis as used in \eqref{eq:Omega_in_ISB}, whose components we collect in the vector $Q$. 
\begin{equation}
\Gamma=p^{I}\alpha_{I}-q_{I}\beta^{I}~,\qquad Q=\begin{pmatrix}
q_{I}\\p^{I}
\end{pmatrix}~.
\end{equation}
The stabilisation equations \eqref{eq:stab} give the following constraint on the holomorphic three-form $\Omega$:
\begin{equation}\label{eq:stab_Omega}
\Gamma=\text{Im}\left[C\,\Omega\right]~,\qquad\text{or in components }\,\, Q=\text{Im}\left[C\,\Pi\right]~.
\end{equation}
Note that the ordering of the electric and magnetic charges is reversed compared to that in \sref{sect:SUGRA_AttMech}, which is purely a matter of changing convention and not conceptually significant. If we fix the charge vector, then the above equation is a constraint on the complex structure moduli of $X$. A Calabi-Yau threefold whose moduli satisfy the above equation for some integral charge vector is said to be an attractor variety. 

If we let the complex constant $C$ have real part $a$ and imaginary part $b$, then the above equation reads
\begin{equation}
Q=a\,\text{Im}\left[\Pi\right]+b\,\text{Re}\left[\Pi\right]~.
\end{equation}
In the Dolbeault decomoposition of $H^{3}(X,\IZ)$, the holomorphic three form generates the $(3,0)$ part. The real and imaginary part of $\Omega$ both belong to $H^{(3,0)}\oplus H^{(0,3)}$, and so the above equation tells us that inside $H^{(3,0)}\oplus H^{(0,3)}$ lives the span of the integral vector $Q$. The statement is that the $(3,0)+(0,3)$ part of $X$'s Dolbeault cohomology contains a 1-dimensional integral lattice.

In some cases, there may be a point $\varphi_{*}$ in the complex structure moduli space of $X$ so that, at this point, the equation \eqref{eq:stab_Omega} is solved for two independent charge vectors $Q_{1},\,Q_{2}$ (with different values of $C$). ``Independence" here means that the symplectic inner product $Q_{1}\Sigma Q_{2}$ does not vanish, and so $Q_{1}$ and $Q_{2}$ are necessarily linearly independent. 

For such moduli values $\varphi_{*}$, $X$'s cohomology has the property that $H^{(3,0)}\oplus H^{(0,3)}$ is the complexification of a rank-two integral lattice in $H^{3}(X,\IZ)$ \cite{Candelas:2019llw}. This gives a splitting of the Hodge structure. We see this at the level of deRham cohomology, however conjecturally this persists into a suitable {\'e}tale cohomology and so restricts the form of the zeta function. We shall look at this more closely in \sref{sect:SUGRA_NewWeight4}.

\section{One-parameter Calabi-Yau attractors in IIA}\label{sect:SUGA_1PCYIIA}
We will only be concerned with IIA compactifications on manifolds $Y$ with $h^{1,1}=1$. If the period vector $\Pi$ solves the attractor equations for two independent charge vectors $Q_{1},\,Q_{2}$, then there must exist independent integral vectors $Q_{3},\,Q_{4}$ such that
\begin{equation}
\Pi\Sigma Q_{3}=\Pi\Sigma Q_{4}=0~.
\end{equation}
This must be so, because the attractor equations imply
\begin{equation}
\text{Span}\left(\text{Re}[\Pi]~,\,\text{Im}[\Pi]\right)=\text{Span}\left(Q_{1}~,\,Q_{2}\right)~. 
\end{equation}
The symplectic complement of $\text{Span}\left(Q_{1}~,\,Q_{2}\right)\subset\IZ^{4}$ is two-dimensional, with generators some $Q_{3}~,\,Q_{4}$, which are by the above relation symplectic-orthogonal to $\text{Re}[\Pi]$ and $\text{Im}[\Pi]$. We shall in what follows make reference to the \emph{orthogonality equation}
\begin{equation}
Q\Sigma\Pi=0~,
\end{equation}
in contrast to the attractor equations \eqref{eq:stab_Omega}~.

This equivalence will be useful when we discuss a set of summation identities based on solutions to the IIA attractor equations, which we now set up. In a one-parameter compactification the IIA prepotential \eqref{eq:IIA_prepotential} reads
\begin{equation}\label{eq:IIA_prepot_1parameter}
\begin{aligned}
\cF&=-\left(z^{0}\right)^{2}\left[\frac{1}{6}Y_{111}t^{3}+\frac{1}{2}Y_{110}t^{2}+\frac{1}{2}Y_{100}t+\frac{1}{6}Y_{000}+\cI\left(t\right)\right]~,\\[5pt]
\cI&=\frac{1}{\left(2\pi\ii\right)^{3}}\sum_{k=1}^{\infty}n^{(0)}_{k}\text{Li}_{3}\left(\ee^{2\pi\ii\,kt}\right)=\frac{1}{\left(2\pi\ii\right)^{3}}\sum_{k=1}^{\infty}\frac{N_{k}}{k^{3}}\ee^{2\pi\ii\,kt}~.
\end{aligned}
\end{equation}
We have written $t=\frac{z^{1}}{z^{0}}$, and in the above expression for the instanton sum $\cI$ we have repackaged the genus-0 instanton numbers into \emph{scaled Gromov-Witten invariants} $N_{k}$. These are computed from the $n^{(0)}_{k}$ and related to the usual Gromov-Witten invariants $N^{GW}_{k}$ \cite{Kontsevich:1994na} by
\begin{equation}
N_{k}=\sum_{d|k}d^{3}n^{(0)}_{d}~,\qquad N_{k}=k^{3}N^{GW}_{k}~.
\end{equation}
The four-component period vector $\Pi$ is
\begin{equation}
\Pi(t)=\begin{pmatrix}
\frac{\partial}{\partial z^{0}}\cF\\
\frac{\partial}{\partial z^{1}}\cF\\
z^{0}\\
z^{1}
\end{pmatrix}=z^{0}\begin{pmatrix}
\frac{1}{6}Yt^{3}-\frac{1}{2}Y_{100}t-\frac{1}{3}Y_{000}-2\cI(t)+t\cI'(t)\\
-\frac{1}{2}Yt^{2}-Y_{110}t-\frac{1}{2}Y_{100}-\cI'(t)\\
1\\
t
\end{pmatrix}~,\qquad Y=Y_{111}~.
\end{equation}
We will now fix a gauge\footnote{Note that this is done only after taking the above derivatives $\frac{\partial}{\partial z^{0}}\cF$}, $z^{0}=1$~. The period vector is seen to be a sum of a quantity $\Pi_{0}$ that is corrected by instantons:
\begin{equation}\label{eq:Pi_decomp}
\begin{gathered}
\Pi=\Pi_{0}+\Delta_{\cI}\Pi~,\\[10pt]
\Pi_{0}=\begin{pmatrix}
\frac{1}{6}Yt^{3}-\frac{1}{2}Y_{100}t-\frac{1}{3}Y_{000}\\
-\frac{1}{2}Yt^{2}-Y_{110}t-\frac{1}{2}Y_{100}\\
1\\
t
\end{pmatrix}~,
\qquad \Delta_{\cI}\Pi=\begin{pmatrix}
-2\cI(t)+t\cI'(t)\\
-\cI'(t)\\
0\\
0
\end{pmatrix}~.
\end{gathered}
\end{equation}
In \cite{Candelas:2021mwz} a method of solving the attractor and orthogonality equations for the above form of period vector was displayed. We provide a brief summary of the solution method. First, the uncorrected equations
\begin{equation}\label{eq:Att_Eq_Pert}
Q=\text{Im}\left[C \Pi_{0}\right]\qquad\text{or}\qquad Q\Sigma\Pi_{0}=0
\end{equation}
are solved, to find a `perturbative' solution $t_{0}$, with real and imaginary part $x_{0}$ and $y_{0}$, so
\begin{equation}
t_{0}=x_{0}+\ii \,y_{0}~.
\end{equation}
Notice that either of \eqref{eq:Att_Eq_Pert} gives algebraic equations for $x_{0}$ and $y_{0}$. In \cite{Shmakova:1996nz} the attractor equations for such uncorrected prepotentials were considered. A full `instanton-corrected' solution $t$ is found by performing perturbation theory in the small parameter 
\begin{equation}
\ee^{-2\pi y_{0}}=|\ee^{2\pi\ii t_{0}}|~.
\end{equation}
The precise form of the solution depends on the charge vector $Q$. The most general form, with electric charges $q_{0},q$ and magnetic charges $p^{0},p$, is
\begin{equation}
Q=\begin{pmatrix}
q_{0}\\q\\p^{0}\\p
\end{pmatrix}=\begin{pmatrix}
q_{\text{D0}}\\
q_{\text{D2}}\\
q_{\text{D6}}\\
q_{\text{D4}}
\end{pmatrix}~,
\end{equation}
where in the second equality we interpret the integral charges as giving the wrapping number of even-dimensional D-branes. We adopt a convention that does not include charges induced on each brane by world-volume curvature coupling. 

We will consider only charge vectors that take one of the two following forms:
\begin{equation}\label{eq:D4_and_D6}
Q_{\text{D4}}=\kappa\begin{pmatrix}
\Lambda\\\Upsilon\\0\\1
\end{pmatrix}~,\qquad Q_{\text{D6}}=\kappa\begin{pmatrix}
\Lambda\\\Upsilon\\1\\0
\end{pmatrix}~.
\end{equation}
The integer $\kappa$ gives the total D4 or D6 charge, while $\Lambda$ and $\Upsilon$ are integer multiples of $\frac{1}{\kappa}$~. 

Our attention will be fixed to solutions of the `D4-D2-D0 orthogonality equations':
\begin{equation}\label{eq:D4_orth}
Q_{\text{D4}}\Sigma\Pi=0~.
\end{equation}
To explain this choice, we note first that if $t$ is a rank-two attractor, then an integral vector of the form $Q_{\text{D4}}$ can be found so that the above equation \eqref{eq:D4_orth} is satisfied. We work with the equations \eqref{eq:D4_orth} and not the attractor equations themselves because, at present, the solution to \eqref{eq:D4_orth} takes a much simpler form where the terms in the perturbative series can be given in closed form (rather than solely by recurrences). 

Neglecting instantons, the equation $Q_{\text{D4}}\Sigma\Pi_{0}=0$ is a simple quadratic,
\begin{equation}\label{eq:orth_D4_pert}
Y_{111}t_{0}^{2}+2\left(Y_{110}+\Upsilon\right)t_{0}+Y_{100}+2\Lambda=0~.
\end{equation} 
We will assume that $\left(Y_{110}+\Upsilon\right)^{2}-Y_{111}\left(Y_{100}+2\Lambda\right)<0$. In that case, the solution to \eqref{eq:orth_D4_pert} is $t_{0}=x_{0}+\ii y_{0}$ with real and imaginary parts 
\begin{equation}
x_{0}=\frac{-Y_{110}-\Upsilon}{Y_{111}}~,\qquad y_{0}=\frac{\sqrt{Y_{111}\left(Y_{100}+2\Lambda\right)-\left(Y_{110}+\Upsilon\right)^{2}}}{Y_{111}}~.
\end{equation}
To reiterate, this algebraic number $x_{0}+\ii y_{0}$ does not solve the attractor  or orthogonality equations, but is termed the \emph{perturbative} solution of the orthogonality equation as it solves the second of the equations \eqref{eq:Att_Eq_Pert}. Then, the full solution of \eqref{eq:D4_orth}, as conjectured\footnote{The conjectural part of this analysis is in identifying the combinatoric functions $a_{\fp}$ and modified Bessel functions in the series coefficients. It is a theorem that some series solution with recursively defined coefficients exists when $y_{0}$ is sufficiently small, see \cite{Candelas:2021mwz}.} in \cite{Candelas:2021mwz}, is
\begin{equation}\label{eq:solution}
t\=t_{0}-\frac{\ii  }{\sqrt{2\pi^{3}Y_{111}}}\sum_{j=1}^{\infty}\ee^{2\pi\ii x_{0} j}\sum_{\fp\in\text{pt}(j)}a_{\fp}N_{\fp}\left(\frac{j}{2\pi y_{0}Y_{111}}\right)^{\ell(\fp)-1/2}K_{\ell(p)-1/2}\left(2\pi j y_{0}\right)~.
\end{equation}
$K_{\nu}(z)$ is the modified Bessel function of the second kind. $\text{pt}(j)$ is the set of partitions of the integer $j$. For such a partition $\fp$, $j$ is partitioned into a set of integers as $j=\sum_{k=1}^{\infty}\mu_{k}k$. That is to say, $\mu_{k}$ is the multiplicity of the integer $k$ in the partition $\fp$ of $j$.

That said, $a_{\fp}$ is a combinatorial factor given by
\begin{equation}
a_{\fp}\=\prod_{k=1}^{\infty}\frac{1}{k^{2\mu_{k}}\mu_{k}!}
\end{equation}
and $N_{\fp}$ is the following product of the enumerative invariants $N_{k}$:
\begin{equation}
N_{\fp}\=\prod_{k=1}^{\infty}N_{k}^{\mu_{k}}~.
\end{equation}
Finally, $\ell(\fp)$ is the length of the partition $\fp$,
\begin{equation}
\ell(\fp)=\sum_{k=0}^{\infty}\mu_{k}~.
\end{equation}

\section{Flux compactifications}\label{sect:SUGRA_FluxComp}
Flux compactifications are string theory compactifications with nontrivial background values for the $(p+1)$-form field strengths. Our account of this topic follows \cite{Douglas:2006es} and \cite{Kachru:2020sio}. The first supersymmetric configurations were given in \cite{Dasgupta:1999ss}. The prospect of realising deSitter vacua was addressed in \cite{Kachru:2003aw}.

Type IIB supergravity's massless bosonic field content includes the even $p$-form Ramond-Ramond fields $C_{0}$, $C_{2}$, and $C_{4}$ which have field strengths
\begin{equation}
F_{i}=\text{d}C_{i-1}~.
\end{equation}
Additionally there is the Kalb-Ramond two-form $B_{2}$ which has field strength
\begin{equation}
H_{3}=\text{d}B_{2}~.
\end{equation}
There is also the dilaton $\phi$, which is packaged along with the axion $C_{0}$ into a single complex scalar, the axiodilaton
\begin{equation}
\tau=C_{0}+\ii\ee^{-\phi}~.
\end{equation}
Another notational change introduces the three-form field $G_{3}$, defined as
\begin{equation}
G_{3}=F_{3}-\tau H_{3}~.
\end{equation}
The utility of these two redefinitions is that it simplifies the action of the $\text{SL}(2,\IZ)$ symmetry of the IIB theory. The axiodilaton $\tau$ and complexified field strength $G_{3}$ transform via
\begin{equation}\label{eq:IIB_modular}
\tau\mapsto\frac{a\tau+b}{c\tau+d}~,\qquad G_{3}\mapsto\frac{1}{c\tau+d}G_{3}~,\qquad\text{with }\begin{pmatrix}
a&b\\c&d
\end{pmatrix}\in\text{SL}(2,\IZ)~.
\end{equation}
In a flux compactification, we dimensionally reduce on a Calabi-Yau threefold $X$, or an orientifold thereof, but do so with nonzero values given to $F_{3}$ and $H_{3}$ supported on the cohomology of $X$. Expanding in an integral symplectic basis, we write the components of $F_{3}$ and $H_{3}$ as
\begin{equation}
F_{3}=(2\pi)^{2}\alpha'\left(f^{a}\alpha_{a}-f_{b}\beta^{b}\right)~,\qquad H_{3}=(2\pi)^{2}\alpha'\left(h^{a}\alpha_{a}-h_{b}\beta^{b}\right)~,
\end{equation}
where $\alpha'$ is the string coupling and $f_{a},\,f^{b},\,h_{a},\,h^{b}$ are all integers. We collect these into the component vectors
\begin{equation}
F=\begin{pmatrix}
f^{a}\\f_{b}
\end{pmatrix}~,\qquad H=\begin{pmatrix}
h^{a}\\h_{b}
\end{pmatrix}~.
\end{equation}
When compactifying with such fluxes turned on, more supersymmetry is broken and the resulting four-dimensional supergravity has $\mathcal{N}=1$ supersymmetry. The geometric moduli of $X$ (both complex structure and K{\"a}hler) and the axiodilaton are scalars in $\mathcal{N}=1$ chiral multiplets \cite{Douglas:2006es}. These scalars are coupled together by a potential term $V$ in the action, which in accordance with $\mathcal{N}=1$ supersymmetry is constructed in a standard manner from a superpotential $W$. Incorporating or neglecting nonperturbative string theory contributions, like D-brane instantons, leads to different superpotentials $W$. We shall work with the uncorrected, classical superpotential which only depends on the complex structure moduli of $X$ and the axiodilaton, reading
\begin{equation}\label{eq:IIB_Superpotential}
W=\int_{X}G_{3}\wedge\Omega=(2\pi)^{2}\alpha'\left(F-\tau H\right)^{T}\Sigma\Pi~,\qquad \Sigma=\begin{pmatrix}
\+0&\mathbb{I}\\-\mathbb{I}&0
\end{pmatrix}~.
\end{equation}
There are two constraints on the choices of flux vectors $F$ and $H$, which bound $F^{T}\Sigma H$ from above and below.
\begin{equation}
0<F^{T}\Sigma H\leq\frac{\chi\left(X_{4}\right)}{24}~.
\end{equation}
The lower bound is necessary to get nontrivial supersymmetric solutions to the equations of motion, while the upper bound is the D3 tadpole condition, which is a consistency condition coming from F-theory. Namely, the different sources of D3 charge must locally cancel out one another. Flux compactifications on an orientifold of $X$ can be lifted to F-theory on an elliptically fibred Calabi-Yau fourfold $X_{4}$ \cite{Sen:1997gv}, and there is a condition
\begin{equation}
\frac{1}{2\kappa_{10}^{2}T_{3}}\int_{X}H_{3}\wedge F_{3}+Q^{\text{D3}}=\frac{\chi\left(X_{4}\right)}{24}~.
\end{equation}
$Q^{\text{D3}}$ is the total charge of any present D3 branes. We can add D3 branes (but not anti-D3 branes) without breaking supersymmetry, from which the upper bound follows.

\section{F-theory lifts}\label{sect:SUGRA_Ftheory}
Sen demonstrated in \cite{Sen:1997gv} that F-theory on an elliptically fibred Calabi-Yau fourfold reduces at weak coupling to an Orientifold compactification of type IIB string theory. We shall briefly review his construction, informed also by \cite{Kachru:2020sio,Kachru:2020abh}.

The modulus $\tau$ of the elliptic fibration is identified with the axiodilaton of the IIB theory. This gives a geometric interpretation to the $\text{SL}(2,\IZ)$ symmetry of IIB. The base of the fibration admits a double covering by a Calabi-Yau threefold, as depicted in \fref{fig:Elliptically_Fibred_Fourfold}. 
\begin{figure}[H]
\begin{center}
\hskip-40pt
	\adjustbox{scale=1,center}{\begin{tikzcd}[row sep=large, column sep=large]			
	\cE_{\tau} \arrow[hookrightarrow]{r} & \arrow[twoheadrightarrow]{d} X_{4}  \\		
	X \arrow[twoheadrightarrow]{r}{2:1} & \cB
	& 
	\end{tikzcd}}
	\vskip10pt
	\capt{6in}{fig:Elliptically_Fibred_Fourfold}{The F-theory fourfold is an elliptic fibration over the base $\cB$. This base is in turn doubly covered by the Calabi-Yau threefold $X$.}	
 \end{center}
\end{figure}
\vskip-25pt 
A method for constructing these fourfolds as intersections in toric varieties, given $X$ as such an intersection, was described in \cite{Collinucci:2008zs,Collinucci:2009uh}~. 

Sen's description gave an explicit relation between the elliptic fibration and the locations of D7 branes and O7 planes in the IIB limit. The most general elliptic fibration over a base $\cB$ has a Weierstrass form
\begin{equation}\label{eq:Weierstrass_general}
y^{2}=x^{3}+f(u)x+g(u)~,
\end{equation}
with $u$ being coordinates on the base $\cB$. As $u$ varies so does the above elliptic curve. For certain values of $u$, given by the vanishing of the discriminant
\begin{equation}
\Delta(u)=4f^{3}+27g(u)^{2}~,
\end{equation}
the elliptic curve becomes singular. Sen computed monodromies of $\tau$ about this singular locus, and recognised the same monodromies as one expects upon circling D7 branes or O7 planes. In so doing, the singular locus $\Delta(u)=0$ on the base $\cB$ is recognised as giving the intersections of spacetime filling 7-dimensional extended objects with the internal orientifold geometry. 

To delineate between the D7 and O7 objects, it is convenient to rewrite $f$ and $g$ in terms of functions $h$, $\eta$, and $\chi$, and a constant $C$ (which is not the $C$ of the previous section):
\begin{equation}\label{eq:Sen_fandg}
\begin{aligned}
f(u)&=C\eta(u)-3h(u)^{2}~,\qquad
g(u)=h(u)\left[C\eta(u)-2h(u)^{2}\right]+C^{2}\chi(u)~.
\end{aligned}
\end{equation}
Having done this, the discriminant is
\begin{equation}\label{eq:Sen_discriminant}
\Delta(u)=C^{2}\left[\eta^{2}\left(4C\eta-9h^{2}\right)+54h\left(C\eta-2h^{2}\right)\chi+27C^{2}\chi^{2}\right]~.
\end{equation}
The $j$-invariant $j_{\cE}$ of an elliptic curve is, in terms of the Weierstrass data $f,g$,
\begin{equation}
j_\cE(u)=\frac{6912f^{3}}{4f^{3}+27g^{2}}~.
\end{equation}
Our prefactor 6912 differs to Sen's convention. Since the elliptic fibration's complex structure parameter $\tau$ (itself some function of the base coordinate $u$) is to be identified with the axiodilaton $\tau$, we should have
\begin{equation}
j_\cE(u)=j\left(\tau(u)\right)~,
\end{equation}
where on the right hand side we have Klein's $j$-invariant
\begin{equation}
j(\tau)=e^{-2\pi\ii\tau}+744+196884\ee^{2\pi\ii\tau}+21493760\ee^{4\pi\ii\tau}+864299970\ee^{6\pi\ii\tau}+\,...~.
\end{equation}
Note that this differs from the function $J(\tau)=\frac{1}{1728}j$, the former being implemented in Mathematica as KleinInvariantJ.

In terms of the functions $f,\,h,\,\chi$, this $j$-invariant is
\begin{equation}
j\left(\tau(u)\right)=\frac{6912\left(C\eta-3h\right)^{2}}{C^{2}\left[\eta^{2}\left(4C\eta-9h^{2}\right)+54h\left(C\eta-2h^{2}\right)\chi+27C^{2}\chi^{2}\right]}~.
\end{equation}
In the limit $C\mapsto0$ the above $j$-invariant diverges like $C^{-2}$. As discussed in \cite{Sen:1997gv} (wherein their $\lambda$ is our $\tau$), up to $\text{SL}(2,\IZ)$ transformations this limit corresponds to sending the axiodilaton $\tau$ to $\ii\infty$ via $\tau=\frac{1}{4\pi\ii}\log(C)+O(C)$~.

In this large $\tau$ limit, taken with constant $h,\,\eta,\,\chi$ and $C\mapsto0$, the discriminant \eqref{eq:Sen_discriminant} becomes
\begin{equation}
\Delta(u)=-C^{2}h^{2}\left(\eta^{2}+12h\chi\right)~.
\end{equation}
This vanishes on the loci
\begin{equation}
h=0~,\qquad \text{ and } \qquad \eta=\pm\sqrt{-12h\chi}~.
\end{equation}
Sen computed the monodromies of the axiodilaton around these loci, and interpreted these as being due to charged objects sourcing the axiodilaton being present. Seven-dimensional extended objects fill the macroscopic four-dimensional spacetime and intersect the base $\cB$ on the loci $\Delta(u)=0$. The locus $h=0$ gives the position of an $O7$ plane, while the loci $\eta=\pm\sqrt{-12h\chi}$ give D-brane positions.

In F-theory, a supersymmetric configuration will be one where the D7 and O7s coincide. By making a choice $\chi=0$ and $h^{2}=\eta$, both objects will be positioned at $\eta=0$.

At the IIB level a maximally supersymmetric configuration should have a spatially constant axiodilaton profile. The sought profile is obtained from Sen's fibration, and the choices made for $\eta,\,h,\,\chi$ give the following $f$ and $g$ as in \eqref{eq:Sen_fandg}:
\begin{equation}
f(u)=(C-3)h(u)^{2}~,\qquad g(u)=(C-2)h(u)^{3}~.
\end{equation}
This gives an elliptic fibration over $\cB$ with Weierstrass model
\begin{equation}
y^{2}=x^{3}+(C-3)h(u)^{2}+(C-2)h(u)^{3}~.
\end{equation}
The singular locus of this curve is $C^{2}(4C-9)h(u)^{6}$. Away from the locus $h(u)=0$ we can make a rescaling of coordinates $y\rightarrow h^{3/2}y~,\,x\rightarrow hx$ so that our model is constant over the base, away from the charged objects.
\begin{equation}
y^{2}=x^{3}+(C-3)x+C-2~.
\end{equation}
The $j$-invariant of this curve is
\begin{equation}
j_{C}=\frac{6912(C-3)^{3}}{C^{2}(4C-9)}~.
\end{equation}
From this follows a cubic equation for $C$, which can be solved in terms of $j(\tau)$. This choice of a $C$ then gives the correct elliptic fibre for the F-theory uplift of the IIB supersymmetric flux vacuum with axiodilaton $\tau$.

\section{The scalar potential}\label{sect:SUGRA_ScalarPot}
Let us temporarily forget the fact that we have an internal \cym, and discuss some general features of these supergravity theories following \cite{Freedman:2012zz}. In any four-dimensional $\mathcal{N}\,{=}\,1$ supergravity coupled to $N$ chiral multiplets, the scalars in those multiplets are coordinates on a projective K\"ahler manifold. The theory is specified by the choice of two functions of the moduli; the K\"ahler potential $\cK$ for the scalar manifold and a superpotential $W$. From these two functions, one calculates the following potential:
\begin{equation} \label{eq:Flux_vacuum_potential}
V \defineas \ee^{\cK}\left(G^{\alpha\bar{\beta}}D_{\alpha}W\overline{D_{\beta}W}-3|W|^{2}\right).
\end{equation}
The sum in the above expression runs over all moduli, so in the case of a Calabi-Yau compactification on $X$, $\alpha,\beta=1,\,...\,,h^{1,1}(X)+h^{1,2}(X)+1$. The metric $G_{\alpha\bar{\beta}}$ is the metric on the scalar manifold, and, by virtue of the K\"ahler condition, $G_{\alpha\bar{\beta}}=\partial_{\alpha}\partial_{\bar{\beta}}\,\cK $. As in \cite{Strominger:1990pd,Candelas:1990pi}, the quantity $D_{\alpha}W$ is the K\"ahler covariant derivative of $W$, which has K\"ahler weight (1,0), so
\begin{equation}\notag
D_{\alpha}W\=\partial_{\alpha}W+\left(\partial_{\alpha}\cK\right)W~.
\end{equation}
In the case of a type IIB Calabi-Yau compactification, the total scalar manifold is the product of the upper half plane (in which the axiodilaton is valued) and the moduli space of the Calabi-Yau manifold~$X$, which factorises, at least locally, into the moduli spaces of complex structures and complexified K\"ahler structures of~$X$. The K\"ahler potential for the total moduli space is then a sum of the K\"ahler potentials for the axiodilaton, complex structure, and K\"ahler structure factors of the moduli space:
\begin{equation}\notag
\cK\=K^\AD + K^\CS + K^\KC~.
\end{equation}
These depend on the moduli as follows:
\begin{equation}\label{eq:K_potentials}
\begin{aligned}
K^\AD =-\log\big({-}\ii\left(\tau-\overline{\tau}\right)\big)~, \quad K^\CS =-\log\left(-\ii\,\Pi^{\dagger}\,\Sigma\,\Pi\right)~, \quad K^\KC =-\log\left(-\ii\,\ip^{\dagger}\Sigma\ip\right)~,
\end{aligned}
\end{equation}
where $\ip$~is the period vector for the mirror manifold, so a function of the K\"ahler moduli of~$X$. The K\"ahler potentials for the axiodilaton and complex structure moduli are exact at tree level, while the potential for complexified K\"ahler structures, $K^{\KC}$, is corrected by fundamental string instantons and $\alpha'$ corrections (this corrected $K^{\KC}$ is the uncorrected $\widetilde{K}^{\CS}$ of the mirror manifold). Additionally, nonperturbative effects can modify $W$ in such a way as to give it a dependence on the K\"ahler moduli of~$X$. If these latter corrections are neglected, then derivatives of $W$ with respect to K\"ahler moduli vanish and the potential \eqref{eq:Flux_vacuum_potential} reduces~to
\begin{equation}\label{eq:flux_vacuum_potential_Wcl}
V_{(W\text{ classical})}\=\ee^{\cK}\bigg(\text{Im}[\tau]^{2}\,|D_{\tau}W|^{2}+g^{i\bar{\jmath}}D_{i}W\overline{D_{j}W}+\left(\widetilde{g}^{\,r\bar{s}}K^{\KC}_{r}K^{\KC}_{\bar{s}}-3\right)|W|^{2}\bigg)~.
\end{equation}
In this expression $g_{i\bar{\jmath}}$ is the metric on the space of complex structures and $\widetilde{g}_{r\bar{s}}$ is the metric on the space of complexified K\"ahler structures. The indices $i,j$ run from 1 to $h^{2,1}(X)$, and $r,s$ run from 1 to $h^{1,1}(X)$.

In supersymmetric configurations the superpotential $W$ vanishes, which greatly simplifies the above expression for the scalar potential $V$, which must vanish\footnote{Note that $V$ does not automatically vanish on a locus in moduli space where $W$ vanishes, because $V$ depends on derivatives of $W$ in each moduli coordinate.} in a vacuum configuration.
\begin{equation}\label{eq:potential_classical}
V\=\ee^{\cK}\bigg(\Im[\tau]^2\, |\partial_{\tau}W|^{2}+g^{i\bar{\jmath}}\,\partial_{\varphi^{i}}W\,\overline{\partial_{\varphi^{j}}W} \bigg)~.
\end{equation} 
This potential is manifestly positive semidefinite, and thus the equations defining a supersymmetric flux vacuum, requiring that both $W$ and $V$ vanish, read:
\begin{equation}\label{eq:sfv_equations}
W\=0~, \qquad \partial_{\tau}W\=0~, \qquad \partial_{\varphi^{i}}W\=0~. 
\end{equation}
Depending on the vacuum expectation values of the field strengths $F_3$ and $H_3$, these equations might be solved for the complex structure moduli $\varphi^i$ as well as the value of the axiodilaton field. The K{\"a}hler moduli are unconstrained, although this ceases to be true when instanton corrections to the action are incorporated~\cite{Kachru_2003}.

\section{Corrections to the potential and a digression on moduli stabilisation}\label{sect:SUGRA_stabilisation}
In what follows, we only consider the case where the superpotential is given by the classical formula \eqref{eq:IIB_Superpotential}, that is, we work in the approximation where we can safely ignore the non-perturbative stringy corrections. Incorporating these corrections to $W$ could alter the space of supersymmetric flux vacua. In particular, the condition of vanishing $W$, which has a cohomological interpretation central to the modularity that we will discuss, only holds when the classical expression \eqref{eq:IIB_Superpotential} for $W$ is used. Additionally, these corrections give $W$ a dependence on the K\"ahler moduli, and so additional derivative terms will appear in the potential. 
 
The nonperturbative corrections to $W$ that we are referring to are the same as those considered in \cite{Kachru_2003}, where it was argued that precisely these corrections allowed for a realisation of metastable de Sitter vacua. One possible source of these corrections is Euclidean D3-instantons, first discussed in \cite{Witten:1996bn}. Additionally, in some setups gluino condensation could occur on spacetime-filling D7-branes which gives a contribution to $W$. Neither of these will be discussed here.

In attempting to build realistic cosmological models from flux compactifications, one can expect to meet the moduli problem. Namely, string theory constructions typically come with a large number of massless scalars. This is phenomenologically undesirable because there is no evidence for the existence such massless scalars. In fact, since these scalars will be coupled to gravity, they would give rise to unobserved fifth-force effects \cite{Cicoli:2023opf,Conlon:2006gv} that would, among other things, affect planetary orbits. Hopes of getting around this problem lie in finding suitable mechanisms of moduli stabilisation. This involves recognising nonperturbative corrections to the scalar potential, so that it acquires a steep local minimum in which the scalars will settle. When the scalars have taken this minimising value, the scalar potential term in the action becomes an effective cosmological constant. To realise a deSitter spacetime necessitates that this minimum be positive. There then remains the problem of giving a natural explanation for the cosmological constant's value being of the order $10^{-122}$ in Planck units.

We will not have anything to say about this problem. However, we will make a short digression now to discuss how the topology of the internal geometry $X$ can affect the large volume behaviour of the potential for the K{\"a}hler moduli. We proceed from the scalar potential \eqref{eq:flux_vacuum_potential_Wcl} that follows from the superpotential \eqref{eq:IIB_Superpotential}, which crucially does not depend on the K{\"a}hler moduli.

The square-bracketed quantities in the first line of \eqref{eq:flux_vacuum_potential_Wcl} are positive definite, so $V$ can be minimised by setting them to zero. This gives $h^{2,1}+1$ equations for the $h^{2,1}+1$ quantities of the axiodilaton $\tau$ and the full set of complex structure moduli $\bm{\varphi}$. Let us assume that these equations all provide independent constraints, so that all complex structure moduli and the axioldilaton are fixed. Let these respectively be fixed to values $\tau_{*}$ and $\bm{\varphi}_{*}$. What remains is
\begin{equation}\label{eq:Vstab}
V(\tau_{*},\bm{\varphi_{*}})\=\frac{\ee^{C(\bm{\varphi}_{*})}}{\text{Im}[\tau_{*}]}|W_{*}|^{2}\cdot\ee^{K}\left(K^{i\bar{j}}K_{i}K_{\bar{j}}-3\right)~,
\end{equation}
which depends on the as-yet-unfixed K{\"a}hler moduli $\bm{t}$ through $K(\bm{t})$.

We want to study the contraction
\begin{equation}
\cC\=K^{i\bar{j}}K_{i}K_{j}~.
\end{equation}
Recall that $K$ can be computed from the prepotential $\cF$ via
\begin{equation}
\ee^{-K}=-\ii\left(z^{a}\bar{\cF}_{a}-\bar{z}^{a}\cF_{a}\right)~.
\end{equation}
$\cF$ depends on the projectivised K{\"a}hler moduli by
\begin{equation}\label{eq:prepotagain}
\cF=-\frac{1}{3!}Y_{abc}\frac{z^{a}z^{b}z^{c}}{z^{0}}-\left(z^{0}\right)^{2}\cI\left(\frac{\bm{z}}{z^{0}}\right)~,\qquad \cI(\bm{t})\=\frac{1}{(2\pi\ii)^{3}}\sum_{\bm{k}\in H_{2}(X,\mathbb{Z})}n_{\bm{k}}\text{Li}_{3}\left(\ee^{2\pi\ii\bm{k}\cdot\bm{t}}\right)~.
\end{equation}
with $a,b,c$ running from 0 to $h^{1,1}$, the above sum includes the perturbative worldsheet corrections to $X$'s quantum volume, in the terms with a zero index. Including or excluding the $Y_{0ij}$ and $Y_{00i}$ terms do not lead to any change in $\cC$, a fact already well-appreciated in the supergravity literature (ultimately because $Y_{0ij}$ and $Y_{00i}$ are real). However the $Y_{000}$ term does play heavily in our discussion, so we take a moment to fix a notation.
\begin{equation}
Y_{000}\equiv-3\frac{\zeta(3)}{(2\pi\ii)^{3}}\chi(X)\=3\frac{\zeta(3)\ii}{8\pi^{3}}\chi\=6\ii \zeta~,\qquad \text{where} \qquad \zeta=\frac{\zeta(3)}{16\pi^{3}}\chi~.
\end{equation}
A salient point, $\zeta$ has the same sign as the Euler characteristic $\chi$. 

We shall write the real and imaginary parts of $\bm{t}$ as $\bm{\mu}$ and $\bm{\rho}$,
\begin{equation}
\bm{t}=\bm{\mu}+\ii\bm{\rho}~.
\end{equation} 
We will denote $\partial_{t^{i}}\cI$ and $\partial_{t^{i}}\partial_{t^{j}}\cI$ by $\cI_{i}$ and $\cI_{ij}$. A computation gives
\begin{equation}
\ee^{-K}=4\left|z^{0}\right|^{2}\bigg[\frac{1}{3}Y_{ijk}\rho^{i}\rho^{j}\rho^{k}-\rho^{i}\,\text{Re}\left[\cI_{i}\right]+\text{Im}\left[\cI\right]+\zeta\bigg]~.
\end{equation}
Differentiating this equation yields
\begin{equation}
\begin{aligned}
\ee^{-K}K_{i}&\=\+2\left|z^{0}\right|^{2}\bigg[\ii Y_{imn}\rho^{m}\rho^{n}+\rho^{m}\cI_{im}-\text{Im}\left[\cI_{i}\right]\bigg]~, \\[5pt]
\ee^{-K}K_{\bar{j}}&\=-2\left|z^{0}\right|^{2}\bigg[\ii Y_{imn}\rho^{m}\rho^{n}-\rho^{m}\cI_{im}+\text{Im}\left[\cI_{i}\right]\bigg]~, \\[5pt]
\ee^{-K}K_{i\bar{j}}-\ee^{-K}K_{i}K_{\bar{j}}&\=-2\left|z^{0}\right|^{2}\bigg[Y_{ijn}\rho^{n}+\text{Im}\left[\cI_{ij}\right]\bigg]~.
\end{aligned}
\end{equation}
By introducing the matrix M with components
\begin{equation}
\text{M}_{ij}=Y_{ijn}\rho^{n}-\ii\,\cI_{ij}
\end{equation}
we can write
\begin{equation}
K_{i\bar{j}}=-\ee^{K}\left(\text{M}+\overline{\text{M}}\right)_{ij}+4\ee^{2K}\bigg[\text{M}_{in}\rho^{n}+\ii\,\text{Im}\left[\cI_{i}\right]\bigg]\bigg[\overline{\text{M}}_{jm}\rho^{m}-\ii\,\text{Im}\left[\cI_{j}\right]\bigg]~.
\end{equation}
Computing $\cC$ necessitates inverting this matrix, so now is a good time to note an elementary formula. For an invertible symmetric matrix A and vectors $\bm{a}$, $\bm{b}$, the matrix B with components
\begin{equation}
\text{B}_{ij}\=\text{A}_{ij}+a_{i}b_{j}
\end{equation}
has inverse
\begin{equation}
\text{B}^{ij}\=\text{A}^{ij}-\frac{1}{1+\bm{a}^{T}\text{A}^{-1}\bm{b}}\text{A}^{im}\text{A}^{jn}a_{m}b_{n}~.
\end{equation}
Let us introduce yet more notation,
\begin{equation}
\text{P}_{ij}=\text{Re}\left[\text{M}_{ij}\right]~, \qquad v_{i}\=\text{M}_{in}\rho^{n}+\ii\,\text{Im}\left[\cI_{i}\right].
\end{equation}
We find
\begin{equation}\label{eq:contractionresult}
\cC\=\frac{2\bm{v}^{T}\text{P}^{-1}\bar{\bm{v}}+4\ee^{K}\bigg(\left[\bm{v}^{T}\text{P}^{-1}\bm{v}\right]\left[\bar{\bm{v}}^{T}\text{P}^{-1}\bar{\bm{v}}\right]-\left[\bm{v}^{T}\text{P}^{-1}\bar{\bm{v}}\right]^{2}\bigg)}{2\bm{v}^{T}\text{P}^{-1}\bar{\bm{v}}-\ee^{-K}}~.
\end{equation}
If all terms in the instanton expansion are set to zero, then $\ee^{-K}=\frac{4}{3}Y_{ijk}\rho^{i}\rho^{j}\rho^{k}+4\zeta$ and $\bm{v}^{T}\text{P}^{-1}\bar{\bm{v}}=\bar{\bm{v}}^{T}\text{P}^{-1}\bar{\bm{v}}=\bm{v}^{T}\text{P}^{-1}\bar{\bm{v}}=Y_{ijk}\rho^{i}\rho^{j}\rho^{k}$. \eqref{eq:contractionresult} then reduces to
\begin{equation}
\cC\=\frac{3Y_{ijk}\rho^{i}\rho^{j}\rho^{k}}{Y_{ijk}\rho^{i}\rho^{j}\rho^{k}-6\zeta}~.
\end{equation}
If the Euler characteristic of the manifold is $0$, then the above expression takes the constant value 3. Then \eqref{eq:Vstab} vanishes, recovering the standard no-scale supergravity potential, for instance appearing in \cite{Kachru_2003}.

However, if the Euler characteristic is not zero, then $\cC$ is not constant and $V$ is less trivial. In fact, $V$ exhibits typical Dine-Seiberg behaviour \cite{Dine:1985he}. The classical value of the manifold's volume is 
\begin{equation}
\upsilon=Y_{ijk}\rho^{i}\rho^{j}\rho^{k}~,
\end{equation}
and for large $\upsilon$, $\cC$ asymptotes to the value 3. The difference $\cC-3$ falls off like $\frac{1}{\upsilon}$. The asymptote is from above if $\chi>0$, and from below if $\chi<0$. In the first case, $\chi>0$, a barrier prevents $\cC$ from becoming smaller than $\upsilon$ (in dimensionless variables) and the potential runs away to infinity. For $\chi<0$, there is a runaway towards $\upsilon=0$. Since this latter point is at small volume, we can expect that mechanisms not incorporated in our analysis become important and so we cannot say much at all about $\cC$ near $\upsilon=0$. 

The denouement of this section is that equation \eqref{eq:contractionresult} allows for one to study how instantons affect the previous paragraph's content. For $\chi\neq0$ there is no substantial change, but the $\chi=0$ scenario is altered. 

If $\chi=0$ but instantons are not neglected, then at large $\rho$ the contraction $\cC$ will asymptote to 3 from above or below depending on the sign of the instanton corrections. What is more, the asymptotic approach will be exponential, rather than the powerlike decay discussed before. This is because, if $\chi=0$, the first nonclassical term in \eqref{eq:prepotagain} that contributes to \eqref{eq:contractionresult} is the first instanton contribution, which falls off exponentially as the K{\"a}hler parameters are increased.

None of that has direct relevance to any useful string models, because (among other problems) we have not stabilised any moduli. We have demonstrated how the values of $\chi$ and the leading instanton number can affect the large volume behaviour of this unrealistic model.

\section{Modularity of supersymmetric vacua in flux compactifications}\label{sect:SUGRA_FluxModularity}
The central equations of this section are those giving the simultaneous vanishing of the potential $V$ and superpotential $W$ in a IIB flux compactification on a CY threefold $X$. We work with the classical superpotential, which has no dependence on the K{\"a}hler moduli. The potential $V$ then has the form \eqref{eq:potential_classical} which we repeat here.
\begin{equation}
W=\int_{X}G_{3}\wedge \Omega=(2\pi)^{2}\alpha'(F-\tau H)\Sigma\Pi~~,\qquad V\=\ee^{\cK}\left(\Im[\tau]^2\, |\partial_{\tau}W|^{2}+g^{i\bar{\jmath}}\,\partial_{\varphi^{i}}W\,\overline{\partial_{\varphi^{j}}W}\right)~.
\end{equation} 
The equations \eqref{eq:sfv_equations} satisfied by supersymmetric vacuum configurations, together with the consistency condition $F^{T}\Sigma H\neq0$, can be recast as the SFV equations
\begin{equation}\label{eq:SFV_equations}
F^{T}\Sigma\Pi=0~,\qquad H^{T}\Sigma\Pi=0~,\qquad F^{T}\Sigma H\neq0~,\qquad\left(F-\tau H\right)\Sigma\left(\partial_{\varphi^{i}}\Pi\right)=0~.
\end{equation}

It is unknown if, for a given manifold $X$, there is always a pair of flux vectors $F$ and $H$ so that within $X$'s moduli space exists a region where the above equations are satisfied. In \cite{Kachru:2020sio} it was conjectured by Kachru, Nally, and Yang that such manifolds, with their moduli on the SFV locus and valued in an algebraic number field, are weight-two modular. They were able to test this conjecture by appealing to two sources of information. One of these was a solution method produced by deWolfe \cite{DeWolfe:2005gy} for solving the SFV equations, valid for $X$ the mirror of a hypersurface in weighted projective space. The other key source of information was the thesis \cite{Kadir:2004zb} of Kadir, in which tables of zeta functions were given, with various choices of moduli, for the mirror of the octic hypersurface in $\IW\IP_{1,1,2,2,2}$. Kachru, Nally, and Yang supported their conjecture by reporting that on deWolfe's solution locus, the tables of Kadir demonstrated weight-two modularity.

We seek to make further tests of this conjecture. This requires more solutions to the SFV equations and more computations of zeta functions. Concerning the first requirement, we determine a method of solving the SFV equations that works on any manifold $X$ whose period vector possesses a $\IZ_{2}$ symmetry. In the thesis \cite{Kuusela:2022hga}, the method of \cite{Candelas:2021tqt} was extended to multiparameter models which, using the solution method that we now describe, allowed for the flux modularity conjecture to be extensively tested, with additional tests carried out in \cite{Candelas:2023yrg}.

\subsection{Solving the SFV equations via permutation symmetry}
It is required that the period vector $\Pi$ has two pairs of entries such that each pair is permuted when two of the moduli are swapped. That is to say, we should be able to identify a pair of moduli $\varphi^{I},\varphi^{J}$ so that
\begin{equation}
\begin{aligned}
\Pi^{1+a}(\varphi^{I},\varphi^{J})&=\Pi^{1+b}(\varphi^{J},\varphi^{I})~,\\[5pt]
\Pi^{2+h^{2,1}+a}(\varphi^{I},\varphi^{J})&=\Pi^{2+h^{2,1}+b}(\varphi^{J},\varphi^{I})~,
\end{aligned}
\end{equation}
where we have suppressed $\Pi$'s dependence on the other moduli. The superscripts indicate components of the vector $\Pi$, with $1\leq a,b\leq h^{2,1}$.

The dependence of the other components of $\Pi$ on any remaining moduli is unimportant. If $\Pi$ has this property, then on the locus $\varphi^{I}=\varphi^{J}$ flux vectors $F$ and $H$ satisfying the first three equations in \eqref{eq:SFV_equations} are
\begin{equation}
F=e_{2+h^{2,1}+I}-e_{2+h^{2,1}+J}~,\qquad H=e_{1+I}-e_{1+J}~,
\end{equation}
where the $e_{k}$ are standard orthonormal basis vectors for $\IR^{2+2h^{2,1}}$, with all components 0 except for the $k^{th}$ component being 1.

With the flux vectors and this restriction on the moduli space $\varphi^{I}=\varphi^{J}$ specified, all bar two of the SFV equations are satisfied for any value of the axiodilaton. The remaining two equations are equivalent and fix the axiodilaton:
\begin{equation}\label{eq:axiodilaton_solution}
\begin{aligned}
\tau(\varphi)&=-\frac{F^{T}\Sigma\left(\partial_{\varphi^{I}}-\partial_{\varphi^{J}}\right)\Pi}{H^{T}\Sigma\left(\partial_{\varphi^{I}}-\partial_{\varphi^{J}}\right)\Pi}\bigg\vert_{\varphi^{I}=\varphi^{J}}\\[10pt]
&=\frac{i}{2\pi}\cdot\frac{\partial_{\varphi^{I}}\left(\varpi_{2,I}-\varpi_{2,J}\right)}{\partial_{\varphi^{I}}\left(\varpi_{1,I}-\varpi_{1,J}\right)}\bigg\vert_{\varphi^{I}=\varphi^{J}}+Y_{0IJ}-Y_{0II}~.
\end{aligned}
\end{equation}
It is perhaps interesting to note that for real moduli in the large complex structure region, the real part of the axiodilaton is 0 or $1/2$ depending on the value of $Y_{0IJ}-Y_{0II}$.

We will now give a select few worked examples. Many more are possible to find than those given here.

\subsection{Examples of supersymmetric flux vacua}
\subsubsection*{The Hulek-Verrill Manifold}
The first example of such a suitable symmetric manifold is the Hulek-Verrill manifold, which is mirror to the CICY
\begin{equation}
\cicy{\IP^1\\\IP^1\\\IP^1\\\IP^1\\\IP^1}{1 & 1\\1 & 1\\1 & 1\\ 1 & 1\\ 1 & 1}^{(5,45)}_{\chi\,=-80~}
\end{equation}
and has the following topological quantities:
\begin{equation}
    Y_{ijk}\=\begin{cases}2\quad i,j,k \text{ distinct.}\\ 0\quad \text{otherwise,}\end{cases}\quad Y_{ij0} \= 0~, \quad Y_{i00} \= -2~, \quad  Y_{000} \= 240 \frac{\zeta(3)}{(2\pi \ii)^3}~.
\end{equation}
Analytic expressions for the periods are available in terms of integrals of Bessel functions, as detailed in \cite{Candelas:2021lkc}. Subject to the condition 
\begin{align} \label{eq:period_region_HV}
\Re\left[\sum_{i=1}^{5}\sqrt{\varphi^{i}}\right]<1~,
\end{align}
we can give the relevant periods as
\begin{equation}\label{eq:periods_Frobenius_HV}
\begin{aligned}
\varpi_{0}(\bm{\varphi})&\=\int_{0}^{\infty}\dd z\;z\,\text{K}_{0}(z)\prod_{i=1}^{5}\text{I}_{0}\left(\sqrt{\varphi^{i}}z\right)~,\\[10pt]
\varpi_{1,j}(\bm{\varphi})&\=-2\int_{0}^{\infty}\dd z\;z\,\text{K}_{0}(z)\text{K}_{0}\left(\sqrt{\varphi^{j}}z\right)\prod_{i\neq j}\text{I}_{0}\left(\sqrt{\varphi^{i}}z\right)~,\\[10pt]
\varpi_{2,j}(\bm{\varphi})&\=8\sum_{\substack{m<n\\m,n\neq j}}\int_{0}^{\infty}\dd z\;z\,\text{K}_{0}(z)\text{K}_{0}\left(\sqrt{\varphi^{m}}z\right)\text{K}_{0}\left(\sqrt{\varphi^{n}}z\right)\prod_{i\neq m,n}\text{I}_{0}\left(\sqrt{\varphi^i}z\right)-4\pi^{2}\varpi_{0}(\bm{\varphi})~.
\end{aligned}
\end{equation}
The five-parameter Hulek-Verrill manifolds support supersymmetric flux vacua on the loci where any two complex structure moduli $\varphi^i$ are equal. For definiteness, let us set two of the $\varphi^{1}$ and $\varphi^2$ equal to a value $\varphi$, and relabel the three remaining $\varphi^{i}$ as $\psi^{1},\,\psi^{2},\,\psi^{3}$. Use of standard Bessel function identities reveals that in the case of Hulek-Verrill manifolds, the axiodilaton given in \eqref{eq:axiodilaton_solution} is
\begin{equation}\label{eq:tau_HV}
\tau\left(\psi^{1},\,\psi^{2},\,\psi^{3}\right)\=\frac{2\ii}{\pi}\, \frac{\int_{0}^{\infty}\dd z
	\;z\,\text{K}_{0}(z)
	\left[\text{K}_{0}\left(\sqrt{\psi^{1}}z \right)\,\text{I}_{0}\left(\sqrt{\psi^{2}}z\right)\,\text{I}_{0}\left(\sqrt{\psi^{3}}z\right)\phantom{\bigg\vert}+\text{cyclic}\right]
}
{\phantom{\bigg\vert}\int_{0}^{\infty}\dd z
	\;z\,\text{K}_{0}(z)
	\,\text{I}_{0}\left(\sqrt{\psi^{1}}z\right)\,\text{I}_{0}\left(\sqrt{\psi^{2}}z\right)\,\text{I}_{0}\left(\sqrt{\psi^{3}}z\right)
}~.
\end{equation}
The dependence of the $j$-invariant can be found numerically by computing the value of $j(\tau)$ on numerous points on the moduli space, and fitting the points on a rational function. It turns out that the $j$-function takes a remarkably simple form
\begin{equation} \label{eq:jtau_HV}
j\left(\tau\left(\psi^{1},\,\psi^{2},\,\psi^{3}\right)\right)\=\frac{\left(\Delta_{F}+16\psi^{1}\psi^{2}\psi^{3}\right)^{3}}{\Delta_{F}\left(\psi^{1}\psi^{2}\psi^{3}\right)^{2}},
\end{equation}
where the polynomial $\Delta_{F}$, which is related to the discriminant \eqref{eq:discriminant_HV}, is given by
\begin{equation}
\begin{aligned}
\Delta_{F}&\=\prod_{\epsilon_{i}=\pm1}\left(1+\epsilon_{1}\sqrt{\psi^{1}}+\epsilon_{2}\sqrt{\psi^{2}}+\epsilon_{3}\sqrt{\psi^{3}}\right)\\[5pt]
&\=\left(\left(1-\psi^{1}-\psi^{2}-\psi^{3}\right)^{2}-4\left(\psi^{1}\psi^{2}+\psi^{2}\psi^{3}+\psi^{3}\psi^{1}\right)\right)^{2}-64\;\psi^{1}\psi^{2}\psi^{3}~.
\end{aligned}
\end{equation}
Note that to derive \eqref{eq:tau_HV} and then subsequently arrive at \eqref{eq:jtau_HV}, we have used the expressions \eqref{eq:periods_Frobenius_HV} for the periods, which are valid only in the region \eqref{eq:period_region_HV}. However, since the expression \eqref{eq:jtau_HV} for the $j$-function is well-defined everywhere outside of the discriminant locus, this is the unique analytic continuation of the left-hand side into this region and the expression \eqref{eq:jtau_HV} is correct throughout moduli space.

It is interesting to note that \eqref{eq:tau_HV} and \eqref{eq:jtau_HV} do not involve the complex structure coordinate $\varphi$, and only depend on the three coordinates $\psi^{1},\,\psi^{2},\,\psi^{3}$. In other examples, as we will see in imminent subsections, $\tau$ does depend on the value to which we set two of the coordinates in our solution method. For this particular example at hand, the elimination of $\varphi$ from these formula was explained geometrically in \cite{Candelas:2023yrg} in terms of the fibred product structure of the Hulek-Verrill manifold. Understanding why this $\varphi$ independence does or does not occur for other examples is an open problem.
\subsubsection*{The mirror bicubic}
This manifold is mirror to the CICY
\begin{equation}
\cicy{\IP^2\\\IP^2}{3 \\ 3}_{\chi\,=-162~,}
\end{equation}
with the constants $Y_{abc}$ given by
\begin{equation}
    Y_{ijk}\=\begin{cases}
        0\quad i=j=k~,\\ 
        3\quad \text{otherwise,}
    \end{cases} \quad
    Y_{ij0}\=\begin{cases}0&i\=j~,\\\frac{1}{2}&i\;\neq\; j~,\end{cases} \,\quad    
    Y_{i00} \= -3~, \quad Y_{000} \= 486 \frac{\zeta(3)}{(2\pi \ii)^3}~.
\end{equation}
The periods can be written in terms of integrals of hypergeometric and Meijer G-functions:
\begin{equation}
\begin{aligned}
\varpi_{0}&\=\+\int_{0}^{\infty}\dd u\;\ee^{-u}\, \zeroFtwo{1}{1}{u^{3}\varphi^{1}}\, \zeroFtwo{1}{1}{u^{3}\varphi^{2}}~,\\[10pt]
\varpi_{1,i}&\=-\int_{0}^{\infty}\dd u\;\ee^{-u}\, G^{2,0}_{0,3}\left(0,0,0\vert u^{3}\varphi^{i}\right)\, \zeroFtwo{1}{1}{u^{3}\varphi^{j}}~,\qquad j \neq i~,\\[10pt]
\varpi_{2,i}&\=3\int_{0}^{\infty}\dd u\, \ee^{-u}\Big[ G^{2,0}_{0,3}\left(0,0,0\vert u^{3}\varphi^{i}\right)\,G^{2,0}_{0,3}\left(0,0,0\vert u^{3}\varphi^{j}\right)\\[5pt]
&\hskip80pt +\zeroFtwo{1}{1}{u^{3}\varphi^{i}}\left(G^{3,0}_{0,3}\left(0,0,0\vert -u^{3}\varphi^{j}\right)+\ii\pi G^{2,0}_{0,3}\left(0,0,0\vert u^{3}\varphi^{j}\right)\right) \Big]\\[5pt]
&\hskip120pt-2\pi^2\varpi_{0}~,\hskip110pt j \neq i~,\\[10pt]
\end{aligned}
\end{equation}
whence it follows that the axiodilaton profile on the $\IZ_2$ symmetric locus $\varphi^1 = \varphi^2$ is given by 
\begin{equation} \notag
\tau(\varphi){=}\frac{3\ii}{2\pi} \, \frac{\int_{0}^{\infty}\dd u\;\ee^{-u}\,\left[\zeroFtwo{1}{1}{u^{3}\varphi}G^{3,0}_{0,3}(0,0,1|{-}u^{3}\varphi)-\zeroFtwo{2}{2}{u^{3}\varphi}G^{3,0}_{0,3}(1,1,1|{-}u^{3}\varphi)\right]}{\int_{0}^{\infty}\dd u\;\ee^{-u}\,\left[\zeroFtwo{1}{1}{u^{3}\varphi}G^{2,0}_{0,3}(0,1,0|u^{3}\varphi)+\zeroFtwo{2}{2}{u^{3}\varphi}G^{2,0}_{0,3}(1,1,1|u^{3}\varphi)\right]}{-}1
~.\end{equation}
Even though it is not immediately obvious, this expression has, for real $\varphi$, a real part equal to $\frac{1}{2}$. One should bear in mind that we have further simplified the ratio of integrals to arrive at the above expression.

Numerical methods strongly suggest that the $j$-invariant $j(\tau(\varphi))$ is given by the following rational function of the moduli:
\begin{align} \label{eq:Mirror_Dicubic_j-invariant}
j(\tau(\varphi)) \= -\frac{(1+24\varphi)^3}{\varphi^3 (1+27\varphi)}~.
\end{align}

\subsubsection*{The mirror maximally split quintic}
This manifold is the mirror of the maximal split of the quintic hypersurface in $\IP^5$, given by the configuration described by the CICY matrix
\begin{equation}
\cicy{\IP^{4}\\\IP^{4}}{\;1\;1\;1\;1\;1\;\\\;1\;1\;1\;1\;1\;}_{\chi=-100}.
\end{equation}
The quantities $Y_{abc}$ of the split quintic are
\begin{equation}
    Y_{ijk}\=\begin{cases}
        5\quad \;\; i=j=k~,\\ 
        10\quad \text{otherwise,}\
    \end{cases} \quad Y_{0ij}\=\begin{cases}\frac{1}{2}&i\=j~,\\0&i\;\neq\;j~,\end{cases} 
    \qquad Y_{i00} \= -\frac{25}{6}~, \quad Y_{000} \= \frac{300\zeta(3)}{(2\pi\ii)^{3}}~.
\end{equation}
Denoting the Meijer G-functions as $G^{a,b}_{c,d}(0,0,0,0,0|x) = G^{a,b}_{c,d}(x)$, and the hypergeometric functions as $_{a}F_{b}(1,1,1,1; x) = {}_{a}F_b(x)$, the periods relevant to our analysis can be written as
% \begin{equation}
% \begin{aligned}
% \varpi^0 &\=\+\int_{0}^{\infty}\text{d}u\;G^{5,0}_{0,5}(u) \; {}_{0}F_4(u \varphi^1) \;{}_{0}F_4(u \varphi^2)~,\\[10pt]
% \varpi^{1} &\=-\int_{0}^{\infty}\text{d}u\;G^{5,0}_{0,5}(u) \; G^{2,0}_{0,5}(u \varphi^{1}) \; {}_{0}F_4(u \varphi^{2})~,\\[10pt]
% \varpi^{2} &\=-\int_{0}^{\infty}\text{d}u\;G^{5,0}_{0,5}(u) {}_{0}F_4(u \varphi^1)G^{2,0}_{0,5}(u \varphi^{2})~,\\[10pt]
% \varpi_{1} &\=5\int_{0}^{\infty}\text{d}u\;G^{5,0}_{0,5}(u)\left[2G^{2,0}_{0,5}(u \varphi^{1})\;G^{2,0}_{0,5}(u \varphi^{2})\right.+\left(G^{3,0}_{0,5}(-u \varphi^1)+\ii\pi G^{2,0}_{0,5}(u \varphi^{1})\right){}_{0}F_4(u \varphi^2)\\[5pt]
% &+2\left(G^{3,0}_{0,5}(-u \varphi^2)+\ii\pi G^{2,0}_{0,5}(u \varphi^{2})\right){}_{0}F_4(u \varphi^1)\left.-\;\frac{5\pi^{2}}{3}\;{}_{0}F_4(u \varphi^1){}_{0}F_4(u \varphi^2)\right] ~,\\[10pt]
% \varpi_{2} &\=5\int_{0}^{\infty}\text{d}u\;G^{5,0}_{0,5}(u)\left[2G^{2,0}_{0,5}(u \varphi^{1})\;G^{2,0}_{0,5}(u \varphi^{2})\right.+2\left(G^{3,0}_{0,5}(-u \varphi^1)+\ii\pi G^{2,0}_{0,5}(u \varphi^{1})\right){}_{0}F_4(u \varphi^2)\\[5pt]
% &+\left(G^{3,0}_{0,5}(-u \varphi^2)+\ii\pi G^{2,0}_{0,5}(u \varphi^{2})\right){}_{0}F_4(u \varphi^1)\left.-\;\frac{5\pi^{2}}{3}\;{}_{0}F_4(u \varphi^1){}_{0}F_4(u \varphi^2)\right] ~.
% \end{aligned}
% \end{equation}
\begin{equation}
\begin{aligned}
\varpi_0 &\=\+\int_{0}^{\infty}\text{d}u\;G^{5,0}_{0,5}(u) \; {}_{0}F_4(u \varphi^1) \;{}_{0}F_4(u \varphi^2)~,\\[10pt]
\varpi_{1,i} &\=-\int_{0}^{\infty}\text{d}u\;G^{5,0}_{0,5}(u) \; G^{2,0}_{0,5}(u \varphi^{i}) \; {}_{0}F_4(u \varphi^{j})~,\hskip156pt i \neq j~,\\[10pt]
\varpi_{2,i} &\=5\int_{0}^{\infty}\text{d}u\;G^{5,0}_{0,5}(u)\left[2G^{2,0}_{0,5}(u \varphi^{i})\;G^{2,0}_{0,5}(u \varphi^j)\right.+\left(G^{3,0}_{0,5}(-u \varphi^i)+\ii\pi G^{2,0}_{0,5}(u \varphi^{i})\right){}_{0}F_4(u \varphi^2)\\[5pt]
&+2\left(G^{3,0}_{0,5}(-u \varphi^j)+\ii\pi G^{2,0}_{0,5}(u \varphi^j)\right){}_{0}F_4(u \varphi^i)\left.-\;\frac{5\pi^{2}}{3}\;{}_{0}F_4(u \varphi^i){}_{0}F_4(u \varphi^j)\right]~,\qquad i \neq j~,
\end{aligned}
\end{equation}
and the axiodilaton is given on the $\IZ_2$ symmetric locus $\varphi^1 = \varphi^2 = \varphi$ by
\begin{equation}
\begin{aligned}
\tau(\varphi)&{=}\frac{5\ii}{2\pi}\frac{\int_{0}^{\infty}\text{d}u\;G^{5,0}_{0,5}(u)\left[G^{3,0}_{0,5}(0,0,1,0,0|-u \varphi){}_{0}F_4(u \varphi){-}G^{3,0}_{0,5}(1,1,1,1,1|-u \varphi)\zeroFfour{2}{2}{2}{2}{u\varphi}\right]}{\int_{0}^{\infty}\text{d}u\;G^{5,0}_{0,5}(u)\left[G^{2,0}_{0,5}(0,1,0,0,0|u \varphi){}_{0}F_4(u \varphi){-}G^{2,0}_{0,5}(1,1,1,1,1|u \varphi)\zeroFfour{2}{2}{2}{2}{u\varphi}\right]} \\&\+\+\+\++2~.
\end{aligned}
\end{equation}
For real $\varphi$, this expression has real part equal to $-\frac{1}{2}$. Attempts to numerically integrate the above combinations of Meijer G functions are met with problems in Mathematica, so instead it is best to expand each of the hypergeometric functions ${}_0 F_{4}$ as a power series in $u$ up to some large order (we used 360). After interchanging the order of summation and integration, each integral in the resulting sum can be evaluated in Mathematica exactly as a single Meijer G function, yielding a series expression amenable to fast evaluation.

The numerical evidence strongly suggests that the $j$-invariant $j(\tau)$ is again a rational function of the complex structure parameter~$\varphi$,
\begin{align}
j(\tau(\varphi)) \= -\frac{\left(1+12\varphi+14\varphi^{2}-12\varphi^{3}+\varphi^{4}\right)^3}{\varphi ^5 (1+11\varphi-\varphi^{2})}~.
\end{align}

\subsection{Speculation on elliptic curves inside the F-theory fourfold}
Consider again the F-theory fourfold, as discussed in \sref{sect:SUGRA_Ftheory}. This is an elliptic fibration, and for the supersymmetric vacuum configurations which we have described, the fibre is the elliptic curve $E_{\tau}$ with $\tau$ the axiodilaton \eqref{eq:axiodilaton_solution} of our solution to the supersymmetric flux vacuum equations.

The base of this fibration, $\cB$, is doubly covered by the threefold $X$, which we have argued is modular of weight two. In line with the conjectures of Kachru, Nally, and Yang in \cite{Kachru:2020sio,Kachru:2020abh}, the weight-two modular form associated to $X$ is itself associated to the elliptic curve $E_{\tau}$ in the sense of the modularity theorem for elliptic curves. 

For the case of the Hulek-Verrill family, it was explained in \cite{Candelas:2023yrg} that the threefold $X$ contained a ruled surface $E_{\tau}\times\IP^{1}$, which was nontrivial in homology. This gave an account of why the threefold's zeta function numerator contained a factor corresponding to the elliptic curve $E_{\tau}$. 

For this Hulek-Verrill example then, the F-theory fourfold that appears in the uplift of the supersymmetric flux vacuum that we describe contains $E_{\tau}$ in two different ways: once as the fibre and again in (a double cover) of the base. It could be interesting to understand whether this happens for every example, and more speculatively, to see if any string duality explains or utilises this twofold appearance of $E_{\tau}$ in the F-theory fourfold.

%\subsection{Additional examples}
%We provide here several other examples, with a briefer presentation. We shall limit ourselves to simply displaying the $j$-invariant, which is constructed in the same way as %in previous sections.

%\begin{equation}
%\cicy{\IP^{3}\\\IP^{3}}{1&1&2\\1&1&2}~,\qquad j\left(\tau(\varphi)\right)=\frac{\left(16+16\varphi+\varphi^{2}\right)^{3}}{\varphi(16+\varphi)}~.
%\end{equation}

%\begin{equation}
%\cicy{\IP^{1}\\\IP^{1}\\\IP^{1}\\\IP^{1}}{2\\2\\2\\2}~,\qquad 
%\end{equation}

\section{New weight four modular manifolds}\label{sect:SUGRA_NewWeight4}
We have seen in \sref{sect:SUGRA_AttMech} that the attractor mechanism fixes the values taken by vector multiplet scalar fields on the horizon of an extremal black hole in $\mathcal{N}=2$ supergravity. In IIB Calabi-Yau compactifications those scalars give the complex structure moduli of the Calabi-Yau $X$, and the fixed points of attractor flows are the solutions of Strominger's equations
\begin{equation}\label{eq:Strominger}
Q=\text{Im}\left[C\Pi\right]~.
\end{equation}
In this section we will describe a method, first worked through in \cite{Candelas:2019llw} and later fully developed in \cite{Candelas:2021tqt}, for finding rank-two attractors. These are complex structure moduli $\varphi_{*}$ that solve the above equation \eqref{eq:Strominger} for two independent charge vectors $Q_{1},\,Q_{2}$. Tables that display the main piece of information that we use, factorisation counts, are collected in \sref{sect:Appendix_Factorisations}. Our search returns a number of known examples, but also some new weight-four manifolds.

We will search among a subset of the one-parameter manifolds. These possess fourth-order Picard Fuchs operators which are of Calabi-Yau type \cite{almkvist2010tables,vanstraten2017calabiyau}. Such operators are tabulated in the database \cite{AESZ-db}, which is attached to the paper \cite{almkvist2010tables}. We shall refer to geometries by the AESZ label of their operator. Note that in the one-parameter setting, rank-two attractors are also solutions to the supersymmetric flux vacuum equations, and indeed all examples of weight-four modular one-parameter manifolds are also weight-two modular.

\subsection*{The search process}
If a rational number $\varphi_{*}$ is such that $X_{\varphi_{*}}$ is weight-four modular, then the numerator of the zeta function $\zeta_{p}\left(X_{\varphi_{*}};T\right)$ should factorise for each prime $p$. Turning this on its head, we will search for rank-two attractors by first tabulating the zeta function for $\varphi^{p}\in\{0,...,p-1\}$, for a number of primes $p$. Fixing a prime $p$, we will count the number of times the zeta function factorises for $\varphi^{p}\in\{0,...,p-1\}$. If this number is `usually' greater than zero (with exceptions expected to correspond to a small number of bad primes), then we inspect the zeta function numerators for the $\varphi^{p}_{*}$ that have factorised numerators and search the LMFDB database \cite{LMFDB} for a weight-four modular form whose $p^{th}$ Fourier coefficient is that read off of the zeta function numerator, to the extent of our tables.

If such a modular form can be found, then we seek a rational $\varphi_{*}$ that reduces modulo $p$ to $\varphi^{p}_{*}$. Having found such a value $\varphi_{*}$, we can numerically check to see if a pair of rational charge vectors $Q_{1},\,Q_{2}$ can be found so that \eqref{eq:Strominger} is solved by $\Pi(\varphi_{*})$.

We perform this check for 61 of the operators on the AESZ list. In Appendix \sref{sect:Appendix_Factorisations} we give bar charts showing the number of factorisations for each prime $5\leq p\leq 131$, for a total of 30 primes. To spot a rational rank-two attractor, we scroll down this list and stop where the bars all have heights greater than 0. In this way we recover a number of known examples: AESZ34 \cite{Candelas:2019llw}, and the pair AESZ4 and AESZ11 \cite{Bonisch:2022mgw}, as well as a number of examples identified previously in \cite{Bonisch:2022slo}. We will highlight those examples in Appendix \sref{sect:Appendix_Factorisations}, and also comment on this choice of 61 operators.

Moreover, we are also able to identify two hitherto-undiscovered attractors. These are in the moduli spaces of AESZ17 and AESZ22, and both at $\varphi_{*}=-1$. 

The operators AESZ17 and AESZ22 possess a property not shared by any other operators for which rank-two attractors are known: 17 and 22 both possess two MUM points, at 0 and infinity. By a change of variables that exchanges the MUM points, the operators AESZ17 and AESZ22 can be respectively transformed to AESZ118 and AESZ290. Indeed, the charts in \sref{sect:Appendix_Factorisations} for the pair (22,118) are identical. We give the associated modular forms in the following table.

\begin{table}[h]
    \centering
    \begin{tabular}{|c|c|c|c|}
    \hline
    \phantom{\bigg(}AESZ no.     & Attractor & weight 2 form  & weight 4 form\\
    \hline & & &\\
     17    & $-1\phantom{^{-5}}$ & 14.2.a.a & 14.4.a.b\\
     & & & \\
     22 & $-1\phantom{^{-5}}$  & 11.2.a.a & 33.4.a.b \\
      & & & \\
     \hline
    \end{tabular}
    \caption{New rank two attractors.}
    \label{tab:new_rks}
\end{table}

One should bear in mind that these are only conjectured to be modular varieties, and that conjecture is supported by tabulating the zeta function for finitely many primes.

We will say more about these geometries in \sref{sect:SUGRA_GWS}, and wrap up this section with the first few terms of the above modular form's $q$-expansions, with $q=\ee^{2\pi\ii\tau}$:
\begin{equation}
\begin{aligned}
f_{14.2.a.a}(\tau)&=q-q^{2}-2q^{3}+q^{4}+2q^{6}+q^{7}-q^{8}+q^{9}+\,...~,\\
f_{14.4.a.b}(\tau)&=q+2q^{2}-2q^{3}+4q^{4}-12q^{5}-4q^{6}+7q^{7}+8q^{8}-23q^{9}+\,...~,\\
&\\
f_{11.2.a.a}(\tau)&=q-2q^{2}-q^{3}+2q^{4}+q^{5}+2q^{6}-2q^{7}-2q^{9}+\,...~,\\
f_{33.4.a.b}(\tau)&=q-q^{2}-3q^{3}-7q^{4}-4q^{5}+3q^{6}-26q^{7}+15q^{8}+9q^{9}+\,...~.
\end{aligned}
\end{equation}
LMFDB \cite{LMFDB} provides the following eta-function expressions for the weight-two forms:
\begin{equation}
f_{14.2.a.a}(\tau)=\eta(\tau)\eta(2\tau)\eta(7\tau)\eta(14\tau)~,\qquad f_{11.2.a.a}(\tau)=\eta(\tau)^{2}\eta(11\tau)^{2}~,
\end{equation}
\hskip150pt with \hskip 20pt $\eta(\tau)=q^{1/24}\prod_{n=1}^{\infty}\left(1-q^{n}\right)~.$

%\subsubsection{Derived equivalence}
%Expanding the genus 0 prepotential for AESZ22 about the MUM point $\varphi=0$ and applying the mirror map as discussed in \sref{} gives the genus 0 invariants for a quotient manifold, the Reye congruence. This geometry was studied by Hosono and Takagi in \cite{}. It can be realised as a $\IZ_{2}$ quotient of a two-parameter CICY that we have already encounted,
%\begin{equation}
%\cicy{\IP^{4}\\\IP^{4}}{1&1&1&1&1\\1&1&1&1&1}_{/\IZ_{2}}~.
%\end{equation}
%The $IZ_{2}$ action amounts to exchanging the ambient $\IP^{4}$ factors. If the complex structure moduli are taken to be such that this $\IZ_{2}$ action is a symmetry of the manifold, then the above quotient can be taken. For generic moduli, this group action is fixed-point free.

\section{Summation identities: L-value ratios from GW invariants}\label{sect:SUGRA_GWS}
Note that all summation identities in this section are at the time of writing conjectures, supported to the various degrees of numerical precision that we give.

In this section, we will specialise the series solution \eqref{eq:solution} to the orthogonality equations to a selection of rank-two attractors. Following \cite{Candelas:2019llw,Bonisch:2022mgw}, we can express the Frobenius periods as linear combinations of the two critical L-values associated to a weight-four modular form. Therefore, the mirror coordinate $t$ is a rational function of those L-values. This value $t$ is a solution to the instanton-corrected IIA attractor equations, and we explained in \sref{sect:SUGA_1PCYIIA} how such a $t$ necessarily solved the orthogonality equations. By expressing $t$ via the solution \eqref{eq:solution}, we arrive at a series identity. Certain combinations of L-values can be expressed as infinite sums whose terms are built out of Gromov-Witten invariants.

In practice, we evaluate no more than the first 75 terms of the series, because the $j^{th}$ term contains a sum over partitions of $j$. Beyond $j=75$ there are too many partitions to work with on a laptop.

These sums converge for small enough values of $y_{0}$, and in those cases we get an honest numerical equality. Shanks transformations \cite{BenderOrszag} can be applied to the series in order to improve the number of figures to which our identities hold. However, for a number of rank-two attractors the appropriate value of $y_{0}$ is too large and the sums formally diverge. 

Building on the work of \cite{Candelas:2021mwz}, where such divergences were not further addressed, we identify Pad\'e resummation as the appropriate scheme for summing these series for larger values of $y_{0}$. We shall write
\begin{equation}
\text{LHS}\Peq \sum_{j=1}\ee^{2\pi\ii x_{0} j} c\left(j,y_{0}\right)
\end{equation}
to indicate that while the series on the right hand side diverges, the Pad\'e approximants about 0 in the variable $X$ to the series $\sum_{j=1}^{\infty}X^{j}c\left(j,y_{0}\right)$ approach (as the order of the approximant is increased) the value on the LHS when the substitution $X=\ee^{2\pi\ii x_{0}}$ is made.

Our examples will feature rank two attractors found in \cite{Candelas:2019llw,Bonisch:2022mgw} for the operators AESZ34, AESZ4, and AESZ11, as well as the new attractors found in \sref{sect:SUGRA_NewWeight4} for AESZ22 and AESZ17.

Note that when we give a $\chi$ among a set of topological data at the start of a subsection, we do so for the B-model geometry.

\subsection*{AESZ4, The mirror of $\IP^{5}[3,3]$}
The Picard-Fuchs equation for this geometry is the following hypergeometric equation:
\begin{equation}
\left[\theta^{4}-729\,z\left(\theta+\frac{1}{3}\right)^{2}\left(\theta+\frac{2}{3}\right)^{2}\right]F(z)\=0~.
\end{equation}
The topological data is
\begin{equation}
Y_{111}\=9~,\qquad Y_{110}\=\frac{1}{2}~,\qquad Y_{100}\=-\frac{9}{2}~,\qquad\chi\=144~.
\end{equation}
It was shown in \cite{Bonisch:2022mgw} that this geometry has a rank two attractor at $z=-2^{-3}\cdot3^{-6}$. The Frobenius periods were shown to be given by the following combinations of critical L-values associated to the weight-4 modular form with LMFDB label 54.4.a.c :
\begin{equation}
\begin{aligned}
\widehat\varpi_{0}&\=-18\,\frac{L(2)}{(2\pi\ii)^{2}}~,\\[5pt] 
\widehat\varpi_{1}&\=-9\,\frac{L(2)}{(2\pi\ii)^{2}}-\frac{9}{4}\,\frac{L(1)}{2\pi\ii}~,\\[5pt]
\widehat\varpi_{2}&\=\+135\,\frac{L(2)}{(2\pi\ii)^{2}}-\frac{81}{8}\,\frac{L(1)}{(2\pi\ii)}~,\\[5pt] 
\widehat\varpi_{3}&\=\+\frac{297}{4}\,\frac{L(2)}{(2\pi\ii)^{2}}+\frac{81}{16}\,\frac{L(1)}{2\pi\ii}+\frac{\chi\,\zeta(3)}{(2\pi\ii)^{3}}\widehat\varpi_{0}~.
\end{aligned}
\end{equation}
These critical L-values have decimal expansions
\begin{equation}
\begin{aligned}
L(1)&\=3.84405587339769322173023\,...~,\\[5pt]
L(2)&\=2.18020008917513916885513\,...~.
\end{aligned}
\end{equation}
In the $t-$plane, the attractor is
\begin{equation}
t\=\frac{1}{2}+\frac{\ii \pi}{4}\frac{L(1)}{L(2)}~.
\end{equation}

The attractor  equations are solved by $z=-2^{-3}\cdot3^{-6}$ for the following charge vectors:
\begin{equation}
Q_{1}\=(0,-5,0,1)^{T},\qquad Q_{2}\=(-3,12,1,0)^{T}~.
\end{equation}
It follows that the integral basis period vector $\Pi$ is symplectic-orthogonal to the charge vectors
\begin{equation}
Q_{3}\=(-3,0,1,0)^{T},\qquad Q_{4}\=(12,-5,0,1)^{T}~.
\end{equation}
$Q_{4}$ is of `D4-type', as the entry corresponding to the D6 charge is zero. This means that we can form an identity as in \cite{Candelas:2021mwz}. The perturbative solution of the orthogonality equation is given by
\begin{equation}
t_{0}=\frac{1}{2}+\frac{\ii\sqrt{69}}{6}~.
\end{equation}
We arrive at 
\begin{equation}
\frac{3\pi}{2}\cdot\frac{L(1)}{L(2)}\=\sqrt{69}-\sqrt{\frac{2}{\pi^{3}}}\,\sum_{j=1}^{\infty}(-1)^{j}\sum_{\fp\in\text{pt}(j)}a_{\fp}N^{\fp}\left(\frac{j}{3\pi\sqrt{69}}\right)^{l(\fp)-1/2}K_{l(\fp)-1/2}\left(\frac{j\,\pi\sqrt{69}}{3}\right)~.
\end{equation}
This sum converges, and there is no need for Pad\'e resummation. The equality should be understood as usual. Using 60 terms of the series, we observe agreement to 73 figures. By performing a Shanks transformation up to 15 times, we observe improvement to 105 figures. Applying further Shanks transformations does not yield further improvement.

\section*{AESZ11, The mirror of $\IW\IP^{5}_{1,1,1,1,1,2}[4,3]$}
This geometry also has a hypergeometric Picard-Fuchs equation:
\begin{equation}
\left[\theta^{4}-1728\,z\left(\theta+\frac{1}{4}\right)\left(\theta+\frac{1}{3}\right)\left(\theta+\frac{2}{3}\right)\left(\theta+\frac{3}{4}\right)\right]F(z)\=0~.
\end{equation}
The topological data is
\begin{equation}
Y_{111}\=6~,\qquad Y_{110}\=0~,\qquad Y_{100}\=-4~,\qquad\chi\=156~.
\end{equation}
It was shown in \cite{Bonisch:2022mgw} that this geometry has a rank two attractor at $z=-2^{-4}\cdot3^{-3}$. The Frobenius periods were again shown to be given by combinations of critical L-values , this time associated to the weight-4 modular form with LMFDB label 180.4.a.e :
\begin{equation}
\begin{aligned}
\widehat\varpi_{0}&\=-18\,\frac{L(2)}{(2\pi\ii)^{2}}~,\\[5pt] 
\widehat\varpi_{1}&\=-9\,\frac{L(2)}{(2\pi\ii)^{2}}-\frac{6}{10}\,\frac{L(1)}{2\pi\ii}~,\\[5pt]
\widehat\varpi_{2}&\=\+45\,\frac{L(2)}{(2\pi\ii)^{2}}-\frac{9}{5}\,\frac{L(1)}{2\pi\ii}~,\\[5pt] 
\widehat\varpi_{3}&\=\+27\,\frac{L(2)}{(2\pi\ii)^{2}}+\frac{3}{5}\,\frac{L(1)}{2\pi\ii}+\frac{\chi\,\zeta(3)}{(2\pi\ii)^{3}}\widehat\varpi_{0}~.
\end{aligned}
\end{equation}
These critical L-values have decimal expansions
\begin{equation}
\begin{aligned}
L(1)&\=10.01558944150908412493925\,...~,\\[5pt]
L(2)&\=\phantom{0}1.99897335900796011552100\,...~.
\end{aligned}
\end{equation}
In the $t-$plane, the attractor this time is
\begin{equation}
t\=\frac{1}{2}+\frac{\ii \pi}{15}\cdot\frac{L(1)}{L(2)}~.
\end{equation}

The value $z=-2^{-4}\cdot3^{-3}$ solves the attractor equations for the following charge vectors:
\begin{equation}
Q_{1}\=(-1,6,1,0)^{T},\qquad Q_{2}\=(1,-3,0,1)^{T}~.
\end{equation}
One computes that the integral basis period vector $\Pi$ is symplectic-orthogonal to the charge vectors
\begin{equation}
Q_{3}\=(-1,1,1,0)^{T},\qquad Q_{4}\=(6,-3,0,1)^{T}~.
\end{equation}
$Q_{4}$ is of D4-type, so we can form an identity. The perturbative solution is given by
\begin{equation}
t_{0}=\frac{1}{2}+\frac{\ii\sqrt{39}}{6}~.
\end{equation}
We thereby find
\begin{equation}
\frac{2\pi}{5}\cdot\frac{L(1)}{L(2)}\Peq\sqrt{39}-\sqrt{\frac{3}{\pi^{3}}}\,\sum_{j=1}^{\infty}(-1)^{j}\sum_{\fp\in\text{pt}(j)}a_{\fp}N^{\fp}\left(\frac{j}{2\pi\sqrt{39}}\right)^{l(\fp)-1/2}K_{l(\fp)-1/2}\left(\frac{j\,\pi\sqrt{39}}{3}\right)~.
\end{equation}
This sum does not converge. As discussed in the overview, the identity is supported numerically by Pad\'e resummation. A diagonal Pad\'e approximant of order 30 gives agreement to 38 figures.

\section*{AESZ34, A quotient of the Hulek-Verrill manifold}
Three rank two attractor points were found for this geometry in \cite{Candelas:2019llw}. One of these, at $z=33-8\sqrt{17}$, lies inside the LCS region and was a central example in \cite{Candelas:2021mwz}, where the first sum of the form considered here was given. We discuss now the other two attractor points, at $z=-7^{-1}$ and $z=33+8\sqrt{17}$, which lie outside of the LCS region and so require Pad\'e resummation. 

The Picard-Fuchs equation is not hypergeometric, and is
\begin{equation}
\begin{aligned}
\left[(1-z)(1-9z)(1-25z)\,\theta^{4}+2z\left(675z^{2}-518z+35\right)\theta^{3}+z\left(2925z^{2}-1580z+63\right)\theta^{2}\right.\\[5pt]\left.
+4z\left(675z^{2}-272z+7\right)\theta+5z\left(180z^{2}-57z+1\right)\right]F(z)\=0~.
\end{aligned}
\end{equation}
The topological data is
\begin{equation}
Y_{111}\=24~,\qquad Y_{110}\=0~,\qquad Y_{100}\=-2~,\qquad\chi\=16~.
\end{equation}
We will divide the remainder of this section into subsections discussing each rank two attractor separately. We choose to work with the `$\kappa=2$' geometry of \cite{Candelas:2019llw}, which has triple intersection number $Y_{111}=24$. The choice does not affect the summation identities, as all factors of $\kappa$ drop out.

\subsection*{$z\=-\frac{1}{7}$}
The periods were conjectured (with 1000 digits of numerical precision supporting) to be given in terms of critical L-values for the modular form 14.4.a.a .
\begin{equation}
\begin{aligned}
\widehat\varpi_{0}&\=-28\,\frac{L(2)}{(2\pi\ii)^{2}}~,\\[5pt]
\widehat\varpi_{1}&\=-14\,\frac{L(2)}{(2\pi\ii)^{2}}-\frac{5}{2}\,\frac{L(1)}{2\pi\ii}~,\\[5pt]
\widehat\varpi_{2}&\=-28\frac{L(2)}{(2\pi\ii)^{2}}-30\,\frac{L(1)}{2\pi\ii}~,\\[5pt]
\widehat\varpi_{3}&\=\+14\,\frac{L(2)}{(2\pi\ii)^{2}}-\frac{11}{2}\frac{L(1)}{2\pi\ii}+\frac{\chi\,\zeta(3)}{(2\pi\ii)^{3}}\widehat\varpi_{0}~.
\end{aligned}
\end{equation}
The decimal expansions of these L-values begin
\begin{equation}
\begin{aligned}
L(1)&\=0.67496319716994177129270\,...~,\\[5pt]
L(2)&\=0.91930674266912115653914\,...~.
\end{aligned}
\end{equation}
The attractor is, in the $t$-plane, 
\begin{equation}
t\=\frac{1}{2}+\frac{5\pi\ii}{28}\cdot\frac{L(1)}{L(2)}~.
\end{equation}
Independent charge vectors for which the attractor equations are solved are
\begin{equation}
Q_{1}\=(-\frac{8}{5},6,1,0)^{T}~,\qquad Q_{2}\=(\frac{16}{5},-12,0,1)^{T}~.
\end{equation}
The integral symplectic period vector is then orthogonal to
\begin{equation}
Q_{3}\=(-\frac{8}{5},\frac{16}{5},1,0)^{T}~,\qquad Q_{4}\=(6,-12,0,1)^{T}~.
\end{equation}
We shall form our identity using $Q_{4}$. The perturbative solution is
\begin{equation}
t_{0}\=\frac{1}{2}+\frac{\ii}{\sqrt{6}}~.
\end{equation}
From these considerations, we can find the identity
\begin{equation}
\frac{15\pi}{14}\cdot\frac{L(1)}{L(2)}\Peq\sqrt{6}-\frac{1}{2}\cdot\sqrt{\frac{3}{\pi^{3}}}\,\sum_{j=1}^{\infty}(-1)^{j}\sum_{\fp\in\text{pt}(j)}a_{\fp}N^{\fp}\left(\frac{j}{8\pi\sqrt{6}}\right)^{l(\fp)-1/2}K_{l(\fp)-1/2}\left(\frac{\pi j \sqrt{6}}{3}\right)~.
\end{equation}
A diagonal Pad\'e approximant of order 30 gives agreement to 34 figures.
\subsection*{$z\=33+8\sqrt{17}$}
This constitutes perhaps our strangest example. The real parts of the critical L-values associated to 34.4.b.a have decimal expansions beginning
\begin{equation}
\begin{aligned}
\lambda(1)&\=0.61300748403501690756896\,...~,\\[5pt]
\lambda(2)&\=0.72053904959503349611019\,...~.
\end{aligned}
\end{equation}
The complete L-values are
\begin{equation}
\begin{aligned}
L(1)&\=\lambda(1)\left(1+\ii\left(\frac{1+\sqrt{17}}{4}\right)^{3}\right)~,\\[5pt]
L(2)&\=\lambda(2)\left(1-\ii\,\frac{1-\sqrt{17}}{4}\right)~.
\end{aligned}
\end{equation}
Introduce symbols for the following numbers:
\begin{equation}
\epsilon_{-}\=4-\sqrt{17}~,\qquad \delta_{+}\=\frac{3+\sqrt{17}}{2}~,\qquad \delta_{-}\=\frac{3-\sqrt{17}}{2}~.
\end{equation}
In terms of the real parts $\lambda(1),\,\lambda(2)$ of the L-values, the periods as computed in \cite{Candelas:2019llw} are (again, conjecturally with 1000 digits of precision found numerically by those authors)
\begin{equation}
\begin{aligned}
\widehat\varpi_{0}&\=-\frac{7\sqrt{17}\epsilon_{-}^{2}\delta_{+}}{2^{3}\pi^{2}}\lambda(2)-\ii\,\frac{3\cdot5\delta_{-}^{3}}{2^{4}\sqrt{17}\pi}\lambda(1)~,\\[5pt]
\widehat\varpi_{1}&\=-\frac{\sqrt{17}\epsilon_{-}^{2}\delta_{+}}{2\pi^{2}}\lambda(2)-\ii\,\frac{5\delta_{-}^{3}}{2^{5}\sqrt{17}\pi}\lambda(1)~,\\[5pt]
\widehat\varpi_{2}&\=-\frac{5^{2}\sqrt{17}\epsilon_{-}^{2}\delta_{+}}{2^{3}\pi^{2}}\lambda(2)+\ii\,\frac{3\cdot5\delta_{-}^{3}}{2^{4}\sqrt{17}\pi}\lambda(1)~,\\[5pt]
\widehat\varpi_{3}&\=-\frac{\sqrt{17}\epsilon_{-}^{2}\delta_{+}}{2\pi^{2}}\lambda(2)+\ii\,\frac{13\delta_{-}^{3}}{2^{5}\sqrt{17}\pi}\lambda(1)+\frac{\chi\zeta(3)}{(2\pi\ii)^{3}}\widehat\varpi_{0}~.
\end{aligned}
\end{equation}
In the $t$-plane, the attractor is
\begin{equation}
t\=\frac{1}{6}+\frac{1156L(2)}{2856L(2)-45\pi\ii\left(9+\sqrt{17}\right)L(1)}~.
\end{equation}
The attractor equations are solved for the charge vectors
\begin{equation}
Q_{1}\=(-\frac{72}{85},\frac{66}{17},1,0)^{T}~,\qquad Q_{2}\=(\frac{296}{85},-\frac{192}{17},0,1)^{T}~.
\end{equation}
The period vector is orthogonal to the charge vectors
\begin{equation}
Q_{3}\=(-\frac{72}{85},\frac{296}{85},1,0)^{T}~,\qquad Q_{4}\=(\frac{66}{17},-\frac{192}{17},0,1)^{T}~.
\end{equation}
$Q_{4}$ shall be used to obtain an identity. The pertubative solution is
\begin{equation}
t_{0}\=\frac{8}{17}+\frac{\ii}{34}\sqrt{\frac{65}{3}}~.
\end{equation}
We arrive at
\begin{equation}\label{eq:aesz34+plus_identity}
\begin{aligned}
&\frac{2^{3}17^{3}\lambda(2)}{2856\lambda(2)-45\pi\ii\left(9+\sqrt{17}\right)\lambda(1)}\Peq\frac{31}{3}+\ii\sqrt{\frac{65}{3}}\\[5pt]
&\hskip30pt-\frac{17\ii}{2\sqrt{3\pi^{3}}}\sum_{j=1}^{\infty}\ee^{\frac{16\pi\ii}{17}\cdot j}\sum_{\fp\in\text{pt}(j)}a_{\fp}N^{\fp}\left(\frac{17j}{8\pi\sqrt{195}}\right)^{l(\fp)-1/2}K_{l(\fp)-1/2}\left(\frac{\pi j }{17}\sqrt{\frac{65}{3}}\right)~.
\end{aligned}
\end{equation}
This identity is much trickier to verify numerically. Using a diagonal Pad\'e approximant of order 30, agreement can be found to six figures. Our conjectural summation identity for this example then is not very well supported. In fact, agreement to six figures can be found using diagonal Pad\'e approximants of orders 23 through 30.

In case our series \eqref{eq:aesz34+plus_identity} does not hold, and in case the other (better evidenced) conjectural series identities from this section should prove true, it would be interesting to understand what sets this example apart.
%We give a plot showing the absolute values of the differences between successive Pad\'e approximants and the anticipated value below.
%\vskip20pt
%\begin{figure}[h]
%\begin{center}
%\scalebox{0.7}{\includegraphics{aesz34_phiplus_pade_plot.pdf}}
%\caption{The absolute value of the difference between the LHS of \eqref{eq:aesz34+plus_identity} and the resummed RHS against the order of the diagonal Pad\'e approximant used. }
%    \label{fig:aesz34_pade_plot}
%\end{center}
%\end{figure}

\section*{The pair AESZ22 and AESZ118}
The following two operators are mapped to one another by a change of variables and scaling transformation
\begin{equation}
z\=\frac{1}{32x}~,\qquad F_{22}(z)=\frac{1}{32x}F_{118}(x)~.
\end{equation}
\begin{equation}
\begin{aligned}
\text{AESZ22:}\phantom{8}&\hskip10pt\left[(7-4z)^{2}\left(1+11z-z^{2}\right)\theta^{4}\right.\\[5pt]&\hskip15pt\left.-2z(7-4z)\left(143+4942z-2084z^{2}+256z^{3}\right)\theta^{3}\right.\\[5pt]&\hskip15pt\left.
-z\left(1638+102261z-72568z^{2}+23024z^{3}-3072z^{4}\right)\theta^{2}\right.\\[5pt]&\hskip15pt\left.
-z\left(637+66094z-30072z^{2}+12896z^{3}-2048z^{4}\right)\theta\right.\\[5pt]&\hskip15pt
\left.-2z\left(49+7868z-1904z^{2}+1472z^{3}-256z^{4}\right)\right]F_{22}(z)\=0~,\\[10pt]
\text{AESZ118:}&\hskip10pt\left[ (1-x) (1-56 x)^2 \left(1-352x-1024 x^2\right)\theta ^4\right.\\
&\hskip15pt-2  x (1-56 x) (9-64 x) \left(33+384x-1792 x^2\right)\theta ^3\\
&\hskip15pt- x \left(431+15136x-335424x^2+4386816x^3 -19267584 x^4\right)\theta ^2\\
   &\hskip15pt-2   x \left(67+7072x-41088x^2 +996532x^3 -6422528x^4 \right)\theta\\
   &\hskip15pt\left.-16 x \left(1+176x-144x^2 +23296x^3 -20070x\right)\right]F_{118}(x)\=0~.
\end{aligned}
\end{equation}
In the first of the above equations, $\theta$ denotes the operator $z \frac{\text{d}}{\text{d}z}$, while in the second $\theta$ denotes $x \frac{\text{d}}{\text{d}x}$.

Every operator in the AESZ list is `adapted' to a MUM point, in that the point $0\in\IC$ is a point of maximal unipotent monodromy for any operator and the exponents for a basis of solutions about this point are $\{0,0,0,0\}$. 

However, as demonstrated by these examples, Calabi-Yau operators can have multiple MUM points. AESZ22 has, in addition to the origin, a MUM point at infinity with exponents $\{1,1,1,1\}$. Under the above change of variables, this point is mapped to 0 while 0 is mapped to AESZ118's secondary MUM point at infinity. The scaling transformation $F(z)\mapsto x F(x)$ is performed so that the exponents about the origin are $\{0,0,0,0\}$.

The $q$-expansion of either the Yukawa coupling or the prepotential about $z=0$ returns integer invariants $n_{d}^{\text{RC}}$ that are interpreted as counts of genus 0 curves with degree $d$ on a mirror geometry, the \textit{Reye Congruence}. This is a $\mathbb{Z}_{2}$ quotient of the complete intersection
\begin{equation}
\cicy{\IP^{4}\\\IP^{4}}{\;1\;1\;1\;1\;1\;\\\;1\;1\;1\;1\;1\;}_{\chi=-100.}
\end{equation}
by the freely acting $\mathbb{Z}_{2}$ quotient that exchanges the two $\IP^{4}$ factors. The first few $n_{d}^{\text{RC}}$ are
\begin{equation}
n_{1}^{\text{RC}}\=50~,\,n_{2}^{\text{RC}}\=325~,\,n_{3}^{\text{RC}}\=1475~,\,n_{4}^{\text{RC}}\=15325~,\,n_{5}^{\text{RC}}\=148575~,\,n_{6}^{\text{RC}}\=1885575~.
\end{equation}
On the other hand, $q$-expanding about $x=0$ returns integers $n_{d}^{\text{QS}}$ that count genus 0 degree $d$ curves on the orthogonal linear section of the double quintic symmetroid \cite{hosono2016double}. 

%\textcolor{red}{Actually, I don't understand many things here. There are two MUM points, and so two sets of curve counts for derived-equivalent families. The Reye congruence was shown by Hosono+Takagi to be derived equivalent to this other space. By fiddling around, I can arrive at an Euler character and triple intersection for AESZ118 that agree with their values, but from the AESZ database I get a value 0 for $c_{2}$ while H+T give 40. This is not explained by a symplectic transformation, by my understanding. Something nontrivial is up.}

The first few of these numbers are
\begin{equation}
n_{1}^{\text{QS}}\=550~,\,n_{2}^{\text{QS}}\=19150~,\,n_{3}^{\text{QS}}\=1165550~,\,n_{4}^{\text{QS}}\=106612400~,\,n_{5}^{\text{QS}}\=12279982850~.
\end{equation}
The formal statement, around which we avoid working, is that these two spaces RC and QS are derived equivalent, having the same derived category of coherent sheaves \cite{Kontsevich:1994dn}. They are not birational, possessing different topological invariants (such as their nonequal triple intersection numbers), but have the same Hodge numbers and the mirror symmetry computation of their curve counts takes place in a common complex structure moduli space for which the Picard-Fuchs operator can be taken to be AESZ22 or AESZ118 by suitable choices of coordinates.

\subsection*{A rational rank-two attractor}
We have identified the point $z=-1$ as a rank-two attractor. Equivalently, this is $x=-1/32$. The zeta function numerator factorises and the ensuing weight-four modular form has LMFDB label 33.4.a.b .

Upon a Mellin transformation, this modular form gives an L-function with special values
\begin{equation}
\begin{aligned}
L(1)&\=-1.0538249565444346000240\,...~,\\
L(2)&\=\+0~.
\end{aligned}
\end{equation}
About either $z=0$ or $x=0$ a Frobenius basis of periods can be computed. It is possible to express the imaginary parts of the period vectors $\widehat\varpi$ as rational multiples of $\frac{L(1)}{2\pi}$ (except that $\widehat\varpi_3$ involves $\frac{\zeta(3)}{(2\pi\ii)^{3}}$). The real parts involve another nonzero number $\frac{\Phi}{(2\pi)^{2}}$, with
\begin{equation}
\Phi\=16.671069275646809663\,...\,~.
\end{equation}Since $L(2)$ is zero, it is not useful to this end. It would be interesting to approach this issue from the perspective of Beilinson's conjecture and give $\Phi$ in terms of the nonzero value $L'(2)$, a problem that would involve calculating a suitable Beilinson regulator \cite{Beilinson}. However, it is possible to express $\Phi$ in terms of the L-values for a different modular form related to 33.4.a.b by a twist. In particular, $\Phi$ can be expressed in terms of a critical value of the Mellin transform of the form with LMFDB label 528.4.a.h:
\begin{equation}
\Phi\=\frac{\pi}{6}L_{\textbf{528.a.h}}(1)~,
\end{equation}
\subsubsection*{The Reye Congruence}
Pertaining to the mirror geometry RC, the topological invariants are
\begin{equation}
Y_{111}\=35~,\qquad Y_{110}\=\frac{1}{2}~,\qquad Y_{100}\=-\frac{25}{6}~,\qquad\chi\=50~.
\end{equation}
at $z=-1$, the Frobenius periods (about $z=0$) evaluate as follows:
\begin{equation}
\begin{aligned}
\widehat\varpi_{0}&\=-\frac{\Phi}{(2\pi\ii)^{2}}-\frac{5}{3}\,\frac{L(1)}{2\pi\ii}~,\\[5pt]
\widehat\varpi_{1}&\=-\frac{1}{2}\,\frac{\Phi}{(2\pi\ii)^{2}}~,\\[5pt]
\widehat\varpi_{2}&\=-\frac{10}{3}\,\frac{\Phi}{(2\pi\ii)^{2}}+\frac{175}{36}\,\frac{L(1)}{2\pi\ii}~,\\[5pt]
\widehat\varpi_{3}-\frac{\chi\,\zeta(3)}{(2\pi\ii)^{3}}\widehat\varpi_{0}&\=-\frac{5}{24}\,\frac{\Phi}{(2\pi\ii)^{2}}+\frac{5}{6}\,\frac{L(1)}{2\pi\ii}~.
\end{aligned}
\end{equation}
The attractor equations are solved for charge vectors given in the integral basis by
\begin{equation}
Q_{1}\=(-\frac{1}{2},5,1,0)^{T}~,\qquad Q_{2}\=(\frac{7}{2},-13,0,1)~.
\end{equation}
The orthogonality equations are solved by 
\begin{equation}
Q_{3}\=(5,-13,0,1)^{T}~,\qquad Q_{4}\=(-\frac{1}{2},\frac{7}{2},1,0)~.
\end{equation}
$Q_{3}$ has zero $D6$ charge. One can determine the perturbative solution to the orthogonality equation to be
\begin{equation}
t_{0}\=\frac{15}{42}+\frac{\sqrt{69}}{42}\ii~.
\end{equation}
These considerations lead to a sum
\begin{equation}
-\frac{3-25\frac{\pi L(1)}{\Phi}\ii}{3+10\frac{\pi L(1)}{\Phi}\ii }\ii\Peq\frac{\sqrt{69}}{6}-\sqrt{\frac{7}{10\pi^{3}}}\sum_{j=1}^{\infty}\ee^{\frac{15\pi\ii}{21}j}\sum_{\fp\in\text{pt}(j)}a_{\fp}N^{\fp}\left(\frac{3j}{5\pi\sqrt{69}}\right)^{l(\fp)-1/2}K_{l(\fp)-1/2}\left(\frac{\pi j \sqrt{69} }{21}\right)~,
\end{equation}
in which the $N_{k}$ are the scaled Gromov-Witten invariants constructed from the invariants $n_{d}^{\text{RC}}$. We use a diagonal Pad\'e approximant of order 30, and find agreement to six figures. This example is also only weakly supported, and our conjectural sum here is not well-evidenced.

\subsubsection*{The Quintic Symmetroid}
This time, for the mirror geometry QS, the topological invariants are
\begin{equation}
Y_{111}\=10~,\qquad Y_{110}\=0~,\qquad Y_{100}\=-\frac{10}{3}~,\qquad\chi\=50~.
\end{equation}
at $x=-1/32$, the Frobenius periods (about $x=0$) evaluate to:
\begin{equation}
\begin{aligned}
\widehat\varpi_{0}&\=-2\,\frac{\Phi}{(2\pi\ii)^{2}}~,\\[5pt]
\widehat\varpi_{1}&\=\phantom{2}-\frac{\Phi}{(2\pi\ii)^{2}}+\frac{10}{3}\,\frac{L(1)}{2\pi\ii}~,\\[5pt]
\widehat\varpi_{2}&\=\+\frac{5}{3}\,\frac{\Phi}{(2\pi\ii)^{2}}+\frac{50}{3}\,\frac{L(1)}{2\pi\ii}~,\\[5pt]
\widehat\varpi_{3}-\frac{\chi\,\zeta(3)}{(2\pi\ii)^{3}}\widehat\varpi_{0}&\=\+\frac{5}{3}\,\frac{\Phi}{(2\pi\ii)^{2}}+\frac{10}{9}\frac{L(1)}{2\pi\ii}~.
\end{aligned}
\end{equation}
The attractor equations are solved for the charge vectors
\begin{equation}
Q_{1}\=(-1,5,1,0)^{T}~,\qquad Q_{2}\=(2,-5,0,1)^{T}~.
\end{equation}
The orthogonality equations are in turn solved by 
\begin{equation}
Q_{3}\=(-1,2,1,0)^{T}~,\qquad Q_{4}\=(5,-5,0,1)^{T}~.
\end{equation}
From $Q_{4}$, one can determine the perturbative solution
\begin{equation}
t_{0}\=\frac{1}{2}+\frac{\ii\sqrt{15}}{6}~.
\end{equation}
Collecting the various pieces, one can reach the sum
\begin{equation}
-20\frac{\pi L(1)}{\Phi}\Peq\sqrt{15}-\frac{1}{\sqrt{10\pi^{3}}}\sum_{j=1}^{\infty}(-1)^{j}\sum_{\fp\in\text{pt}(j)}a_{\fp}N^{\fp}\left(\frac{j}{60\pi\sqrt{15}}\right)^{l(\fp)-1/2}K_{l(\fp)-1/2}\left(\pi j \sqrt{\frac{5}{3}}\right)~,
\end{equation}
in which the scaled Gromov-Witten invariants are constructed from the instanton numbers $n_{d}^{\text{QS}}$. We use a diagonal Pad\'e approximant of order 36, and find agreement to 29 figures.

\section*{The pair AESZ17 and AESZ290}
These are another pair of operators with two MUM points that are exchanged by a coordinate transformation and scaling
\begin{equation}
z\=z\mapsto-\frac{1}{3^{6}x}~,\qquad F_{17}(z)=-\frac{1}{3^{6}x}F_{290}(x)~.
\end{equation} 
\begin{equation}
\begin{aligned}
\text{AESZ17:}&\hskip10pt\big[(5-9 z)^2 (1-27 z) \left(1+27 z^2\right)\theta ^4 
\\&\hskip15pt-36  z (5-9z) \left(7-15z+621z^{2}-729z^{3}\right)\theta ^3
\\&\hskip15pt-6  z \left(59049 z^4-91935z^3+39591 z^2-541 z+180\right)\theta ^2
\\&\hskip15pt-6   z \left(39366 z^4-64233 z^3+34155 z^2-155z+75\right)\theta
\\&\hskip15pt-3 z \left(19683 z^4-32562 z^3+21060 z^2-30 z+25\right)\big]F_{17}(z)\=0~,\\[10pt]
\text{AESZ290:}&\hskip10pt\big[(1+27x)(1+405x)^{2}\left(1+19683x^{2}\right)\theta^{4}
\\&\hskip15pt-108x(1+405x)\left(7-729x-177147x^{2}-7971615x^{3}\right)\theta^{3}
\\&\hskip15pt-6x\left(80-37017x-8155323x^{2}-1506635235x^{3}-87169610025x^{4}\right)\theta^{2}
\\&\hskip15pt-6x\left(17-17415x-3720087x^{2}-789189885x^{3}-58113073350x^{4}\right)\theta
\\&\hskip15pt-9x\left(1-1998x-454896x^{2}-111602610x^{3}-9685512225x^{4}\right)\big]F_{290}(x)\=0~.
\end{aligned}
\end{equation}
The geometric picture is markedly different to the pair AESZ22/118, in that $q$-expansions about $x=0$ do not return the BPS invariants of a manifold because there is no mirror manifold. Indeed, we will see that any putative mirror should have $Y_{111}=-30/13$, which defies an interpretation as a triple intersection number. 

The MUM point $z=0$ is mirror the the large volume point of a one-parameter family of threefolds constructed as the $\IZ_{3}$ quotient of the split bicubic:
\begin{equation}\label{eq:Z3Quotient}
\cicy{\IP^{2}\\\IP^{2}\\\IP^{2}}{\;1\; 1\; 1\;\;\\\;1\; 1\; 1\;\;\\\;1\; 1\; 1\;\;}_{/\IZ_{3}}
\end{equation}

The above quotient geometry has Euler characteristic $\chi=-30$. Since $h^{1,1}=1$, we have that $h^{2,1}=16$ for the above space. AESZ17 is the Picard Fuchs operator for a mirror $X$ to \eqref{eq:Z3Quotient}. From the Hodge diamond reflection, we have that $h^{1,1}(X)=16$ and $=h^{2,1}(X)=1$. If the mum point $x=0$ (which is $z=\infty$) in the complex structure moduli space of $X$ has a mirror $Y$, then $Y$'s Hodge numbers must read $h^{1,1}(Y)=1~,h^{2,1}(Y)=16$. That is to say, $Y$ should have the same Hodge numbers as \eqref{eq:Z3Quotient}. As a consequence, $Y$ and \eqref{eq:Z3Quotient} must have the same Euler characteristic. 

Now we consider monodromies \cite{vanStratenMonodromy} for the operator AESZ290. This has conifold singularities at $x=-1/27,\,x=\pm\frac{\sqrt{3}\ii}{343}$. As usual, since the operator's local exponents at $x=0$ are $(0,0,0,0)$ we can construct a Frobenius basis 
\begin{equation}
\begin{aligned}
\varpi_{0}(x)&=f_{0}(x)~,\\
\varpi_{1}(x)&=\frac{1}{2\pi\ii}\left(f_{0}(x)\log(x)+f_{1}(x)\right)~,\\
\varpi_{2}(x)&=\frac{Y_{111}}{2(2\pi\ii)^{2}}\left(f_{0}(x)\log(x)^{2}+2f_{1}(x)\log(x)+f_{2}(x)\right)~,\\
\varpi_{3}(x)&=\frac{Y_{111}}{6(2\pi\ii)^{3}}\left(f_{0}(x)\log(x)^{3}+3f_{1}(x)\log(x)^{2}+3f_{2}(x)\log(x)+f_{3}(x)\right)~,
\end{aligned}
\end{equation}
where all of the $f_{i}$ are power series, with $f_{0}(0)=1$ and $f_{1}(0)=f_{2}(0)=f_{3}(0)=0$. Now we attempt to construct an integral symplectic period vector
\begin{equation}
\Pi=\begin{pmatrix}
-\frac{\chi(Y)\zeta(3)}{(2\pi\ii)^{3}}&-\frac{1}{2}Y_{100}&0&1\\
-\frac{1}{2}Y_{100}&-Y_{110}&-1&0\\
1&0&0&0\\
0&1&0&0
\end{pmatrix}\varpi~.
\end{equation}
The task at hand then is to find real values of $Y_{111},\,Y_{100}$ and $Y_{110}$ so that the monodromy matrices about each conifold are integral. Note that we have fixed the value $\chi(Y)=30$ by our earlier argument. In practice we compute monodromy matrices for a modified version of the Frobenius vector that does not include the factors of $Y_{111}$, and then restore these symbolically with a diagonal change of basis.

It is instructive to check the monodromy matrix that we compute at $x=\frac{\ii\sqrt{3}}{343}$ for the above $\Pi$. The $(1,1)$ component is
\begin{equation}
-\frac{7}{8}+\frac{9Y_{100}}{Y_{111}}+\frac{9\ii\left(30+13Y_{111}\right)\zeta(3)}{4Y_{111}\pi^{3}}~.
\end{equation}
In order for this to be real, we must choose $Y_{111}$ so that the imaginary part vanishes. So we learn that the triple intersection number of our manifold $Y$ should be 
\begin{equation}
Y_{111}=-\frac{30}{13}~.
\end{equation}
On these grounds, we dispense with the idea that there is a mirror manifold at $x=0$. We cannot by any choice of $Y_{100}$ and $Y_{110}$ render the conifold matrix integral, so we press on without fixing any value for these.

\subsection*{A rational rank-two attractor}
Here we are able to identify a rank-two attractor at $z=-1$, which is $x=\frac{1}{3^{6}}$, with the corresponding weight-four modular form having LMFDB label 14.4.a.b . Mellin transforming this gives the critical L-values
\begin{equation}
\begin{aligned}
L(1)&\=0.81476235013261396706625\,...~,\\[5pt]
L(2)&\=1.13620338571911095621858\,...~.
\end{aligned}
\end{equation}
\subsubsection*{AESZ17}
The topological invariants are
\begin{equation}
Y_{111}\=30~,\qquad Y_{110}\=0~,\qquad Y_{100}\=-3~,\qquad\chi\=30~.
\end{equation}
The Frobenius periods about $z=0$ can be evaluated at $z=-1$, giving
\begin{equation}
\begin{aligned}
\widehat\varpi_{0}&\=-14\,\frac{L(2)}{(2\pi\ii)^{2}}~,\\[5pt]
\widehat\varpi_{1}&\=-7\,\frac{L(2)}{(2\pi\ii)^{2}}-\frac{3}{4}\,\frac{L(1)}{2\pi\ii}~,\\[5pt]
\widehat\varpi_{2}&\=-42\,\frac{L(2)}{(2\pi\ii)^{2}}-\frac{45}{4}\,\frac{L(1)}{2\pi\ii}~,\\[5pt]
\widehat\varpi_{3}-\frac{\chi\,\zeta(3)}{(2\pi\ii)^{3}}\widehat\varpi_{0}&\=-\frac{7}{2}\,\frac{L(2)}{(2\pi\ii)^{2}}-\frac{19}{8}\frac{L(1)}{2\pi\ii}~.
\end{aligned}
\end{equation}
Independent charge vectors for which the attractor equations are solved are
\begin{equation}
Q_{1}\=\left(-\frac{4}{3},6,1,0\right)^{T}~,\qquad Q_{2}\=\left(\frac{14}{3},-15,0,1\right)~,
\end{equation}
while the orthogonality equations are solved for 
\begin{equation}
Q_{3}\=\left(6,-15,0,1\right)^{T}~,\qquad Q_{4}\=\left(-\frac{4}{3},\frac{14}{3},1,0\right)~.
\end{equation}
Using $Q_{3}$ one can find the perturbative solution
\begin{equation}
t_{0}\=\frac{1}{2}+\frac{\ii\sqrt{5}}{10}~.
\end{equation}
We obtain the sum
\begin{equation}
\frac{3\pi}{14}\frac{L(1)}{L(2)}\Peq\frac{1}{\sqrt{5}}-\frac{1}{\sqrt{15\pi^{3}}}\sum_{j=1}^{\infty}(-1)^{j}\sum_{\fp\in\text{pt}(j)}a_{\fp}N^{\fp}\left(\frac{j}{6\pi\sqrt{5}}\right)^{l(\fp)-1/2}K_{l(\fp)-1/2}\left(\frac{\pi j}{\sqrt{5}} \right)~.
\end{equation}
A diagonal Pad\'e approximant of order 36 provides agreement to 10 figures.
\subsubsection*{AESZ290}
We lack a geometric interpretation for a mirror at $x=0$, but proceed with the computation using the topological invariants 
\begin{equation}
Y_{111}\=-\frac{30}{13}~,\qquad Y_{110}\=Y_{110}~,\qquad Y_{100}\=Y_{100}~,\qquad\chi\=30~.
\end{equation}
The rank two attractor $z=-1$ maps to $x=3^{-6}$, and we find that at this point the Frobenius periods for AESZ290 evaluate to
\begin{equation}
\begin{aligned}
\widehat\varpi_{0}&\=\frac{9\sqrt{3}}{2}\frac{L(1)}{(2\pi)}~,\\[5pt]
\widehat\varpi_{1}&\=21\ii\sqrt{3}\frac{L(2)}{(2\pi)^{2}}~,\\[5pt]
\widehat\varpi_{2}&\=\frac{45\sqrt{3}}{8}\frac{L(1)}{2\pi}~,\\[5pt]
\widehat\varpi_{3}-\frac{\chi\,\zeta(3)}{(2\pi\ii)^{3}}\widehat\varpi_{0}&\=\frac{315\ii\sqrt{3}}{52}\frac{L(2)}{(2\pi)^{2}}~.
\end{aligned}
\end{equation}
The attractor equations are solved for the charge vectors
\begin{equation}
Q_{1}\=\left(\frac{15}{52}-\frac{Y_{100}}{2},\,-Y_{110},\,0,\,1\right)^{T}~,\qquad Q_{2}\=\left(0,\,-\frac{5}{4}-\frac{Y_{100}}{2},\,1,\,0\right)^{T}~.
\end{equation}
The orthogonality equations are solved for charges
\begin{equation}
Q_{3}\=\left(0,\,\frac{15}{52}-\frac{Y_{100}}{2},\,1,\,0\right)^{T}~,\qquad Q_{4}\=\left(-\frac{5}{4}-\frac{Y_{100}}{2},\,-Y_{110},\,0,1\right)^{T}~.
\end{equation}
The orthogonality equation for the charge vector $Q_{4}$ is solved at the perturbative level by
\begin{equation}
t_{0}=\frac{\ii}{2}\sqrt{\frac{13}{3}}~.
\end{equation}
From this information, we construct the sum
\begin{equation}
\frac{14}{3\pi}\frac{L(2)}{L(1)}\=\sqrt{\frac{13}{3}}-\sqrt{\frac{13}{15\pi^{3}}}\sum_{j=1}^{\infty}\sum_{\fp\in\text{pt}(j)}a_{\fp}N^{\fp}(-1)^{l(\fp)}\left(\frac{\sqrt{\frac{13}{3}}j}{10\pi}\right)^{l(\fp)-1/2}K_{l(\fp)-1/2}\left(\pi j\sqrt{\frac{13}{3}} \right)~.
\end{equation}
There is no Pad\'e resummation here, as this sum converges (by the same arguments as appeared in Appendix F of \cite{Candelas:2021mwz}). By computing  60 terms in the series, we find agreement to 50 figures. By taking two Shanks transformations \cite{BenderOrszag}, this agreement can be improved to 55 figures. Three Shanks transformations also yields 55 figures of agreement, and further Shanks transformations worsen the agreement.

%now enable appendix numbering format and include any appendices
\appendix
\chapter{Coordinates on complex structure moduli space}\label{sect:Appendix_coordinates}
\setlength\epigraphwidth{.9\textwidth}
\epigraph{Smith : Because of you, I'm no longer an Agent of this system. Because of you, I've\\\phantom{Smith : } changed. I'm unplugged. A new man, so to speak. Like you, apparently, free.\\
Neo \+\+\;: Congratulations.\\
Smith : Thank you. But, as you well know, appearances can be deceiving, which brings\\\phantom{Smith : } me back to the reason why we're here.}{Lana and Lilly Wachowski, \textit{The Matrix Reloaded}}
In this appendix, we explain our choice of coordinates for the complex structure moduli spaces of the mirrors of complete intersections. We follow the presentations of \cite{Hosono:1994ax,Hosono:1995bm}, whose choice of coordinates we use. This involves a combinatoric problem in toric geometry, a subject that we do not go into detail on, instead referring to the textbook \cite{CoxBook}. The construction of mirror manifolds via operations on polyhedrons that encode intersections in toric varieties goes back to work of Batyrev, Borisov, and Nill \cite{borisov1993mirror,Batyrev:1994pg,Batyrev:2023vaa,BatyrevNill1,BatyrevNill2}.

The presentation that we offer is solely given to define a choice of coordinates. We will not discuss the proofs underlying these methods.

As in the main body of the thesis, $Y$ will be a family of complete intersections. For this appendix we shall allow the ambient space to be a product of weighted projective spaces, which is a mild generalisation of the CICYs discussed in \sref{sect:Essen_CICYs}. So the data of $Y$ is
\begin{equation}
\cicy{\IW\IP^{n_{1}}_{\left(w_{1}^{(1)}~,\,...~,\,w_{n_{1}+1}^{(1)}\right)}\\\\\vdots\\\\\IW\IP^{n_{K}}_{\left(w_{1}^{(K)}~,\,...~,\,w_{n_{K}+1}^{(K)}\right)}}{d^{\,(1)}_{1}~,\,...~,\,d^{\,(1)}_{C}\\\\\vdots\\\\d^{\,(K)}_{1}~,\,...~,\,d^{\,(K)}_{C}}~.
\end{equation}
We assume, following \cite{Hosono:1994ax}, that the intersection $Y$ does not intersect with the singularities of the ambient variety $\cA=\IW\IP^{n_{1}}_{\left(w_{1}^{(1)}~,\,...~,\,w_{n_{1}+1}^{(1)}\right)}\times\,...\,\times\IW\IP^{n_{K}}_{\left(w_{1}^{(K)}~,\,...~,\,w_{n_{K}+1}^{(K)}\right)}$, and without loss of generality take $w_{n_{i}+1}^{(i)}=1$ for each $i$. We will also assume a favourability condition: $h^{1,1}=K$. We will display here formulae relevant for such intersections in products of weighted projective spaces. Results for CICYs in products of (unweighted) projective spaces, as in all our main lines of discussion in this thesis, can be obtained by setting all $w^{(k)}_{i}$ equal to 1.

Each weighted projective space $\IW\IP^{n_{i}}_{\bm{w}^{(i)}}$ is a toric variety, associated to a reflexive simplicial polyhedron $\Delta_{i}$. $\Delta_{i}$ has integer vertices in $\IR^{n_{i}}$. The ambient space $\cA$ is also a toric variety, and the associated polyhedron is 
\begin{equation}
\Delta=\Delta_{1}\times\,...\,\times\Delta_{K}\subset \IR^{n_{1}}\times\,...\,\times\IR^{n_{K}}~.
\end{equation}
The Batyrev-Borisov procedure produces a mirror $X$ to an intersection in a toric variety, like $Y$ above, from the data of the polyhedron $\Delta^{*}$, which is the polar dual of $\Delta$. We intentionally gloss over the specifics of $\Delta$ and its association to $\cA$, and skip to the presentation of the vertices $\nu_{i,j}^{*}$ of $\Delta^{*}$. For each fixed $i$, there are vertices $\nu_{i,j}^{*}$ that have components $0$ in each factor $\IR^{n_{j}}$ of $\IR^{n_{1}}\times\,...\,\times\IR^{n_{K}}$ with $j\neq i$. The components of $\nu_{i,j}^{*}$ in the factor $\IR^{n_{i}}$ are
\begin{equation}\begin{gathered}
\nu_{i,1}^{*}=(1,0,\,...\,,0)~,\qquad...~,\qquad\nu_{i,n_{i}}^{*}=(0,\,...\,,0,1)~,\\
\nu_{i,n_{i}+1}^{*}=\left(-w_{1}^{(1)}~,\,...~,\,-w_{n_{i}}^{(i)}\right)~.
\end{gathered}\end{equation}
Denote the set of all the vertices by $E$, and partition $E$ into distinct sets $E_{m}$, $1\leq m \leq C$. For each fixed $i$, each set $E_{m}$ contains $d^{\,(i)}_{m}$ vertices from the set $\nu^{*}_{i,j}$. 

Now for each fixed $m$, each vertex $\nu_{i,j}^{*}\in E_{m}$ is extended to 
\begin{equation}
\overline{\nu}_{i,j}^{*}=\left(e^{(m)},\nu_{i,j}^{*}\right)\in\IR^{C}\times\IR^{n_{1}}\times\,...\,\times\IR^{n_{K}}~.
\end{equation} 
By $e^{(m)}$ we mean the unit vector in the $m^{th}$ direction of $\IR^{C}$. We then form additional vectors
\begin{equation}
\overline{\nu}_{0,p}^{*}=\left(e^{(p)}~,\bm{0}\right)\in\IR^{C}\times\IR^{n_{1}}\times\,...\,\times\IR^{n_{K}}~.
\end{equation}
Importantly, it turns out that there are $K$ independent linear relations between the vectors $\overline{\nu}_{i,j}^{*}$,
\begin{equation}\label{eq:relations}
\sum_{i,j}l^{(s)}_{i,j}\,\overline{\nu}_{i,j}^{*}=\bm{0}~,\qquad 1\leq s\leq K~.
\end{equation}
The components of $l^{(s)}_{i,j}$ are
\begin{equation}\label{eq:relationcomponents}\begin{aligned}
l^{(s)}_{0,j}&=-d_{j}^{\,(s)}~,\quad &1\leq j\leq C~,&\quad\\[5pt]
l^{(s)}_{i,j}&=\delta_{i}^{\,(s)}w_{j}^{(s)}~,\quad & 1\leq i\leq K~,&\quad 1\leq j\leq n^{(s)}_{n_{s}+1}~,\\[5pt]
\delta_{i}^{(s)}&=\begin{cases}0&i\neq s~,\\1&i=s~.\end{cases}&&
\end{aligned}\end{equation}
The choice of overall sign, with minus signs in the components of $l^{(s)}_{0,j}$, is one of convention. We fix this choice in our work, so that MUM points are located at the origin. We remark that for the more general case of intersections in toric varieties, relations of the form \eqref{eq:relations} hold. It is only in our simplified case of complete intersections in weighted projective spaces that the particular form \eqref{eq:relationcomponents}, with the Kronecker delta, hold.

Let $\Delta_{i}^{*}$ denote the convex hull of $E_{i}$ and the origin. The mirror family of $Y$ consists of varieties $X$ birational to the vanishing loci of $C$ equations $P_{r}=0$ in the toric variety $\IP_{\Delta_{1}^{*}+\,...\,+\Delta_{C}^{*}}$, where $+$ denotes the Minkowski sum of polyhedra. The ambient space coordinates are $X_{m,n}$, with $1\leq m\leq K$ and $1\leq n \leq n_{m}$. Those equations read
\begin{equation}\label{eq:mirror_equation}
P_{r}\equiv a_{0r}-\sum_{\nu_{i,j}^{*}\in E_{r}}a_{i,j}X^{\nu_{i,j}^{*}}=0~.
\end{equation}
Let us clarify the notation here: $X^{\nu_{i,j}^{*}}$ means $\prod_{j=1}^{n_{i}}X_{i,j}^{\nu_{i,j}^{*}}$.

The set of $a_{ij}$ that appear in \eqref{eq:mirror_equation} outnumber the set of complex structure moduli of $X$. Coordinate redefinitions remove the redundancies, and these are encoded in the relations $l^{s}_{i,j}$. We work with coordinates
\begin{equation}\label{eq:varphi_defined}
\varphi^{s}=\frac{\prod_{j=1}^{n_{s}+1}a_{s,j}}{\prod_{j=1}^{C}a_{0,j}^{d_{j}^{\,(s)}}}\equiv a^{l^{(s)}}~.
\end{equation}
Inspecting this process, one sees that there is some superficial ambiguity in the choice of sets $E_{m}$ which can lead to different polynomials $P_{m}$. The set of coordinates obtained from \eqref{eq:varphi_defined} does not depend on this choice. The apparent ambiguity in the polynomials is removed when these varieties are desingularised, but we do not discuss that in this appendix.

\chapter{Zeta function factorisation counts}\label{sect:Appendix_Factorisations}
\setlength\epigraphwidth{.4\textwidth}
\epigraph{We're not banging rocks together here.}{Cave Johnson}
This appendix contains bar charts that display for each labelled operator, with $p$ on the $x$-axis, the number of times that the zeta function $\zeta_{p}(\varphi;T)$ factorises as $\varphi$ runs over $\IF_{p}$. Our tables are for primes $5\leq p \leq131$.

It should be understood that this list is not exhaustive: other operators have been studied elsewhere (see \cite{Candelas:2019llw,Bonisch:2022mgw,Bonisch:2022slo}), and many more remain uninvestigated. We depend on the digital database \cite{AESZ-db}. The computation that obtains these factorisation counts needs the topological numbers $Y_{111}$ and $\chi$, which are not given for every operator in \citep{AESZ-db}. Further, a number of the operators listed in \cite{AESZ-db} are typeset in such a way that posed a problem for our code that recovered the operators from the website. Rather than fix these surmountable problems, we omit those examples. We also did not make an effort to work through the whole set for which there was sufficient data available for our purposes. This choice of 61 is a matter of simplistic practicality and does not belie anything interesting.

These tables were generated using the methods of \cite{Candelas:2021tqt}, and are available in electronic form \cite{mcgovern2023a}.\\[20pt]

\includegraphics[scale=0.4]{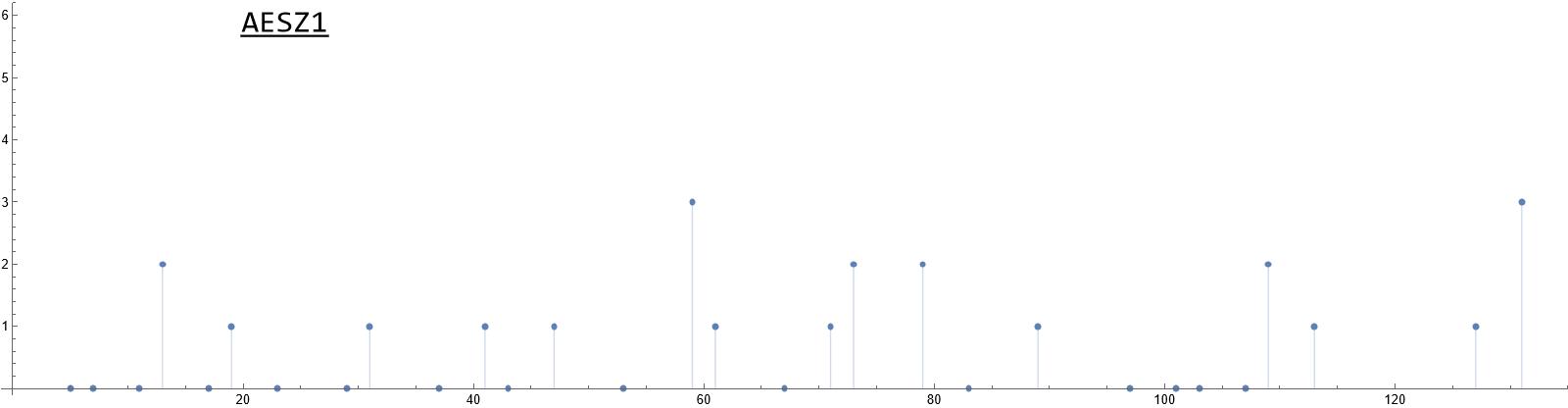}
\includegraphics[scale=0.4]{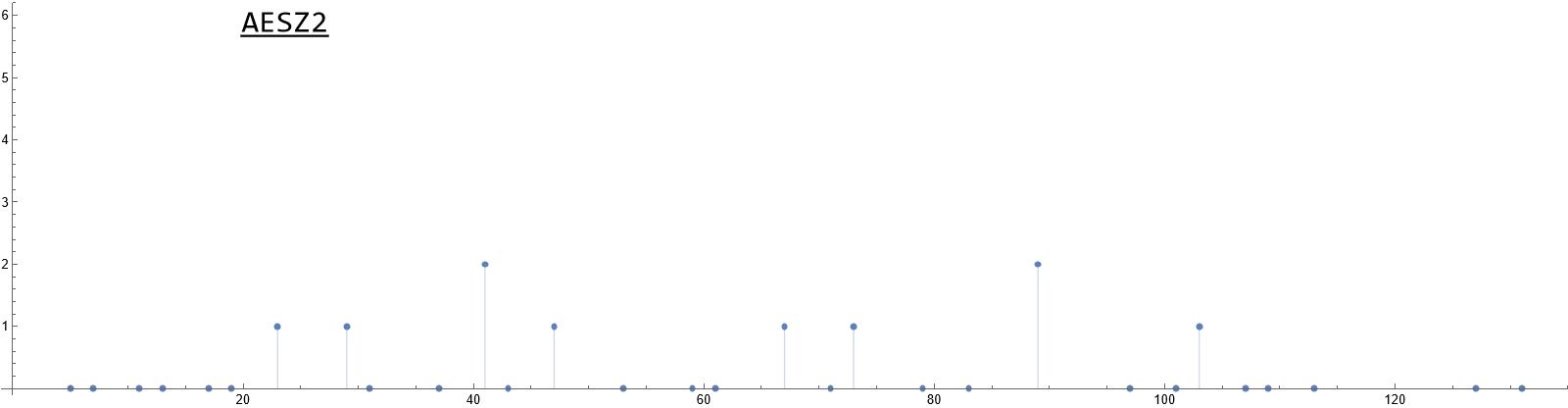}
\includegraphics[scale=0.4]{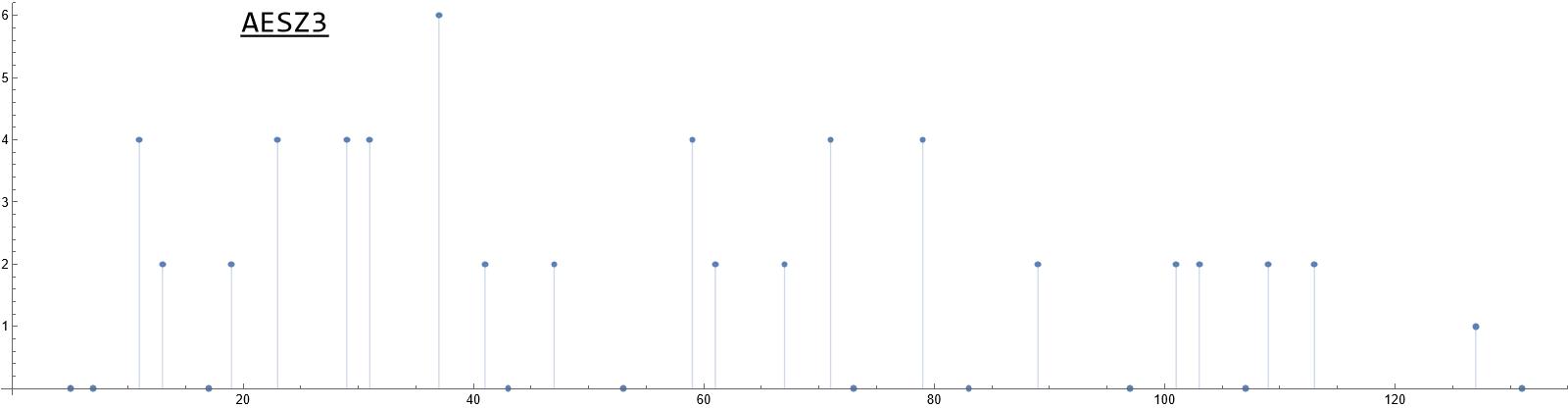}
\includegraphics[scale=0.4]{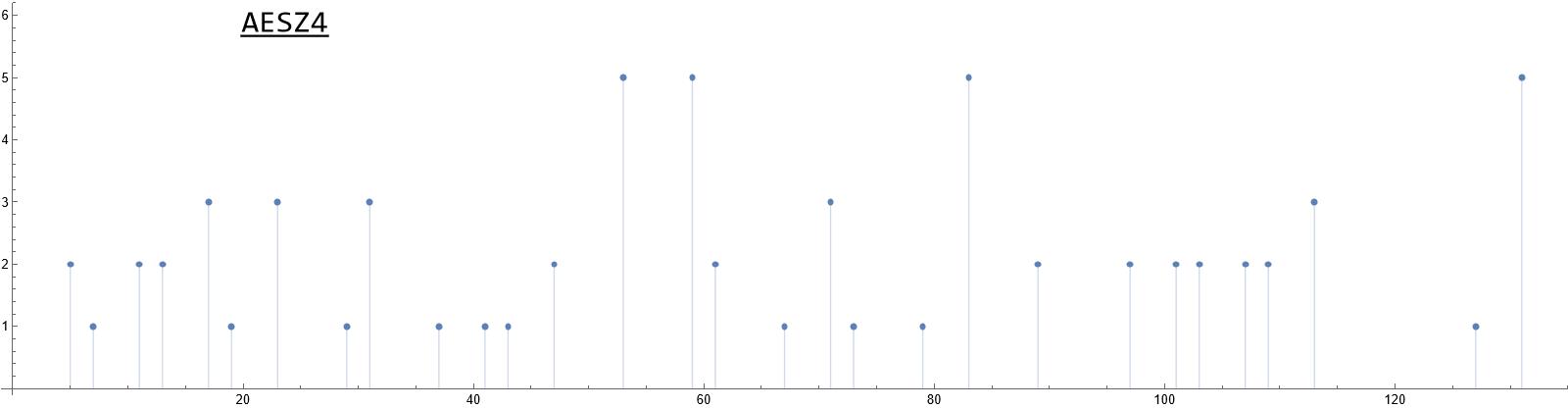}
AESZ4 was studied in \cite{Bonisch:2022mgw,Bonisch:2022slo} and is known to possess a rational rank two attractor.\\[20pt]
\includegraphics[scale=0.4]{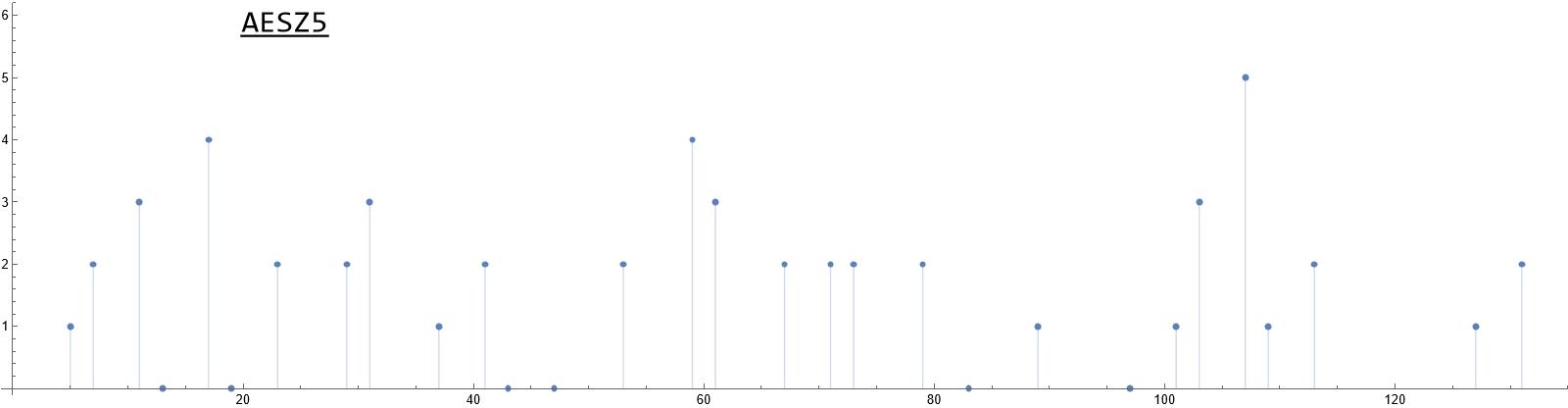}
\includegraphics[scale=0.4]{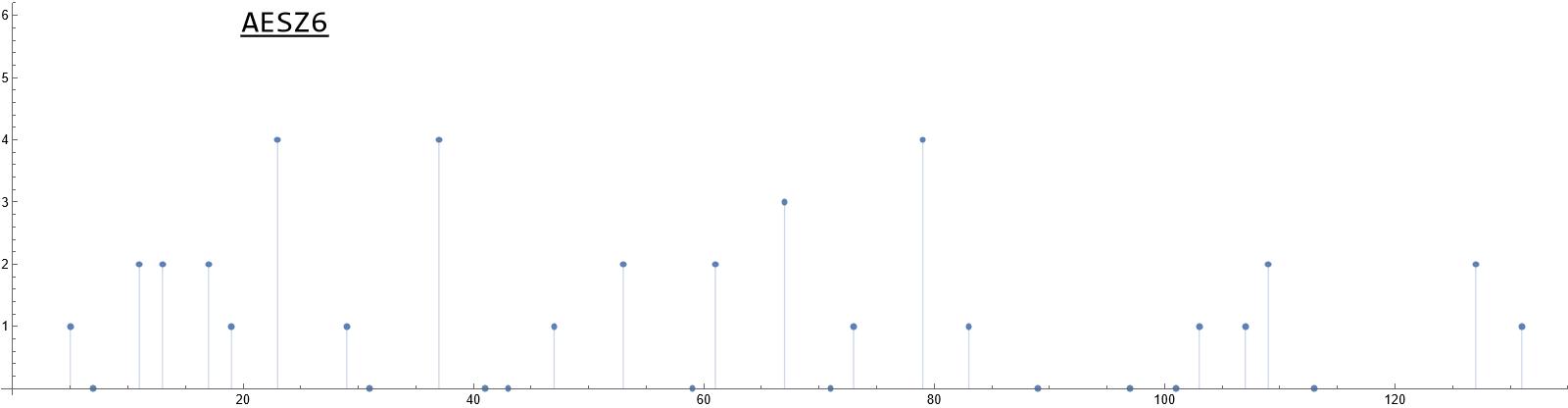}
\includegraphics[scale=0.4]{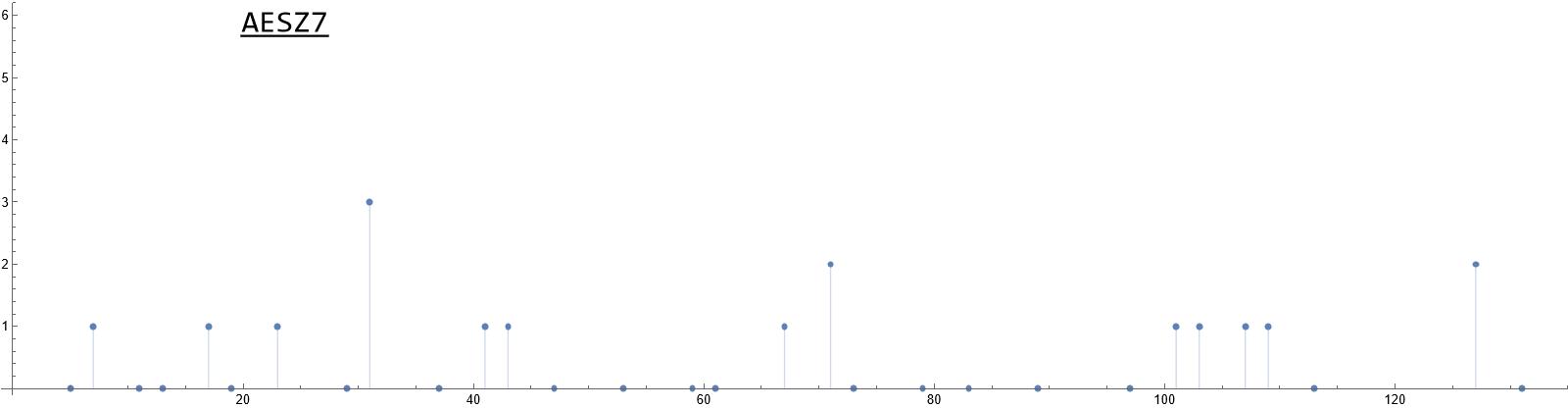}
\includegraphics[scale=0.4]{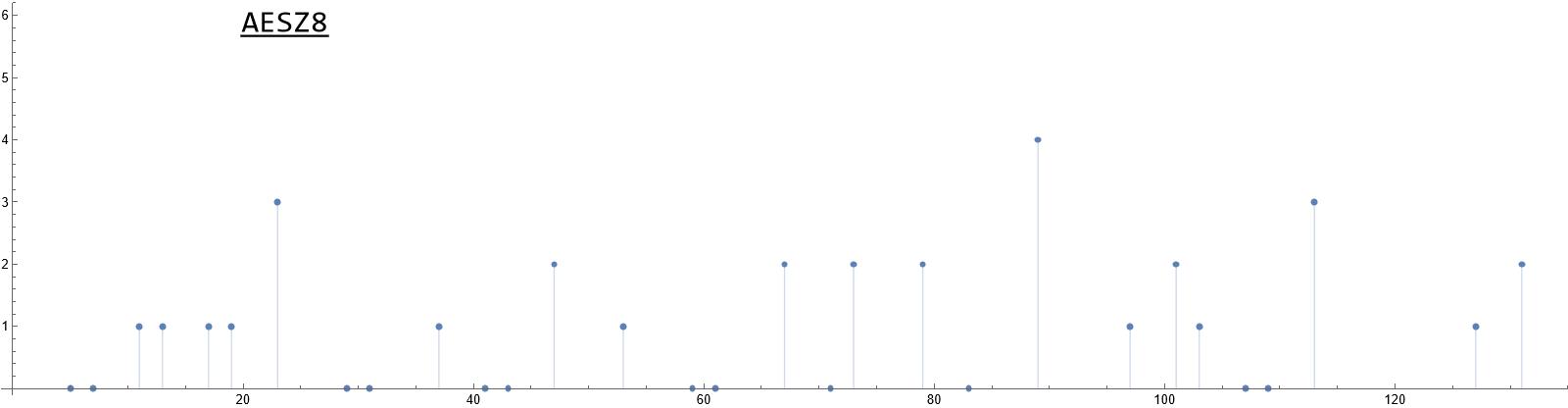}
\includegraphics[scale=0.4]{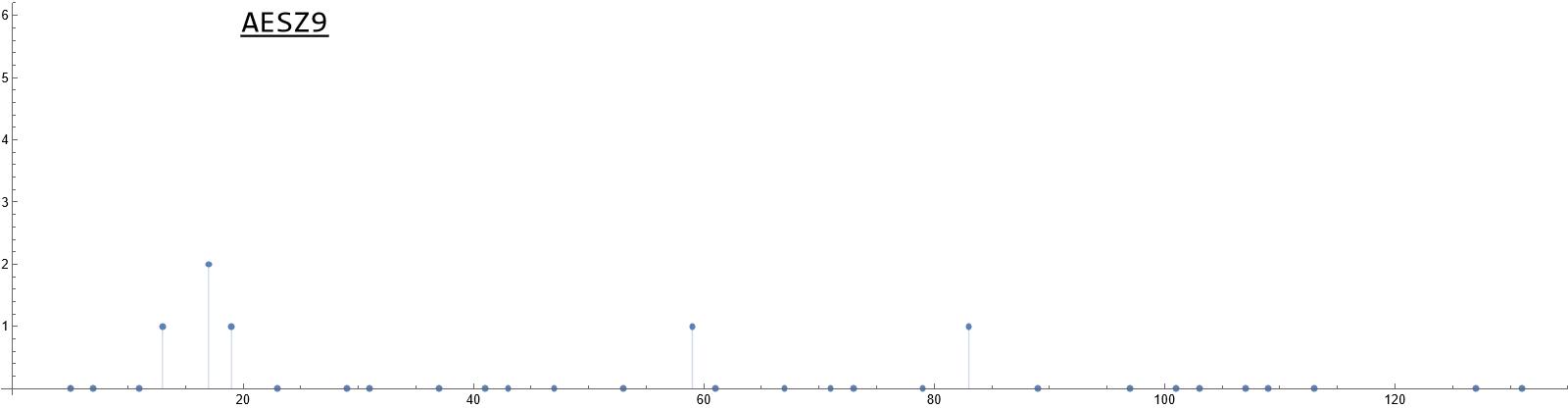}
\includegraphics[scale=0.4]{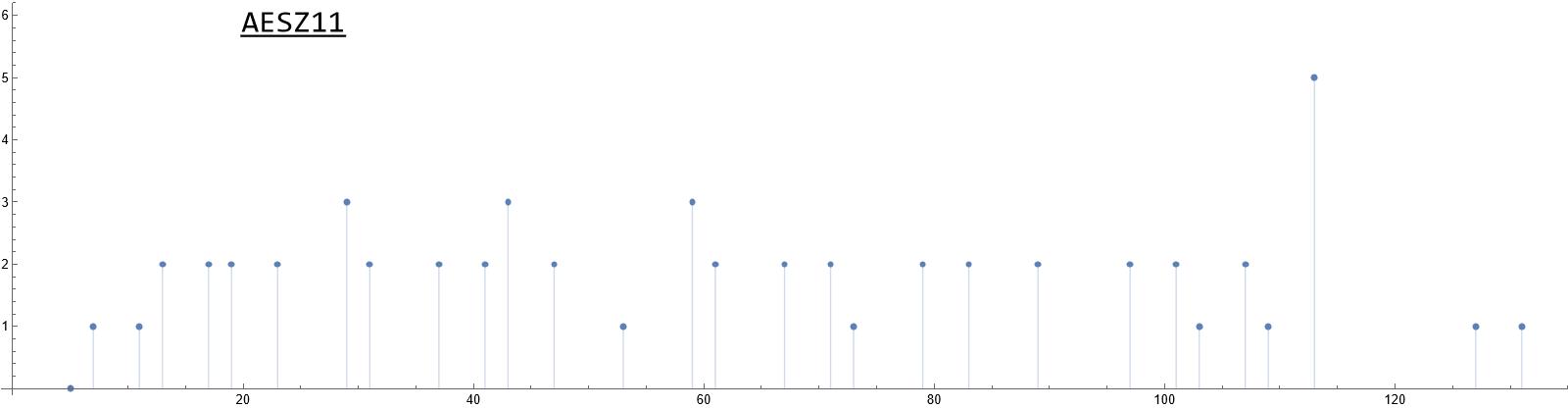}
AESZ11 was studied in \cite{Bonisch:2022mgw,Bonisch:2022slo} and is known to possess a rational rank two attractor.\\[20pt]
\includegraphics[scale=0.4]{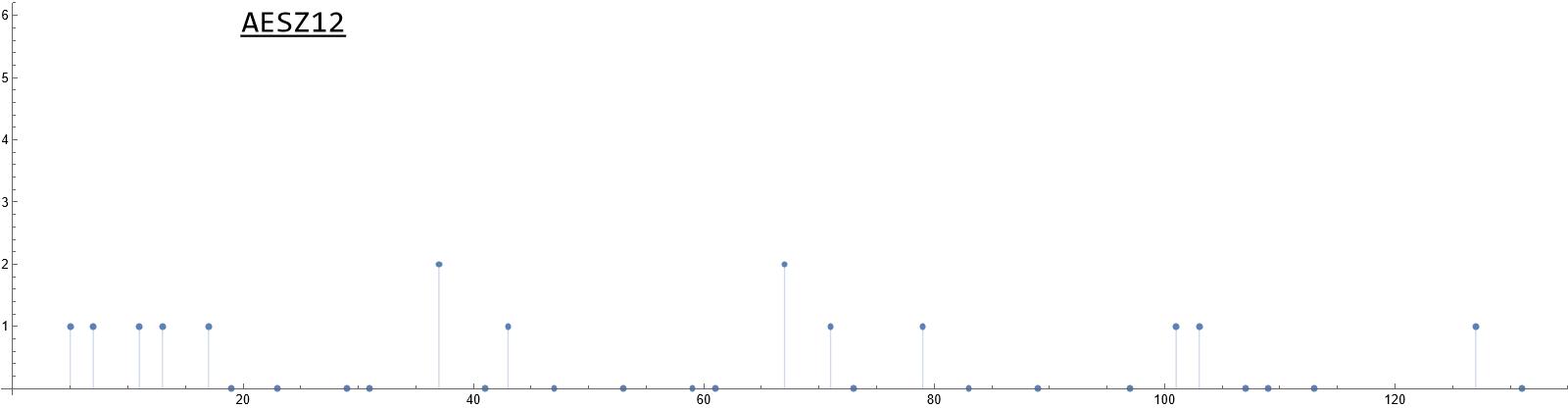}
\includegraphics[scale=0.4]{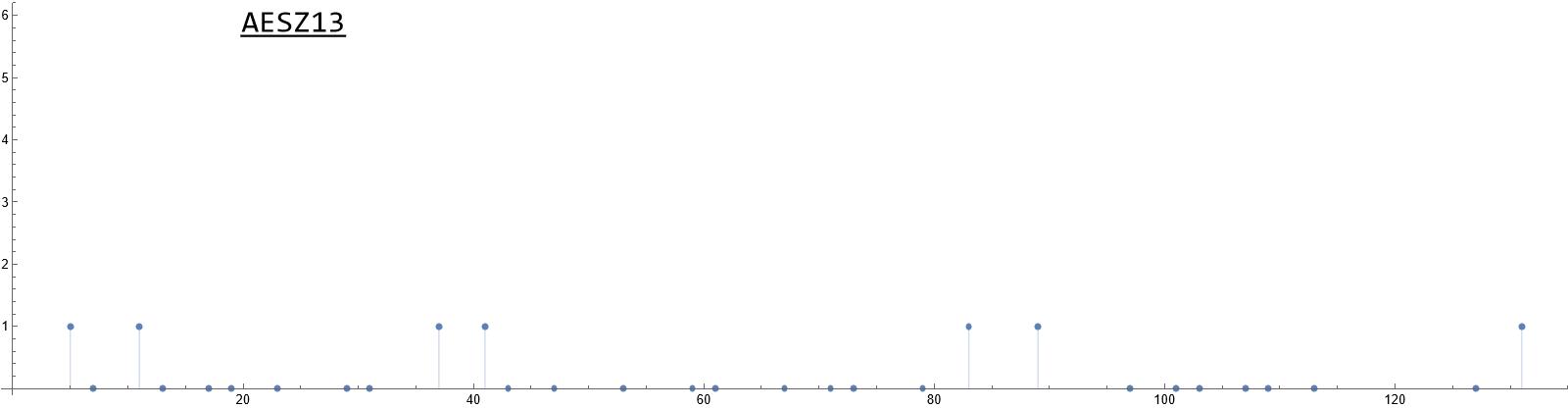}
\includegraphics[scale=0.4]{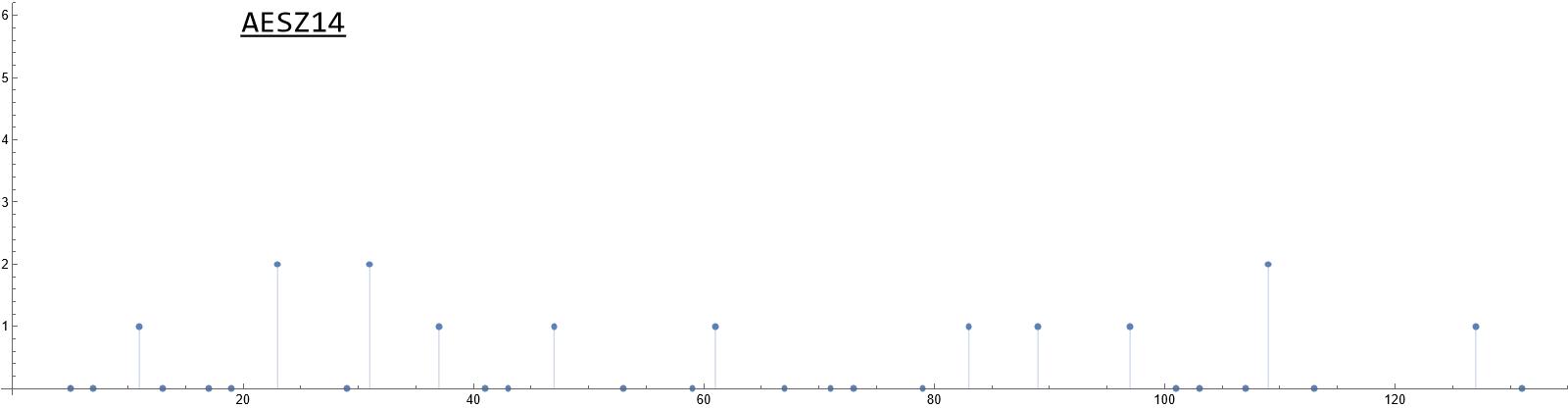}
\includegraphics[scale=0.4]{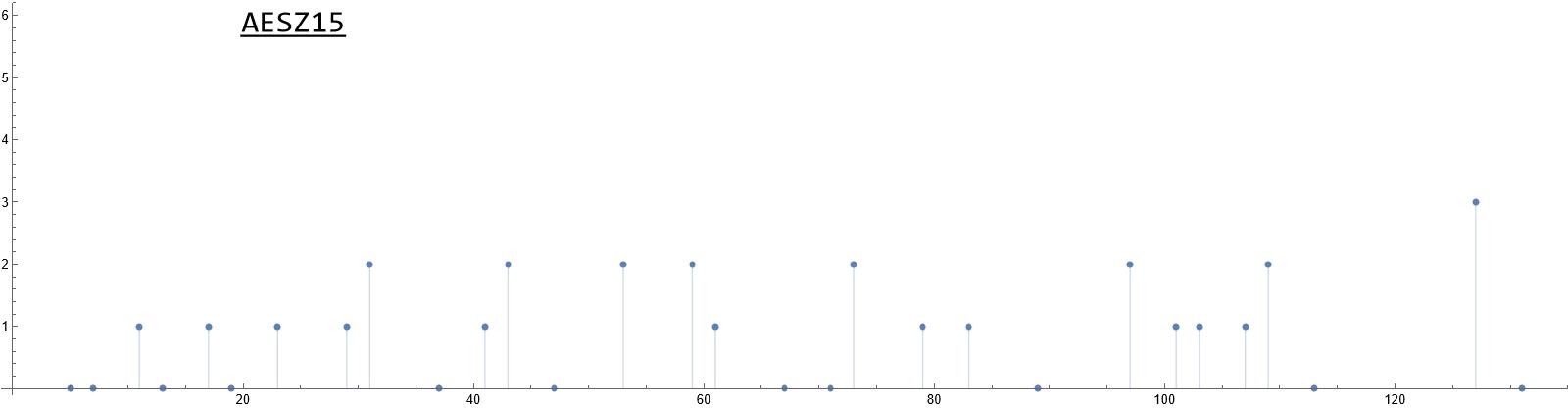}
\includegraphics[scale=0.4]{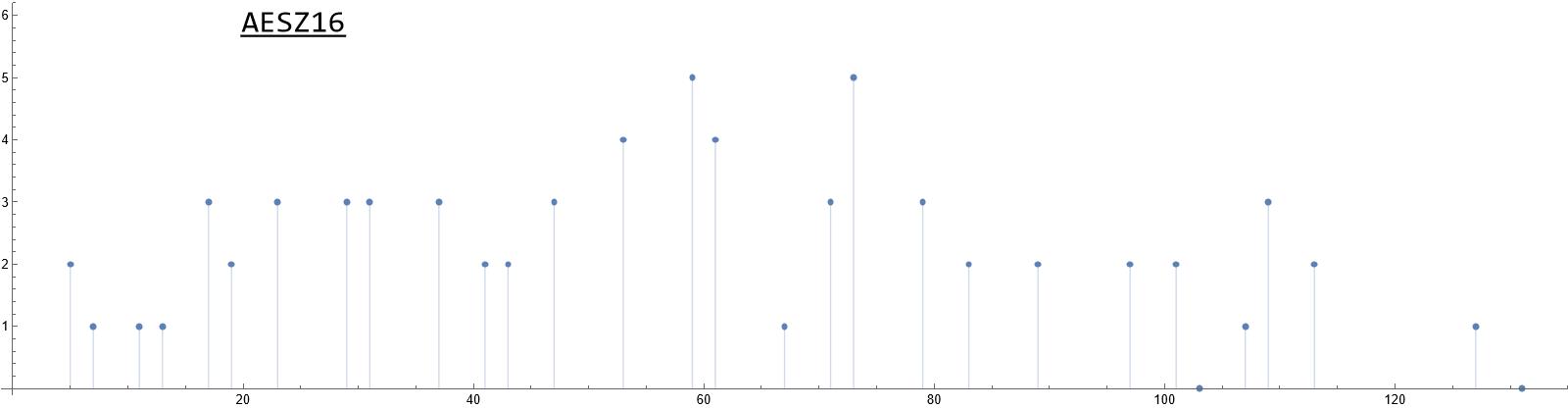}
\includegraphics[scale=0.4]{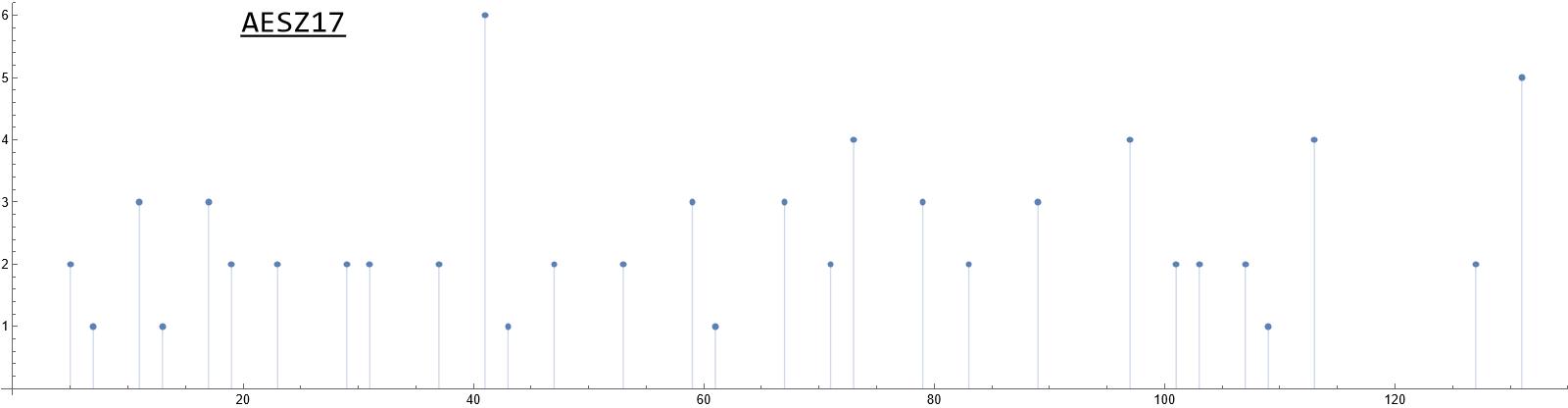}
The existence of a rational rank two attractor for AESZ17 is a new result of this thesis.\\[20pt]
\includegraphics[scale=0.4]{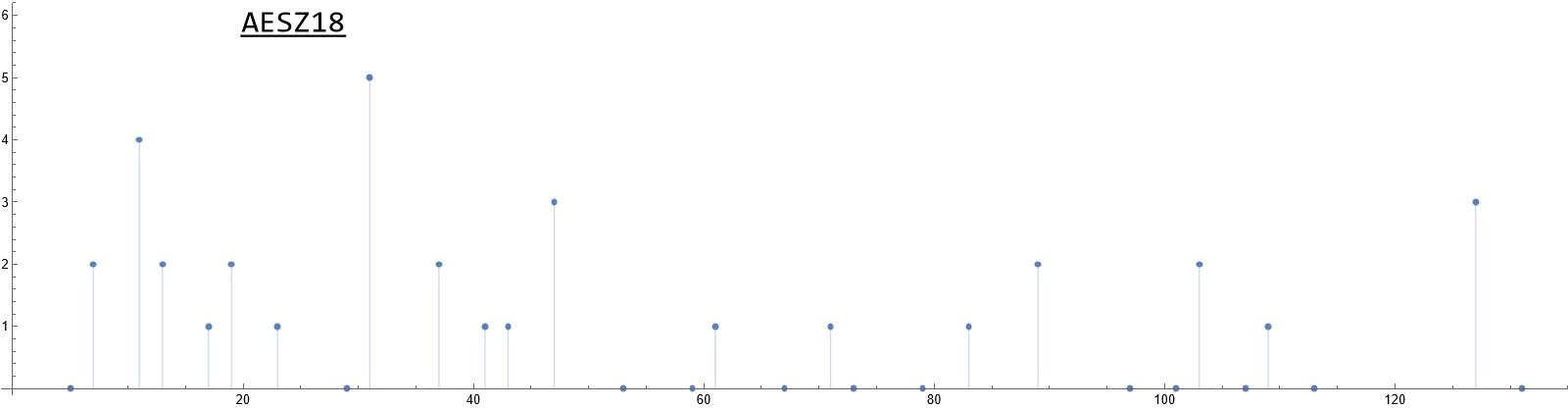}
\includegraphics[scale=0.4]{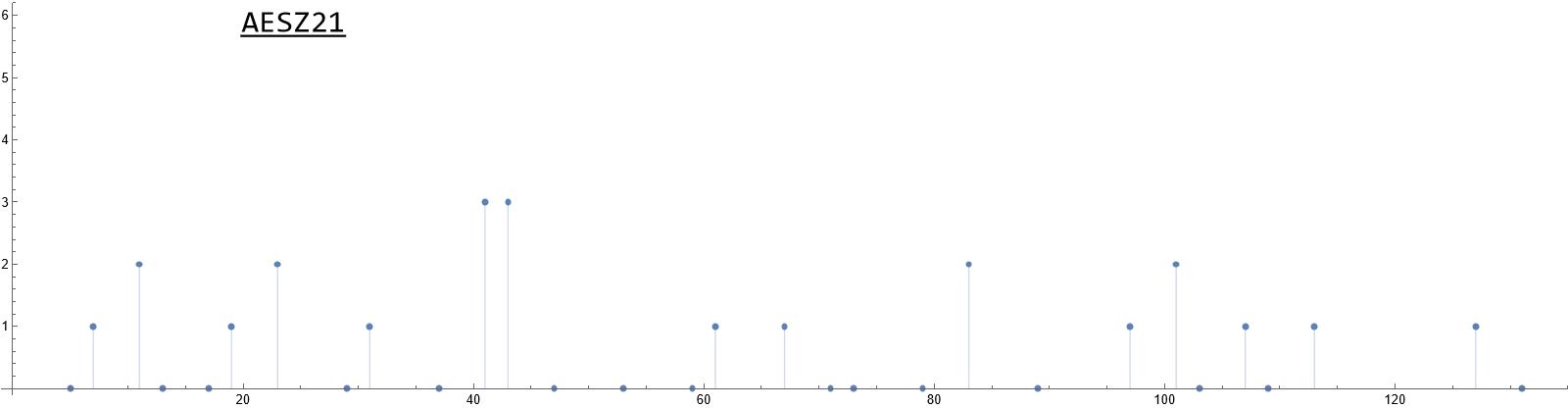}
\includegraphics[scale=0.4]{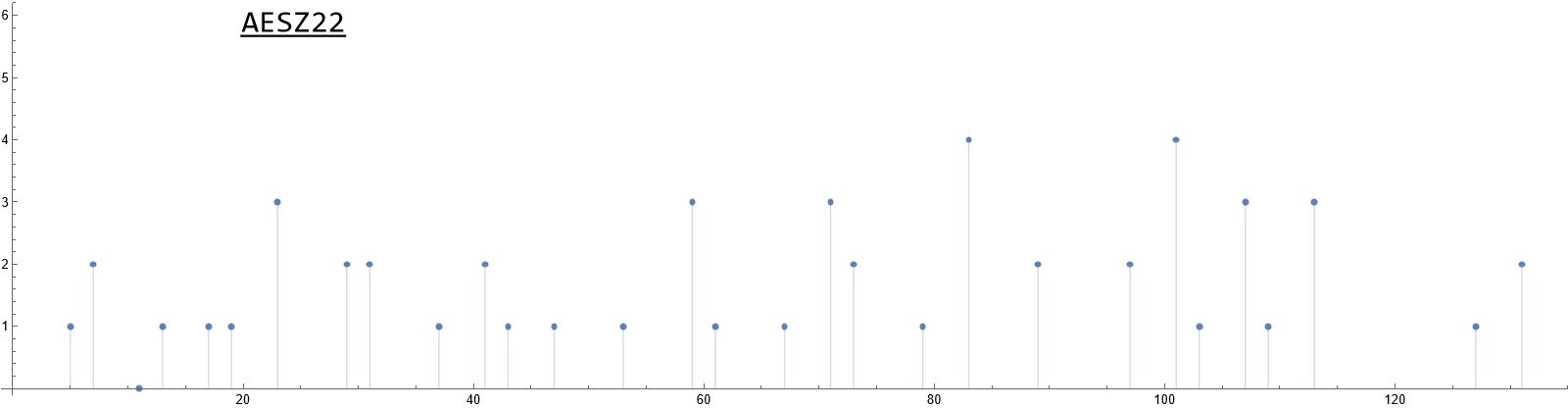}
The existence of a rational rank two attractor for AESZ22 is a new result of this thesis.\\[20pt]
\includegraphics[scale=0.4]{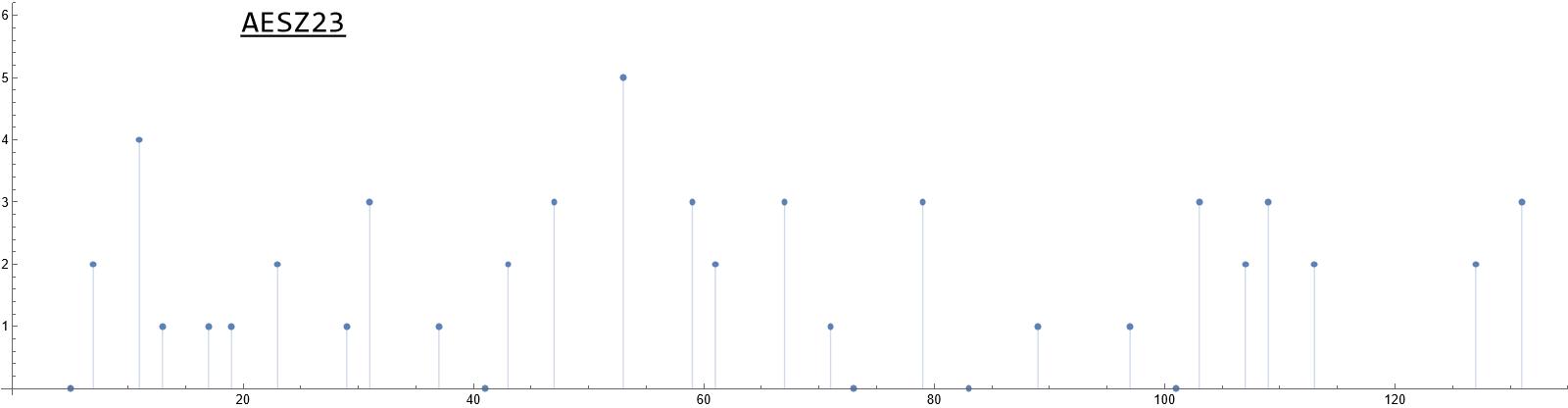}
\includegraphics[scale=0.4]{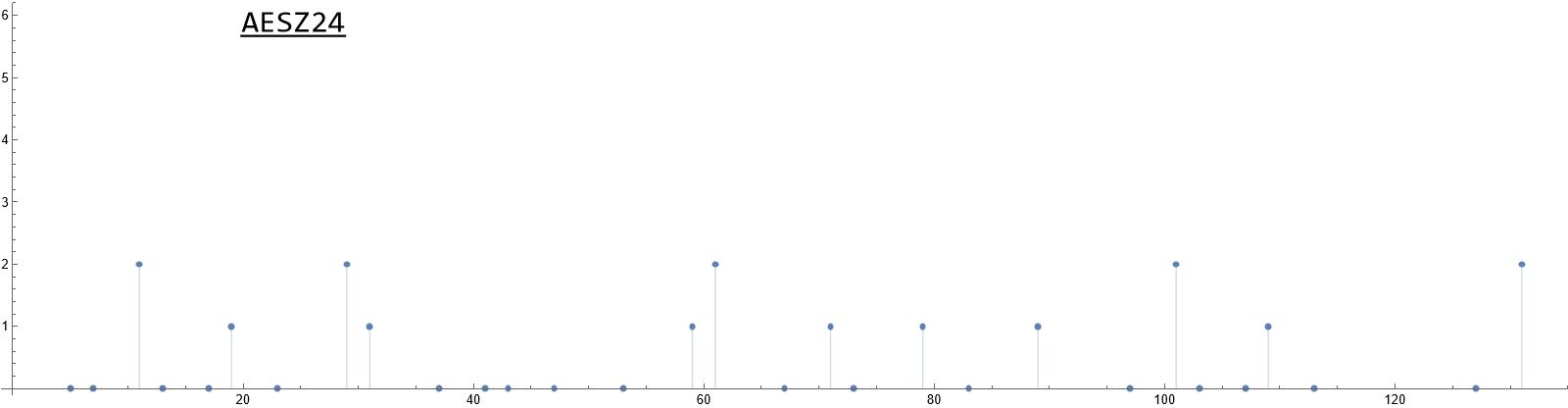}
\includegraphics[scale=0.4]{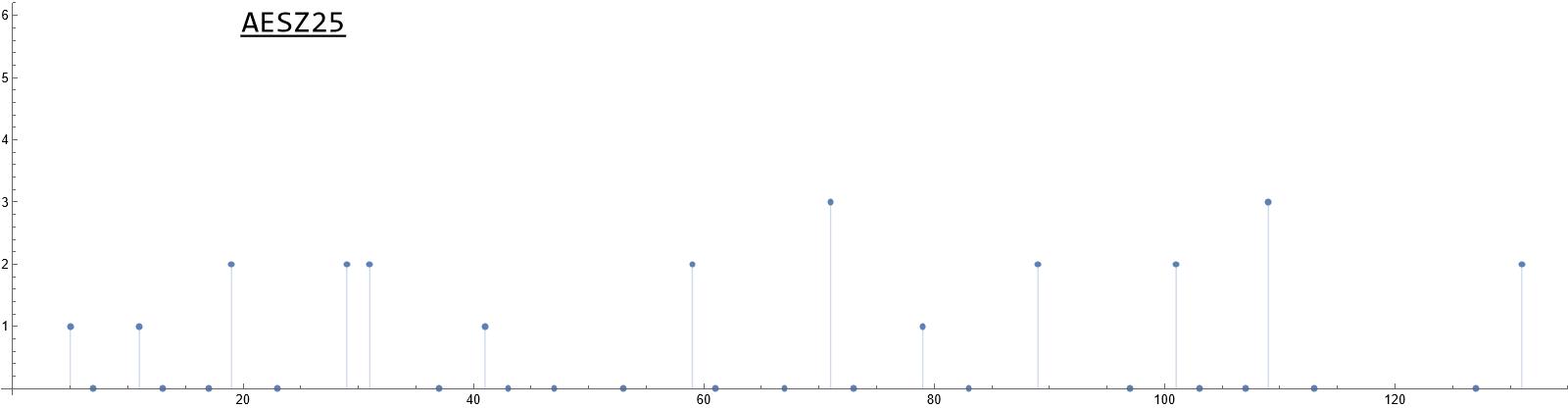}
\includegraphics[scale=0.4]{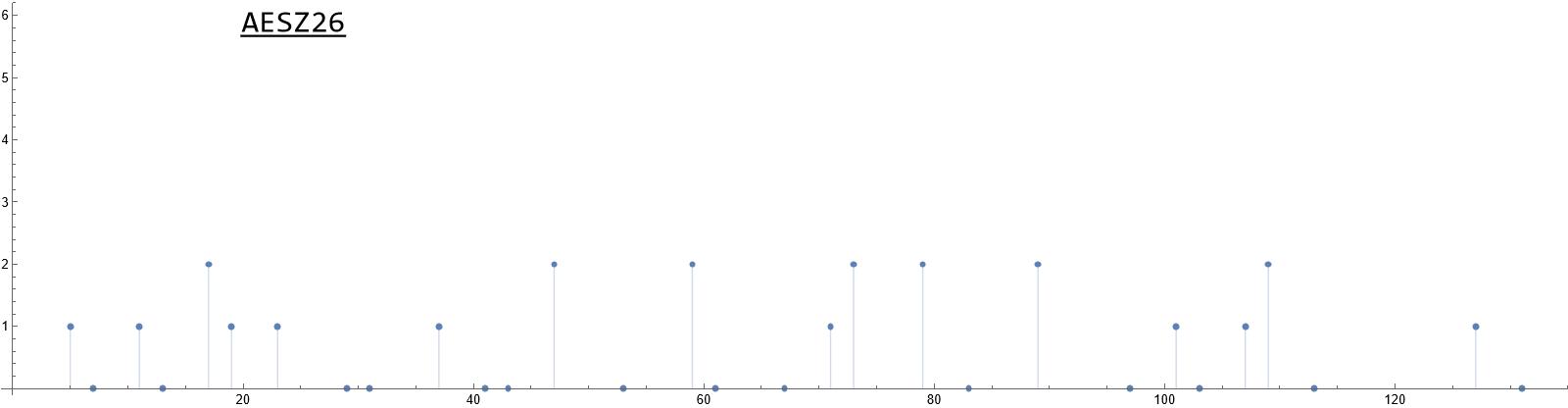}
\includegraphics[scale=0.4]{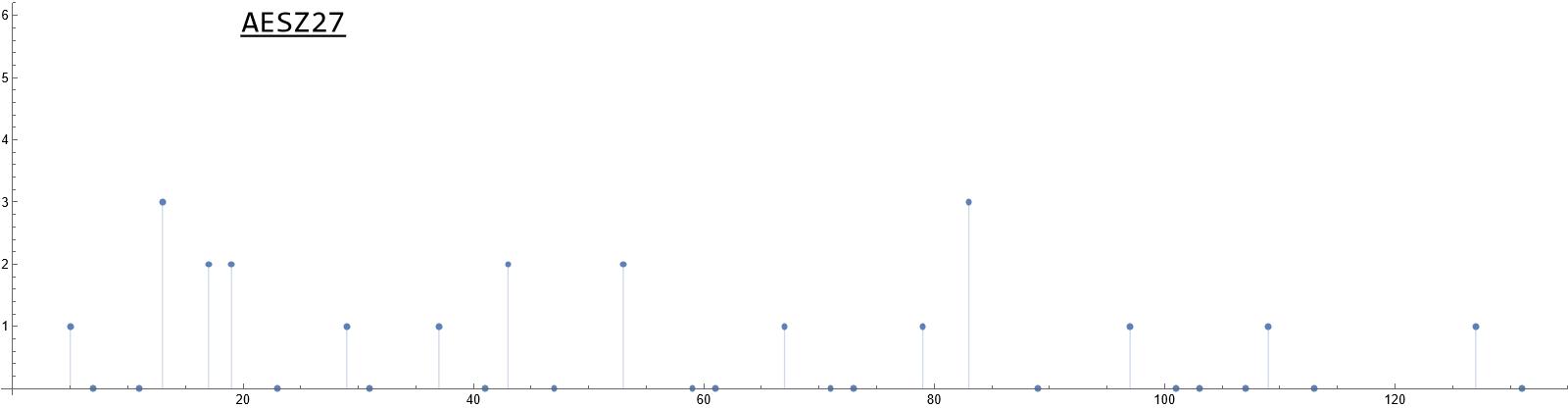}
\includegraphics[scale=0.4]{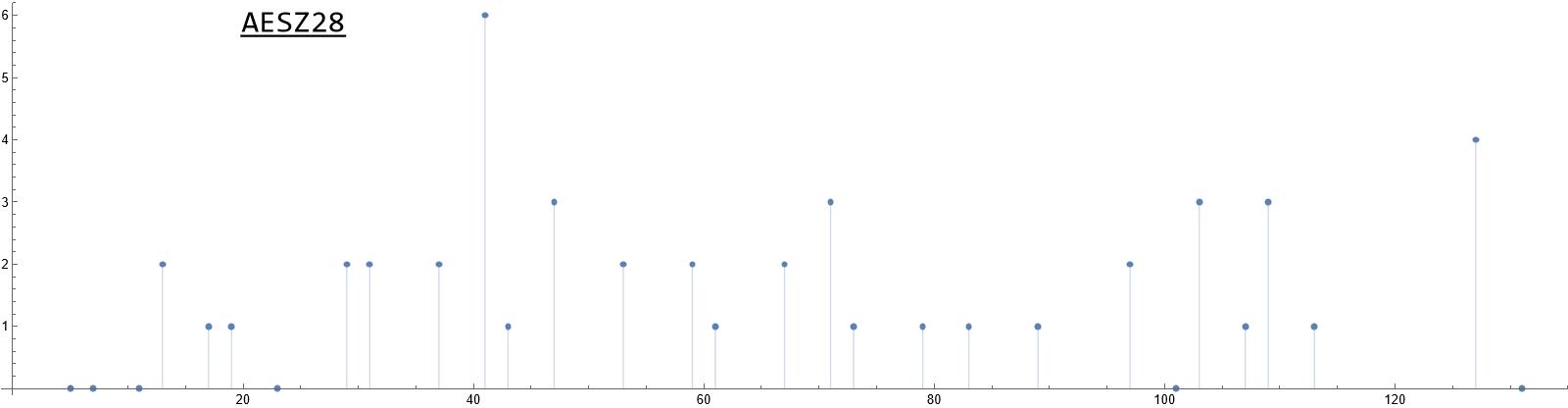}
\includegraphics[scale=0.4]{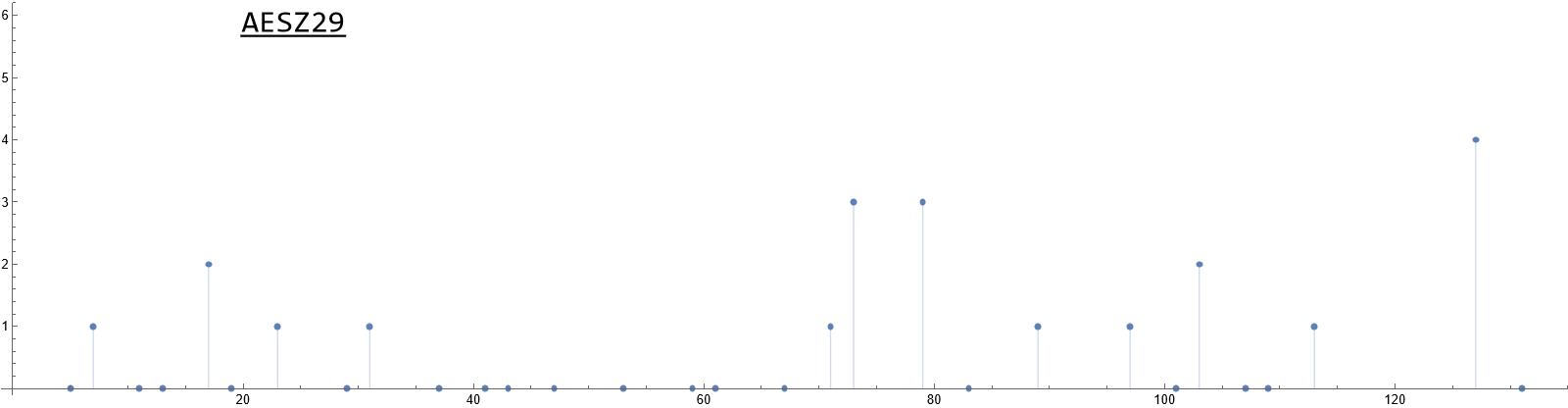}
\includegraphics[scale=0.4]{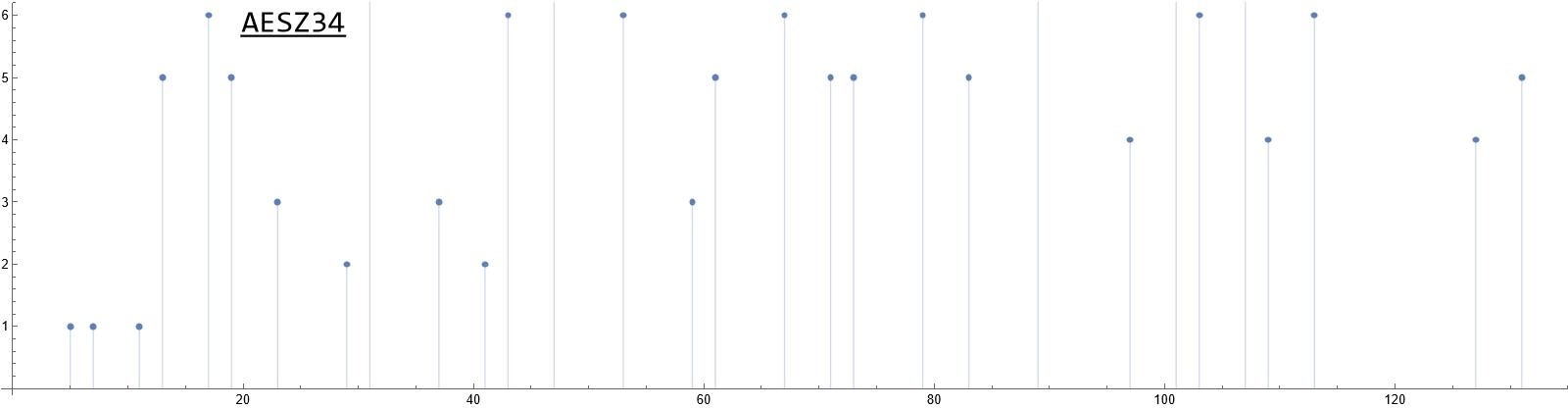}
AESZ34 was studied in \cite{Candelas:2019llw}. It is known to possess one rational and two quadratic rank two attractors.\\[20pt]
\includegraphics[scale=0.4]{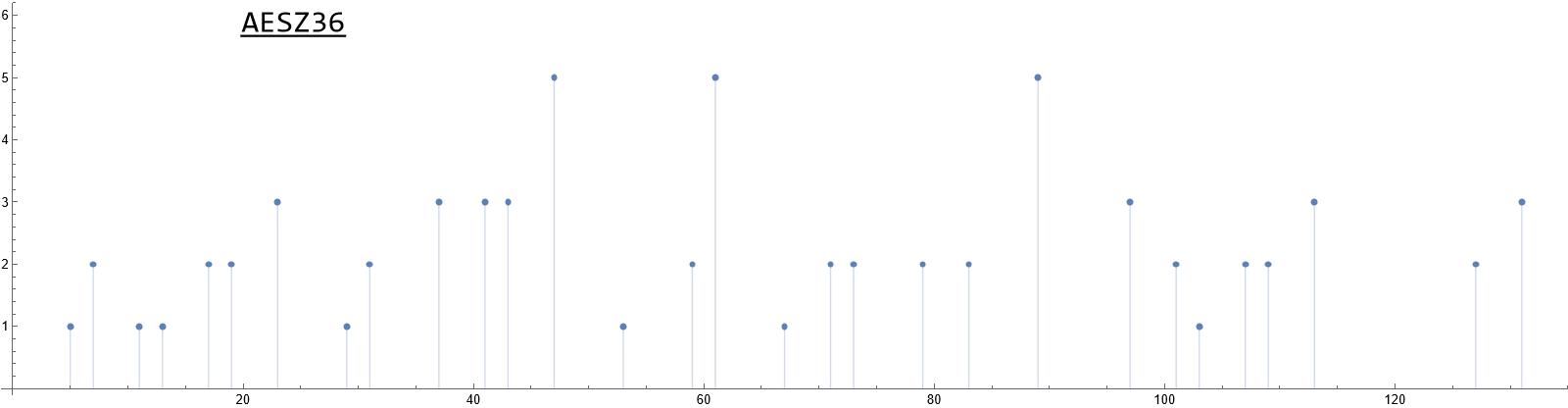}
AESZ36 was studied in \cite{Bonisch:2022slo} and is known to possess a rational rank two attractor.\\[20pt]
\includegraphics[scale=0.4]{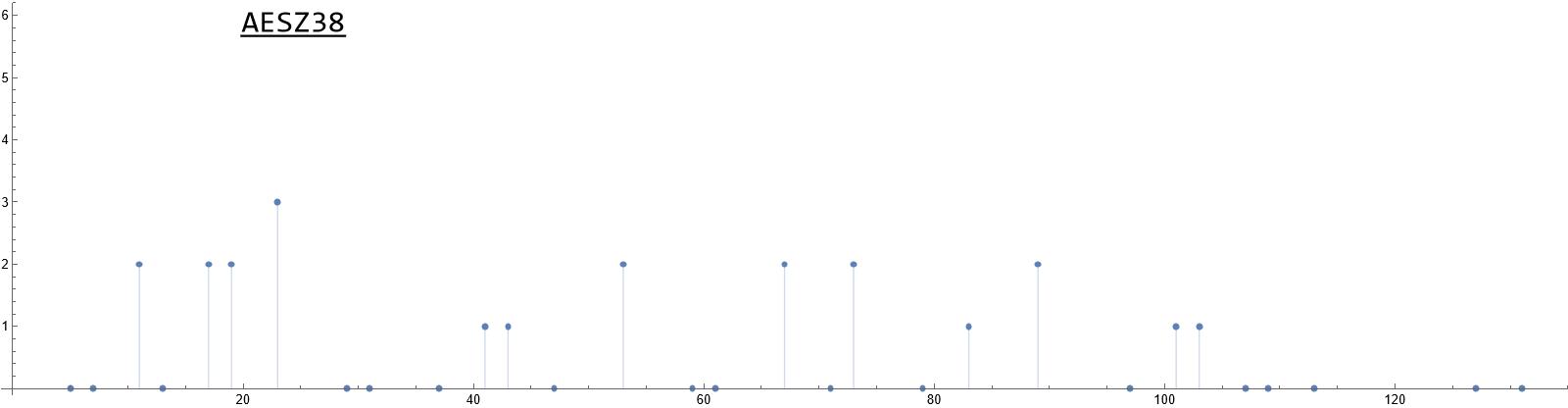}
\includegraphics[scale=0.4]{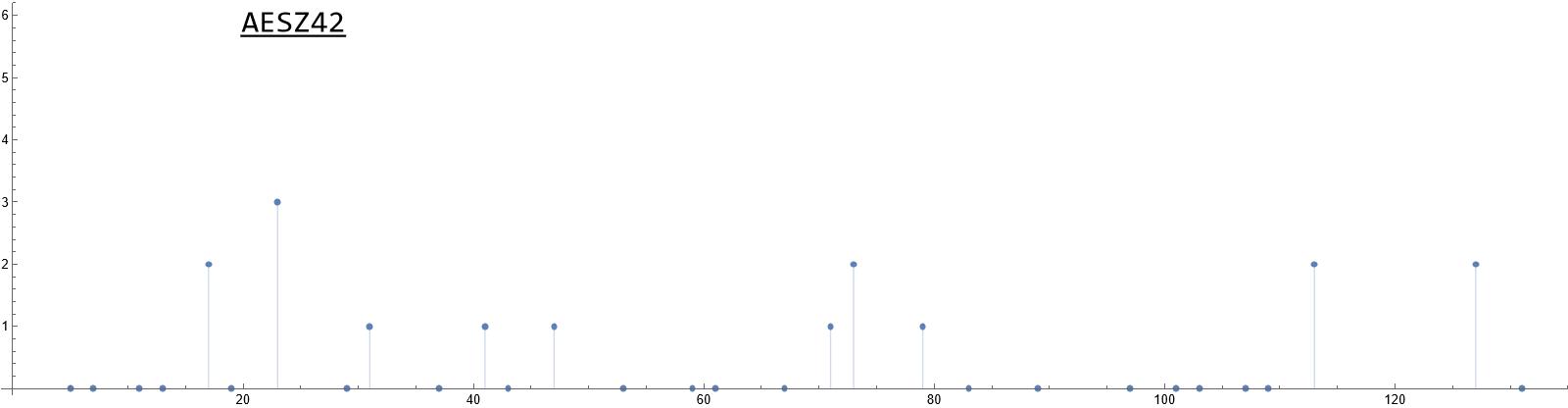}
\includegraphics[scale=0.4]{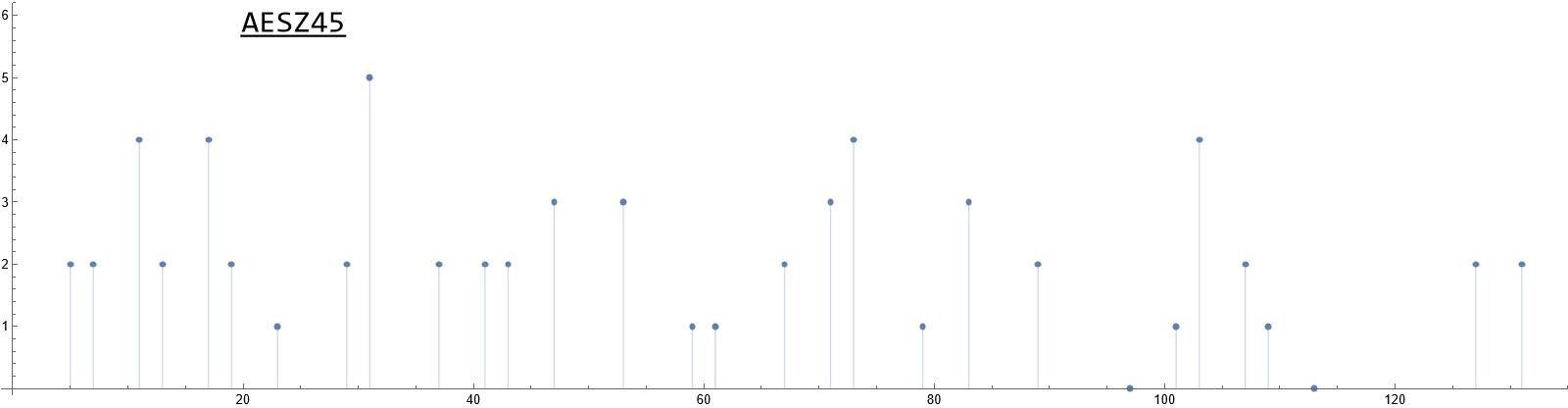}
\includegraphics[scale=0.4]{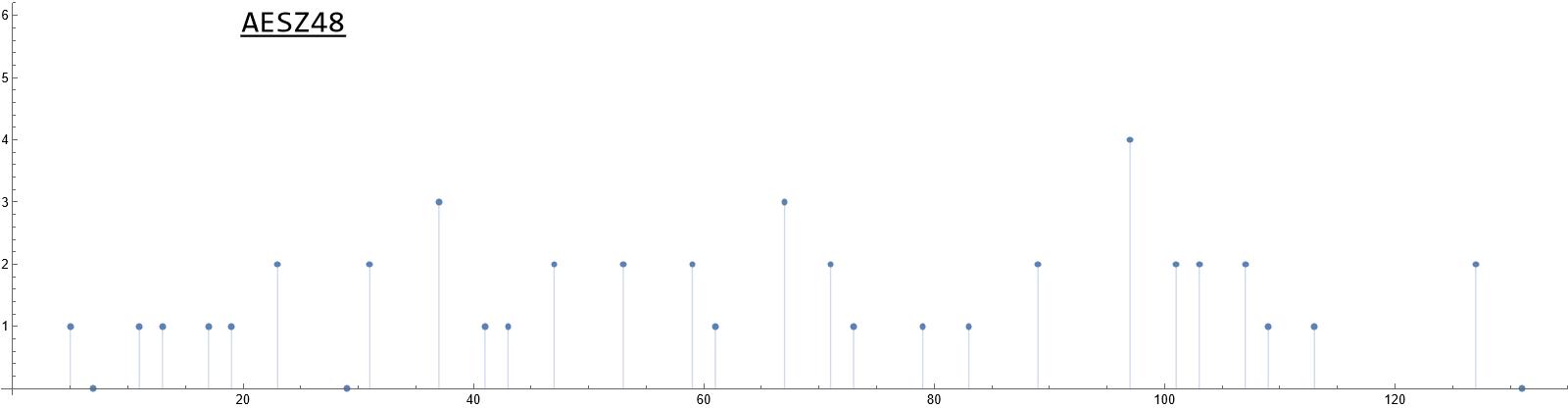}
\includegraphics[scale=0.4]{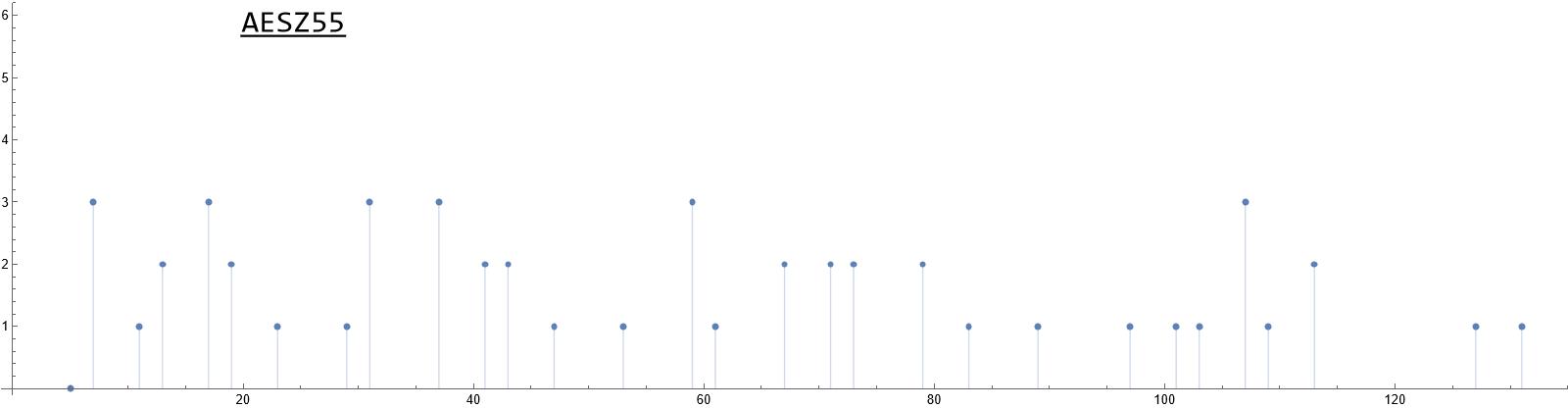}
AESZ55 was studied in \cite{Bonisch:2022slo} and is known to possess a rational rank two attractor.\\[20pt]
\includegraphics[scale=0.4]{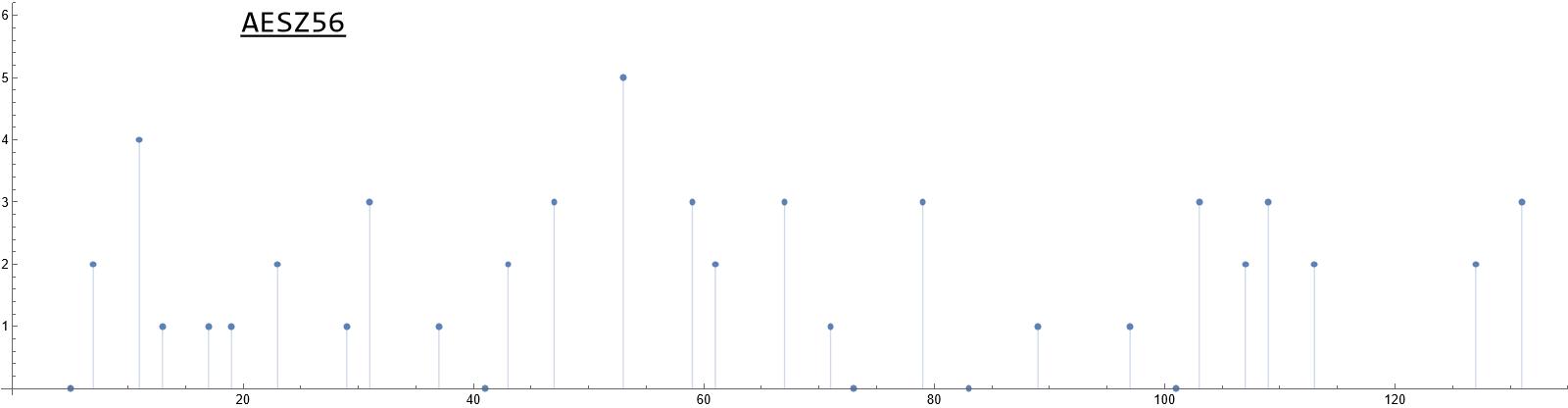}
\includegraphics[scale=0.4]{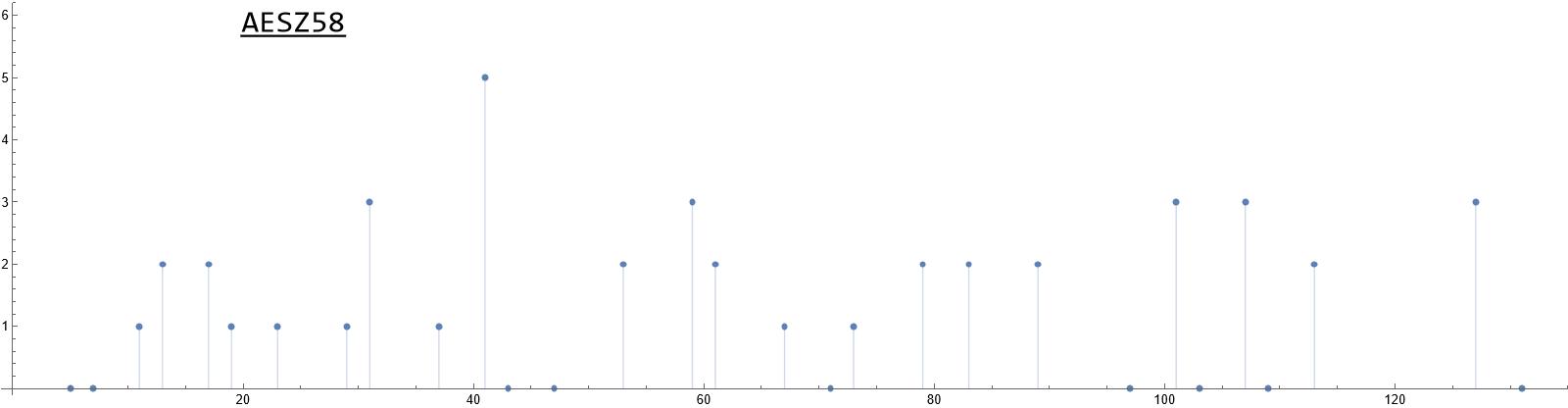}
\includegraphics[scale=0.4]{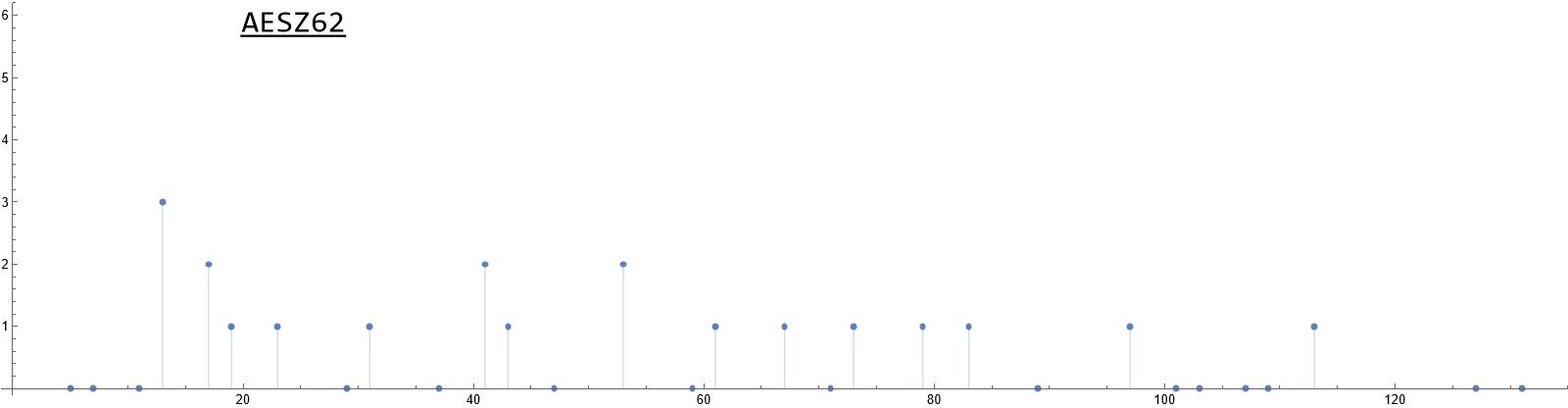}
\includegraphics[scale=0.4]{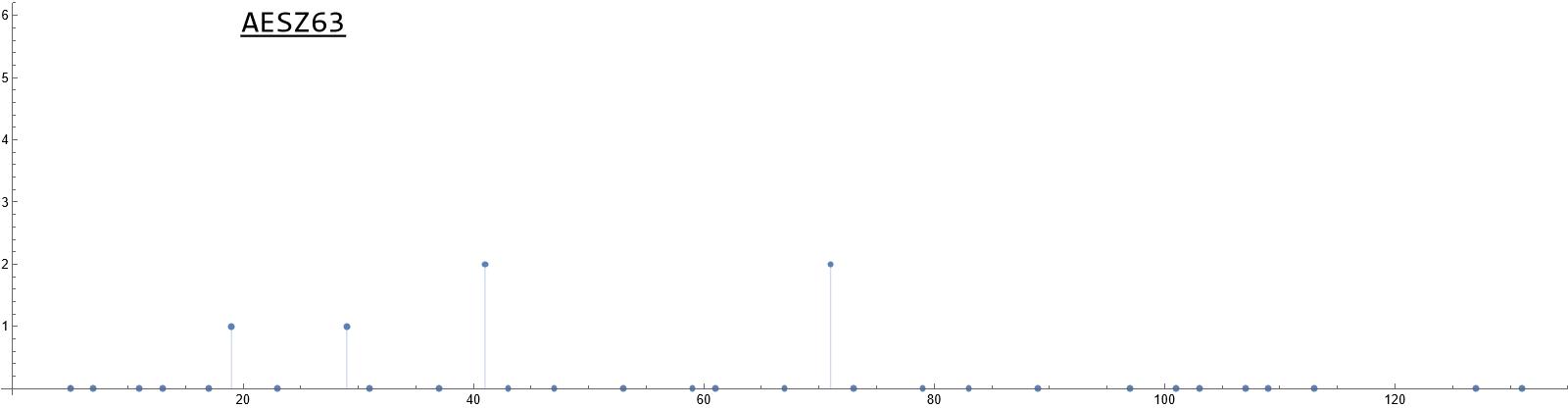}
\includegraphics[scale=0.4]{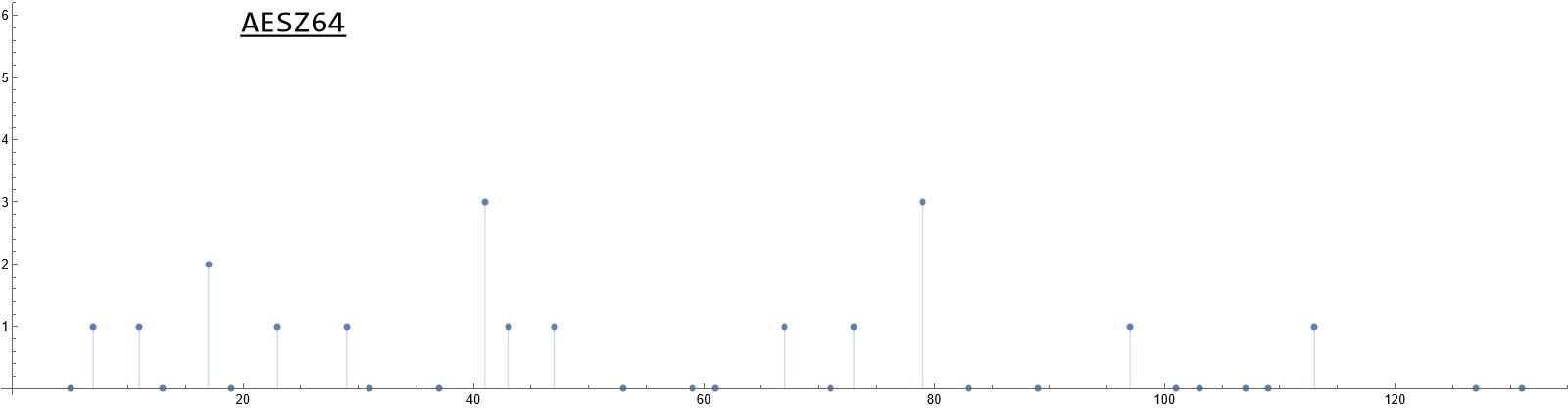}
\includegraphics[scale=0.4]{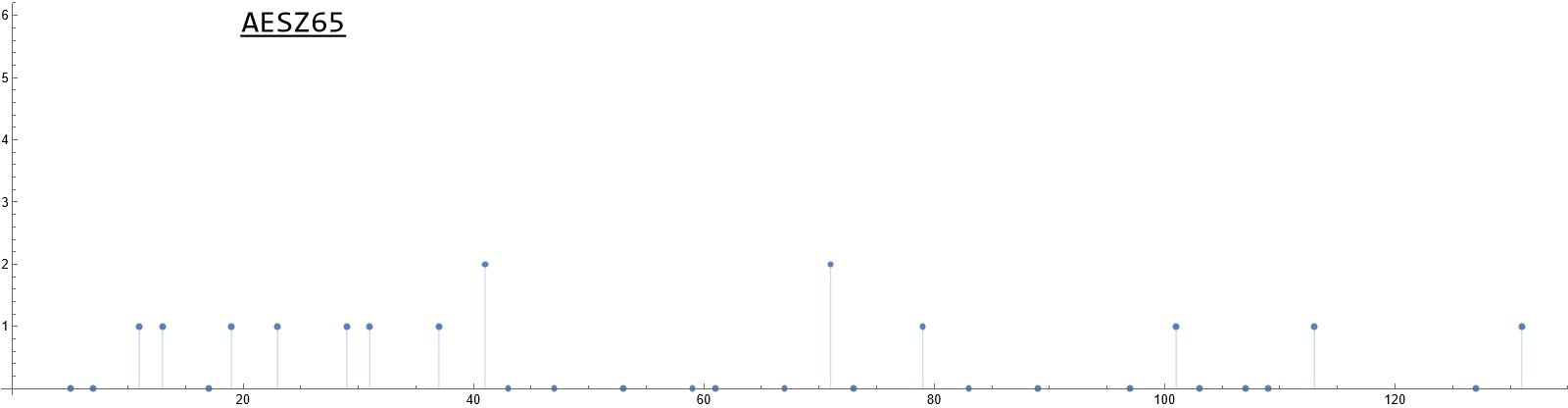}
\includegraphics[scale=0.4]{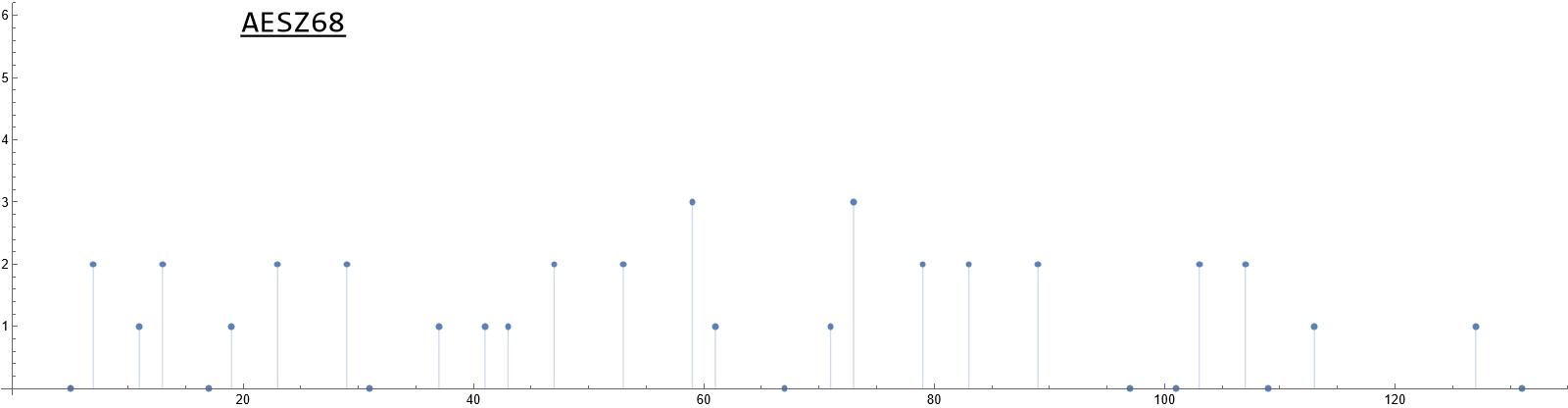}
\includegraphics[scale=0.4]{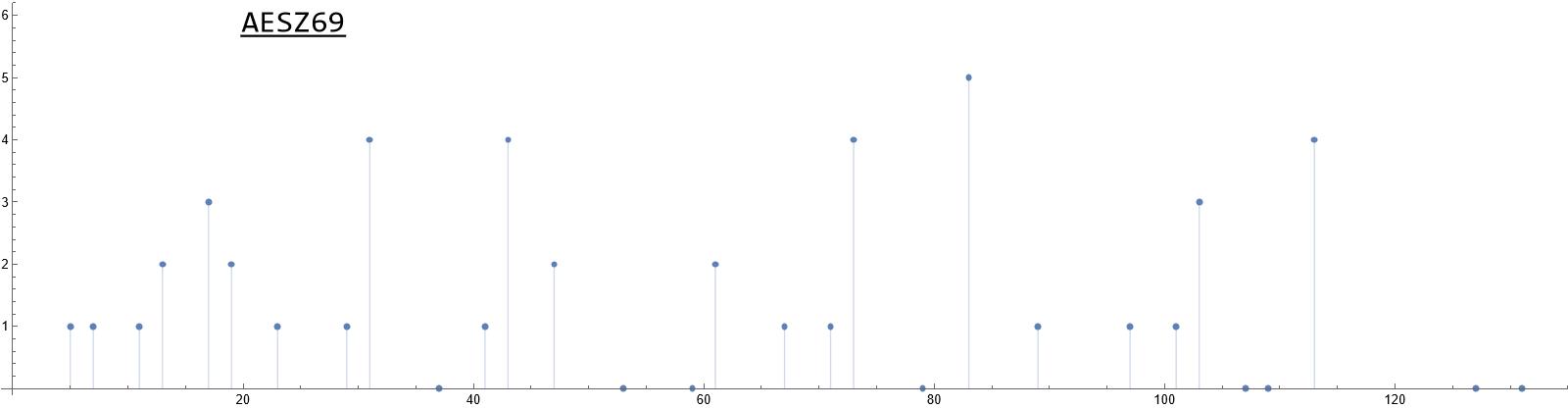}
\includegraphics[scale=0.4]{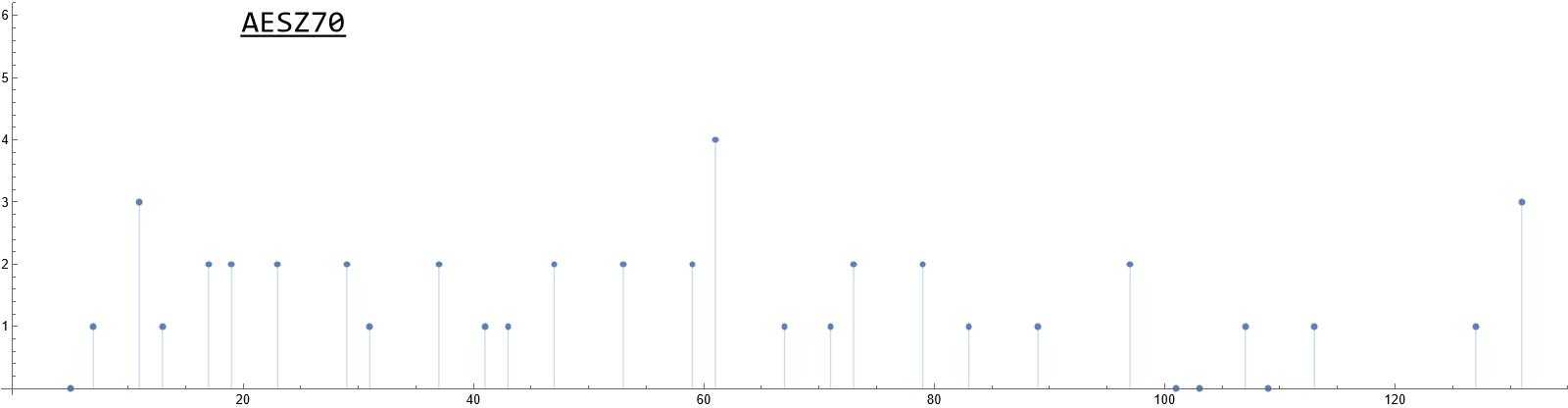}
\includegraphics[scale=0.4]{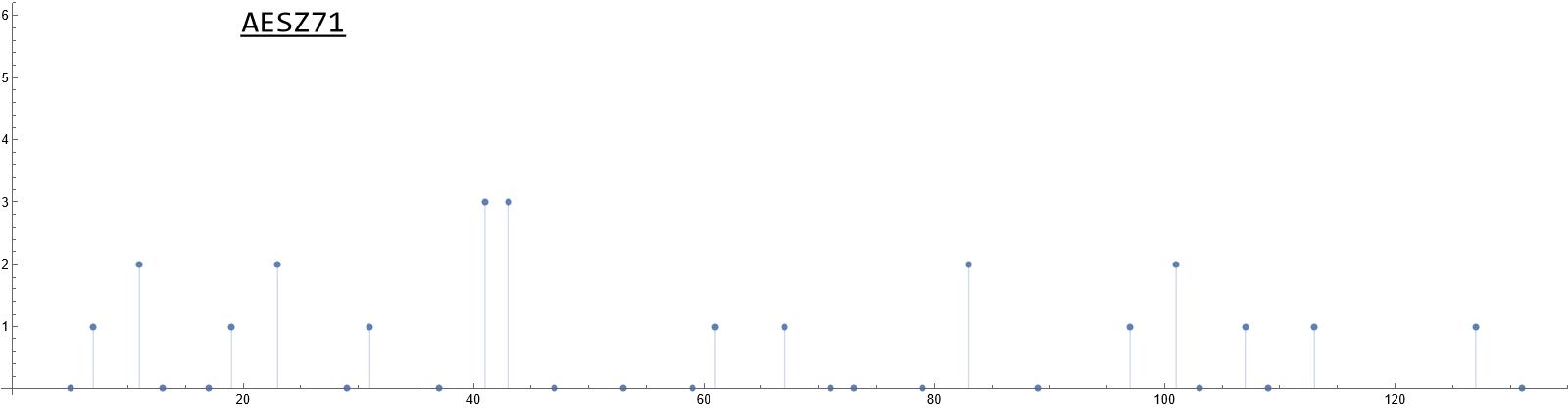}
\includegraphics[scale=0.4]{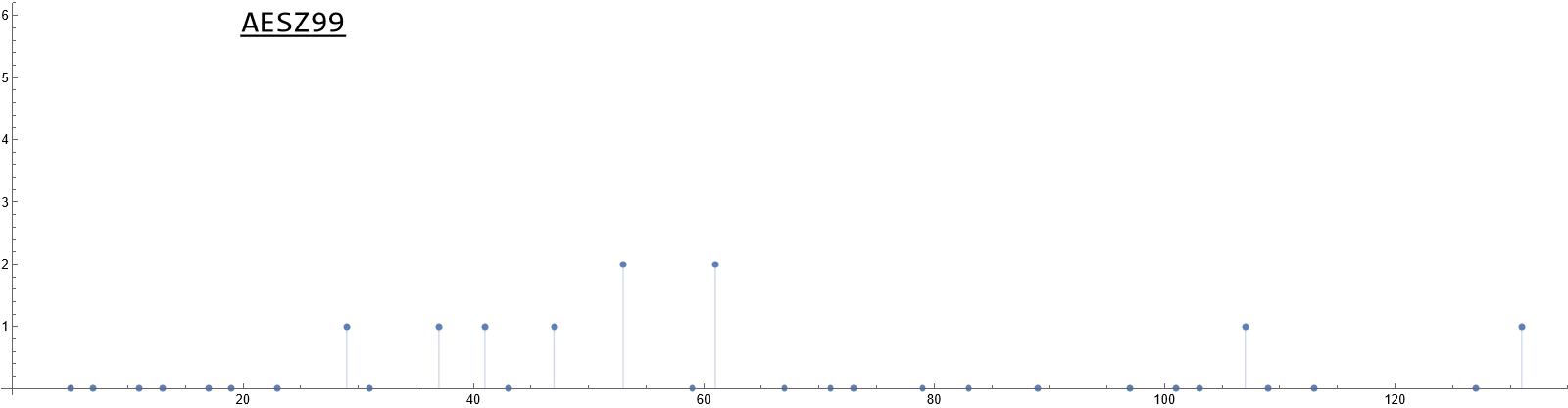}
\includegraphics[scale=0.4]{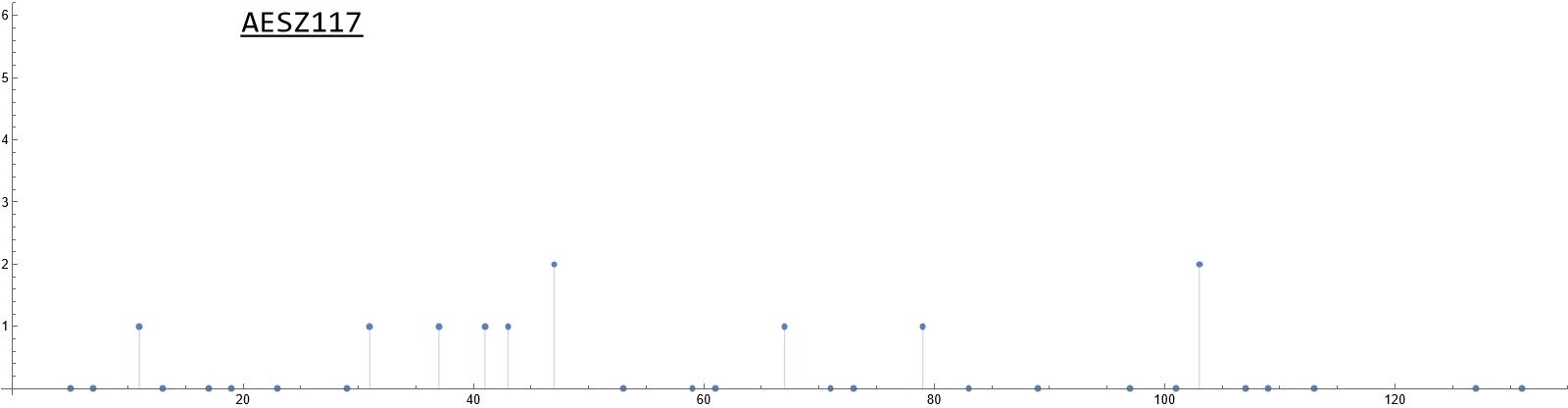}
\includegraphics[scale=0.4]{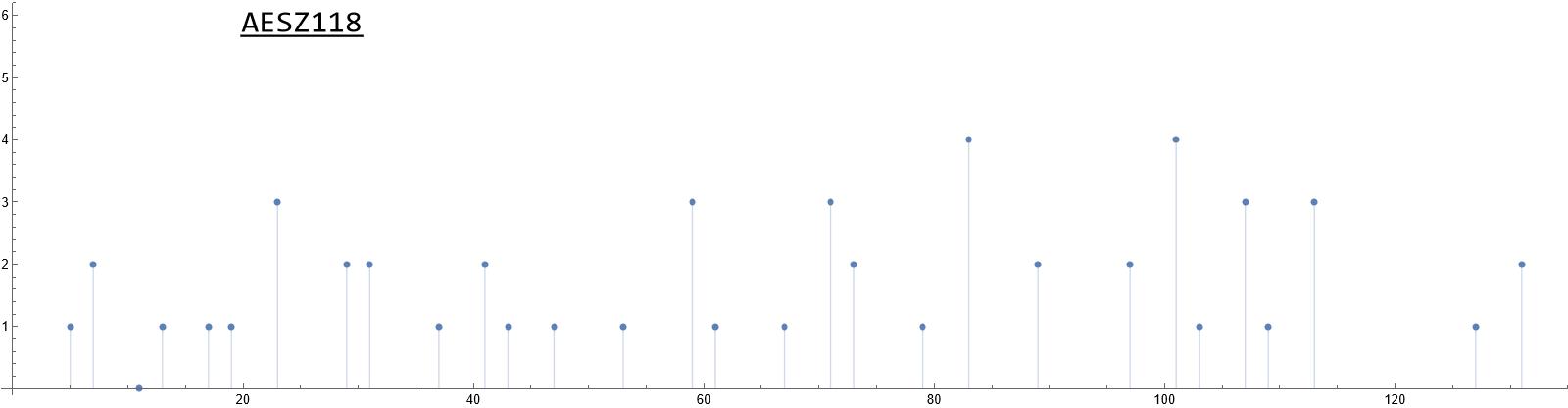}
AESZ118 is equivalent to AESZ22 after a change of variables (see the AESZ22 plot). \\[20pt]
\includegraphics[scale=0.4]{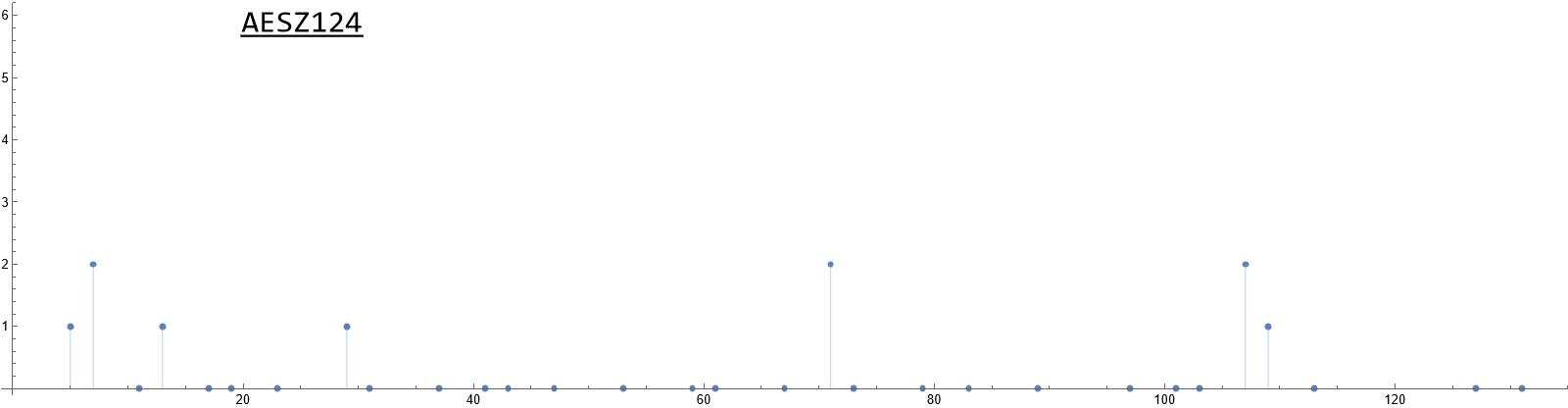}
\includegraphics[scale=0.4]{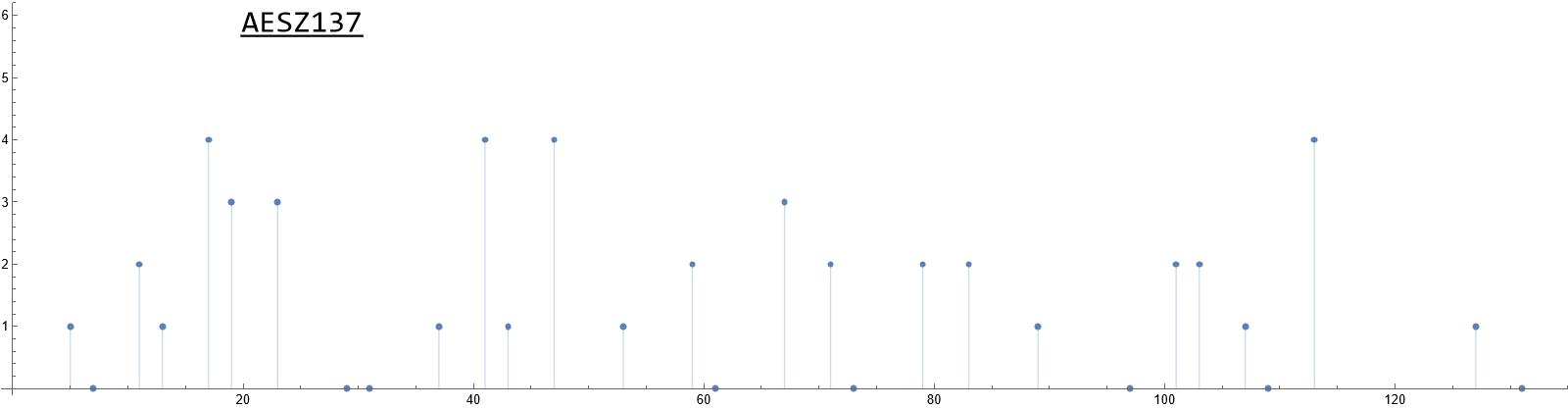}
\includegraphics[scale=0.4]{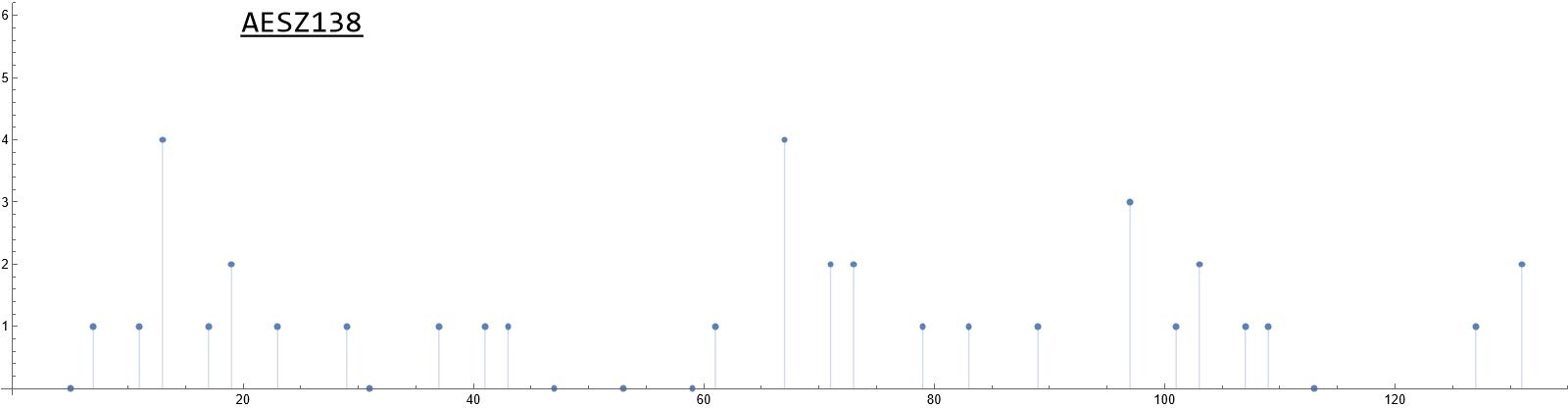}
\includegraphics[scale=0.4]{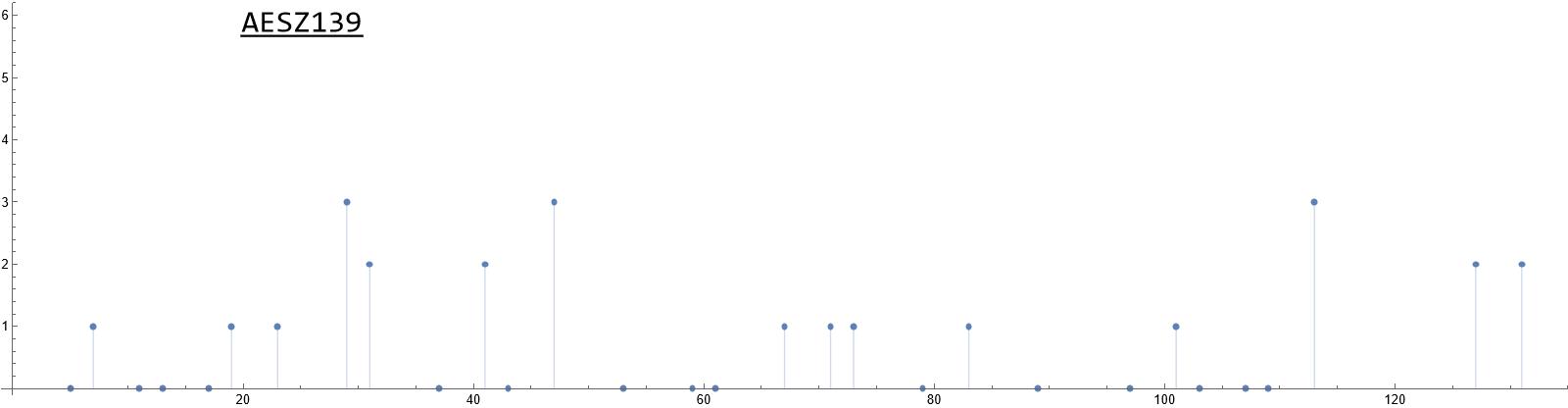}
\includegraphics[scale=0.4]{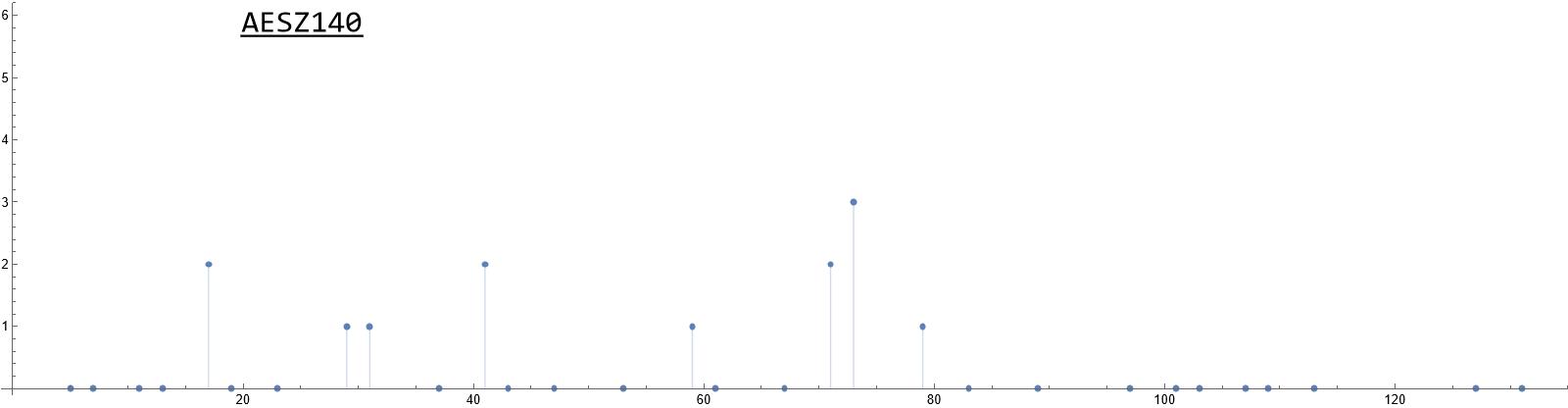}
\includegraphics[scale=0.4]{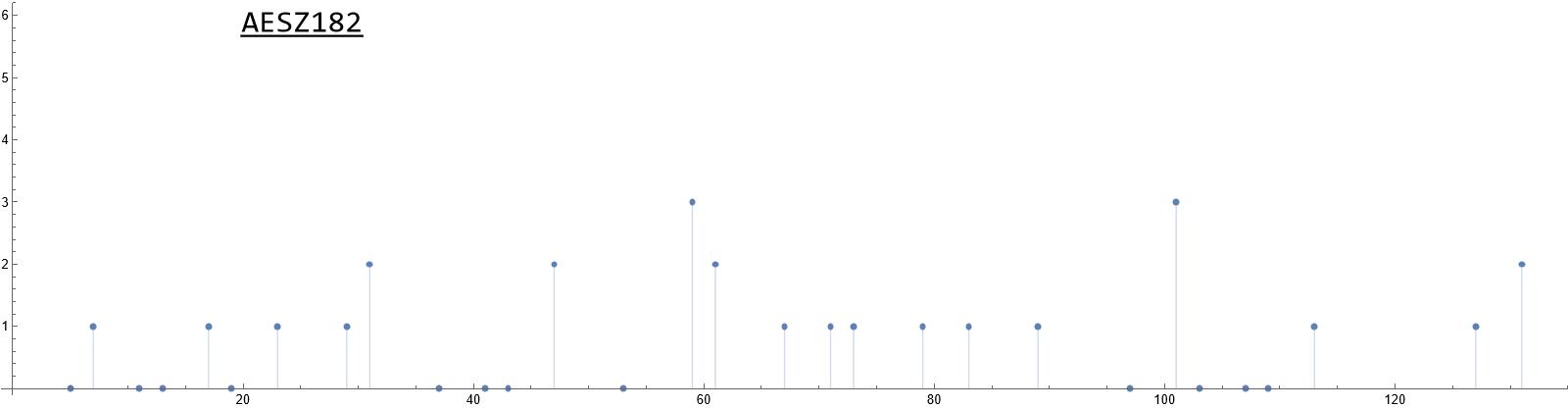}
\includegraphics[scale=0.4]{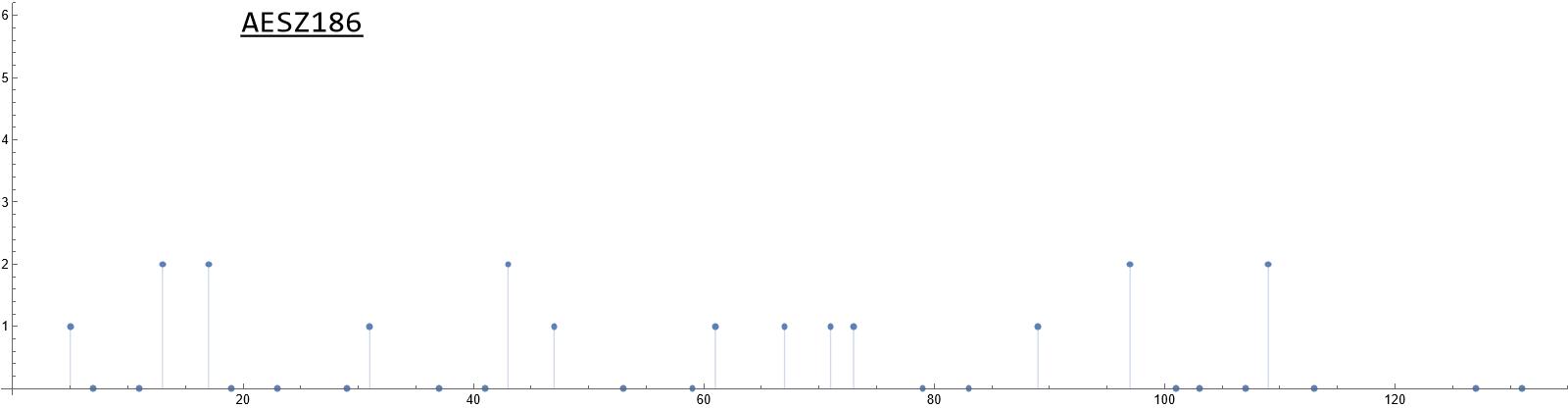}
\includegraphics[scale=0.4]{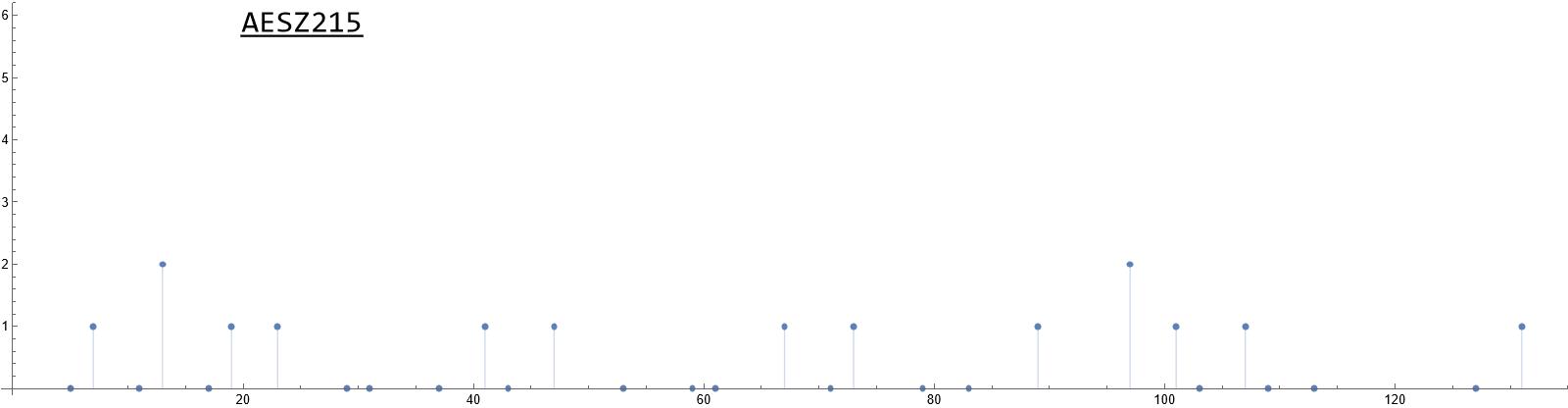}
\includegraphics[scale=0.4]{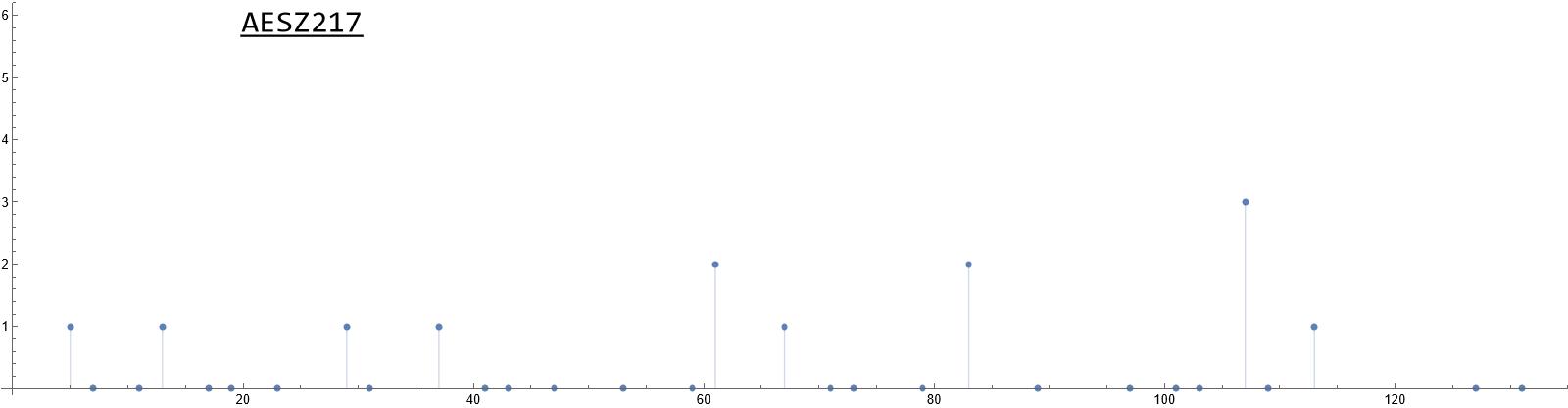}
\includegraphics[scale=0.4]{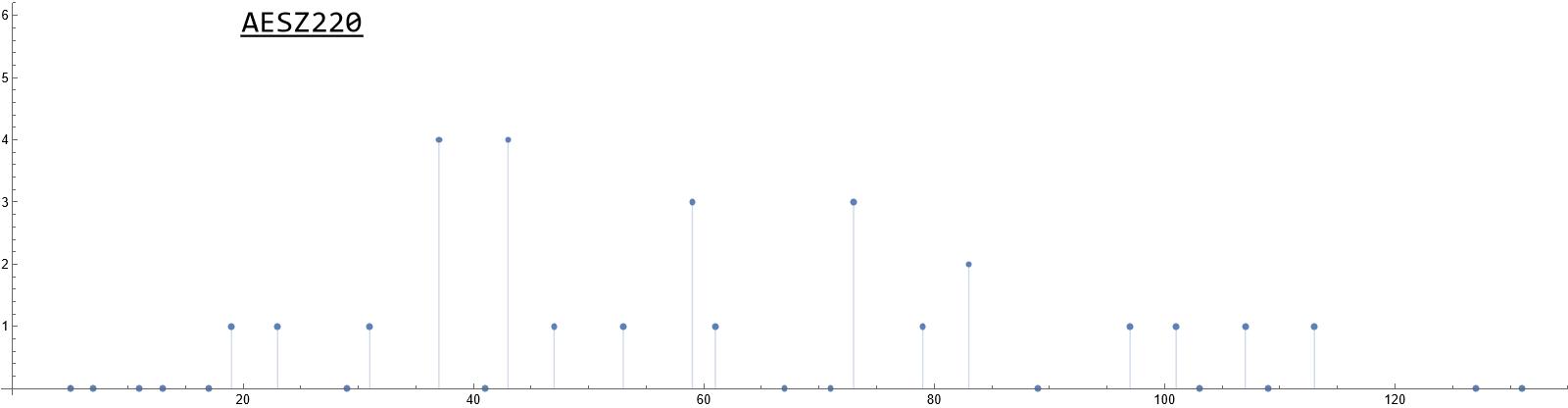}
\includegraphics[scale=0.4]{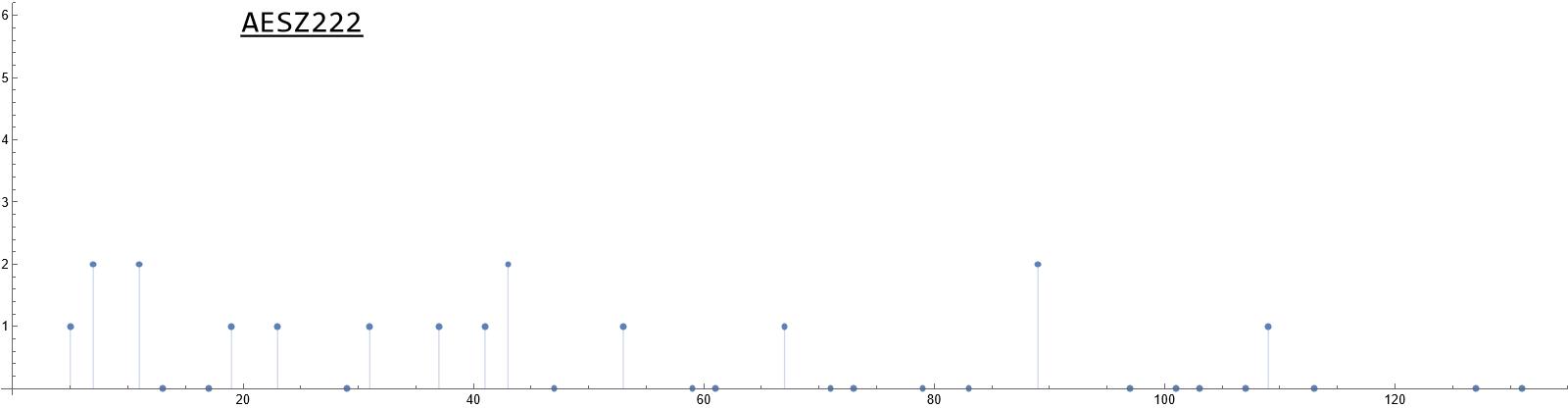}
\includegraphics[scale=0.4]{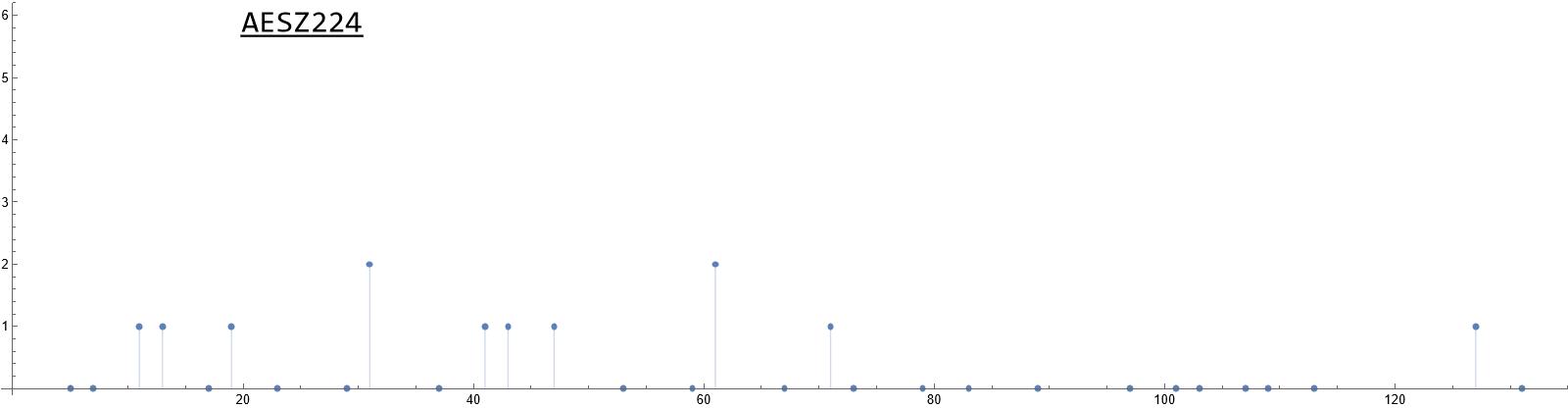}
\includegraphics[scale=0.4]{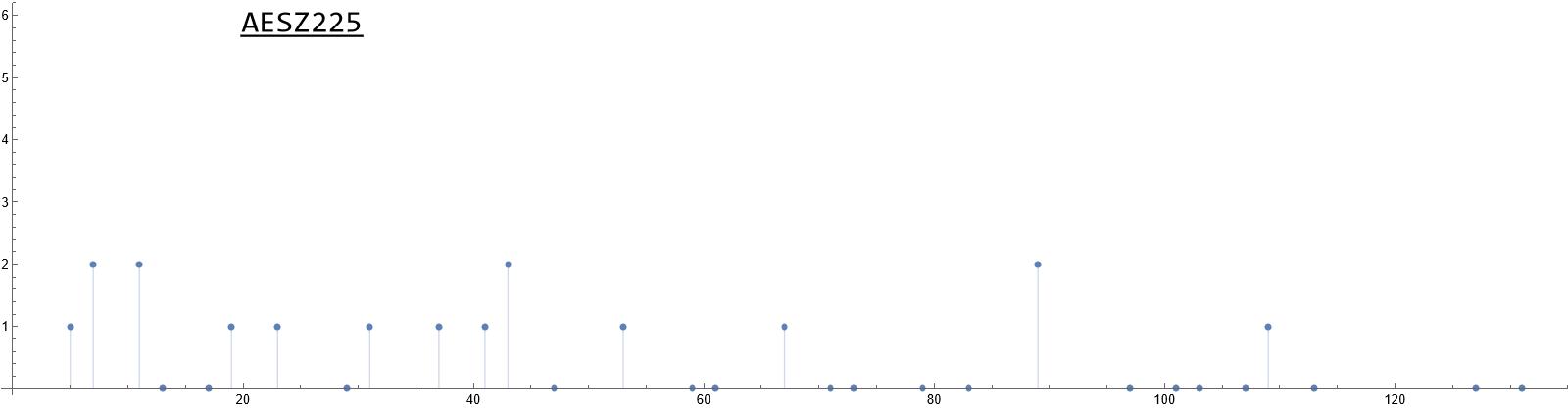}
\includegraphics[scale=0.4]{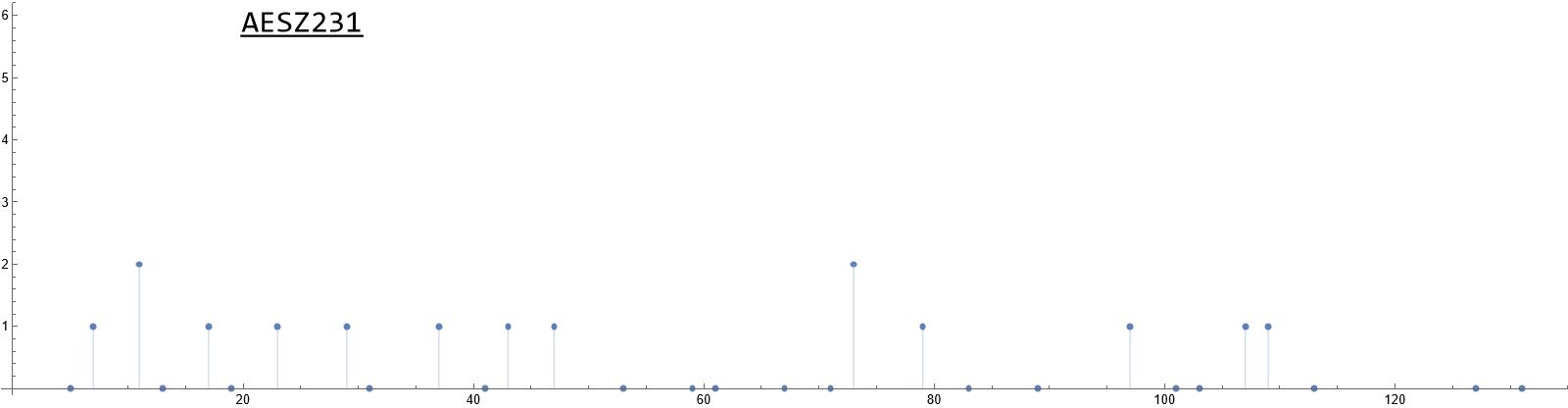}
\includegraphics[scale=0.4]{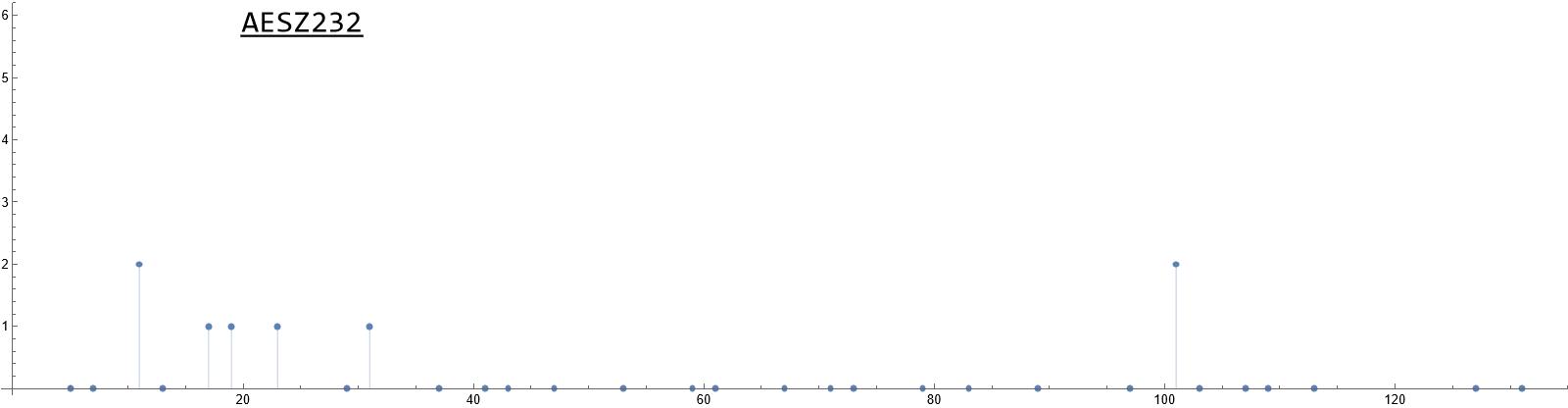}

%\include{appendix1}
%\include{appendix2}

%next line adds the Bibliography to the contents page
\addcontentsline{toc}{chapter}{Bibliography}
%uncomment next line to change bibliography name to references
%\renewcommand{\bibname}{References}

\bibliography{THESIS}        %use a bibtex bibliography file refs.bib
\bibliographystyle{JHEP}  %use the plain bibliography style

\end{document}